\newif\ifabstract
\abstracttrue
\newif\iffull
\ifabstract \fullfalse \else \fulltrue \fi
\documentclass[11pt]{article}
\usepackage{amsfonts}
\usepackage{amssymb}
\usepackage{amstext}
\usepackage{amsmath}
\usepackage{xspace}
\usepackage{ntheorem}
\usepackage{graphicx}
\usepackage{url}
\usepackage{graphics}
\usepackage{colordvi}
\usepackage{hyperref}
\usepackage{subfigure}
\usepackage{xcolor}
\usepackage{bm}

\advance \oddsidemargin by -0.8in
\newcommand{\myparskip}{3pt}
\parskip \myparskip
\newcommand{\neww}{\tiny{\mathsf{new}}}

\newif\ifnocomments
 \nocommentsfalse %Uncommenting this  gives the comments version
\ifnocomments

\newcommand{\mynote}[1]{}
\newcommand{\znote}[1]{}
\newcommand{\snote}[1]{}

\else
\newcommand{\mynote}[1]{\textcolor{purple}{\sc\bf{[JC: #1]}}}
\newcommand{\znote}[1]{\textcolor{red}{\sc\bf{[ZT: #1]}}}
\newcommand{\snote}[1]{\textcolor{blue}{\sc\bf{[SM: #1]}}}

\fi

\newcommand{\MCN}{{\sf Minimum Crossing Number}\xspace}
\newcommand{\CP}{{\sf Cluster Placement}\xspace}
\newcommand{\hE}{\hat E}
\newcommand{\hH}{\hat H}
\newcommand{\hG}{\hat \Gamma}
\newcommand{\hC}{\hat C}
\newcommand{\hcset}{\hat{\mathcal{C}}}
\newcommand{\hGamma}{\hat \Gamma}

\newcommand{\desc}{\operatorname{Desc}}
\newcommand{\algsc}{\ensuremath{{\mathcal{A}}_{\mbox{\textup{\scriptsize{ARV}}}}}\xspace}

\newcommand{\alphasc}{\ensuremath{\beta_{\mbox{\tiny{\sc ARV}}}}}
\newcommand{\eg}{\mathsf{eg}}
\newcommand{\cro}{\mathsf{cr}}
\newcommand{\out}{\mathsf{out}}

\newcommand{\mvp}{\mathsf{mvp}}
\newcommand{\mep}{\mathsf{mep}}

\newcommand{\ok}{acceptable}
\newcommand{\ir}{\mathsf{IRG}}
\newcommand{\mcn}{\textsf{Minimum Crossing Number}\xspace}
\newcommand{\MP}{\textsf{Minimum Planarization}\xspace}
\newcommand{\MVP}{\textsf{Minimum Vertex Planarization}\xspace}
\newcommand{\CNwRS}{\textsf{MCNwRS}\xspace}

\newcommand{\procdraw}{\ensuremath{\sf ProcDraw}\xspace}

\newcommand{\procface}{\ensuremath{\sf ProcessFace}\xspace}

\newcommand{\bGamma}{{\mathbf{\Gamma}}}
\newcommand{\cms}{{\aset_{\text{CMS}}}}
\newcommand{\tcms}{{\aset^3_{\text{CMS}}}}

\newcommand{\mincycle}{\operatorname{MinCycle}}
\newcommand{\dnot}{D^{\circ}}

\newenvironment{prog}[1]{
	\begin{minipage}{5.8 in}
	\begin{center}
	{\sc #1}
	\end{center}
	\begin{enumerate}}
	{
	\end{enumerate}
	\end{minipage}}

\newcommand{\program}[2]{\fbox{\vspace{2mm}\begin{prog}{#1} #2 \end{prog}\vspace{2mm}}}

\newcommand{\tG}{\textbf{G}}
\newcommand{\tH}{\textbf{H}}
\newcommand{\tE}{\textbf{E}'}

\newcommand{\tphi}{\bm{\phi}}

\newcommand{\tB}{\tilde B}
\newcommand{\dout}{D_{\mbox{\tiny{out}}}}
\newcommand{\notF}{\overline{F}}
\newcommand{\NP}{\mbox{\sf NP}}

\newcommand{\opt}{\mathsf{OPT}}
\newcommand{\optcro}{\mathsf{OPT}_{\mathsf{cr}}}
\newcommand{\optcrors}{\mathsf{OPT}_{\mathsf{cnwrs}}}
\newcommand{\set}[1]{\left\{ #1 \right\}}
\newcommand{\sse}{\subseteq}

\newcommand{\tset}{{\mathcal T}}
\newcommand{\uset}{{\mathcal U}}

\newcommand{\pset}{{\mathcal{P}}}
\newcommand{\nset}{{\mathcal{N}}}
\newcommand{\dset}{{\mathcal{D}}}
\newcommand{\discset}{{\mathcal{D}}}

\newcommand{\qset}{{\mathcal{Q}}}

\newcommand{\lset}{{\mathcal{L}}}
\newcommand{\bset}{{\mathcal{B}}}
\newcommand{\tbset}{\tilde{\mathcal{B}}}
\newcommand{\aset}{{\mathcal{A}}}
\newcommand{\cset}{{\mathcal{C}}}
\newcommand{\fset}{{\mathcal{F}}}
\newcommand{\mset}{{\mathcal M}}
\newcommand{\jset}{{\mathcal{J}}}
\newcommand{\xset}{{\mathcal{X}}}

\newcommand{\gset}{{\mathcal{G}}}
\newcommand{\oset}{{\mathcal{O}}}
\newcommand{\yset}{{\mathcal{Y}}}
\newcommand{\rset}{{\mathcal{R}}}

\newcommand{\hset}{{\mathcal{H}}}
\newcommand{\sset}{{\mathcal{S}}}
\newcommand{\zset}{{\mathcal{Z}}}

\newcommand{\nots}{\overline S}
\newcommand{\eint}{E^{\tiny\mbox{int}}}
\newcommand{\event}{{\cal{E}}}
\newcommand{\floor}[1]{\ensuremath{\left\lfloor#1\right\rfloor}}
\newcommand{\ceil}[1]{\ensuremath{\left\lceil#1\right\rceil}}
\newtheorem{theorem}{Theorem}[section]
\newtheorem{lemma}[theorem]{Lemma}

\newtheorem{observation}[theorem]{Observation}
\newtheorem{corollary}[theorem]{Corollary}
\newtheorem{claim}[theorem]{Claim}
\newtheorem{proposition}[theorem]{Proposition}

\newtheorem*{definition}{Definition.}
\newenvironment{proof}{\par \smallskip{\bf Proof:}}{\hfill\stopproof}
\def\stopproof{\square}
\def\square{\vbox{\hrule height.2pt\hbox{\vrule width.2pt height5pt \kern5pt
\vrule width.2pt} \hrule height.2pt}}

\newenvironment{proofof}[1]{\noindent{\bf Proof of #1.}}
{\hspace*{\fill}\stopproof}

\renewcommand{\phi}{\varphi}
\newcommand{\eps}{\epsilon}

\newcommand{\bdg}{bounded-degree graph }

\newcommand{\half}{\ensuremath{\frac{1}{2}}}
\newcommand{\poly}{\operatorname{poly}}
\newcommand{\dist}{\mbox{\sf dist}}

\newcommand{\indicator}{{\mathbf{1}}}

\newcommand{\expect}[2][]{\text{\bf E}_{#1}\left [#2\right]}
\newcommand{\prob}[2][]{\text{\bf Pr}_{#1}\left [#2\right]}

\newenvironment{properties}[2][0]
{
\begin{enumerate} \setcounter{enumi}{#1}}{\end{enumerate}}

\renewcommand{\cong}{\operatorname{cong}}

\newcommand{\explain}{{\sc \color{red}\bf{[Explain]}}}
\newcommand{\tGamma}{\tilde \Gamma}

\newcommand{\eac}{\chi}
\begin{document}
\bibliographystyle{alpha}

\begin{titlepage}

\title{Towards Better Approximation of Graph Crossing Number}
\author{Julia Chuzhoy\thanks{Toyota Technological Institute at Chicago. Email: {\tt cjulia@ttic.edu}. Supported in part by NSF grant CCF-1616584.}   \and Sepideh Mahabadi\thanks{Toyota Technological Institute at Chicago. Email: {\tt mahabadi@ttic.edu}.} \and Zihan Tan\thanks{Computer Science Department, University of Chicago. Email: {\tt zihantan@uchicago.edu}. Supported in part by NSF grant CCF-1616584.}}
\maketitle

	\thispagestyle{empty}
\begin{abstract}
	Graph Crossing Number is a fundamental and extensively studied problem with wide ranging applications. In this problem, the goal is to draw an input graph $G$ in the plane so as to minimize the number of crossings between the images of its edges. The problem is notoriously difficult, and despite extensive work, non-trivial approximation algorithms are only known for bounded-degree graphs. Even for this special case, the best current algorithm %[Kawarabayashi, Sidiropoulos, FOCS '17, STOC 19]\znote{Should we delete citation in the abstract?}
	achieves a $\tilde O(\sqrt n)$-approximation, while the best current negative results do not rule out constant-factor approximation. 
	All current approximation algorithms for the problem build on the same paradigm, which is also used in practice: compute a set $E'$ of edges (called a \emph{planarizing set}) such that $G\setminus E'$ is planar; compute a planar drawing of $G\setminus E'$; then add the drawings of the edges of $E'$ to the resulting drawing. Unfortunately, there are examples of graphs $G$, in which any implementation of this method must incur $\Omega  (\opt^2 )$ crossings, where $\opt$ is the value of the optimal solution. % with crossing number $k$, and a planarizing set $E'$ of $k$ edges, such that, if we add the images of edges in $E'$ to any planar drawing of $G\setminus E'$, then any resulting drawing must have $\Omega(k^2)$ crossings. 
	This barrier seems to doom the only currently known approach to designing approximation algorithms for the problem, and to prevent it from yielding a better than $O(\sqrt n)$-approximation.
	
	In this paper we propose a new paradigm that allows us to overcome this barrier. %, and show that it can be used to obtain better approximation algorithms for the problem. 
	We show an algorithm, that, given a bounded-degree graph $G$ and a planarizing set $E'$ of its edges, computes another planarizing edge set $E''$ with $E'\subseteq E''$, such that $|E''|$ is relatively small, % small compared to $|E'|$ and the crossing number of $G$, 
	and there exists a near-optimal drawing of $G$ in which no edges of $G\setminus E''$ participate in crossings. This allows us to reduce the Crossing Number problem to \emph{Crossing Number with Rotation System} -- a variant of the Crossing Number problem, in which the ordering of the edges incident to every vertex is fixed as part of input. In our reduction, we obtain an instance $G'$ of this problem, where $|E(G')|$ is roughly bounded by the crossing number of the original graph $G$. We show a randomized algorithm for this new problem, that allows us to obtain  an $O(n^{1/2-\eps})$-approximation for Graph Crossing Number on bounded-degree graphs, for some constant $\eps>0$.
\end{abstract}
\end{titlepage}

\pagenumbering{gobble}
\tableofcontents
\newpage 
\pagenumbering{arabic}

\section{Introduction}\label{sec: intro}

%\znote{Should we remove all the space-saving lines (like vspace -xpt) now?}

Given a graph $G=(V,E)$, a \emph{drawing} of $G$ is an embedding of the graph into the plane, that maps every vertex to a point in the plane, and every edge to a continuous curve connecting the images of its endpoints. We require that the image of an edge may not contain images of vertices of $G$ other than its two endpoints, and no three curves representing drawings of edges of $G$ may intersect at a single point, unless that point is the image of their shared endpoint. We say that two edges \emph{cross} in a given drawing of $G$ iff their images share a point other than the image of their common endpoint.
In the \MCN problem, the goal is to compute a drawing of the input $n$-vertex graph $G$ with minimum number of crossings. We denote the value of the optimal solution to this problem, also called the \emph{crossing number} of $G$, by $\optcro(G)$. %We denote by $n$ the number of vertices in the input graph $G$.

The \MCN problem naturally arises in several areas of Computer Science and Mathematics. The problem was initially introduced by Tur\'an \cite{turan_first}, who considered the question of computing the crossing number of complete bipartite graphs, and since then it has been a subject of extensive studies (see, e.g., \cite{turan_first, chuzhoy2011algorithm, chuzhoy2011graph, chimani2011tighter, chekuri2013approximation, KawarabayashiSidi17, kawarabayashi2019polylogarithmic}, and also \cite{richter_survey, pach_survey, matousek_book, vrto_biblio} for surveys on this topic).  The problem is known to be NP-hard \cite{crossing_np_complete}, and it remains NP-hard, and APX-hard on cubic graphs \cite{Hlineny06a,cabello2013hardness}. The \MCN problem appears to be notoriously difficult from the approximation perspective. All currently known algorithms achieve approximation factors that depend polynomially on $\Delta$ -- the maximum vertex degree of the input graph, and, to the best of our knowlgedge, no non-trivial approximation algorithms are known for graphs with arbitrary vertex degrees. We note that the famous Crossing Number Inequality \cite{ajtai82,leighton_book} shows that, for every graph $G$ with $|E(G)|\geq 4n$, the crossing number of $G$ is  $\Omega(|E(G)|^3/n^2)$. Since the problem is most interesting when the crossing number of the input graph is low, and since our understanding of the problem is still extremely poor, it is reasonable to focus on designing algorithms for low-degree graphs. Throughout, we denote by $\Delta$ the maximum vertex degree of the input graph. While we do not make this assumption explicitly, it may be convenient to think of $\Delta$ as being bounded by a constant or by $\poly\log n$.

The first non-trivial algorithm for \mcn, due to Leighton and Rao \cite{leighton1999multicommodity}, combined their algorithm for balanced separators with the framework of \cite{bhatt1984framework}, to compute a drawing of input graph $G$ with $O((n+\optcro(G)) \cdot \Delta^{O(1)}\cdot \log^4n)$ crossings. This bound was later improved to $O((n+\optcro(G)) \cdot \Delta^{O(1)}\cdot \log^3n)$ by \cite{even2002improved}, and then to $O((n+\optcro(G)) \cdot \Delta^{O(1)}\cdot \log^2n)$   as a consequence of the improved algorithm of \cite{ARV} for Balanced Cut. All these results provide an $O(n\poly(\Delta \log n))$-approximation for \MCN (but perform much better when $\optcro(G)$ is high). For a long time, this remained the  best approximation algorithm for \mcn, while the best inapproximability result, to this day, only rules out the existence of a PTAS, unless NP has randomized subexponential time algorithms \cite{crossing_np_complete, Ambuhl07}. We note that it is highly unusual that achieving an $O(n)$-approximation for an unweighted graph optimization problem is so challenging. However, unlike many other unweighted graph optimization problems, the value of the optimal solution to \mcn may be as large\footnote{This can be seen by applying the Crossing Number Inequality to the complete $n$-vertex graph.} as $\Omega(n^4)$.

A sequence of papers~\cite{chuzhoy2011graph,chuzhoy2011algorithm} was the first to break the barrier of $\tilde \Theta(n)$-approximation, providing a $\tilde O \left (n^{9/10}\cdot \Delta^{O(1)}\right )$-approximation algorithm. Recently, a breakthrough sequence of works \cite{KawarabayashiSidi17, kawarabayashi2019polylogarithmic} has led to an improved  $\tilde O\left(\sqrt{n}\cdot \Delta^{O(1)}\right )$-approximation for \mcn.
All the above-mentioned algorithms exploit the same algorithmic paradigm, that builds on  the connection of \mcn~to the  \MP~problem, that we discuss next.

\noindent{\bf Minimum Planarization.}
In the \MP~problem, the input is an $n$-vertex graph $G=(V,E)$, and the goal is to compute a minimum-cardinality subset $E^*$ of its edges (called a \emph{planarizing set}), such that graph $G\setminus E^*$ is planar. This problem and its close variant \MVP~(where we need to compute a minimum-cardinality subset $V'$ of vertices such that $G\setminus V'$ is planar) are of independent interest and have been studied extensively (see, e.g., \cite{chuzhoy2011graph, KawarabayashiSidi17, kawarabayashi2019polylogarithmic}). 
It is immediate to see that, for every graph $G$, $\opt_{\mvp}(G) \leq \opt_{\mep}(G)\leq \optcro(G)$, where $\opt_{\mvp}(G)$ and $\opt_{\mep}(G)$ are the optimal solution values of the \MVP and the \MP problems on $G$, respectively.
A simple application of the Planar Separator Theorem of \cite{planar-separator} was shown to give an  $O(\sqrt{n\log n}\cdot \Delta)$-approximation algorithm for both problems \cite{chuzhoy2011graph}.  
Further, \cite{chekuri2013approximation} provided an $O(k^{15}\cdot \poly(\Delta\log n)$-approximation algorithm for  \MP and \MVP,  where $k$ is the value of the optimal solution.
The more recent breakthrough result of Kawarabayashi and Sidiropoulus \cite{KawarabayashiSidi17, kawarabayashi2019polylogarithmic} provides an $O(\Delta^3\cdot \log^{3.5}n)$-approximation algorithm for \MVP, and an $O(\Delta^4\cdot \log^{3.5}n)$-approximation algorithm for \MP. %Studying \MP~ has played an important role in solving \mcn \cite{kawarabayashi2019polylogarithmic}. 

Returning to the \MCN problem, all currently known approximation algorithms for the problem rely on the same paradigm, which is also used in heuristics (see, e.g. \cite{crossing-survey}). For convenience of notation, we call it Paradigm $\Pi$.

%$\ $

%\vspace{-0mm}
%\centering
%\begin{centering}
\program{Paradigm $\Pi$}{
		\item compute a planarizing set $E'$ of edges for $G$; \label{enum-paradigm-item-1}
		
		%\vspace{-2mm}
		
		\item compute a planar drawing of $G\setminus E'$; \label{enum-paradigm-item2}
		%\vspace{-2mm}
		
		\item add the images of the edges of $E'$ to the resulting drawing.\label{enum-paradigm-item-3}
}
%\end{centering}
%\vspace{-0mm}

$\ $

\iffalse
\paragraph{Paradigm $\Pi$}

\begin{enumerate}
	\item compute a planarizing set $E'$ of edges for $G$; \label{enum-paradigm-item-1}
	\item compute a planar drawing of $G\setminus E'$; \label{enum-paradigm-item2}
	\item add the images of the edges in $E'$ to the resulting drawing.\label{enum-paradigm-item-3}
\end{enumerate}
\fi

We note that graph $G\setminus E'$ may not be $3$-connected and thus it may have several planar drawings; there are also many ways in which the edges of $E'$ can be added to the drawing.
It is therefore important to understand the following questions: 

%\vspace{-3mm}
\begin{quote}
\emph{Can this paradigm be implemented in a way that provides a near-optimal drawing of $G$? What is the best approximation factor that can be achieved when using paradigm $\Pi$?}
\end{quote}
%\vspace{-3mm}

These questions were partially answered in previous work. Specifically, \cite{chuzhoy2011graph} provided an efficient algorithm, that, given an input graph $G$, and a planarizing set $E'$ of $k$ edges for $G$, draws the graph with $O\left(\Delta^3\cdot k \cdot (\optcro(G) + k)\right)$ crossings.
Later, Chimani and Hlin{\v{e}}n{\`y} \cite{chimani2011tighter} improved this bound via a different efficient algorithm to 
$O\left(\Delta\cdot k \cdot (\optcro(G) +\log k)+k^2\right)$.
Both works can be viewed as an implementation of the above paradigm. 
Combining these results with the $O(\poly(\Delta\log n))$-approximation algorithm  for \MP of \cite{kawarabayashi2019polylogarithmic} in order to compute the initial planarizing edge set $E'$  with $|E'|\leq O(\optcro(G)\cdot \poly(\Delta\log n))$, yields an implementation of Paradigm $\Pi$ that produces a drawing of the input graph $G$ with $O\left ((\optcro(G))^2\cdot\poly(\Delta\log n)\right )$ crossings. Lastly, combining this with the  $O(n\poly(\Delta\log n))$-approximation algorithm of \cite{leighton1999multicommodity,bhatt1984framework,even2002improved} leads to the best currently known    $O(\sqrt{n}\cdot\poly(\Delta\log n))$-approximation algorithm of \cite{kawarabayashi2019polylogarithmic} for \MCN.

The bottleneck in using this approach in order to obtain a better than $O(\sqrt n)$-approximation for \mcn is the bounds of \cite{chuzhoy2011graph} and \cite{chimani2011tighter}, whose algorithms produce a drawing of the graph $G$ with $O\left(k \cdot \optcro(G) + k^2\right)$ crossings when $\Delta=O(1)$, where $k$ is the size of the given planarizing set. The quadratic dependence of this bound on $k$ and the linear dependence on $k\cdot \optcro(G)$ are unacceptable if our goal is to obtain better approximation using this technique. A natural question therefore is:

%\vspace{-3mm}
\begin{quote}
	\emph{Can we obtain a stronger bound, that is linear in $(\optcro(G)+|E'|)$, using Paradigm $\Pi$? }
\end{quote}
%\vspace{-3mm}

 Unfortunately, Chuzhoy, Madan and Mahabadi \cite{chuzhoy-lowerbound} have answered this question in the negative; see Section \ref{sec: appx-lower} for details. Their results show that the bounds of \cite{chuzhoy2011graph} and \cite{chimani2011tighter} are almost tight, and so we cannot hope to break the $\Theta(\sqrt{n})$ approximation barrier for \mcn  using the above paradigm. This seems to doom the only currently known approach for designing approximation algorithms for the problem.

In this paper, we propose a new paradigm towards overcoming this barrier, and show that it leads to a better approximation of \MCN. Specifically, we show an efficient algorithm, that, given a planarizing set $E'$ of edges, augments $E'$ in order to obtain a new planarizing set $E''$, whose cardinality is $O\left ((|E'|+\optcro(G))\poly(\Delta \log n)\right )$. Moreover, we show that there exists a drawing $\phi$ of the graph $G$, with at most $O\left ((|E'|+\optcro(G))\poly(\Delta \log n)\right )$ crossings, where the edges of $G\setminus E''$ do not participate in any crossings. In other words, the drawing $\phi$ of $G$ can be obtained by first computing a planar drawing of $G\setminus E''$, and then inserting the images of the edges of $E''$ into this drawing. 
This new paradigm can be summarized as follows:

%\vspace{-0mm}

\program{Paradigm $\Pi'$}{
	\item compute a planarizing set $E'$ of edges for $G$; \label{new-paradigm-item-1}
	
	%\vspace{-2mm}
	
	\item compute an augmented planarizing edge set $E''$ with $E'\subseteq E''$ that has some additional useful properties; \label{new-paradigm-item-1.5}
	
	%\vspace{-2mm}
	
	\item compute a planar drawing of $G\setminus E''$; \label{new-paradigm-item2}
	%\vspace{-2mm}
	
	\item add the images of the edges in $E''$ to the resulting drawing.\label{new-paradigm-item-3}
}
%\vspace{-0mm}

%This allows us to break the $n^{1/2}$ barrier for approximating the \MCN problem: we present an algorithm that achieves $\tilde O(n^{1/2-\eps}\cdot \Delta^{O(1)})$ number of crossings, showing that the new paradigm can indeed be used to overcome the $\Omega(\sqrt{n})$ barrier. 
%We hope that this paradigm will lead to better approximation algorithms. 

$\ $

Our result, combined with the $O(\poly(\Delta\log n))$-approximation algorithm  for \MP of \cite{kawarabayashi2019polylogarithmic}, provides an efficient implementation of Steps (\ref{new-paradigm-item-1}) and (\ref{new-paradigm-item-1.5}) of Paradigm $\Pi'$, such that there exists an implementation of Steps (\ref{new-paradigm-item2}) and (\ref{new-paradigm-item-3}), that produces a drawing of $G$ with $O(\optcro(G)\cdot\poly(\Delta\log n))$ crossings. This still leaves open the following question:

%\vspace{-3mm}
\begin{quote}
	\emph{Can we implement Steps (\ref{new-paradigm-item2}) and (\ref{new-paradigm-item-3}) of Paradigm $\Pi'$ efficiently in near-optimal fashion? }
\end{quote}
%\vspace{-3mm}

One way to address this question is by designing algorithms for the following problem: given a graph $G$ and a planarizing set $E^*$ of its edges, compute a drawing of $G$, such that the corresponding induced drawing of $G\setminus E^*$ is planar, while minimizing the number of crossings; in other words, every crossing in the drawing must be between an edge of $E^*$ and another edge of $E(G)$. This problem was considered by Chimani and Hlin{\v{e}}n{\`y} \cite{chimani2011tighter} who showed an efficient algorithm, that computes a drawing of $G$ with $\optcro^{E^*}(G)+\tilde O(k^2 )$ crossings, where $k=|E^*|$, and $\optcro^{E^*}(G)$ is the optimal solution value for this problem. Unfortunately, if our goal is to break the $\Theta(\sqrt n)$ barrier on the approximation factor for \MCN, the quadratic dependence of this bound on $k$ is prohibitive.

We propose a different approach in order to 
implement Steps (\ref{new-paradigm-item2}) and (\ref{new-paradigm-item-3}) of Paradigm $\Pi'$. We provide an efficient algorithm that exploits our algorithm for Steps (\ref{new-paradigm-item-1}) and (\ref{new-paradigm-item-1.5})  in order to reduce the \MCN problem to another problem, called {\sf Minimum Crossing Number with Rotation System} (\CNwRS). In this problem, the input is a multi-graph $G$ with arbitrary vertex degrees. Additionally, for every vertex $v$ of $G$, we are given a circular ordering $\oset_v$ of the edges that are incident to $v$ in $G$. The goal is to compute a drawing of $G$ in the plane with minimum number of crossings, that respects the orderings $\set{\oset_v}_{v\in V(G)}$ of the edges incident to each vertex (but we may choose whether the ordering is clock-wise or counter-clock-wise in the drawing). We denote by $\Sigma=\set{\oset_v}_{v\in V}$ the collection of all these orderings, and we call $\Sigma$ a \emph{rotation system for graph $G$}. Given an instance $(G,\Sigma)$ of \CNwRS, we denote by $\optcrors(G,\Sigma)$ the value of the optimal solution for this instance.
We show a reduction, that, given an instance $G$ of \MCN with maximum vertex degree $\Delta$, produces an instance $(G',\Sigma)$ of \CNwRS, such that $|E(G')|\leq O\left(\optcro(G)\cdot \poly(\Delta\log n)\right)$, and moreover,  $\optcrors(G',\Sigma)\leq O\left(\optcro(G)\cdot \poly(\Delta\log n)\right )$. In particular, our reduction shows that, in order to obtain an $O(\alpha \poly(\Delta\log n))$-approximation for \MCN, it is sufficient to obtain an $\alpha$-approximation algorithm for \CNwRS. We also show an efficient randomized algorithm, that, given an instance $(G,\Sigma)$ of \CNwRS, with high probability computes a solution to the problem with at most $\tilde O\left (\left(\optcrors(G,\Sigma)+|E(G)|\right )^{2-\eps}\right )$ crossings, for $\eps=1/20$. Combining this result with our reduction, we obtain an efficient algorithm that computes a drawing of an input graph $G$ with maximum vertex degree $\Delta$, with at most $\tilde O((\optcro(G))^{2-\eps}\cdot \poly(\Delta\log n))$ crossings. We note that this algorithm can be viewed as implementing Steps (\ref{new-paradigm-item2}) and (\ref{new-paradigm-item-3}) of Paradigm $\Pi'$. The resulting algorithm, in turn, leads to a $\tilde O(n^{1/2-\eps'}\poly(\Delta))$-approximation algorithm for \MCN, for some fixed constant $\eps'>0$. While this only provides a modest improvement in the approximation factor, we view this as a proof of concept that our new method can lead to improved approximation algorithms, and in particular, this result breaks the barrier that the previously known methods could not overcome. 

%For simplicity of notation, given an input graph $G$ for the \MCN problem, denote $\cro=\optcro(G)$. Our results also immediately imply an $O(\poly(\Delta\log n))$-approximation algorithm for \MCN, with running time  $2^{O(\cro (\log \cro) \poly(\log n))}$. This shows that it is unlikely that the problem of obtaining a $O(\poly(\Delta \log n))$-approximation for \MCN in bounded-degree graphs is \NP-hard in instances $G$ with $\optcro(G)<n^{\eps}$ for any constant $0<\eps<1$.

%In other words, if we
%Using our paradigm we essentially reduce the crossing number on bounded-degree graphs to the following problem: the graph is no longer bounded-degree, but for every vertex we are given a circular ordering of its incident edges, and we are guaranteed that there is a drawing that respects all these orderings whose crossing number is at most $\tilde O(|E(G)|)$, while the optimal crossing number is $\tilde\Omega(|E(G)|)$. Alternatively, by replacing each such vertex by $d_v\times d_v$ grid ( $d_v$ denotes the degree of the vertex $v$), we obtain a problem on bounded-degree graphs where the optimal crossing number is $\Omega(\sqrt{|E(G)|})$. Our algorithm shows that this is indeed simpler than the general problem as we can get a better approximation algorithm for it.
%%%%%%%%%%%%%%%%%%%%%%%%%%%%%%%%%%%%%%%%%%
%%%%%%%%%%%%%%%%%%%%%%%%%%%%%%%%%%%%%%%%%%
%%%%%%%%%%%%%%%%%%%%%%%%%%%%%%%%%%%%%%%%%%

%\znote{We need to also mention the following paper on crossings number with rotation system somewhere: 
%Crossing Numbers of Graphs with Rotation Systems 
%by Pelsmajer et al.}

\textbf{Other related work.}
There has been a large body of work on FPT algorithms for several variants and special cases of the \MCN problem (see, e.g. \cite{Grohe04, KawarabayashiR07, pelsmajer2007crossing, bannister2014crossing, hlinveny2015crossing, didimo2019survey}).
In particular, Grohe \cite{Grohe04} obtained an algorithm for solving the problem exactly, in time $f(\optcro(G))\cdot n^{2}$, where function $f$ is at least doubly exponential. In his paper, he conjectures that there exists an FPT algorithm for the problem with running time $2^{O(\optcro(G))}\cdot n$. The dependency on $n$ in the algorithm of \cite{Grohe04} was later improved by \cite{KawarabayashiR07} from $n^2$ to $n$.

For cubic (that is, $3$-regular) graphs, the \CNwRS problem is equivalent to the \mcn problem. 
Hlin{\v{e}}n{\`y}~\cite{Hlineny06a} proved that \mcn is NP-hard, and Cabello~\cite{cabello2013hardness} proved that it is APX-hard on cubic graphs.
Therefore, the \CNwRS problem is also APX-hard on cubic graphs.
Pelsmajer et al.~\cite{pelsmajer2011crossing} studied a variation of the \CNwRS problem, where for each vertex $v$, both the circular ordering $\oset(v)$ of its incident edges, and the orientation of this ordering (say clock-wise) are fixed. They showed that this variant of the problem is also NP-hard,  and provided an $O(n^4)$-approximation algorithm with running time $O(m^n\log m)$ for it, where $m$ is the number of edges in the graph\footnote{Note that, since the input graph $G$ in both \CNwRS and this variant is allowed to be a multi-graph, it is possible that $m\gg n^2$, and the optimal solution value may be much higher than $n^4$.}.
They also provided additional  approximation algorithms for some special cases. 

We now proceed to describe our results more formally.

%\vspace{-4mm}
\subsection{Our Results}
%\vspace{-2mm}

%\item Check that the drawing that the algorithm of \cite{chuzhoy2011graph} finds draws $G\setminus E'$ in a planar way. This is crucially used by our current proof. \znote{This is true only when the input graph is $3$-connected (Theorem 7). A short explanation for why this is not true for non-$3$-connected graphs is that the drawing produced by Theorem 13 does not guarantee that the induced drawing of $G\setminus e$ is planar.}
%\mynote{We need to add a theorem statement saying that if $G$ is $3$-connected then  \cite{chuzhoy2011graph} finds a drawing where $G\setminus E^*$ is planar (and need to check this very carefully). Thm 13 in \cite{chuzhoy2011graph}  uses a result for drawing a planar graph $G$ plus one edge $e$. I believe that results draws $G$ in a planar way and all intersections are with $e$. So when $G$ is not 3-connected we'll need to delete edges incident to $2$-cuts (where both sides are non-planar) and make sure that at the end all new intersections are between these edges.}
%\znote{I think this can be handled, but need to make sure everything is correct in the process of writing Section 8.}

%In this paper, first we show the limits of the explained paradigm, witnessed by the following theorem.
%\begin{lemma}
%There exists a graph $G$ composed of a planar graph $H$ and the set of additional edges $E'$ such that adding the edges of $E'$ to the panar drawing of $H$ causes $\Omega(|E'|(|E'|+\optcro(G)))$ crossings.
%\end{lemma}

Our main technical result is summarized in the following theorem.

%\vspace{-0.1cm}
\begin{theorem}\label{thm: main}
There is an efficient algorithm, that, given an $n$-vertex graph $G$ with maximum vertex degree $\Delta$ and a planarizing set $E'$ of its edges, computes another planarizing edge set $E''$ for $G$, with $E'\subseteq E''$, such that
$|E''|\leq O\left((|E'|+\optcro(G))\cdot\poly(\Delta\log n)\right)$, and, moreover, there is a drawing $\phi$ of $G$ with at most $O\left((|E'|+\optcro(G))\cdot\poly(\Delta\log n)\right)$ crossings, such that the edges of $G\setminus E''$ do not participate in any crossings in $\phi$.
\end{theorem}
%\vspace{-0.2cm}

Recall that Kawarabayashi and Sidiropoulos~\cite{kawarabayashi2019polylogarithmic} provide an efficient $O(\poly(\Delta\log n))$-approximation algorithm for the \MP problem. Since, for every graph $G$, there is a planarizing set $E^*$ containing at most $\optcro(G)$ edges, we can use their algorithm in order to compute, for an input graph $G$, a planarizing edge set of cardinality $O(\optcro(G)\cdot\poly(\Delta\log n))$. Combining this with Theorem \ref{thm: main}, we obtain the following immediate corollary.

%\begin{theorem}[Corollary 1.2 of~\cite{kawarabayashi2019polylogarithmic}]
%There exists a polynomial-time algorithm, that, given a graph $G$, computes a set $E'\subseteq E(G)$ of $\tilde O(\optcro(G))$ edges, such that the graph $G\setminus E'$ is planar.
%\end{theorem}
%Combining the above two theorems, we have the following corollary.

\begin{corollary}\label{cor:intro-planarizing-set}
There is an efficient algorithm, that, given an $n$-vertex graph $G$ with maximum vertex degree $\Delta$, computes a planarizing set $E'\subseteq E(G)$ of $O(\optcro(G)\cdot \poly(\Delta\log n))$ edges, such that there is a drawing $\phi$ of $G$  with $O(\optcro(G)\cdot \poly(\Delta\log n))$ crossings, and the edges of $E(G)\setminus E'$ do not participate in any crossings in $\phi$.
\end{corollary}
%\vspace{-0.2cm}

\iffalse
The following corollary then easily follows from Corollary \ref{cor:intro-planarizing-set}.
\begin{corollary}\label{cor:sub-exp-alg}
	There is an algorithm, that, given an $n$-vertex graph $G$ with maximum vertex degree $\Delta$, computes a drawing of $G$ in the plane with $O(\optcro(G)\cdot \poly(\Delta\log n))$ crossings, in time  $2^{O(\cro (\log \cro) \poly(\Delta\log n))}$, where $\cro=\optcro(G)$.
\end{corollary}
%\vspace{-0.2cm}

The proof of Corollary \ref{cor:sub-exp-alg} appears in Section \ref{sec: apx-sub-exp-alg} of Appendix.
Notice that, from the above corollary, it is unlikely that the problem of obtaining an $O(\poly\log n)$ approximation for \MCN in bounded-degree graphs is \NP-hard in instances where the value of the optimal solution $\cro<n^{\eps}$ for any constant $\eps$.
\fi

Next, we show a reduction from \MCN to the \CNwRS problem. 

%\vspace{-0.2cm}
\begin{theorem}\label{thm: reduction}
There is an efficient algorithm, that, given an $n$-vertex graph $G$ with maximum vertex degree $\Delta$, computes an instance $(G',\Sigma)$ of the \CNwRS problem, such that the number of edges in $G'$ is at most $O\left(\optcro(G)\cdot \poly(\Delta\log n)\right)$, and   $\optcrors(G',\Sigma)\leq O\left(\optcro(G)\cdot \poly(\Delta\log n)\right )$. Moreover, there is an efficient algorithm that, given any solution to instance $(G',\Sigma)$ of \CNwRS of value $X$, computes a drawing of $G$ with at most  $O\left ((X+\optcro(G))\cdot \poly(\Delta\log n)\right )$ crossings.
\end{theorem}
Notice that the above theorem shows that an $\alpha$-approximation algorithm for \CNwRS immediately gives an $O(\alpha\poly(\Delta\log n))$-approximation algorithm for the \MCN problem. In fact, even much weaker guarantees for \CNwRS suffice: if there is an algorithm that, given an instance $(G,\Sigma)$ of \CNwRS, computes a solution of value at most $\alpha(\optcrors(G,\Sigma)+|E(G)|)$, then there is an $O(\alpha\poly(\Delta\log n))$-approximation algorithm for \MCN. Recall that \cite{leighton1999multicommodity,even2002improved} provide an algorithm for the  \MCN problem that draws an input graph $G$ with $\tilde O((n+\optcro(G)) \cdot \Delta^{O(1)})$ crossings. While it is conceivable that this algorithm can be adapted to the \CNwRS problem, it only gives meaningful guarantees when the maximum vertex degree in $G$ is low, while in the instances of \CNwRS produced by our reduction this may not be the case, even if the initial instance of \MCN had bounded vertex degrees. %While the reduction is currently only limited to $3$-connected graphs, there are techniques, including those presented in the current paper, that allow to generalize algorithms for \MCN in $3$-connected graphs to algorithms that achieve similar guarantees in general graphs.
Our next result provides an algorithm for the \CNwRS problem.
%\vspace{-0.2cm}
\begin{theorem}\label{thm: main_rot}
There is %a universal constant $0<\eps<1$, and
 an efficient randomized algorithm, that, given an instance $(G,\Sigma)$ of \CNwRS, with high probability computes a solution of value at most $\tilde O\left (\left(\optcrors(G,\Sigma)+|E(G)|\right )^{2-\eps}\right )$ for this instance, for $\eps=1/20$.
%\znote{sorry this theorem needs to be changed to $\tilde O((\optcrors(G,\Sigma)+|E(G)|)^{2-\eps})$}
\end{theorem}
%\vspace{-0.2cm}

% graph $G$ and a rotation system $\Sigma$ on $G$, computes a drawing with $\tilde O(\optcro(G,\Sigma)+\optcro(G,\Sigma)^{1-\epsilon}\cdot |E(G)|)$ crossings, for $\eps=1/\con$, that respects the rotation system $\Sigma$. 

By combining Theorem \ref{thm: reduction} with  Theorem \ref{thm: main_rot}, we obtain the following immediate corollary.

%\vspace{-0.1cm}
\begin{corollary}\label{cor: main_algo}
There is %a universal constant $0<\eps<1$, and 
an efficient randomized algorithm, that, given any $n$-vertex graph $G$ with maximum vertex degree $\Delta$, with high probability computes a drawing of $G$ in the plane with at most $O\left ((\optcro(G))^{2-\epsilon}\cdot\poly(\Delta\log n)\right )$ crossings, for $\eps=1/20$.
%\znote{this needs to be changed to $(\optcro(G))^{2-\epsilon/2}$...}
\end{corollary}
%\vspace{-0.2cm}

\iffalse
\begin{proof}
Let $\tG$ be the input bounded-degree graph. We apply the algorithm of~\cite{kawarabayashi2019polylogarithmic} to compute a planarizing set $\tE\subseteq E(\tG)$ with $|\tE|=\tilde O(\optcro(\tG))$. 
Then we apply Theorem~\ref{thm: main} to $\tG$ and $\tE$ to obtain a new planarizing set $\tE'$ of edges.
We now contract each connected component of $\tG\setminus \tE'$ into a node, and denote the resulting graph by $\tG'$.
Note that the input $\tG'$ may be a multi-graph, where parallel edges and self-loops are allowed. We first preprocess the graph $\tG'$ by splitting each edge with a new vertex. Let $\tG''$ be the new graph, then it is clear that (i) $\tG''$ is a simple graph; 
(ii) $|E(\tG'')|=2|E(\tG')|$;
(iii) $\sum_{v}\deg_{\tG''}(v)^2=\sum_{v}\deg_{\tG'}(v)^2+\Theta(|E(\tG')|)$; and
(iv) any drawing of $\tG''$ can be easily transformed into a drawing of $\tG'$ without creating or eliminating any crossings, and vice versa.
Then we apply Theorem~\ref{thm: main_rot} to $\tG''$ and eventually obtain a drawing of $\tG$.
\end{proof}
\fi
 
Lastly, by combining the algorithm from Corollary~\ref{cor: main_algo} with the algorithm of Even et al.~\cite{even2002improved}, we obtain the following corollary, whose proof appears in Section \ref{sec: cor main algo} of Appendix.

%\vspace{-0.2cm}
\begin{corollary}\label{cor: approx alg}
There is an efficient $O(n^{1/2-\eps'}\cdot\poly(\Delta))$-approximation algorithm
for \mcn, for some universal constant $\eps'$.
\end{corollary}
%\vspace{-0.6cm}

\subsection{Our Techniques}
%\vspace{-3mm}

In this subsection, we provide a high-level intuitive overview of the main technical result of our paper -- the proof of Theorem \ref{thm: main}. As in much of previous work, we focus on the special case of \MCN, where the input graph $G$ is $3$-connected. This special case seems to capture the main difficulty of the problem, and the extension of our algorithm to the general case is somewhat standard. We use the standard graph-theoretic notion of well-linkedness: we say that a set $\Gamma$ of vertices of $G$ is \emph{well-linked} in $G$ iff for every pair $\Gamma',\Gamma''$ of disjoint equal-cardinality subsets of $\Gamma$, there is a collection of $|\Gamma'|$ paths in $G$, connecting every vertex of $\Gamma'$ to a distinct vertex of $\Gamma''$, such that every edge of $G$ participates in at most $\poly\log n$ such paths. Given a sub-graph $C$ of $G$, we let $\Gamma(C)$ be the set of its \emph{boundary vertices}: all vertices of $C$ that are incident to edges of $E(G)\setminus E(C)$. The following observation can be shown to follow from arguments in \cite{chuzhoy2011algorithm}: Suppose we are given a collection $\cset=\set{C_1,\ldots,C_r}$  of disjoint sub-graphs of $G$, such that each subgraph $C_i$ has the following properties:

%\vspace{-0.4cm}
\begin{itemize}
	\item {\bf 3-Connectivity:} graph $C_i$ is $3$-connected;
	%\vspace{-0.2cm}
	\item {\bf Well-Linkedness:} the vertex set $\Gamma(C_i)$ is well-linked in $C_i$; and
	%\vspace{-0.2cm}
	\item {\bf Strong Planarity:} graph $C_i$ is planar, and the vertices of $\Gamma(C_i)$ lie on the boundary of a single face in the unique planar drawing of $C_i$.
\end{itemize}

%\vspace{-0.3cm}

Arguments from \cite{chuzhoy2011algorithm} can then be used to show that there is a near-optimal drawing of $G$, in which the edges of $\bigcup_{i=1}^rE(C_i)$ do not participate in any crossings. Therefore, in order to prove Theorem \ref{thm: main}, it is enough to show that there is a small planarizing set $E''$ of edges in $G$, with $E'\subseteq E''$, such that every connected component $C$ of $G\setminus E''$ has the above three properties. We note that the Well-Linkedness property is easy to achieve: there are standard algorithms (usually referred to as ``well-linked decomposition''), that allow us to compute a set $E''$ of $O(|E'|)$ edges with $E'\subseteq E''$, such that every connected component of $G\setminus E''$ has the Well-Linkedness property. 
Typically, such an algorithms starts with $E''=E'$, and then iterates, as long as some connected component $C$ of the current graph $G\setminus E''$ does not have the Well-Linkedness property. The algorithm then computes a cut $(X,Y)$ of $C$ that is sparse with respect to the set $\Gamma(C)$ of vertices, adds the edges of $E(X,Y)$ to the set $E''$ and continues to the next iteration. One can show, using standard arguments, that, once the algorithm terminates, $|E''|\leq O(|E'|)$ holds.

\begin{figure}[h]
	\centering
	\subfigure[Bad example for the Strong Planarity property. The vertices of $\Gamma_1\cup \Gamma_2$ and the edges of $E_1\cup E_2$ are shown in red.]{\scalebox{0.4}{\includegraphics{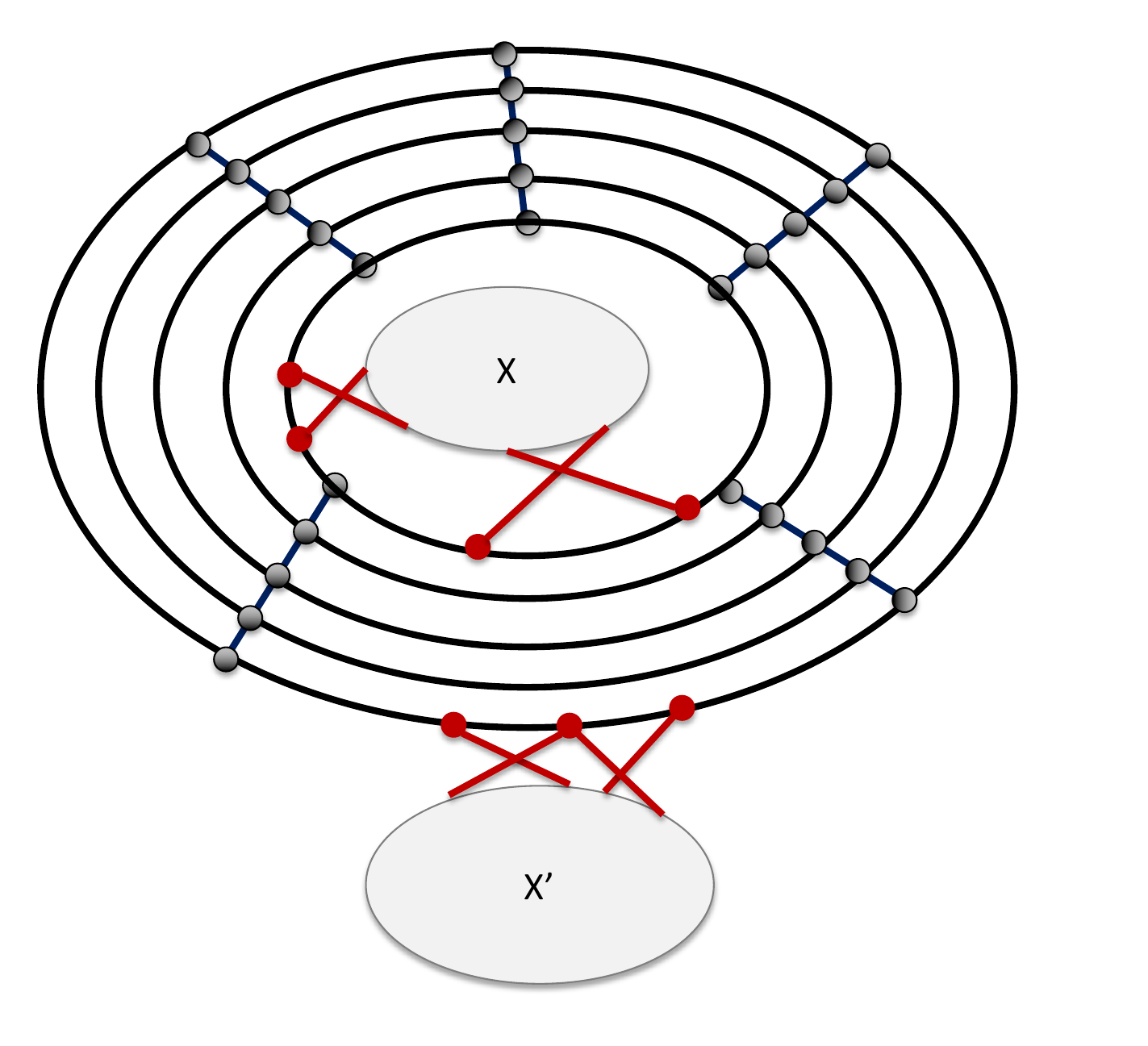}}\label{fig: bad example for strong planarity}}
	\hspace{1cm}
	\subfigure[Deleting the blue edges in this figure ensures that all vertices of $\Gamma_1\cup \Gamma_2$ lie on the boundary a single face. Note that this deletion may cause $C$ to violate the $3$-connectivity property.]{
		\scalebox{0.4}{\includegraphics{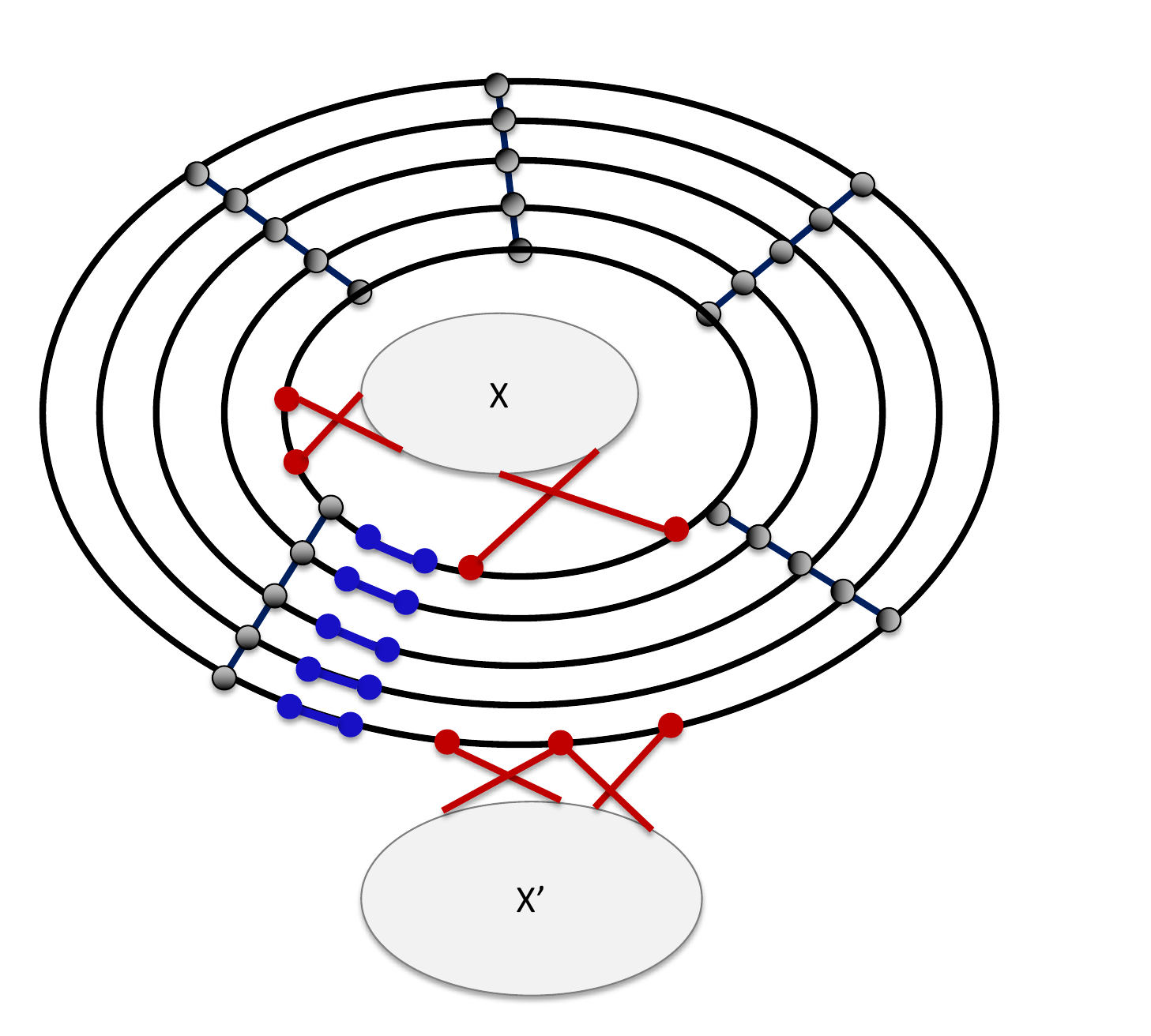}}\label{fig: bad example for strong planarity fixed}}
	%\hspace{1cm}
	%\subfigure[caption subfigure 3]{\scalebox{0.2}{\includegraphics{file3.pdf}}\label{fig: label3}}
	\caption{Obtaining  the Strong Planarity property\label{fig: strong planarity}}
\end{figure}

The remaining two properties, however, seem impossible to achieve, if we insist that the set $E''$ of edges is relatively small. For example, consider a cylindrical grid $C$, that consists of a large number $N$ of disjoint cycles, and a number of additional paths that intersect each cycle in a way that forms a cylindrical grid (see Figure \ref{fig: bad example for strong planarity}). Consider two additional graphs $X$ and $X'$. Let $G$ be a graph obtained from the union of $C,X,X'$, by adding two sets of edges: set $E_1$, connecting some vertices of $X$ to some vertices of the innermost cycle of $C$ (the set of endpoints of these edges that lie in $C$ is denoted by $\Gamma_1$), and set $E_2$, connecting some vertices of $X'$ to some vertices of the outermost cycle of $C$ (the set of endpoints of these edges that lie in $C$ is denoted by $\Gamma_2$). Assume that the given  planarizing set of edges for $G$ is $E'=E_1\cup E_2$, so that $\Gamma(C)=\Gamma_1\cup \Gamma_2$. Then, in order to achieve the strong planarity property for the graph $C$, we are forced to add $N$ edges to set $E''$: one edge from every cycle of the cylindrical grid, in order to ensure that the vertices of $\Gamma_1\cup \Gamma_2$ lie on the boundary of a single face (see Figure \ref{fig: bad example for strong planarity fixed}).

In order to overcome this difficulty, we weaken the Strong Planarity requirement, as follows. Informally, given a graph $G$ and a sub-graph $C$ of $G$, a \emph{bridge} for $C$ is a connected component of $G\setminus C$. If $R$ is a bridge for $C$, then we let $L(R)$ be the set of \emph{legs} of this bridge, that contains all vertices of $C$ that have a neighbor in $R$. We replace the Strong Planarity requirement with the following weaker property:

%\vspace{-0.4cm}
\begin{itemize}
	\item {\bf Bridge Property:} for every bridge $R$ for $C_i$, all vertices of $L(R)$ lie on the boundary of a single face in the unique planar drawing of $C_i$.
\end{itemize}

%\vspace{-0.2cm}

We illustrate the intuition for why this weaker property is sufficient on the example from Figure \ref{fig: bad example for strong planarity}. Let $G_1=C\cup E_1\cup X$, and let $G_2=C\cup E_2\cup X'$. Notice that $C$ has the strong planarity property in both $G_1$ and $G_2$. Therefore, from the results of \cite{chuzhoy2011algorithm}, there is a drawing $\phi_1$ of $G_1$, and a drawing $\phi_2$ of $G_2$, such that the edges of $C$ do not participate in any crossings in either drawing. In particular, the drawings of the graph $C$ induced by $\phi_1$ and $\phi_2$ must be identical to the unique planar drawing of $C$, and hence to each other. 
Moreover, in drawing $\phi_1$ of $G_1$, all edges and vertices of $X$ must be drawn inside a disc whose boundary is the innermost cycle of $C$, and in the drawing $\phi_2$ of $G_2$, all edges and vertices of $X'$ must be drawn outside of the disc whose boundary is the outermost cycle of $C$.
Therefore, we can ``glue'' the drawings $\phi_1$ and $\phi_2$ to each other via the drawing of $C$ in order to obtain the final drawing $\phi$ of $G$, such that the edges of $C$ do not participate in crossings in $\phi$.

Assume now that the Bridge Property does not hold for $C$ (for example, assume that $X$ and $X'$ are connected by an edge). Then we can show that in any drawing of $G$, at least $\Omega(N)$ edges of $C$ must participate in crossings. We can then add $N$ edges of $C$ to $E''$ -- one edge per cycle, as shown in Figure \ref{fig: bad example for strong planarity fixed}, in order to ensure that all vertices of $\Gamma_1\cup \Gamma_2$ lie on the boundary of a single face of the resulting drawing. This increase in the cardinality of $E''$ can be charged to the crossings of the optimal drawing of $G$ in which the edges of $C$ participate.

Intuitively, if our goal were to only ensure the Well-Linkedness and the Bridge properties, we could start with $E''=E'$, and then gradually add edges to $E''$, until every connected component of $G\setminus E''$ has the bridge property, using reasonings that are similar to the ones above. After that we could employ the well-linked decomposition in order to ensure the Well-Linkedness property of the resulting clusters. One can show that, once the Bridge Property is achieved, it continues to hold over the course of the algorithm that computes the well-linked decomposition.

Unfortunately, it is also critical that we ensure the $3$-connectivity property. Assume that a connected component $C$ of $G\setminus E''$ is $3$-connected, let $\psi$ be its unique planar drawing, and let $\phi^*$ be the optimal drawing of $G$. Then one can show that drawing $\psi$ of $C$ is ``close'' to the drawing $\phi^*_C$ of $C$ induced by $\phi^*$. We measure the ``closeness'' between two drawings using the notion of irregular vertices and edges, that was introduced in \cite{chuzhoy2011graph}. A vertex $v$ of $C$ is irregular, if the ordering of the edges incident to $v$, as they enter $v$, is different in the drawings $\psi$ and $\phi^*_C$ (ignoring the orientation). The notion of irregular edges is somewhat more technical and, since we do not need it, we will not define it here. It was shown in \cite{chuzhoy2011graph}, that, if $C$ is $3$-connected, then the total number of irregular vertices and edges of $C$ is roughly bounded by the number of crossings in which the edges of $C$ participate in $\phi^*$. Therefore, if we think of the number of crossings in $\phi^*$ as being low, then the two drawings $\psi$ and $\phi^*_C$ are close to each other. However, if graph $C$ is not $3$-connected, and, for example, is only $2$-connected, then the number of irregular vertices in $C$ may be much higher. Let $S_2(C)$ denote the set of all vertices of $C$ that participate in $2$-separators, that is, a vertex $v$ belongs to $S_2(C)$ iff there is another vertex $v'\in V(C)$, such that graph $C\setminus\set{v,v'}$ is not connected. 
It is easy to see that, even if the drawing $\phi^*_C$ is planar, it is possible that there are as many as $|S_2(C)|$ irregular vertices in $C$. Since we do not know what the optimal drawing $\phi^*$ looks like, it seems impossible to fix a planar drawing of $C$ that is close to $\phi^*_C$.

Ensuring the $3$-connectivity property for the connected components of $G\setminus E''$, however, seems a daunting task. As edges are added to $E''$, some components $C$ may no longer be $3$-connected. Even if we somehow manage to decompose them into $3$-connected subgraphs, while only adding few edges to $E''$, the addition of these new edges may cause the well-linkedness property to be violated. Then we need to perform the well-linked decomposition from scratch, which in turn can lead to the violation of the 3-connectivity property, and it is not clear that this process will terminate while $|E''|$ is still sufficiently small.

In order to get around this problem, we slightly weaken the $3$-connectivity property. We first observe that, even if graph $C$ is not $3$-connected, and is instead $2$-connected, but the number of vertices participating in $2$-separators (vertices in set $S_2(C)$) is low, then this is sufficient for us. Intuitively, the reason is that, 
the results of  \cite{chuzhoy2011graph} show that the number of irregular vertices in such a graph $C$ is roughly bounded by $|S_2(C)|$ plus the number of crossings in $\phi^*$ in which the edges of $C$ participate. Another intuitive explanation is that, when $|S_2(C)|$ is low, there are fewer possible planar drawings of $C$, so we may think of all of them as being ``close'' to $\phi^*_C$. Unfortunately, even this weaker property is challenging to achieve, since we need to ensure that it holds simultaneously with the Well-Linkedness and the Bridge properties, for all connected components of $G\setminus E''$. In order to overcome this obstacle, we allow ourselves to add a small set $A$ of ``fake'' edges to the graph $G$, whose addition ensures that each component of $(G\setminus E'')\cup A$ is a $2$-connected graph with few $2$-separators, for which the Well-Linkedness and the Bridge properties hold. Intuitively, we show that the fake edges of $A$ can be embedded into the graph $G$, so that, in a sense, we can augment the optimal drawing $\phi^*$ of $G$ by adding the images of the edges of $A$ to it, without increasing the number of crossings (though we note that this is an over-simplification that is only intended to provide intuition). The proof of Theorem \ref{thm: main} can then be thought of as consisting of two parts. In the first part, we present an efficient algorithm that computes the set $E''$ of edges of $G$ with $E'\subseteq E''$, and the collection $A$ of fake edges, such that, for every connected component $C$ of graph $(G\setminus E'')\cup A$, one of the following holds: either $|\Gamma(C)|\leq \poly(\Delta \log n)$, or $C$ has the Well-Linkedness and the Bridge properties, together with  the weakened $3$-Connectivity property. We also compute an embedding of the fake edges in $A$ into $G$ in this part. In the second part, we show that there exists a near-optimal drawing of $G$ in which the edges of $G\setminus E''$ do not participate in crossings. The latter part formalizes and significantly generalizes ideas presented in \cite{chuzhoy2011algorithm}.

\noindent {\bf Organization.}
We start with some basic definitions and notation in Section \ref{sec:prelim}. We then present a high-level overview of the proofs of all our results in Section \ref{sec:overview}, with most details deferred to subsequent sections.
Section \ref{sec:block_decompos} contains definitions and general results regarding block decompositions of graphs (mostly from previous work). The proof of Theorem \ref{thm: main} is completed in Sections \ref{sec:thm1}--\ref{sec:non_3}, and Section \ref{sec:cr_rotation_system} completes the proof of Theorem \ref{thm: main_rot}.

\section{Preliminaries}
\label{sec:prelim}

\iffalse
------------------

To be used later for locally non-interfering/non-interfering paths (for this the definition should be adapted). We may also keep it in prelims as a definition of crossing paths. Maybe we should also define a drawing of a path as a concatenation of images of its edges. We use it quite a bit.

\begin{definition}
	Let $\gamma,\gamma'$ be two curves in the plane or on the sphere. We say that $\gamma$ and $\gamma'$ \emph{cross},  iff there is a disc $D$, whose boundary is a simple closed curve that we denote by $\beta$, such that: 
	
	\begin{itemize}
		\item  $\gamma\cap D$ is a simple open curve, whose endpoints we denote by $a$ and $b$; 
		\item $\gamma'\cap D$ is a simple open curve, whose endpoints we denote by $a'$ and $b'$; and
		
		\item $a,a',b,b'\in \beta$, and they appear on $\beta$ in this circular order.
	\end{itemize}
\end{definition}

Given a graph $G$ embedded in the plane or on the sphere, we say that two paths $P,P'$ in $G$ cross iff their images cross. Similarly, we say that a path $P$ crosses a curve $\gamma$ iff the image of $P$ crosses $\gamma$.

------------------------------
\fi

By default, all logarithms are to the base of $2$. All graphs are finite, simple and undirected. Graphs with parallel edges are explicitly referred to as multi-graphs.

We follow standard graph-theoretic notation. Assume that we are given a graph $G=(V,E)$. 
%We denote by $n$ the number of vertices and by $\Delta$ the maximum vertex degree in the graph which may not necessarily be a constant.
For a vertex $v\in V$, we denote by $\delta_G(v)$ the set of all  edges of $G$ that are incident to $v$.
For two disjoint subsets $A,B$ of vertices of $G$, we denote by $E_G(A,B)$ the set of all edges with one endpoint in $A$ and another in $B$.
For a subset $S\subseteq V$ of vertices, we denote by $E_G(S)$ the set of all edges with both endpoints in $S$, and we denote by $\out_G(S)$ the subset of edges of $E$ with exactly one endpoint in $S$, namely $\out_G(S)=E_G(S, V\!\setminus\! S)$. 
We denote by $G[S]$ the subgraph of $G$ induced by $S$.
We sometimes omit the subscript $G$ if it is clear from the context. 
%We will also refer to connected subgraphs of $G$ by \emph{clusters}. \mynote{do we actually do it for subgraphs or connected components?}
%For a connected subgraph $C$ of the graph $G$, we denote $\out(C)=\out(V(C))$, and we use $\Gamma(C)$ to denote the set of all vertices of $C$ that serve as endpoints of edges in $\out(C)$.
%
%Given a set $\pset$ of paths in a graph $G$, we denote by $E(\pset)=\bigcup_{P\in\pset}E(P)$ the set of all edges participating in the paths of $\pset$.
%For each edge $e\in E(G)$, we denote by $c_{\pset}(e)$ the number of paths in $\pset$ that contain the edge $e$. We also denote $c(\pset)=\max_{e\in E(G)}\set{c_{\pset}(e)}$, and we refer to $c(\pset)$ as the \emph{congestion} caused by the paths in $\pset$.
%
%
%\textbf{Connectivity.}
 We say that a graph $G$ is $\ell$-connected for some integer $\ell>0$, if there are $\ell$ vertex-disjoint paths between every pair of vertices in $G$. %Throughout the paper, we use connectivity to denote vertex-connectivity, unless specified otherwise.

Given a graph $G=(V,E)$, a \emph{drawing}  $\phi$ of $G$ is an embedding of the graph into the plane, that maps every vertex to a point and every edge to a continuous curve that connects the images of its endpoints. We require that the interiors of the curves representing the edges do not contain the images of any of the vertices. We say that two edges \emph{cross} at a point $p$, if the images of both edges contain $p$, and $p$ is not the image of a shared endpoint of these edges. We require that no three edges cross at the same point in a drawing of $\phi$. 
We say that $\phi$ is a \emph{planar drawing} of $G$ iff no pair of edges of $G$ crosses in $\phi$.
For a vertex $v\in V(G)$, we denote by $\phi(v)$ the image of $v$, and for an edge $e\in E(G)$, we denote by $\phi(e)$ the image of $e$ in $\phi$. For any subgraph $C$ of $G$, we denote by $\phi(C)$ the union of images of all vertices and edges of $C$ in $\phi$. For a path $P\subseteq G$, we sometimes refer to $\phi(P)$ as the \emph{image of $P$ in $\phi$}. Note that a drawing of $G$ in the plane naturally defines a drawing of $G$ on the sphere and vice versa; we use both types of drawings.
Given a graph $G$ and a drawing $\phi$ of $G$ in the plane, we use $\cro(\phi)$ to denote the number of crossings in $\phi$. Let $\phi'$ be the drawing of $G$ that is a mirror image of $\phi$. We say that $\phi$ and $\phi'$ are \emph{identical} drawings of $G$, and that their \emph{orientations} are different. We sometime say that $\phi'$ is obtained by \emph{flipping} the drawing $\phi$. If $\gamma$ is a simple closed curve in $\phi$ that intersects $G$ at vertices only, and $S$ is the set of vertices of $G$ whose images lie on $\gamma$, with $|S|\geq 3$, then we say that the circular orderings of the vertices of $S$ along $\gamma$ in $\phi$ and $\phi'$ are identical, but the orientations of the two orderings are different, or opposite.

Whitney \cite{whitney1992congruent} proved that every $3$-connected planar graph has a unique planar drawing.
Throughout, for a $3$-connected planar graph $G$, we denote by $\rho_G$ the unique planar drawing of $G$.

\textbf{Problem Definitions.} 
The goal of the \mcn~problem is to compute a drawing of the input graph $G$ in the plane with smallest number of crossings.
The value of the optimal solution, also called the \emph{crossing number} of $G$, is denoted by $\optcro(G)$.

We also consider a closely related problem called Crossing Number with Rotation System (\CNwRS). In this problem, we are given a multi-graph $G$, and, for every vertex $v\in V(G)$, a circular ordering $\oset_v$ of its incident edges. We denote by $\Sigma=\set{\oset_v}_{v\in V(G)}$ the collection of all such orderings, and we refer to $\Sigma$ as a \emph{rotation system} for graph $G$.  We say that a drawing $\phi$ of $G$ \emph{respects} the rotation system $\Sigma$ if the following holds: For every vertex $v\in V(G)$, let $\eta(v)$ be an arbitrarily small disc around $v$ in $\phi$. Then the images of the edges of $\delta_G(v)$ in $\phi$ must intersect the boundary of $\eta(v)$ in a circular order that is identical to $\oset_v$ (but the orientation of this ordering may be arbitrary). In the \CNwRS problem, the input is a {\bf multi-graph} $G$ with a rotation system $\Sigma$, and the goal is to compute a drawing of $G$ in the plane that respects $\Sigma$ and minimizes the number of crossings.

{\bf Faces and Face Boundaries.}
Suppose we are given a planar graph $G$ and a drawing $\phi$ of $G$ in the plane. The set of faces of $\phi$ is the set of all connected regions of $\mathbb{R}^2\setminus \phi(G)$. We designate a single face of $\phi$ as the ``outer'', or the ``infinite'' face. 
The \emph{boundary} $\delta(F)$ of a face $F$ is a subgraph of $G$ consisting of all vertices and edges of $G$ whose image is incident to $F$. Notice that, if graph $G$ is not connected, then boundary of a face may also be not connected. Unless $\phi$ has a single face, the boundary $\delta(F)$ of every face $F$ of $\phi$ must contain a simple cycle $\delta'(F)$ that separates $F$ from the outer face. We sometimes refer to graph $\delta(F)\setminus\delta'(F)$ as the \emph{inner boundary} of $F$. Lastly, observe that, if $G$ is $2$-connected, then the boundary of every face of $\phi$ is a simple cycle. %, and its inner boundary is empty.

\textbf{Bridges and Extensions of Subgraphs.}
Let $G$ be a graph, and let $C\subseteq G$ be a subgraph of $G$. A \emph{bridge} for $C$ in graph $G$ is either (i) an edge $e=(u,v)\in E(G)$ with $u,v\in V(C)$ and $e\not \in E(C)$; or (ii) a connected component of $G\setminus V(C)$. %, together with all edges (and their endpoints) between the component and $C$.
%{\znote{this is slightly different from the previous version, only for brevity in notations}}
We denote by $\rset_G(C)$ the set of all bridges for $C$ in graph $G$. For each bridge $R\in \rset_G(C)$, we define the set of vertices $L(R)\subseteq V(C)$, called the \emph{legs of $R$}, %to contain every vertex $v\in V(C)$ that has a neighbor in $R$.
 as follows. If $R$ consists of a single edge $e$, then $L(R)$ contains the endpoints of $e$. Otherwise, $L(R)$ contains all vertices $v\in V(C) $, such that $v$ has a neighbor that belongs to $R$. %, such that some edge connecting $v$ to a vertex of $R$ belongs to $G$.

Next, we define an \emph{extension} of the subgraph $C\subseteq G$, denoted by $X_G(C)$. The extension contains, for every bridge $R\in \rset_G(C)$, a tree $T_R$, that is defined as follows. If $R$ is a bridge consisting of a single edge $e$, then the corresponding tree $T_R$ only contains the edge $e$. Otherwise, 
let $R'$ be the subgraph of $G$ consisting of the graph $R$, the vertices of $L(R)$, and all edges of $G$ connecting vertices of $R$ to vertices of $L(R)$.
We let $T_R\subseteq R'$ be a tree, whose leaves are precisely the vertices of $L(R)$. %, that contains the smallest number of edges among all such trees. 
Note that such a tree exists because graph $R$ is connected, and it can be found efficiently. % and thus contains a spanning tree $T$ as a subgraph. %Finally we can recursively remove all leaves of the tree that are not in $L(R)$. %This gives one such tree, illustrating the existence of $T_R$. 
We let the extension of $C$ in $G$ be $X_G(C)=\set{T_R\mid R\in \rset_G(C)}$.

\textbf{Sparsest Cut and Well-Linkedness.} %We use a standard notion of well-linkedness:
%
%\vspace{-3mm}
%\begin{definition}
%	Given a graph $G$, a subset $T$ of its vertices, and a parameter $\alpha > 0$, we say that $T$ is \emph{$\alpha$-well-linked} in $G$, iff for any partition $(A,B)$ of $T$, $|E(A, B)|\ge \alpha\cdot\min \{|T\cap A|, |T\cap B|\}$. %\snote{shouldn't this be $|E(A, B)|\ge \alpha\cdot\min \{|\out(T)\cap \out(A)|, |\out(T)\cap \out(B)|\}$?}
%\end{definition}
%\vspace{-0.3cm}
%
Suppose we are given a graph $G=(V,E)$, and a subset $\Gamma\sse V$ of its vertices. We say that a cut $(X,Y)$ in $G$ is a valid $\Gamma$-cut iff $X\cap \Gamma,Y\cap \Gamma\neq \emptyset$. The \emph{sparsity} of a valid $\Gamma$-cut $(X,Y)$ is $\frac{|E(X,Y)|}{\min\set{|X\cap \Gamma|, |Y\cap \Gamma|}}$. %, and the value of the sparsest cut in $G$ is defined to be:
%$\Phi(G)=\min_{S\subsetneq V}\set{\Phi(S)}$.
In the Sparsest Cut problem, given a graph $G$ and a subset $\Gamma$ of its vertices, the goal is to compute a valid $\Gamma$-cut of minimum sparsity. Arora, Rao and Vazirani~\cite{ARV} have shown an $O(\sqrt {\log n})$-approximation algorithm for the sparsest cut problem. We denote this algorithm by \algsc, and its approximation factor by $\alphasc(n)=O(\sqrt{\log n})$. 

We say that a set $\Gamma$ of vertices of $G$ is \emph{$\alpha$-well-linked} in $G$, iff the value of the sparsest cut in $G$ with respect to $\Gamma$ is at least $\alpha$.

\section{High-Level Overview}
\label{sec:overview}

In this section we provide a high-level overview of the proofs of our main results, and state the main theorems from which the proofs are derived. 
As in previous work, we start by considering a special case of the \MCN problem, where the input graph $G$ is $3$-connected. This special case seems to capture the main technical challenges of the whole problem, and the extension to non-$3$-connected graphs is relatively easy and follows the same framework as in previous work \cite{chuzhoy2011graph}.
We start by defining several central notions that our proof uses.

%\mynote{An explanation why the things are set up this way and not the way we discussed. Our plan was to define an acceptable cluster like this:  a type-1 acceptable cluster has $\poly\log n$ terminals. A type-2 acceptable cluster is $3$-connected, good (no path connecting terminals lying on different faces), and terminals are well-linked. To get a planarizing set $\hat E$ of edges so that in $H=G\setminus \hat E$ every cluster is acceptable, we would follow these 3 steps:
%(1) make all clusters good, under a suitable definition of good.
%(2) do a well-linked decomposition.
%(3) cut-off 2-cuts to get 3-connected components.
%However, the third step does not actually work. If we start cutting off 2-vertex cuts we may end up creating new 2-vertex cuts that weren't there before. So type-2 acceptable clusters need to be defined differently. We need to allow to add fake edges that will replace the 2-vertex cuts we cut off, and the cluster itself with the fake edges should be 3-connected. This motivates the setup in this section.}

%\subsection*{Special Case: Input Graph $G$ is $3$-Connected}

\subsection{Acceptable Clusters and Decomposition into Acceptable Clusters}
In this section we define acceptable clusters and decomposition into acceptable clusters. These definitions are central to all our results. 
Let $G$ be an input graph on $n$ vertices of maximum degree at most $\Delta$; we assume that $G$ is $3$-connected. Let $\hat E$ be any planarizing set of edges for $G$, and let $H=G\setminus \hat E$. Let $\Gamma\subseteq V(G)$ be the set of all vertices that serve as endpoints of edges in $\hat E$; we call the vertices of $\Gamma$ \emph{terminals}. We will define a set $A$ of \emph{fake edges}; for every fake edge $e\in A$, both endpoints of $e$ must lie in $\Gamma$. We emphasize that the edges of $A$ do not necessarily lie in $H$ or in $G$; in fact we will use these edges in order to augment the graph $H$. 

We denote by $\cset$ the set of all connected components of graph $H\cup A$, and we call elements of $\cset$ \emph{clusters}.  For every cluster $C\in \cset$, we denote by $\Gamma(C)=\Gamma\cap V(C)$ the set of all terminals that lie in $C$. We also denote by $A_C=A\cap C$ the set of all fake edges that lie in $C$.

%Next, we define some desirable properties of these clusters. 

%\vspace{-2mm}
\begin{definition}
	We say that a cluster $C\in \cset$ is a \emph{type-1 acceptable cluster} iff:
	
	%\vspace{-2mm}
	\begin{itemize}
		\item $A_C=\emptyset$; and
		
		%\vspace{-2mm}
		\item  $|\Gamma(C)|\leq \mu$ for $\mu=512\Delta\alphasc(n)\log_{3/2} n=O(\Delta\log^{1.5}n)$  (recall that $\alphasc(n)=O(\sqrt{\log n})$ is the approximation factor of the algorithm $\algsc$ for the sparsest cut problem).
	\end{itemize}
\end{definition}
%\vspace{-0.2cm}

Consider now some cluster $C\in \cset$, and assume that it is $2$-connected. For a pair $(u,v)$ of vertices of $C$, we say that $(u,v)$ is a $2$-separator for $C$ iff the graph $C\setminus\set{u,v}$ is not connected. We denote by $S_2(C)$ the set of all vertices of $C$ that participate in $2$-separators, that is,  a vertex $v\in C$ belongs to $S_2(C)$ iff there is another vertex $u\in C$ such that $(v,u)$ is a $2$-separator for $C$.
Next, we define type-2 acceptable clusters.

%Consider a cluster $C\in \cset$, and assume that it is planar. Let $\psi'_C$ be any planar drawing of $C$ on the sphere. 
%\znote{it is not clear at this point why $C$ is planar, since $C$ may contains edges in $A$. should we claim beforehand that $H\cup A$ is planar?}
%Then $\psi'_C$ induces a planar drawing $\psi_{C\setminus A_C}$ of $C\setminus A_C$ on the sphere. Next, we define type-2 acceptable clusters.

		%\vspace{-2mm}
\begin{definition}
We say that a cluster $C\in \cset$ is a \emph{type-2 acceptable cluster} with respect to its drawing $\psi'_C$ on the sphere if the following conditions hold:

		%\vspace{-2mm}
\begin{itemize}
\item {\bf (Connectivity):} $C$ is a simple $2$-connected graph, and $|S_2(C)|\leq O(\Delta|\Gamma(C)|)$. Additionally, graph $C\setminus A_C$ is a $2$-connected graph. %\znote{does this implies that $C$ is 2-connected?} \mynote{yes but I thought it's good to spell it out because we are bounding $|S_2(C)|$}.

		%\vspace{-2mm}
\item {\bf (Planarity):} $C$ is a planar graph, and the drawing $\psi'_C$ is planar. We denote by  $\psi_{C\setminus A_C}$ the drawing of $C\setminus A_C$ is induced by $\psi'_C$.

		%\vspace{-2mm}

%	or it is isomorphic to $K_3$. In particular, $C$ has a unique planar drawing, that is denoted by $\rho_{C}$. Let $\rho'_C$ be the drawing of $C\setminus A_C$ induced by $\rho_C$.
\item {\bf (Bridge Consistency Property):} for every bridge $R\in \rset_{G}(C\setminus A_C)$, there is a face $F$ in the  drawing $\psi_{C\setminus A_C}$ of $C\setminus A_C$, %(the drawing of $C\setminus A_C$ that is induced by the drawing $\psi'_C$ of $C$), 
such that all vertices of $L(R)$ lie on the boundary of $F$; and

		%\vspace{-2mm}

%{\color{blue} Note that as $C$ is no longer $3$-connected, its boundary is not a simple cycle and here we consider the full boundary of it, i.e., all edges $e$ such that every point on the drawing $\rho'_C(e)$ is adjacent to $F$.}
\item {\bf (Well-Linkedness of Terminals):} the set $\Gamma(C)$ of terminals is $\alpha$-well-linked in $C\setminus A_C$, for $\alpha=\frac{1}{128\Delta \alphasc(n)\log_{3/2} n}=\Theta \left (\frac{1}{\Delta \log^{1.5}n}\right )$.
%	\item if $Q$ is any path in $G$ whose endpoints lie in $C$, such that $Q$ is internally disjoint from $C$, then the endpoints of $Q$ are terminals that lie on the boundary of a single face of the drawing $\rho_{C\cup A}$.
\end{itemize}
\end{definition}	
%\vspace{-0.2cm}

%We emphasize that graph $C\setminus A_C$ for cluster $C\in \cset_2$ may not be connected.

Let $\cset_1\subseteq \cset$ denote the set of all type-1 acceptable clusters. % and $\cset_2\subseteq \cset$ denote the set of all type-2 acceptable clusters. (This is only done for the embedding, we don't define $\cset_2$ here yet).
For a fake edge $e=(x,y)\in A$, an \emph{embedding} of $e$ is a path $P(e)\subseteq G$ connecting $x$ to $y$. We will compute an embedding of all fake edges in $A$ that has additional useful properties summarized below.

		%\vspace{-2mm}
\begin{definition}
	A \emph{legal embedding} of the set $A$ of fake edges is a collection $\pset(A)=\set{P(e)\mid e\in A}$ of paths in $G$, such that the following hold.
	
			%\vspace{-2mm}
	
	\begin{itemize}
		\item For every edge $e=(x,y)\in A$, path $P(e)$ has endpoints $x$ and $y$, and moreover, there is a type-1 acceptable cluster $C(e)\in \cset_1$ such that $P(e)\setminus\set{x,y}$ is contained in $C(e)$; and
				%\vspace{-2mm}
		
		\item For any pair $e,e'\in A$ of distinct edges, $C(e)\neq C(e')$; 
	\end{itemize}
\end{definition}
%\vspace{-0.2cm}

Note that from the definition of the legal embedding, all paths in $\pset(A)$ must be mutually internally disjoint.
%
%		
%
%
%\subsection{Main Theorems}
%\label{subsec: main technical theorems}
%
%
Finally, we define a decomposition of a graph $G$ into acceptable clusters; this definition is central for the proof of our main result.

		%\vspace{-2mm}
\begin{definition} 
	A \emph{decomposition of a graph $G$ into acceptable clusters} consists of:
	
			%\vspace{-2mm}
	\begin{itemize}
		\item a planarizing set $\hat E\subseteq E(G)$ of edges of $G$;
				%\vspace{-2mm}
		\item a set $A$ of fake edges (where the endpoints of each fake edge are terminals with respect to $\hat E$);
				%\vspace{-2mm}
		\item a partition $(\cset_1,\cset_2)$ of all connected components (called clusters) of the resulting graph $(G\setminus \hat E)\cup A$ into two subsets, such that every cluster $C\in \cset_1$ is a type-1 acceptable cluster;
				%\vspace{-2mm}
		\item for every cluster $C\in \cset_2$, a planar drawing $\psi'_C$ of $C$ on the sphere, such that $C$ is a type-2 acceptable cluster with respect to $\psi'_C$; and
		
				%\vspace{-2mm}
		\item a legal embedding $\pset(A)$ of all fake edges.
		\end{itemize}
	
			%\vspace{-4mm}
We denote such a decomposition by $\dset=\left (\hat E,A,\cset_1,\cset_2,\set{\psi'_C}_{C\in \cset_2},\pset(A)\right )$.
\end{definition}
%\vspace{-0.2cm}

\iffalse
The proof of Theorem~\ref{thm: main} is divided into two parts.
In the first part (Section~\ref{sec:thm1} and Section~\ref{sec:canonical drawing}), we will show the proof of Theorem~\ref{thm: main}, assuming that the input \bdg $G$ is $3$-connected. 
In the second part (Section~\ref{sec:non_3}), we will complete the proof of Theorem~\ref{thm: main} by showing how to handle non-$3$-connected graphs.
We now state the two main theorems of the first part.
\fi

Our first result is the following theorem, whose proof appears in Section \ref{sec:thm1}, that allows us to compute a decomposition of the input graph $G$ into acceptable clusters. This result is one of the main technical contributions of our work.
%\vspace{-0.1cm}
\begin{theorem}\label{thm: can find edge set w acceptable clusters}
There is an efficient algorithm, that, given a $3$-connected $n$-vertex graph $G$ with maximum vertex degree at most $\Delta$ and a planarizing set $E'$ of edges for $G$, computes a decomposition  $\dset=\left ( E'',A,\cset_1,\cset_2,\set{\psi'_C}_{C\in \cset_2},\pset(A)\right )$ of $G$ into acceptable clusters, such that $E'\subseteq  E''$ and $| E''|\leq O((| E'|+\optcro(G))\cdot\poly(\Delta\log n))$.
\end{theorem}
%\vspace{-0.2cm}

\subsection{Canonical Drawings}

In this subsection, we assume that we are given a $3$-connected $n$-vertex graph $G$ with maximum vertex degree at most $\Delta$, and a decomposition  $\dset=\left ( E'',A,\cset_1,\cset_2,\set{\psi'_C}_{C\in \cset_2},\pset(A)\right )$ of $G$ into acceptable clusters. Next, we define drawings of $G$ that are ``canonical'' with respect to the clusters in the decomposition. 
For brevity of notation, we refer to type-1 and type-2 acceptable clusters as type-1 and type-2 clusters, respectively.

%Note that, for a type-2 acceptable cluster $C\in \cset_2$, graph $C\setminus A_C$ is not necessarily connected. We denote by $\sset(C)$ the set of all connected components of $C\setminus A_C$, and we let $\cset_2'=\bigcup_{C\in \cset_2}\sset(C)$.
%Note that $\cset_1\cup \cset_2'$ is precisely the set of all connected components of $G\setminus E''$.

Intuitively, in each such canonical drawing, we require that, for every type-2  cluster $C\in \cset_2$, the edges of $C\setminus A_C$ do not participate in any crossings, and for every type-1 acceptable cluster $C\in \cset_1$, the edges of $C$ only participate in a small number of crossings (more specifically, we will define a subset $E^{**}(C)$ of edges for each cluster $C\in \cset_1$ that are allowed to participate in crossings). We then show that any drawing of $G$ can be transformed into a drawing that is canonical with respect to all clusters in $\cset_1\cup \cset_2$, while only slightly increasing the number of crossings. This is sufficient in order to complete the proof of Theorem \ref{thm: main}, by adding to $E''$ the set $\bigcup_{C\in \cset_1}E^{**}(C)$ of edges. However, in order to be able to reduce the problem to the \CNwRS problem, as required in Theorem \ref{thm: reduction}, we need stronger properties. We will define, for every type-1 cluster $C\in \cset_1$, a fixed drawing $\psi_C$, and we will require that, in the final drawing of $G$, the induced drawing of each such cluster $C$ is precisely $\psi_C$. For every type-2 cluster $C\in \cset_2$, we have already defined a drawing $\psi_{C\setminus A_C}$ of $C\setminus A_C$ -- the drawing of $C\setminus A_C$ that is induced by the drawing $\psi'_C$ of $C$. We will require that the drawing of $C\setminus A_C$ that is induced by the final drawing of $G$ is precisely $\psi_{C\setminus A_C}$.
Additionally, for each cluster $C\in \cset_1$, and for each bridge $R\in \rset_G(C)$, we will define a disc $D(R)$ in the drawing $\psi_C$ of $C$, and we will require that all vertices and edges of $R$ are drawn inside $D(R)$ in the final drawing of $G$. 
Similarly, for each type-2 acceptable cluster $C\in \cset_2$, for every bridge $R\in \rset_G(C\setminus A_C)$, we define a disc $D(R)$  in the drawing $\psi_{C\setminus A_C}$ of $C\setminus A_C$, and we will require that all vertices and edges of $R$ are drawn inside $D(R)$ in the final drawing of $G$. 
%Note that for every component $C'\in \sset(C)$, drawing $\psi(C\setminus A_C)$ of $C\setminus A_C$ induces a unique drawing of $C'$, that we denote by $\psi(C')$.
%
This will allow us to fix the locations of the components of $\cset_1\cup \cset_2$ with respect to each other (that is, for each pair $C,C'\in \cset_1\cup \cset_2$ of clusters, we will identify a face $F$ in the drawing $\psi_{C\setminus A_C}$ of $C\setminus A_C$, and a face $F'$ in the drawing $\psi_{C'\setminus A_{C'}}$ of $C'\setminus A_{C'}$, such that, in the final drawing $\phi$ of the graph $G$, graph $C'\setminus A_{C'}$ is drawn inside the face $F$ (of the drawing of $C\setminus A$ induced by $\phi$, which is identical to $\psi_{C\setminus A_C}$), and similarly graph $C\setminus A_C$ is drawn inside the face $F'$).

%Before we proceed, it would be convenient for us to ensure that for every type-1 acceptable cluster $C\in \cset_1$, for every pair of distinct bridges $R,R'\in \rset_G(C)$, the sets $L(R)$ and $L(R')$ of vertices are distinct. In order to achieve this, 

Before we continue, it would be convenient for us to ensure that, for every type-1 cluster $C\in \cset_1$, the vertices of $\Gamma(C)$ have degree $1$ in $C$, and degree $2$ in $G$; it would also be convenient for us to ensure that no edge of $ E''$ connects two vertices that lie in the same cluster. 
In order to ensure these properties, we subdivide some edges of $G$. Specifically, if $e=(u,v)\in E''$ is an edge with $u,v\in C$, for some cluster $C\in \cset_1\cup \cset_2$, then we subdivide the edge $(u,v)$ with two vertices, replacing it with a path $(u,u',v',v)$. The edges $(u,u')$ and $(v',v)$ are then added to set $E''$ instead of the edge $(u,v)$, and we add a new cluster to $\cset_1$, that consists of the vertices $u',v'$, and the edge $(u',v')$. This transformation ensures that no edge of $E''$ connects two vertices that lie in the same cluster. 
Consider now any type-1 cluster $C\in \cset_1$. For every edge $e=(u,v)\in E''$ with $u\in V(C)$ and $v\not\in V(C)$, we subdivide the edge with a new vertex $u'$, thereby replacing the edge with the path $(u,u',v)$. Vertex $u'$ and edge $(u,u')$ are added to the cluster $C$, while edge $(u',v)$ replaces the edge $(u,v)$ in set $E''$. Note that $u'$ now becomes a terminal, and, once all edges of $E''$ that are incident to the vertices of $C$ are processed, $u$ will no longer be a terminal. Abusing the notation, the final cluster that is obtained after processing all edges of $E''$ incident to $V(C)$ is still denoted by $C$. Notice that now the number of terminals that lie in $C$ may have grown by at most a factor $\Delta$, and so $|\Gamma(C)|\leq \mu\Delta$ must hold. Abusing the notation, we will still refer to $C$ as a type-1 acceptable cluster, and we will continue to denote by $\cset_1$ the set of all such clusters in the decomposition. Observe that this transformation ensures that every vertex of $\Gamma(C)$ has degree $1$ in $C$ and degree $2$ in $G$. Once every cluster $C\in \cset_1$ is processed in this manner, we obtain the final graph $G'$. Observe that $|E''|$ may have increased by at most a constant factor.
 Notice also that, for every fake edge $e=(x,y)\in A$, the endpoints of $e$ remain terminals in $\Gamma$, and the path $P(e)\in \pset$ that was used as a legal embedding of the edge $e$ can be converted into a path $P'(e)$ embedding $e$ in the new graph $G'$, by possibly subdividing the first and the last edge of $P(e)$ if needed. If $C(e)\in \cset_1$ is the type-$1$ cluster with $P(e)\setminus\set{x,y}\subseteq C(e)$, then the new vertices that (possibly) subdivide the first and the last edge of $P(e)$ lie in the new cluster $C'(e)$ corresponding to $C(e)$, so $P'(e)\setminus\set{x,y}\subseteq C'(e)$ continues to hold. The resulting path set $\pset'=\set{P'(e)\mid e\in A}$ is a legal embedding of the set of fake edges into $G'$.  
 Lastly, observe that any drawing of $G'$ on the sphere immediately gives a drawing of $G$ with the same number of crossings.
 Therefore, to simplify the notation, we will denote the graph $G'$ by $G$ and $\pset'$ by $\pset$ and we will assume that the decomposition $\dset$ of $G$ into acceptable clusters has the following two additional properties:

\begin{properties}{P}
	\item For every edge $e\in E''$, the endpoints of $e$ lie in different clusters of $\cset_1\cup \cset_2$; \label{prop: no edge with endpoints in same cluster} and
	\item For every type-1 cluster $C\in \cset_1$, for every terminal $t\in \Gamma(C)$, the degree of $t$ in $C$ is $1$, and its degree in $G$ is $2$. \label{prop: degrees of terminals for type 1}
\end{properties}

We now proceed to define canonical drawings of the graph $G$ with respect to the clusters of $\cset_1\cup \cset_2$.

\subsubsection{Canonical Drawings for Type-2 Acceptable Clusters}
Consider any type-2 cluster $C\in \cset_2$. Recall that the decomposition $\dset$ into acceptable clusters defines a planar drawing $\psi'_C$ of $C$ on the sphere, that induces a planar drawing $\psi_{C\setminus A_C}$ of $C\setminus A_C$ on the sphere. Recall that the Bridge Consistency Property of  type-2 acceptable clusters ensures that, for every bridge $R\in \rset_{G}(C\setminus A_C)$, there is a face $F$ of the drawing $\psi_{C\setminus A_C}$, such that the vertices of $L(R)$ lie on the boundary of $F$ (we note that face $F$ is not uniquely defined; we break ties arbitrarily). Since graph $C\setminus A_C$ is 2-connected, the boundary of face $F$ is a simple cycle, whose image is a simple closed curve. We denote by $D(R)$ the disc corresponding to the face $F$, so the boundary of $D(R)$ is the simple closed curve that serves as the boundary of the face $F$.
%: consider the drawing $\psi_{C\setminus A_C}$ of $C\setminus A_C$ on the sphere. If we remove all points participating in the drawings of the edges and the vertices of $C\setminus A_C$, then we obtain a collection of connected regions, each of which defines a face of $\psi_{C\setminus A_C}$. Since $C\setminus A_C$ is $2$-connected, each such region is a simple disc. \znote{I think you mean that the boundary of each such region is a simple cycle? I think a region on a plane is always homotopic to a simple disc.}\mynote{not sure what this means. it just defines the simple disc $D$, that corresponds to the region of the plane defined by the face.}
%We let $D(R)$ be the disc that corresponds to the face $F$. An important property that will be useful for us later is that for every pair $R,R'\in \rset_G(C)$ of discs, either $D(R)\cap D(R')=\emptyset$, or $D(R)=(R')$. Moreover, for every bridge $R
%Notice that for bridges $R\neq R'$, the discs $D(R),D(R')$ may not be disjoint.% Lastly, we set $\psi(C\setminus A_C)=\rho'_C$.
%As mentioned earlier, for every component $C'\in \sset(C)$, the drawing $\psi(C\setminus A_C)$ defines a drawing of $C'$, that we denote by $\psi(C')$.
%
Notice that the resulting set $\set{D(R)}_{R\in \rset_G(C\setminus A_C)}$ of discs has the following properties:
\begin{properties}{D}
	\item If $R\neq R'$ are two distinct bridges in $\rset_G(C\setminus A_C)$, then either $D(R)=D(R')$, or $D(R)\cap D(R')$ only contains points on the boundaries of the two discs; \label{prop: disjointness of discs} and

\item For every bridge $R\in \rset_G(C\setminus A_C)$, the vertices of $L(R)$ lie on the boundary of $D(R)$ in the drawing $\psi_{C\setminus A_C}$.\label{prop: legs on boundary of disc}
\end{properties}

We are now ready to define canonical drawings with respect to type-$2$ clusters.

\begin{definition}
	Let $\phi$ be any drawing of the graph $G$ on the sphere. We say that the drawing $\phi$ is \emph{canonical} with respect to a type-$2$  cluster $C\in \cset_2$ iff:
	\begin{itemize}
		\item the drawing of $C\setminus A_C$ induced by $\phi$ is identical to $\psi_{C\setminus A_C}$ (but its orientation may be different);
	\item the edges of $C\setminus A_C$ do not participate in any crossings in $\phi$; and
	\item for every bridge $R\in \rset_{G}(C\setminus A_C)$, all vertices and edges of $R$ are drawn in the interior of the disc $D(R)$ (that is defined with respect to the drawing $\psi_{C\setminus A_C}$ of $C\setminus A_C$). 
	\end{itemize}
\end{definition}
%\vspace{-0.3cm}

\subsubsection{Canonical Drawings for Type-1 Acceptable Clusters}
%Consider a cluster $C\in \cset_1$. %Recall that, from the definition of type-1 acceptable clusters, we are guaranteed that $|\Gamma(C)|\leq \mu$. 
%
%
%
%In this subsection, we show that we can find a drawing of $C\cup X_G(C)$ with relatively few crossings, and we will exploit this fact in the proof of Theorem~\ref{thm: from acceptable clusters to drawing} when dealing with type-1 acceptable clusters.
For convenience, we denote $\cset_1=\set{C_1,\ldots,C_q}$. 
We fix an arbitrary optimal drawing $\phi^*$ of the graph $G$. For each $1\leq i\leq q$, we denote by $\chi_i$ the set of all crossings $(e,e')$ such that either $e$ or $e'$ (or both) are edges of $E(C_i)$.   
The following observation is immediate.

\begin{observation}\label{obs: sum of chis}
$\sum_{i=1}^q|\chi_i|\leq 2\cdot\cro(\phi^*)=2\cdot\optcro(G)$. 
\end{observation}

We use the following theorem in order to fix a drawing of each type-$1$ acceptable cluster $C_i$; the proof appears in Section \ref{sec: fixing drawing of type 1 clusters}.

\begin{theorem}\label{thm: drawing of cluster extension-type 1}
	There is an efficient algorithm that, given a type-$1$ cluster $C_i\in \cset_1$, computes a drawing $\psi_{C_i}$ of $C_i$ on the sphere with $ O\left((|\chi_i|+1)\cdot \poly(\Delta\log n)\right)$ crossings, together with a set $E^*(C_i)\subseteq E(C_i)$ of at most $ O\left((|\chi_i|+1)\cdot \poly(\Delta\log n)\right)$ edges, such that graph $C_i\setminus E^*(C_i)$ is connected, and the drawing of $C_i\setminus E^*(C_i)$ induced by $\psi_{C_i}$ is planar.  Additionally, the algorithm computes, for every bridge $R\in \rset_{G}(C_i)$, a closed disc $D(R)$, such that:
	\begin{itemize}

		\item the vertices of $L(R)$ are drawn on the boundary of $D(R)$ in $\psi_{C_i}$;
		
		\item the interior of $D(R)$ is disjoint from the drawing $\psi_{C_i}$; and
		
%		\item the image of every edge of $C_i$  is disjoint from $D(R)$ in $\psi_{C_i}$, except possibly for an endpoint that is drawn on the boundary of $D(R)$;  and
		\item for every pair $R,R'\in \rset_G(C_i)$ of bridges, either $D(R)=D(R')$, or $D(R)\cap D(R')=\emptyset$.
%		\item for every edge of $\tilde E(R)$, its image intersects the boundary of the disc $D(R)$ at a single point; and

%		\item the image of the edge $e_i$ has a non-empty intersection with the boundary of the disc $D(R_i)$, and this intersection is a contiguous curve; it does not intersect the interior of $D(R_i)$, and it does not intersect any other disc $D(R')$ for any $R'\in \rset_G(C_i)\setminus\set{R_i}$.
	\end{itemize}
\end{theorem}

Note that in particular, Properties (\ref{prop: disjointness of discs}) and (\ref{prop: legs on boundary of disc}) also hold for the discs in $\set{D(R)}_{R\in \rset_G(C)}$. %Using standard uncrossing procedure, we can assume w.l.o.g. that for every pair of edges of $C_i$, their images cross at most once in $\psi_{C_i}$, while preserving all other properties of the drawing $\psi_{C_i}$.

For each type-1 cluster $C_i\in \cset_1$, let $E^{**}(C_i)\subseteq E(C_i)$ be the set of all edges of $C_i$ that participate in crossings in  $\psi_{C_i}$. Clearly, $|E^{**}(C_i)|\leq O(\cro(\psi_{C_i}))\leq O((|\chi_i|+1)\poly(\Delta\log n))$.
Let $E^*=\bigcup_{C_i\in \cset_1}E^{**}(C_i)$. Then, from Observation \ref{obs: sum of chis} and Theorem \ref{thm: can find edge set w acceptable clusters}: %, and since $\mu=O(\Delta\log^{1.5}n)$:

\[\begin{split}
|E^*|& \leq \sum_{C_i\in \cset_1}O\!\left((|\chi_i|+1)\cdot \poly(\Delta\log n)\right)\\
&\leq O\left ((\optcro(G)+|E''|) \poly(\Delta\log n)\right )\\
&\leq O\left ((\optcro(G)+|E'|) \poly(\Delta\log n)\right ).
\end{split}
\]

%We denote, for each type-$1$ acceptable cluster $C_i\in \cset_1$, by $\psi_{C_i}$ the drawing of $C_i$ on the sphere that is induced by the drawing $\psi_i$ of $C_i^+$ (note that this drawing excludes the edge $e_i$).

We now define canonical drawings with respect to type-1 clusters. 
\begin{definition}
	Let $\phi$ be any drawing of the graph $G$ on the sphere, and let $C_i\in \cset_1$ be a type-$1$ cluster. We say that $\phi$ is a \emph{canonical drawing} with respect to $C_i$, iff:
	\begin{itemize}
		\item the drawing of $C_i$ induced by $\phi$ is identical to $\psi_{C_i}$ (but orientation of the two drawings may be different); and
		\item for every bridge $R\in \rset_{G}(C_i)$, all vertices and edges of $R$ are drawn in the interior of the disc $D(R)$ (that is defined with respect to the drawing $\psi_{C_i}$ of $C_i$).
	\end{itemize}
\end{definition}

Notice that the definition implies that the only edges of $C_i$ that participate in crossings of $\phi$ are the edges of $E^{**}(C_i)$.

\subsubsection{Obtaining a Canonical Drawing}

%Recall that we are given a partition $(\cset_1,\cset_2)$ of all connected components of $H$ into type-1 and type-2 acceptable clusters, respectively. 
%\mynote{we need to be careful in that the boundary of $F$ may not be $2$-connected. We need to write this more carefully.}
%For a type-1 acceptable cluster $C\in \cset_1$, we denote by $\rho'_C$ the drawing of $C$ that is induced by the drawing of corresponding graph $C^+$ given by Corollary \ref{cor: drawing of cluster extension-type 1}. Recall that for every tree $T_R\in X_G(C)$, there is a disc $D_{T_R}$, such that, in the drawing of $C^+$ given by Corollary~\ref{cor: drawing of cluster extension-type 1}, $T_R\setminus \tilde E(R)$ is drawn inside the disc $D_{T_R}$ and graph $C$ is drawn outside $D_{T_R}$.

Our next result shows that there exists a near-optimal drawing of the graph $G$ that is canonical with respect to all clusters.
The proof of the following theorem appears in Section \ref{sec:canonical drawing}.

\begin{theorem}\label{thm: canonical drawing}
There is an efficient algorithm, that, given, as input:
\begin{itemize}
\item an $n$-vertex graph $G$ of maximum vertex degree at most $\Delta$;
\item an arbitrary drawing $\phi$ of $G$; 
\item a decomposition $\dset=\left ( E'',A,\cset_1,\cset_2,\set{\psi_{C}}_{C\in \cset_2},\pset(A)\right )$ of $G$ into acceptable clusters for which Properties (\ref{prop: no edge with endpoints in same cluster}) and  (\ref{prop: degrees of terminals for type 1}) hold;
\item a drawing $\psi_{C_i}$ and an edge set $E^*(C_i)$ for each cluster $C_i\in \cset_1$ as defined above; and
\item for each cluster $C\in \cset_1\cup \cset_2$,  a collection $\set{D(R)}_{R\in \rset_G(C\setminus A_C)}$ of discs on the sphere with Properties (\ref{prop: disjointness of discs}) and (\ref{prop: legs on boundary of disc});
\end{itemize}
computes a drawing $\phi'$ of $G$ on the sphere with $O\left ((|E''|+\cro(\phi))\cdot\poly(\Delta \log n)\right)$ crossings, such that $\phi'$ is canonical with respect to every cluster $C\in \cset_1\cup \cset_2$. 
\end{theorem}

We note that for our purposes, an existential variant of the above theorem, that shows that a drawing $\phi'$ with the required properties exists, is sufficient. We provide the proof of the stronger constructive result in case it may be useful for future work on the problem.

\subsection{Completing the Proof of Theorem \ref{thm: main} for $3$-Connected Graphs}
Notice that Theorem \ref{thm: canonical drawing} concludes the proof of Theorem \ref{thm: main} for the special case where $G$ is a $3$-connected graph. Indeed, given a $3$-connected graph $G$ and a planarizing set $E'$ of its edges, we use Theorem  \ref{thm: can find edge set w acceptable clusters} to compute a decomposition  $\dset=\left ( E'',A,\cset_1,\cset_2,\{\psi_C\}_{C\in \cset_2},\pset(A)\right )$ of $G$ into acceptable clusters, such that $E'\subseteq  E''$ and $| E''|\leq O((| E'|+\optcro(G))\cdot\poly(\Delta\log n))$. Next, we apply Theorem \ref{thm: drawing of cluster extension-type 1} to each type-1 cluster $C_i\in \cset_1$, to obtain the set $E^{**}(C_i)$ of edges, the drawing $\psi_{C_i}$ of $C_i$, and the discs $D(R)$ for all bridges $R\in \rset_G(C_i)$. Let $E^*=\bigcup_{C_i\in \cset_1}E^{**}(C_i)$, so $|E^*|\leq O\left ((\optcro(G)+|E'|) \poly(\Delta\log n)\right )$, as observed above. The final output of the algorithm is the set $E''\cup E^*$ of edges. 
Observe that  $|E''\cup E^*|\leq O((| E'|+\optcro(G))\cdot\poly(\Delta\log n))$, as required. Moreover, by using Theorem \ref{thm: canonical drawing} with the optimal drawing $\phi^*$ of $G$, we conclude that there exists a drawing $\phi'$ of $G$ with $O\left ((|E'|+\optcro(G))\cdot\poly(\Delta \log n)\right )$ crossings, that is canonical with respect to all clusters in $\cset_1\cup \cset_2$. In particular, the only edges that may participate in crossings in $\phi'$ are edges of $E''\cup E^*$. %By deleting the images of the fake edges from this drawing, we obtain the desired drawing of $G$.

\subsection{Extension to General Graphs}
In Section \ref{sec:non_3} we show how to extend the above proof of Theorem \ref{thm: main}  to general graphs, that are not necessarily $3$-connected. The extension builds on techniques that were introduced in \cite{chuzhoy2011graph}. Additionally, in Section \ref{sec:non_3}, we prove the following theorem, that provides a black-box reduction from the problem of approximating \MCN in general graphs using paradigm $\Pi'$, to the problem of approximating \MCN in $3$-connected graphs, using the same paradigm.

\begin{theorem}
\label{thm:main_non_3}
Suppose there exists an efficient (possibly randomized) algorithm, that, given a $3$-connected $n$-vertex graph $G$ with maximum vertex degree $\Delta$, and a planarizing set $E'$ of its edges, computes a drawing of $G$ with at most $f(n,\Delta)\cdot(\optcro(G)+|E'|)$ crossings, for any function $f$ that is monotonously increasing in both $n$ and $\Delta$. %, and satisfies $f(n,\Delta)\ge \Delta$ \mynote{why we need this?} for all $n,\Delta>0$. 
Then there exists an efficient (possibly randomized) algorithm that, given a (not necessarily 3-connected) graph $\hat{G}$ on $n$ vertices with maximum vertex degree $\Delta$,  and a planarizing set $\hat{E}'$ of its edges, computes a drawing of $\hat G$ with the number of crossings bounded by  $O\left( f(n,\Delta)\cdot (\optcro(\hat G)+|\hat E'|)\cdot \poly(\Delta\log n)\right)$.
\end{theorem}
%\vspace{-0.5cm}

\subsection{Reduction to Crossing Number with Rotation System -- Proof of Theorem \ref{thm: reduction}}\label{subsec: reduction}

In this section we provide a reduction from \MCN in $3$-connected graphs to \CNwRS, completing the proof of Theorem \ref{thm: reduction} in the case where the input graph $G$ is $3$-connected.
We extend this proof to general graphs in Section \ref{sec:non_3}.
Recall that Kawarabayashi and Sidiropoulos~\cite{kawarabayashi2019polylogarithmic} provide an efficient $O(\poly(\Delta\log n))$-approximation algorithm for the \MP problem. Since, for every graph $G$, there is a planarizing set $E^*$ containing at most $\optcro(G)$ edges, we can use their algorithm in order to compute, for the input graph $G$, a planarizing edge set $E'$ of cardinality $O(\optcro(G)\cdot\poly(\Delta\log n))$. We then use Theorems \ref{thm: can find edge set w acceptable clusters} and \ref{thm: drawing of cluster extension-type 1} to compute another  planarizing edge set $E''$ of cardinality $O(\optcro(G)\cdot\poly(\Delta\log n))$ for $G$, the collection $\cset=\cset_1\cup \cset_2$ of clusters, together with their drawings $\psi_C$, and we use families $\set{D(R)}_{R\in \rset_G(C)}$ of discs for all $C\in \cset$ that we have computed. We will not need fake edges anymore, so for every type-2 cluster $C\in \cset_2$, we let $C'=C\setminus A_C$, and we let $\psi_{C'}$ be the planar drawing of $C'$ that was used in the definition of the canonical drawing. For a type-1 cluster $C\in \cset_1$, consider the drawing $\psi_C$ of $C$ given by Theorem \ref{thm: drawing of cluster extension-type 1}. We let $C'$ be a graph obtained from $C$, by placing a vertex on every crossing of a pair of edges in $\psi_C$. Therefore, graph $C'$ is planar, and we denote by $\psi_{C'}$ its planar drawing that is induced by $\psi_C$. We still denote by $\Gamma(C')$ the set of all vertices of $C'$ that serve as endpoints of the edges of $E''$. Consider the graph $G'$, that is obtained by taking the union of all clusters in $\set{C'\mid C\in \cset}$ and the edges in $E''$. Suppose we compute a drawing $\phi$ of $G'$ with $z$ crossings, such that the only edges that participate in the crossings are the edges of $E''$, and for every cluster $C\in \cset$, the drawing of $C'$ induced by $\phi$ is identical to $\psi_{C'}$. Then we can immediately obtain a drawing $\phi'$ of $G$ with $O(z+\optcro(G)\poly(\Delta \log n))$ crossings, where the additional crossings arise because we replace, for every type-1 cluster $C\in \cset_1$, the planar drawing $\psi_{C'}$ of $C'$ with the (possibly non-planar) drawing $\psi_C$ of $C$. 
Let $\cset_1'=\set{C'\mid C\in \cset_1}$, $\cset_2'=\set{C'\mid C\in \cset_2}$, and $\cset'=\cset_1'\cup \cset_2'$. Since we do not use the original clusters in $\cset$ in the remainder of this subsection, for simplicity of notation, we denote $\cset_1',\cset_2'$ and $\cset'$ by $\cset_1,\cset_2$ and $\cset$, and we will use notation $C\in \cset$ instead of $C'$. Recall that for every cluster $C\in \cset$ we are now given a fixed planar drawing $\psi_C$. In order to reduce the problem to \CNwRS, we use the \CP problem, that we define next, as an intermediate problem.

\paragraph{Cluster Placement Problem.}
In the \CP problem, we are given a collection $\hcset$ of disjoint connected planar graphs (that we call clusters). For every cluster $\hC\in \hcset$, we are also given a planar drawing  $\psi_{\hC}$ of $\hC$ on the sphere. We denote by $\fset_{\hC}$ the set of all faces of this drawing. Additionally, for every ordered pair $(\hC_1,\hC_2)\in \hcset$ of clusters, we are given a face $F_{\hC_1}(\hC_2)\in \fset_{\hC_1}$. The goal is to compute a planar drawing $\phi$ of $\bigcup_{\hC\in \hcset}\hC$ on the sphere such that, for every cluster $\hC_1\in \hcset$, the drawing of $\hC_1$ induced by $\phi$ is identical to $\psi_{\hC_1}$, and moreover, for every cluster $\hC_2\in \hcset\setminus\set{\hC_1}$, the drawing of $\hC_2$ in $\phi$ is contained in the face $F_{\hC_1}(\hC_2)$ of the drawing of $\hC_1$ in $\phi$. 

The proof of the following simple theorem is deferred to Section \ref{sec: solving cluster placement problem} of Appendix.

\begin{theorem}\label{thm: solve cluster placement}
	There is an efficient algorithm, that, given an instance of the \CP problem, finds a feasible solution for the problem, if such a solution exists.
	\end{theorem}

%------------------------------
%-----------------------------

We note that the current collection $\cset=\cset_1\cup \cset_2$ of clusters that we obtained for the instance $G$ of the \MCN problem naturally defines an instance of the \CP problem. For every cluster $C\in \cset$, we have already defined a fixed planar drawing $\psi_C$ of $C$ on the sphere. 
Consider now some ordered pair $(C_1,C_2)\in \cset$ of clusters. Then there must be some bridge $R\in \rset_{G'}(C_1)$, such that $C_2\subseteq R$. Recall that we have defined a disc $D(R)$ corresponding to the bridge $R$ in the drawing $\psi_{C_1}$ of $C_1$, and that the images of the edges and vertices of $C_1$ in $\psi_{C_1}$ are disjoint from the interior of the disc $D(R)$. Therefore, there is some face $F$ in the drawing $\psi_{C_1}$ of $C_1$ with $D(R)\subseteq F$. We then set $F_{C_1}(C_2)$ to be this face $F$. This defines a valid instance of the \CP problem. Moreover, since Theorem \ref{thm: canonical drawing} guarantees the existence of a canonical drawing of the graph $G$, this problem has a feasible solution. 
In fact Theorem \ref{thm: canonical drawing} provides the following stronger guarantees:

\begin{observation}\label{obs: drawing consistent with Cluster Placement}
	There is a drawing $\phi$ of graph $G'$ with $O(\optcro(G)\poly(\Delta\log n))$ crossings, such that for every cluster $C\in \cset$, the edges of $C$ do not participate in any crossings, and the drawing of $C$ induced by $\phi$ is identical to $\psi_C$ (but the orientation may be arbitrary). Moreover, for any ordered pair $(C_1,C_2)\in \cset$ of clusters, the image of $C_2$ in $\phi$ is contained in the interior of the face $F_{C_1}(C_2)$ of the drawing of $C_1$ in $\phi$.
\end{observation}

We use the algorithm from Theorem \ref{thm: solve cluster placement} in order to compute a feasible solution to this instance of the \CP problem, obtaining a drawing $\tilde \phi$ of $\bigcup_{C\in \cset}C$ on the sphere.

In order to compute a final drawing of $G'$ (and hence of $G$), it is enough to add the drawings of the edges of $E''$ into $\tilde \phi$. We do so by defining an instance of the \CNwRS problem.

\paragraph{Defining Instances of \CNwRS.}
Let $\fset$ be the set of all faces in the drawing $\tilde \phi$ of the graph $\bigcup_{C\in\cset}C$. For every face $F\in \fset$, let $\hset(F)\subseteq \cset$ be the set of all clusters $C\in \cset$, such that at least one terminal of $\Gamma(C)$ (that is, endpoint of an edge of $E''$ that lies in $C$) lies on the boundary of the face $F$.

We associate, with each face $F\in \fset$, an instance $(G^F,\Sigma^F)$ of the \CNwRS problem, as follows.
Let $E^F\subseteq E''$ be the set of all edges whose both endpoints lie on the boundary of the face $F$ in $\tilde \phi$.
In order to construct the graph $G^F$, we start with the union of the clusters $C\in \hset(F)$, and add the edges of $E^F$ to the resulting graph. Then we contract every cluster $C\in \hset(F)$ into a vertex $v(C)$, keeping parallel edges and deleting self-loops. This concludes the definition of the graph $G^F$. We now define a rotation system for $G^F$. Consider any vertex $v=v(C)\in V(G^F)$, and let $\delta(v)$ be the set of all edges that are incident to $v(C)$ in $G^F$. If $C\in \cset_1$, then $|\delta(v)|\leq |\Gamma(C)|\leq \poly(\Delta\log n)$. We define $\oset_v$ to be an arbitrary ordering of the edges in $\delta(v)$. Assume now that $C\in \cset_2$. Recall that graph $C$ must be $2$-connected, so the intersection of the boundary of $F$ with $C$ is a simple cycle, that we denote by $K^F(C)$.
We denote by $\Gamma^F(C)$ the set of all vertices of $\Gamma(C)$ that lie on the cycle $K^F(C)$.
From Observation \ref{obs: drawing consistent with Cluster Placement},  the vertices of $\Gamma(C)$ that serve as endpoints of the edges of $\delta(v)$ must belong to $\Gamma^F(C)$. We let $\tilde \oset^F(C)$ denote the circular ordering of the vertices of $\Gamma^F(C)$ along the cycle $K^F(C)$. The ordering $\oset_v$ of the edges of $\delta(v)$ is determined by the ordering of their endpoints in $\tilde \oset^F(C)$: edges that are incident to the same vertex of $K^F(C)$ appear consecutively in $\oset_v$ in an arbitrary order. The ordering of the edges that are incident to different vertices of $\Gamma^F(C)$ follows the ordering $\tilde \oset^F(C)$. We then let $\Sigma^F=\set{\oset_v}_{v\in V(G^F)}$.
 
This completes the definition of instance $(G^F,\Sigma^F)$ of \CNwRS. Since Theorem  \ref{thm: reduction} calls for a single instance of \CNwRS, we let $G''$ be the disjoint union of the graphs in $\set{G^F}_{F\in \fset}$, and we let $\Sigma=\bigcup_{F\in \fset}\Sigma^F$. This defines the 
final instance $(G'',\Sigma)$ of the \CNwRS problem. Notice that $E(G'')\subseteq E''$ and so $|E(G'')|\leq O(\optcro(G)\poly(\Delta\log n))$. We need the following observation.
\begin{observation}\label{obs: cover all E''}
$E''=\bigcup_{F\in \fset}E^F$.
\end{observation}
\begin{proof}
From the definition of edge sets $E^F$, $\bigcup_{F\in \fset}E^F\subseteq E''$, so it is enough to show that $E''\subseteq \bigcup_{F\in \fset}E^F$. Assume for contradiction that this is not the case. Then there is some edge $e=(u,v)\in E''$, such that no face of $\fset$ contains both $u$ and $v$ on its boundary. 
Let $C'$ be the cluster containing $u$ and $C''$ the cluster containing $v$; recall that $C'\neq C''$ must hold. 
Then there must be some cluster $C$ and a cycle $K$ in $C$, such that $u,v\not \in K$, and the image of $K$ in $\tilde \phi$ separates the images of $u$ and $v$. If $C=C'$, then we let $F_1$ be any face in the drawing of $C$ that is incident to $u$; otherwise, we let $F_1=F_C(C')$. Similarly, if $C=C''$, then  we let $F_2$ be any face in the drawing of $C$ that is incident to $v$; otherwise, we let $F_2=F_C(C')$. Notice that $F_1\neq F_2$, and moreover, since $K$ separates $u$ from $v$, if $C=C'$ then $u$ does not lie on the boundary of $F_2$ (and similarly, if $C=C''$, then $v$ does not lie on the boundary of $F_1$). But that means that, in the drawing $\phi$ of graph $G'$ that is given by Observation \ref{obs: drawing consistent with Cluster Placement}, there is some cycle $K'\subseteq C$ that separates the image of $u$ from the image of $v$ in $\phi$. But then the image of edge $e$ must cross the image of some edge of $C$ in $\phi$, which is impossible.
\end{proof}

Next, we show that the final instance $(G',\Sigma)$ of \CNwRS has a sufficiently cheap solution. The proof of the next lemma appears in Section \ref{sec: cheap soln for CNRS} of Appendix.
\begin{lemma}\label{lem: cheap solution}
There is a solution to instance $(G'',\Sigma)$ of \CNwRS of value $O(\optcro(G)\cdot \poly(\Delta\log n))$.
\end{lemma}
Lastly, in order to complete the proof of Theorem \ref{thm: reduction}, it is enough to show an efficient algorithm, that, given  any solution to instance $(G'',\Sigma)$ of \CNwRS of value $X$, computes a drawing of $G'$ with $O\left ((X+\optcro(G))\cdot \poly(\Delta\log n)\right)$ crossings.
 
\paragraph{Obtaining the Final Drawing of $G'$.}
We assume that we are given a solution $\hat \phi$ to instance $(G'',\Sigma)$ of \CNwRS, whose value is denoted by $X$. Since graph $G''$ is the disjoint union of graphs $\set{G^F}_{F\in \fset}$, we can use $\hat \phi$ to obtain, for each face $F\in \fset$, a solution $\hat \phi^F$ to instance $(G^F,\Sigma^F)$ of \CNwRS, of value $X^F$, such that $\sum_{F\in \fset}X^F\leq X$. Intuitively, ideally, we would like to start with the drawing $\tilde \phi$ of $\bigcup_{C\in \cset}C$ given by the solution to the \CP problem, and then consider the faces $F\in \fset$ one-by-one. For each such face, we would like to use the drawing $\hat \phi^F$ of graph $G^F$ in order to insert the images of the edges of $E^F$ into the face $F$. Since, from Observation \ref{obs: cover all E''}, $E''=\bigcup_{F\in \fset}E^F$, once every face of $\fset$ is processed, all the edges of $E''$ are inserted into the drawing, and we obtain a valid drawing of graph $G'$. There is one difficulty with using this approach. Recall that, a solution $\hat \phi^F$ to instance $(G^F,\Sigma^F)$ guarantees that for every vertex $v\in V(G^F)$, the images of the edges of $\delta(v)$ enter $v$ in an ordering identical to $\oset_v$. However, the \emph{orientation} of this ordering may be arbitrary. In other words, in order to insert the edges of $E^F$ into the face $F$ of the drawing of $\phi'$, by copying their drawings in $\hat \phi^F$, we may need to \emph{flip} the drawings of some clusters $C\in \hset(F)$. Since each cluster may belong to several sets $\hset(F)$, we need to do this carefully. 

Consider the drawing $\tilde \phi$ of $\bigcup_{C\in \cset}C$. 
Consider any cluster $C\in \cset$ and any face $F\in \fset$, such that $C\in \hset(F)$. 
As before, we let $\Gamma^F(C)\subseteq \Gamma(C)$ be the set of terminals that belong to $C$ and lie on the boundary of the face $F$. Next, we define a disc $D^F(C)$, as follows. If $C\in \cset_1$, then let $\gamma^F(C)$ be a simple closed curve that contains every terminal in $\Gamma^F(C)$, and separates the drawing of $C\setminus \Gamma^F(C)$ from the drawing of every cluster $C'\in \hset(F)\setminus\set{C}$. If $C\in \cset_2$, then we let $\gamma^F(C)$
be the image of the cycle $K^F(C)$ in $\tilde \phi$. We then let $D^F(C)$ be a disc, whose boundary is $\gamma^F(C)$, that contains the drawing of $C$ in $\tilde \phi$. Notice that for every cluster  $C'\in \hset(F)\setminus\set{C}$, the drawing of $C'$ in $\tilde \phi$ is disjoint from $D^F(C)$. Notice also that, if $C\in \cset_2$, then the ordering of the terminals in $\Gamma^F(C)$ on the boundary of $D^F(C)$ is identical to $\tilde \oset^F(C)$.
 
We now proceed as follows. First, we describe a procedure \procface, that, intuitively, will allow us to insert, for a given face $F\in \fset$, all edges of $E^F$ into the drawing; this may require flipping drawings in some discs $D^F(C)$, for $C\in \hset(F)$. We then show an algorithm that builds on this procedure in order to compute a drawing of $G'$.

\paragraph{Procedure \procface.}
The input to procedure \procface is a face $F\in \fset$, and a collection $\set{D^F(C)}_{C\in \hset(F)}$ of disjoint discs (intuitively, disc $D^F(C)$ already contains a drawing of some parts of the graph $G$, as defined above, but in this procedure we do not modify parts of the graph drawn inside the disc, and consider these parts as being fixed; we may however flip the drawing that is contained in $D(C)$). Additionally, for every cluster $C\in \hset(F)$, we are given a drawing of the terminals in $\Gamma^F(C)$ on the boundary of the disc $D^F(C)$. We require that, if $C\in \cset_2$, then the circular ordering of the terminals in $\Gamma^F(C)$ on the boundary of the disc $D^F(C)$ is identical to $\tilde \oset^F(C)$. We are also given a cluster $C^*\in \hset(F)$, whose orientation is fixed (that is, we are not allowed to flip the disc $D^F(C^*)$). The procedure inserts the edges of $E^F$ into this drawing, while possibly flipping some discs $D^F(C)$.

In order to execute the procedure, we start with the solution $\hat \phi^F$ to instance $(G^F, \Sigma^F)$. For every vertex $v\in V(G^F)$, we consider a small disc $\eta(v)$ around the drawing of $v$ in $\hat \phi^F$. We also define a smaller disc $\eta'(v)\subseteq \eta(v)$ that is contained in the interior of $\eta(v)$ and contains the image of $v$. For every edge $e=(v,v')$, we truncate the image of $e$, so that it originates at some point $p_v(e)$ on the boundary of $\eta(v)$ and terminates at some point $p_{v'}(e)$  on the boundary of $\eta(v')$. 

Consider now some vertex $v(C)\in V(G^F)$, whose corresponding cluster $C$ lies in $\cset_2$. Recall that for every terminal $t\in \Gamma^F(C)$, 
 there must be a contiguous segment $\sigma(t)$ on the boundary of $\eta(v)$ that contains all points $\set{p_{v(C)}(e)}$, where $e\in \delta(v)$ is an edge that is incident to $t$, so that all resulting segments in $\set{\sigma_t\mid t\in \Gamma^F(C)}$ are disjoint. The circular ordering of these segments along the boundary of $\eta(v)$ is identical to $\tilde \oset^F(C)$. We place images of the terminals in $\Gamma^F(C)$ on the boundary of $\eta'(v)$, in the circular order that is identical to $\tilde \oset^F(C)$, whose orientation is the same as the orientation of the ordering of the segments in $\set{\sigma_t\mid t\in \Gamma^F(C)}$. If the orientation of this ordering is identical to the orientation of the ordering of terminals of $ \Gamma^F(C)$ on the boundary of the disc $D^F(C)$, then we say that cluster $C$ \emph{agrees} with the orientation of $\hat \phi^F$, and otherwise we say that it disagrees with it. We can assume without loss of generality that, if cluster $C^*\in \cset_2$, then it agrees with the orientation of $\hat \phi^F$, since otherwise we can flip the drawing $\hat \phi^F$. 

In order to insert the edges of $E^F$ into the current drawing, we will do the opposite: we will ``insert'' the discs $D^F(C)$ into the discs $\eta'(v(C))$ in the drawing $\hat \phi^F$. Specifically, for every cluster $C\in \cset$, if $C$ agrees with the orientation of $\hat \phi^F$, then we insert the disc $D^F(C)$ into the disc $\eta'(v(C))$  in the current drawing $\hat \phi^F$, so that the images of the terminals of $\Gamma^F(C)$ coincide (it may be convenient to think of the disc $D^F(C)$ as containing a drawing of $C$ and maybe some additional subgraphs of $G'$). If $C$ disagrees  with the orientation of $\hat \phi^F$, then we first create a mirror image of the disc $D^F(C)$ (that will result in flipping whatever drawing currently appears in $D^F(C)$), and then insert the resulting disc into the disc $\eta(v(C))$ in the current drawing $\hat \phi^F$, so that the images of the terminals of $\Gamma^F(C)$ coincide. 
In either case, we can extend the drawings of the edges of $\delta(v(C))$ inside $\eta(v(C))\setminus\eta'(v(C))$, so that for every terminal $t\in \Gamma^F(C)$, the drawing of every edge $e$ that is incident to $t$ terminates at the image of $t$, without introducing any crossings.
Lastly, if $C\in \cset_1$, then we simply insert the disc $D^F(C)$ into the disc $\eta'(C)$.  We extend the drawings of the edges of $\delta(v(C))$ inside $\eta(v(C))\setminus\eta'(v(C))$, so that for every terminal $t\in \Gamma^F(C)$, the drawing of every edge $e$ that is incident to $t$ terminates at the image of $t$, while introducing at most $|\Gamma^F(C)|^2$ new crossings. This completes the description of Procedure \procface.

We are now ready to complete the drawing of the graph $G'$. Our algorithm performs a number of iterations. In each iteration $i$ we will fix an orientation of some subset $\cset^i\subseteq \cset$ of clusters. 
We maintain the invariant that for every cluster $C\in \cset^i$, if $F\in \fset$ is any face with $C\in \hset(F)$ that has not been processed yet, then no cluster of $(\cset^1\cup \cset^2\cup\cdots\cup \cset^i)\setminus\set{C}$ lies in $\hset(F)$.
We let $\cset^0$ consist of a single arbitrary cluster $C_0\in \cset$. 

In order to execute the first iteration, we let $F\in \fset$ be any face with $C_0\in \hset(F)$. We run Procedure \procface on face $F$, with cluster $C^*=C_0$. Notice that the outcome of this procedure can be used in order to insert the edges of $E^F$ into the current drawing $\tilde \phi$ of $\bigcup_{C\in \cset}C$, after possibly flipping the images inside some discs in $\set{D(C)}_{C\in \hset(F)\setminus\set{C_0}}$. We then let $\cset^1$ contain all clusters in $\hset(F)$. Notice that the invariant holds for this definition of set $\cset^1$. For each  cluster $C\in \cset^1$, its orientation is fixed from now on, and the drawing of $C$ will never be modified again.
 
In order to execute the $i$th iteration, we start with $\cset^i=\emptyset$. We consider each cluster $C\in \cset_{i-1}$, one-by-one. For each such cluster $C$, for every face $F\in \fset$ with $C\in \hset(F)$, that has not been processed yet, we apply Procedure \procface to face $F$, with $C=C^*$. As before, this procedure can be used in order to insert all edges of $E^F$ into the current drawing, after, possibly, flipping the images contained in some discs $D^F(C')$, for $C'\in \hset(F)\setminus\set{C}$. Notice, however, that from our invariant, none of the clusters corresponding to these discs may belong to $\cset^1\cup \cdots\cup \cset^{i-1}$. We then add, to set $\cset^i$, all clusters of $\hset(F)\setminus\set{C}$. It is easy to verify that the invariant continues to hold. Once every face in $\fset$ is processed, we have inserted all edges of $E''$ into $\phi'$, and obtain a final drawing $\phi''$ of the graph $G'$.

We now bound the number of crossings in $G'$. In addition to the crossings that were present in the drawings $\hat \phi^F$, for $F\in \fset$, we may have added, for every cluster $C\in \cset_1$, at most $O(\Delta^2 |\Gamma(C)|^2)$ new crossings of edges that are incident to the terminals of $C$ (this bound follows the same reasonings as those in the proof of Lemma \ref{lem: cheap solution}). Since, for every cluster $C\in \cset_1$, $|\Gamma(C)|\leq O(\poly(\Delta\log n))$, we get that the total number of crossings in the drawing $\phi''$ of $G'$ is at most:
\[
\begin{split}
 X+O\Big(\sum_{C\in \cset_1} \Delta |\Gamma(C)|^2\Big)& \leq  X+O(|\Gamma|\poly(\Delta\log n)) \\
 &\leq X+O(|E''|\poly(\Delta\log n))\\
 &\leq X+O(\optcro(G) \poly(\Delta\log n)).
 \end{split}\]
Note that drawing $\phi''$ of $G'$ immediately induces a drawing of the original graph $G$, where the number of crossings is bounded by $\cro(\phi'')$ plus the sum, over all original type-1 clusters $C\in \cset_1$, of the number of crossings in the original drawing $\psi_C$ of $C$ (recall that we have replaced all such crossings with vertices in $G'$). However, the total number of all such additional crossings, as shown already, is bounded by $O(\optcro(G) \poly(\Delta\log n))$, and so overall, the total number of crossings in the final drawing of $G$ is bounded by 
 $X+O(\optcro(G) \poly(\Delta\log n))$. This completes the proof of Theorem \ref{thm: reduction} for the special case where the input graph $G$ is $3$-connected. We extend the proof to general graphs in Section \ref{sec:non_3}.

 In Section \ref{sec:cr_rotation_system} we provide an algorithm for the \CNwRS problem,  proving Theorem \ref{thm: main_rot}.
\section{Block Decompositions and Embedding of Fake Edges}
\label{sec:block_decompos}
\paragraph{Blocks.}
Let $G$ be a $2$-connected graph. A \emph{$2$-separator} for $G$ is a pair $(u,v)$ of vertices, such that graph $G\setminus\set{u,v}$ is not connected. 
 Most of the definitions in this section are from \cite{chuzhoy2011graph}.
 
\begin{definition}
Let $G=(V,E)$ be a $2$-connected graph. 
A subgraph $B=(V',E')$ of $G$ is called a \emph{block} iff: 
\begin{itemize}
\item $V\setminus V'\neq\emptyset$ and $|V'|\geq 3$;
\item There are two distinct vertices $u,v\in V'$, called \emph{block end-points} and denoted by $I(B)=(u,v)$, such that there are no edges from $V\setminus V'$ to $V'\setminus\set{u,v}$ in $G$. All other vertices of $B$ are called \emph{inner vertices}; 
\item  $B$ is the subgraph of $G$ induced by $V'$, except that it {\bf does not} contain the edge $(u,v)$ even if it is present in $G$.
\end{itemize}
\end{definition}
Notice that, if $G$ is a $2$-connected graph, then every $2$-separator $(u,v)$ of $G$ defines at least two internally disjoint blocks $B',B''$ with $I(B'),I(B'')=(u,v)$.
If $B$ is a block with endpoints $u$ and $v$, then the \emph{complement} of $B$, denoted by $B^c$, is the sub-graph of $G$ induced by the vertices of $(V(G)\setminus V(B))\cup \set{u,v}$. Notice that $B^c$ is itself a block, unless edge $e=(u,v)$ belongs to $B^c$; in the latter case, $B^c\setminus\set{e}$ is a block.

\subsection{Block Decomposition of $2$-Connected Graphs.}

Let $\lset$ be a collection of sub-graphs of $G$. We say that $\lset$ is \emph{laminar} iff for every pair $H,H'\in \lset$ of subgraphs, either $H\cap H'=\emptyset$, or $H\subseteq H'$, or $H'\subseteq H$ hold. Given a laminar collection $\lset$ of subgraphs of $G$ with $G\in \lset$, we can associate a tree $\tau=\tau(\lset)$ with it, called a \emph{decomposition tree}, as follows. For every graph $H\in \lset$, there is a vertex $v(H)$ in $ \tau$. The tree is rooted at the vertex $v(G)$. For every pair $H,H'\in \lset$ of subgraphs, such that $H\subsetneq H'$, and there is no other graph $H''\in \lset\setminus\set{H,H'}$ with $H\subsetneq H''\subsetneq H'$, there is an edge $(v(H),v(H'))$ in the tree, and $v(H')$ is the parent of $v(H)$ in $\tau$.

Let $G$ be a $2$-connected graph, and
assume that we are given a laminar family of sub-graphs of $G$ with $G\in \lset$. Let $\tau=\tau(\lset)$ be the decomposition tree associated with $\lset$. Assume further that every graph  $B\in \lset\setminus\set{G}$ is a block. For each such graph $B\in \lset$,  we define a new graph, $\tilde {B}$; this definition is used throughout the paper. The edges of $\tilde B$ will be classified into ``fake'' edges and ``real'' edges, where every real edge of $\tilde B$ is an edge of $B$.
In order to obtain graph $\tilde B$, we start with the graph $B$. We then consider every child vertex $v(B')$ of $v(B)$ in $\tau(\lset)$ one-by-one. For each such child vertex $v(B')$, we delete all edges and vertices of $B'$ from $B$, except for the endpoints $I(B')$. 
%\snote{it is vague in the sense that it is not clear whether the edge between the endpoints $I(B')$ is deleted or not.}
If the current graph $\tilde B$ does not contain an edge connecting the endpoints of $B'$, then we add such an edge as a fake edge. Consider now the graph $\tilde B$ obtained after processing every child vertex of $v(B)$ in tree $\tau(\lset)$. If $B\neq G$, then we add a fake edge connecting endpoints of $B$ to $\tilde B$ (notice that, from the definition of a block, $\tilde B$ may not contain a real edge connecting the endpoints of $B$). This completes the definition of the graph $\tilde B$. Observe that by our construction, $\tilde B$ has no parallel edges.
The definition of $\tilde B$ depends on the family $\lset$, so when using it we will always fix some such family. We denote by $A_{\tB}$ the set of all fake edges in $\tilde B$. We also denote by $e^*_{\tB}$ the unique fake edge connecting the endpoints of $B$; if no such edge exists, then $e^*_{\tB}$ is undefined. We refer to $e^*_{\tB}$ as the \emph{fake parent-edge of $\tB$}.

We let $\nset(B)$ be a collection of sub-graphs of $G$ that contains the graph $B^c$ -- the complement of the block $B$, and additionally, for every child vertex $v(B')$ of  $v(B)$ in the tree $\tau$, the block $B'$. Observe that every fake edge $e=(u,v)$ of $\tilde B$ is associated with a distinct graph $B_e\in \nset(B)$, where $I(B_e)=(u,v)$.
We are now ready to define a block decomposition.

\begin{definition}
	Let $G$ be a $2$-connected graph, 
	let $\lset$ be a laminar family of sub-graphs of $G$ with $G\in \lset$, and let $\tau=\tau(\lset)$ be the decomposition tree associated with $\lset$. We say that $\lset$ is a \emph{block decomposition} of $G$, iff:
	
	\begin{itemize}
		\item every graph $B\in \lset\setminus\set{G}$ is a block;
		
		\item for each graph $B\in \lset$, either $\tilde{B}$ is $3$-connected, or $\tilde B$ is isomorphic to $K_3$ -- a clique graph on $3$ vertices; and
		\item if a vertex $v(B)\in V(\tau)$ has exactly one child vertex  $v(B')$, then $I(B) \neq I(B')$.
	\end{itemize} 
\end{definition}

For convenience, if $\lset$ is a block decomposition of $G$, then we call the elements of $\lset$ \emph{pseudo-blocks}. Note that each pseudo-block is either a block of $G$, or it is $G$ itself. 
The following theorem was proved in \cite{chuzhoy2011graph}.

\begin{theorem}\label{thm: block decomposition} [Theorem 12 in the arxiv version of \cite{chuzhoy2011graph}]
There is an efficient algorithm, that, given a $2$-connected graph $G=(V,E)$ with $|V|\geq 3$, computes a block decomposition $\lset$ of $G$, such that, for each vertex $v\in V$ that participates in some $2$-separator $(u,v)$ of $G$, either (i)  $v$ is an endpoint of a block $B\in\lset$, or (ii) $v$ has exactly two neighbors in $G$, and there is an edge $(u',v)\in E$, such that $u'$ is an endpoint of a block $B\in \lset$.
\end{theorem}

%old version below:
%\begin{theorem}\label{thm: block decomposition} [Theorem 12 in the arxiv version of \cite{chuzhoy2011graph}]
%There is an efficient algorithm, that, given a $2$-connected graph $G=(V,E)$ with $|V|\geq 3$, computes a block decomposition $\lset$ of $G$, such that, for each vertex $v\in V$ that participates in some $2$-separator $(u,v)$ of $G$: (i) either it is an endpoint of a block $B\in\lset$, or it has degree $2$ in $G$; and (ii) there is an edge $(u,v)\in E$, such that $u$ is an endpoint of a block $B\in\lset$.
%\end{theorem}

We need the following two simple observations.

\begin{observation}\label{obs: paths for fake edges}
	Let $G$ be a $2$-connected graph, let $\lset$ be a block decomposition of $G$, and let $B\in \lset$ be a pseudo-block in the decomposition. Consider the corresponding graph $\tilde B$. Then, for every fake edge $e\in A_{\tilde B}$, there is a path $P(e)$ in $G$, connecting its endpoints, that is internally disjoint from
	$V(\tilde B)$ and is completely contained in some graph $B_e\in \nset(B)$. Moreover, if $e\neq e'$ are two distinct fake edges in $A_{\tilde B}$, then $B_e\neq B_{e'}$, and so paths $P(e)$ and $P(e')$ are internally disjoint.
\end{observation}
%Let $v(B')$ be the child block of $v(B)$ that corresponds to the fake edge $e=(u,w)$. Then $B'$ should be connected otherwise $\tilde {B'}$ cannot be $3$-connected or equal to $K_3$. Therefore, we can use the path inside $B'$ to connect $u$ to $w$. Note that all such paths are internally disjoint as $u,w$ disconnects $B'$ from the rest of the graph which includes $V(\tilde B)$ and the vertices of other paths. Similarly, if $B\neq G$, and there is a fake edge $e=(u,w)$ connecting the endpoints of $B$, then let $B'$ be such that $v(B')$ is the parent node of $v(B)$. Then since $\tilde {B'}$ is $3$-connected or equal to $K_3$, there should be a path connecting $u$ to $w$ in the graph $\tilde {B'}\setminus{e}$ which gives a path in the graph $B'\setminus B$.
\begin{observation}\label{obs: block in 2-connected graph is planar}
	Let $G$ be a $2$-connected planar graph, let $\lset$ be a block decomposition of $G$, and let $B\in \lset$ be a pseudo-block. Then $\tilde B$ is a planar graph, and it has a unique planar drawing.
\end{observation}
\iffalse
\begin{proof}
{\color{blue} We get a planar drawing of $\tilde B$ using a planar of $G$. We draw every real edge of $\tilde B$ along its drawing in $G$. Moreover, by Observation \ref{obs: paths for fake edges}, there is a path corresponding to every fake edge of $\tilde B$ that is internally disjoint from other paths and the vertices of $\tilde B$. Therefore, we can draw the fake edge along this path without introducing any crossings. Therefore, the graph $\tilde B$ is also planar. Using the definition of block decomposition, we know that either $\tilde B$ is $3$-connected or equal to $K_3$, meaning that it has a unique planar drawing.
}
\end{proof}
\fi

\subsection{Block Decomposition of General Graphs}
So far we have defined a block decomposition for $2$-connected graphs. We now extend this notion to general graphs, that may not even be connected, and introduce some useful notation.
Let $G$ be any graph. We denote by $\cset(G)$ the set of all connected components of $G$.
Consider now some connected component $C\in \cset(G)$. Let $\zset(C)$ denote the collection of all maximal $2$-connected sub-graphs of $C$ (that is, $Z\subseteq C$ belongs to $\zset(C)$ iff $Z$ is $2$-connected, and it is not strictly contained in any other $2$-connected subgraph of $C$). %\snote{it might be good to confirm whether a single edge connecting two 2-connected components is not in any block, and what we will do with them} 
It is easy to verify that $\zset(C)$ is uniquely defined and can be computed efficiently. %{\color{blue} Note that two-connected components of a graph can be computed using a single DFS algorithm.} 
For convenience, we call the elements in $\zset(C)$ \emph{super-blocks}. %\mynote{a better name?}
We also denote by $\zset(G)=\bigcup_{C\in \cset(G)}\zset(C)$ the collection of all resulting super-blocks. 

Finally, for every super-block $Z\in \zset(G)$, we let $\lset(Z)$ be the block decomposition of $Z$ given by Theorem \ref{thm: block decomposition}.
Recall that $\lset(Z)$ contains the graph $Z$, and all other graphs in $\lset(Z)$ are blocks of $Z$.
We denote by $\bset(C)=\bigcup_{Z\in \zset(C)}\lset(Z)$ the collection of all pseudo-blocks in the block decompositions of the subgraphs $Z\in \zset(C)$, and we denote $\bset(G)=\bigcup_{C\in \cset(G)}\bset(C)$. We will refer to the collection $\bset(G)$ of pseudo-blocks a \emph{block decomposition of $G$}. Observe that this generalizes the definition of block decompositions of $2$-connected graphs to general graphs.

Consider now some super-block $Z\in \zset(G)$, and some pseudo-block $B\in \lset(Z)$. If $B=Z$, then the complement block $B^c$ is empty. Otherwise, the complement block $B^c$ is defined exactly as before, with respect to the graph $Z$. In other words, $B^c$ is the sub-graph of $Z$ induced by the set $(V(Z)\setminus V(B))\cup I(B)$ of vertices. We define the set $\nset(B)$ of graphs as before: we add $B^c$ to $\nset(B)$, and additionally, for every child vertex $v(B')$ of $v(B)$, we add the block $B'$ to $\nset(B)$. %Informally, we think of the collection $\nset(B)$ of graphs as the ``neighbors'' of the block $B$.

\subsection{Embedding of Fake Edges}
We will repeatedly use the following lemma, whose proof appears in Section \ref{subsec:block_endpoints_routing}.
\begin{lemma}
\label{lem:block_endpoints_routing}
Let $G$ be a graph, and let $\bset(G)$ be its block decomposition.
Denote $\tilde\bset(G)=\set{\tilde B\mid B\in \bset(G)}$, and let $\tilde \bset^*(G)\subseteq \tilde\bset(G)$ contain all graphs $\tilde B$ that are not isomorphic to $K_3$. Then we can efficiently compute, for each graph $\tilde{B}\in \tilde\bset^*(G)$, a collection $\pset_{\tilde B}=\set{P_{\tilde B}(e)\mid e\in A_{\tilde B}}$ of paths in $G$, such that:
\begin{itemize}
\item for each fake edge $e=(u,v)\in A_{\tilde B}$, the path $P_{\tilde B}(e)$ connects $u$ to $v$ in $G$ and it is internally disjoint from $\tilde B$; 
\item all paths in $\pset_{\tilde B}$ are mutually internally disjoint; and
\item if we denote $\pset=\bigcup_{\tilde B\in \tilde\bset^*(G)}\left(\pset_{\tilde B}\setminus \set{P_{\tilde B}(e^*_{\tilde B})} \right)$, then every edge of $G$ participates in at most $6$ paths in $\pset$.
\end{itemize} 
\end{lemma}

\subsection{Monotonicity of Blocks under Edge Deletions}
Intuitively, our algorithm for Theorem \ref{thm: main}  starts with some planarizing set $E'$ of edges for the input graph $G$. We denote $H=G\setminus E'$, and consider the block decomposition $\bset(H)$ of $H$. As the algorithm progresses, more edges are added to $E'$, graph $H$ evolves, and its block decomposition $\bset(H)$ changes. The following lemma, whose proof is deferred to Appendix~\ref{apd:block_containment},  allows us to keep track of these changes.

\begin{lemma}\label{lem: block containment}
	Let $G$ be a graph, and let $E_1,E_2$ be two planarizing edge sets for $G$, with $E_1\subseteq E_2$. Denote $H_1=G\setminus E_1$, $H_2=G\setminus E_2$, and let $\bset_1=\bset(H_1),\bset_2=\bset(H_2)$ be block decompositions of $H_1$ and of $H_2$, respectively. Then for every pseudo-block $B_2\in \bset_2$ with $|V(\tilde B_2)|>3$, there is a pseudo-block $B_1\in \bset_1$, such that: %(i) every vertex of $\tilde B_2$ lies in $\tilde B_1$; and (ii) every real edge of $\tilde B_2$ belongs to $\tilde B_1$ as a real edge, except possibly for the edge connecting the endpoints of $B_1$.

	\begin{itemize}
		\item $V(\tilde B_2)\subseteq V(\tilde B_1)$;
		\item every real edge of $\tilde B_2$ belongs to $\tilde B_1$ as a real edge, except (possibly) for an edge whose endpoints are the endpoints of $B_1$; 
		\item for every fake edge $e=(u,v)$ in $\tilde B_2$, there is a path $P(e)$ in $\tilde B_1$ that connects $u$ to $v$ (and may contain fake edges of $\tilde B_1$), that is internally disjoint from $V(\tilde B_2)$. Moreover, if $e$ and $e'$ are distinct fake edges of $\tilde B_1$, then their corresponding paths $P(e)$ and $P(e')$ are internally disjoint; and
		\item if graph $\tilde B_2$ contains a real edge connecting the endpoints of $B_1$, then for every fake edge $e\in E(\tilde B_2)$, the path $P(e)$ does not contain the edge connecting the endpoints of $B_1$.
		
	\end{itemize}
\end{lemma}

\section{Computing a Decomposition into Acceptable Clusters: Proof of Theorem~\ref{thm: can find edge set w acceptable clusters}}
\label{sec:thm1}

This section is dedicated to the proof of Theorem \ref{thm: can find edge set w acceptable clusters}. 
\iffalse
The proof consists of two stages. In the first stage, we compute a subset $E_1$ of edges of $G$ with $E'\subseteq E_1$ and 
$|E_1|\leq O\left ( (|E'|+\optcro(G))\cdot \poly(\Delta\log n)\right )$. We will ensure that for every connected component $C$ of the resulting graph $G\setminus E_1$, either $C$ contains at most $\mu$ terminals (endpoints of edges of $E_1$) and therefore can be considered a type-1 acceptable cluster, or it has another nice property that is somewhat similar to the Bridge Consistency Property. In the second part of the proof, we augment the edges in $E_1$ to another set $E''$ of edges, and complete the construction of the decomposition. Intuitively, the first stage deals with the Bridge Consistency property, while the second stage ensures the Terminal Well-Linkedness and $3$-Connectivity properties of the type-2 acceptable clusters.
\fi
%-----------------------------
%-----------------------------
%-----------------------------
%-----------------------------
%-----------------------------
%-----------------------------
%-----------------------------
%-----------------------------
%-----------------------------
%-----------------------------
We start by introducing some notation and defining the notions of good pseudo-blocks.
		
\subsection{Good Pseudo-Blocks}
\label{sec: good blocks and acceptable clusters}

Throughout this subsection, we assume that we are given some planarizing set $\hat E$ of edges for the input graph $G$, that is, $G\setminus \hat E$ is planar. We note that $\hat E$ is not necessarily the same as the original planarizing set $E'$, since, as the algorithm progresses, we may add edges to the planarizing set. The definitions in this subsection refer to any planarizing set $\hat E$ that may arise over the course of the algorithm.

Given a planarizing set $\hat E$ of edges for $G$, let $H=G\setminus \hat E$. Recall that we say that a vertex $v$ of $H$ is a \emph{terminal} iff it is incident to some edge in $\hat E$, and we denote the set of terminals by $\Gamma$.

Recall that we have defined a block decomposition of a graph $H$ as follows. We denoted by $\cset=\cset(H)$ the set of all connected components of $H$, and we refer to the elements of $\cset$ as \emph{clusters}. For each cluster $C\in \cset$, we have defined a collection $\zset(C)$ of super-blocks (maximal $2$-connected subgraphs) of $C$, and we denoted $\zset(H)=\bigcup_{C\in \cset}\zset(C)$. Lastly, for each superblock $Z\in \zset(H)$, we let $\lset(Z)$ be the block decomposition of $Z$ given by Theorem \ref{thm: block decomposition}, and we denoted by $\bset(C)=\bigcup_{Z\in \zset(C)}\lset(Z)$ the resulting collection of pseudo-blocks for cluster $C$. The final block decomposition of $H$ is defined to be $\bset(H)=\bigcup_{C\in \cset}\bset(C)$. %, and the elements of $\bset(H)$ are called pseudo-blocks.

Consider some pseudo-block $B\in \bset(H)$, and let $\tilde B$ be the corresponding graph that is either a $3$-connected graph or isomorphic to $K_3$. 
Recall that $\tilde B$ has a unique planar drawing, that we denote by $\rho_{\tilde B}$. Throughout this section, for any pseudo-block $B$, we denote by $\tilde B'\subseteq \tilde B$ the graph obtained from $\tilde B$ by deleting all its fake edges. Note that $\tilde B'$ may be not $3$-connected, and it may not even be connected. The drawing $\rho_{\tilde B}$ of $\tilde B$ naturally induces a drawing of $\tilde B'$, that we denote by $\rho'_{\tilde B'}$. %\snote{why dont we use the same $\rho_{\tilde B'}$? $\rho_{\tilde B'}$ is the same drawing as $\rho'_{\tilde B'}$ but for a sub-graph.}
%Consider now the set $\rset_G(\tilde B')$ of all bridges for $\tilde B'$ in graph $G$. 

\begin{definition}
We say that a pseudo-block $B\in \bset(H)$ is a \emph{good pseudo-block} iff there is a planar drawing $\hat{\psi}_B$ of $B$, that we call the \emph{associated drawing}, such that, for each bridge $R\in \rset_G(B)$, there is a face $F$ in $\hat{\psi}_B$, whose boundary contains all vertices of $L(R)$. If $B$ is not a good pseudo-block, then it is called a \emph{bad pseudo-block}.
\end{definition}

 Note that, if $B$ is a bad pseudo-block, then for every planar drawing $\hat \psi_B$ of $B$, there is some bridge $R\in \rset_G(B)$, such that no face of $\hat{\psi}_B$ contains all vertices of $L(R)$. We call such a bridge $R$ a \emph{witness} for $B$ and $\hat \psi_B$. We note that for each bridge $R\in \rset_G(B)$ and for every vertex $v\in L(R)$, either $v$ is a terminal of $\Gamma$, or it is a separator vertex for the connected component $C$ of $H$ that contains $B$. %\znote{make it an observation?}

\iffalse
\paragraph{Good Clusters.}
Finally, we define good clusters.

\begin{definition}
	We say that a cluster $C\in \cset$ is \emph{good} iff pseudo-block $B\in \bset(C)$ is good.
\end{definition}
\fi

\iffalse
Our algorithm starts with some planarizing set $E'$ of edges, and then, over the course of several steps, gradually adds additional edges to set $E'$. Suppose that at some point we compute a planarizing set $\hat E$ of edges, such that, in the block decomposition given by Theorem~\ref{thm: block decomposition} of every cluster of $G\setminus \hat E$, all pseudo-blocks are good. As the algorithm progresses, and more edges are added to $\hat E$, we would like to claim that this property continues to be preserved. The following lemma allows us to do so. Its proof is deferred to Appendix~\ref{apd:good_cluster_preserving}.

\begin{lemma}\label{lem: good clusters cant be killed}
Suppose we are given two planarizing sets $\hat E,\hat E'$ of edges, with $\hat E\subseteq \hat E'$. Let $H=G\setminus\hat E$ and $H'=G\setminus \hat E'$, so that $H'\subseteq H$. Consider some pseudo-block $B_1\in \bset(H')$, and let $B_2\in \bset(H)$ be the corresponding block given by Lemma \ref{lem: block containment}. Assume further that $\tilde B_2$ does not contain a real edge connecting the endpoints of $B_1$. Then, if $B_2$ is a good pseudo-block, so is $B_1$. % Assume that every pseudo-block in $\bset_H$ is good. Then every pseudo-block in $\bset_{H'}$ is good as well. 
\end{lemma}
\fi

The remainder of the proof of Theorem \ref{thm: can find edge set w acceptable clusters} consists of two stages. In the first stage, we augment the initial planarizing set $E'$ of edges to a new edge set $E_1$, such that 
for every connected component $C$ of $G\setminus E_1$, either $C$ is a type-$1$ acceptable cluster, or every pseudo-block in the block decomposition of $C$ is good. In the second stage, we further augment $E_1$ in order to obtain the final edge set $E''$ and a decomposition of $G$ into acceptable clusters. We now describe each of the two stages in turn.

%-------------------------------------------
%-------------------------------------------
%-------------------------------------------
%-------------------------------------------

%-------------------------------------------
%-------------------------------------------
%-------------------------------------------
%-------------------------------------------
\subsection{Stage 1: Obtaining Type-1 Acceptable Clusters and Good Pseudo-Blocks}
\label{sec:stage1}

%-------------------------------------------
%-------------------------------------------
%-------------------------------------------
%-------------------------------------------

%-------------------------------------------
%-------------------------------------------
%-------------------------------------------
%-------------------------------------------

The main result of this stage is summarized in the following theorem.

\begin{theorem}\label{thm: getting good pseudoblocks}
There is an efficient algorithm, that, given a $3$-connected $n$-vertex graph $G$ with maximum vertex degree $\Delta$ and a planarizing set $E'$ of edges for $G$, computes a planarizing edge set $E_1$ with $E'\subseteq E_1$, such that $|E_1|\leq O\left ( (|E'|+\optcro(G))\cdot \poly(\Delta\log n)\right )$, and, if we denote $H_1=G\setminus E_1$, and let $\Gamma_1$ be the set of all endpoints of edges in $E_1$, then for every connected component $C$ of $H_1$, either $|V(C)\cap \Gamma_1|\leq \mu$, or every pseudo-block in the block decomposition $\bset(C)$ of $C$ is a good pseudo-block.
Moreover, for each connected component $C$ of the latter type, for each pseudo-block $B$ in $\bset(C)$, the algorithm also computes its associated planar drawing $\hat{\psi}_B$.
\end{theorem}

The remainder of this subsection is dedicated to proving Theorem \ref{thm: getting good pseudoblocks}. 
%
%
%The goal of this stage is to compute a planarizing edge set $E_1$ with $E'\subseteq E_1$, such that $|E_1|\leq O\left ( (|E'|+\optcro(G))\cdot \poly(\Delta\log n)\right )$, and, if we denote $H=G\setminus E_1$, and by $\Gamma_1$ the set of all endpoints of edges in $E_1$, then for every connected component $C$ of $H$, either $|C\cap \Gamma_1|\leq \mu$, or $C$ has another nice property that is somewhat similar to bridge consistency property, and is defined below. 
We start with a high-level intuition in order to motivate the next steps in this stage. Consider any pseudo-block $B\in \bset(H)$ in the block decomposition of the graph $H$. We would like to construct a small set $E^*(B)$ of edges of $\tilde B'$, such that, for every bridge $R\in \rset_G(\tilde B')$, all vertices of $L(R)$ lie on the boundary of a single face in the drawing of $\tilde B'\setminus E^*(B)$ induced by $\rho_{\tilde B}$. In general, we are able to find such an edge set $E^*(B)$, while ensuring that its size is small, compared with the following two quantities. The first quantity is the size of the vertex set $\Gamma'(B)$, that is defined to be the union of (i) all terminals (that is, endpoints of edges in $E'$) lying in $\tilde B$; (ii) all endpoints of the fake edges of $\tilde B$; and (iii) all separator vertices of $C$ that lie in $\tilde B$, where $C\in \cset$ is the component of $H$ that contains $B$. The second quantity is, informally, the number of crossings in which the edges of $\tilde B'$ participate in the optimal drawing $\phi^*$ of $G$ (we need a slightly more involved definition of the second quantity that we provide later).
We would like to augment $E'$ with the edges of $\bigcup_{B\in \bset(H)}E^*(B)$ to obtain the desired set $E_1$. 
Unfortunately, we cannot easily bound the size of $|E_1|$, as we cannot directly bound $\sum_{B\in \bset(H)}|\Gamma'(B)|$. For example, consider a situation where the decomposition tree $\tau_Z$ associated with some maximal $2$-connected subgraph $Z$ of $H$ contains a long induced path $P$. Then for every vertex $v(B)$ on path $P$, graph $\tilde B$ contains exactly two fake edges, one corresponding to its parent, and the other corresponding to its unique child. Note that, it is possible that many of the graphs $\tilde B$ with $v(B)\in P$ do not contain any terminals or separator vertices of the component of $H$ that contains $B$, so $\sum_{v(B)\in P}|\Gamma'(B)|$ is very large, and it may be much larger than $\optcro(G)+|E'|$. To vercome this difficulty, we carefully decompose all such  paths $P$, such that, after we delete a small number of edges from the graph, we obtain a collection of type-1 acceptable clusters, and for each block $B$ with $v(B)\in P$, graph $\tilde B$ is contained in one of these clusters. Our proof proceeds as follows. First, we bound the cardinality of the set $U$ of the separator vertices of $H$ by comparing it to the size of $\Gamma$. 
%Then we ``mark'' some vertices $v(B)$ in the forest associated with the block decomposition $\bset(H)$ of $H$. We say that a block $B$ is marked if its corresponding vertex $v(B)$ is marked.
Then we mark some blocks $B$ in the block decomposition $\bset(H)$ of $H$. 
We will ensure that, on the one hand, we can suitably bound the total cardinality of the vertex sets $\Gamma'(B)$ for all marked blocks $B$, while on the other hand, in the forest associated with the block decomposition $\bset(H)$ of $H$, if we delete all vertices corresponding to the marked blocks in $\bset(H)$, we obtain a collection of paths, each of which can be partitioned into subpaths, which we use in order to define type-1 acceptable clusters. Let $\bset'$ denote the set of all marked blocks. For each block $B\in \bset'$, we define a collection $\chi(B)$ of crossings in the optimal drawing $\phi^*$ of $G$, such that $\sum_{B\in \bset'}|\chi(B)|$ can be suitably bounded. We then process each such block $B\in \bset'$ one by one, computing the edge set $E^*(B)$ of $\tilde B$, such that, 
for every bridge $R\in \rset_G(\tilde B')$, all vertices of $L(R)$ lie on the boundary of a single face in the drawing of $\tilde B'\setminus E^*(B)$ induced by $\rho_{\tilde B}$.  The cardinality of $E^*(B)$ will be suitably bounded by comparing it to $|\Gamma'(B)|+|\chi(B)|$. 

We now proceed with the formal proof. 
Throughout the proof, we denote $H=G\setminus E'$ and denote by $\cset$ the set of connected components of $H$ that we call clusters. The set $\Gamma$ of terminals contains all vertices that are endpoints of the edges of $E'$. 
For every cluster $C\in \cset$, we denote by $U(C)$ the set of all separator vertices of $C$; that is, vertex $v\in U(C)$ iff graph $C\setminus\set{v}$ is not connected. Let $U=\bigcup_{C\in \cset}U(C)$. We start by proving the following observation.

\begin{observation}\label{obs: few sep vertices}
	$|U|\leq O(|\Gamma|)$.
\end{observation}
\begin{proof}
It suffices to show that, for every cluster $C\in \cset$, $|U(C)|\leq O(|\Gamma\cap V(C)|)$. From now on we fix a cluster $C\in \cset$. Let $\zset(C)$ be the decomposition of $C$ into super-blocks. 
We can associate a graph $T$ with this decomposition as follows. The set $V(T)$ of vertices is defined to be the union of (i) the set $U(C)$ of separator vertices,  that we also refer to as regular vertices; and (ii) the set $U'=\set{v_Z\mid Z\in \zset(C)}$ of vertices called supernodes, representing the super-blocks of the decomposition.
%We call the vertices in $U'$ \emph{super-nodes}. 
For every super-block $Z\in \zset(C)$ and every separator vertex $u\in U(C)$ such that $u\in V(Z)$, we add the edge $(u,v_Z)$ to the graph.
For every pair of distinct separator vertices $u,u'\in U(C)$, such that $(u,u')\in E(C)$ and there is no super-block $Z\in \zset(C)$ that contains both $u$ and $u'$, we add the edge $(u,u')$ to the graph.
It is easy to verify that graph $T$ is a tree. 

We partition the set $V(T)$ of vertices into the following three subsets: (i) the set $V_1$ contains all vertices that have degree $1$ in $T$, namely $V_1$ is the set of all leaf vertices of $T$; (ii) the set $V_{2}$ contains all vertices that have degree $2$ in $T$; and (iii) the set $V_{\ge 3}$ contains all vertices that have degree at least $3$ in $T$.
We further partition the set $V_{2}$ into three subsets: the set $U'_2=U'\cap V_2$ containing all supernodes of $V_2$; the set $\hat U_{2}$ containing all regular vertices  $u\in V_2$, such that  both neighbors of $u$ are regular vertices; and the set $\tilde U_{2}=(U(C)\cap V_2)\setminus \hat U_{2}$ containing all remaining vertices.
%Let $L\subseteq V(T)$ be the set of all leaves of $T$.
We use the following observation.
\begin{observation}
$|V_1|\le |\Gamma\cap V(C)|$,
$|U'_2|\le |\Gamma\cap V(C)|$, 
and $|\hat U_2| \le |\Gamma\cap V(C)|$.
\end{observation}
\begin{proof}
Observe that $V_1\subseteq U'$, and moreover, if some node $v_Z\in V_1$ corresponds to the super-block $Z\in \zset(C)$, then there must be at least one terminal vertex in $\Gamma\cap V(Z)$ that does not belong to $U(C)$. This follows from the fact that $G$ is $3$-connected, and $|V(Z)\cap U(C)|=1$. Therefore, $|V_1|\le |\Gamma\cap V(C)|$.
Similarly, we can deduce that, for each vertex $v_Z\in U'_2$, its corresponding block $Z$ must contain a terminal in $\Gamma\cap V(Z)$ that does not belong to $U(C)$. Therefore, $|U'_2|\le |\Gamma\cap V(C)|$.
From the definition of $\hat U_{2}$ and the fact that $G$ is $3$-connected, we get that every node in $\hat U_2$ has to be a terminal in $\Gamma\cap V(C)$. Therefore, $|\hat U_2| \le |\Gamma\cap V(C)|$.
\end{proof}

From the definition of the sets $V_1$ and $V_{\ge 3}$, $|V_{\ge 3}|\le |V_1|\le |\Gamma\cap V(C)|$.
Moreover, if we denote by $ E^*$ the set of all edges of the tree $T$ that are incident to a vertex of $V_{\ge 3}$, then $| E^*|\le O(|V_1|)$.
%Observe that each vertex of $U(C)$ has degree at least $2$ in $T$. Note that, if a vertex $u\in U(C)$ has degree $2$ in $T$, and both neighbors of $u$ also belong to $U(C)$, then $u$ has to be a terminal in $\Gamma\cap V(C)$.

Consider a vertex $u\in \tilde U_2$. Recall that $u$ has exactly two neighbors in $T$, that we denote by $x$ and $y$, and $x$ and $y$ are not both regular vertices. If $x\in V_{\geq 3}$ or $y\in V_{\geq 3}$, then $u$ is an endpoint of an edge in $E^*$. If either of the vertices $x$ or $y$ lies in $V_1$, then $u$ is the unique neighbor of that vertex in $T$. Assume now that neither of the two vertices lies in $V_1\cup V_{\geq 3}$, that is, both vertices lie in $V_2$. Assume w.l.o.g. that $x\not\in U$, so $x$ is a supernode. Then $u$ is one of the two neighbors of a supernode $x\in U_2'$. To summarize, if $u\in \tilde U_2$, then either (i) $u$ is an endpoint of an edge of $ E^*$; or (ii) $u$ is a unique neighbor of a vertex of $V_1$; or (iii) one of the two neighbors of a vertex of $U'_2$. Therefore, $|\tilde U_2|\le O(| E^*|+|V_1|+|U'_2|)\le O(|V_1|+|U'_2|)\le O(|\Gamma\cap V(C)|)$.
%Also observe that, the number of vertices of $U(C)$ that has degree at least $3$ is at most $|L|$. 
Altogether: 
\[
|U(C)|= |U(C)\cap V_1|+|U(C)\cap V_2|+|U(C)\cap V_{\ge 3}|
\le 0+ |\hat U_2|+|\tilde U_2|+|V_{\ge 3}|
\le O(|\Gamma\cap V(C)|).\]
Summing over all clusters $C\in \cset$, we get that $|U|\le O(|\Gamma|)$.
%Let $S_2\subseteq V(T)$ be the set of all nodes in $U'$ that have degree $2$ in $T$. Similarly, for every vertex $v_Z\in S_2$, there must be at least one terminal vertex of $\Gamma\cap V(Z)$ in the corresponding superblock $Z$ that does not belong to $U(C)$. Therefore, $|S_1|+|S_2|\leq |\Gamma\cap V(C)|$.
%Let $T'$ be the tree obtained from $T$ by suppressing every vertex $u\in U(C)$ whose degree in $T$ is $2$. Then $T'$ is a tree, and $|V(T')|\leq O(|S_1|+|S_2|)\leq O(|\Gamma\cap V(C)|)$. Since for every vertex $u\in U(C)$ that has degree $2$ in $T$, both its endpoints are in $U'$, we get that $|V(T)|\leq |V(T')|+|E(T')|\leq O(|V(T')|)\leq O(|\Gamma\cap V(C)|)$. Therefore, $|U(C)|\leq O(|\Gamma\cap V(C)|)$.
\end{proof}

Let $C\in \cset$ be any cluster of $H$, and let  $\zset(C)$ be the set of all super-blocks of $C$. For each super-block $Z\in \zset(C)$, we let $\lset(Z)$ be the block decomposition of $Z$, given by Theorem \ref{thm: block decomposition}. Recall that this block decomposition is associated with a decomposition tree, that we denote for brevity by $\tau_Z$. As before, we denote by $\bset(C)=\bigcup_{Z\in \zset(C)}\lset(Z)$ and by $\bset(H)=\bigcup_{C\in \cset}\bset(C)$ the block decompositions of $C$ and $H$, respectively. Let $\tset$ be the forest that consists of the trees $\tau_Z$ for all $Z\in \bigcup_{C\in\cset}\zset(C)$. Recall that every vertex $v(B)\in \tset$ corresponds to a pseudo-block $B\in \bset(H)$ and vice versa.

Consider now some tree $\tau_Z\in \tset$. We \emph{mark} a vertex $v(B)$ of $\tau_Z$ iff, either (i) vertex $v(B)$ has at least two children in the tree $\tau_Z$, or (ii) graph $\tilde B$ contains at least one vertex of $\Gamma\cup U$ that is not an endpoint of a fake edge of $\tilde B$. 
We denote by $\bset'\subseteq \bset(H)$ the set of all pseudo-blocks $B$ whose corresponding vertex $v(B)$ was marked. For each pseudo-block $B\in \bset'$, let $\Gamma'(B)$ be the set of vertices of $\tilde B$ that contains all terminals of $\Gamma$ that lie in $\tilde B$, the vertices of $U$ that lie in $\tilde B$, and all endpoints of the fake edges that belong to $\tilde B$.
We need the following simple observation.

\begin{observation}\label{obs: bound gamma for marked blocks}
	$\sum_{B\in \bset'}|\Gamma'(B)|\leq O(\Delta\cdot|\Gamma|)$.
\end{observation}
\begin{proof}
Consider a pseudo-block $B\in \bset'$. We partition the set $\Gamma'(B)$ of vertices into three subsets: set $\Gamma'_1(B)$ contains all endpoints of fake edges of $\tilde B$; set $\Gamma'_2(B)$ contains all vertices of $\Gamma\setminus U$ that lie in $\tilde B$ and do not serve as endpoints of fake edges, and set $\Gamma'_3(B)$ contains all vertices of $U$ that lie in $\tilde B$ and do not serve as endpoints of fake edges.
	
Notice that the sets $\set{\Gamma'_2(B)}_{B\in \bset'}$ are mutually disjoint, since for any pair $B_1,B_2\in \bset(H)$ of pseudo-blocks in the decomposition, the only vertices of $\tilde B_1$ that may possibly lie in $\tilde B_2$ are vertices of $U$ and vertices that serve as endpoints of fake edges in $\tilde B_1$. 
 Therefore, $\sum_{B\in \bset'}|\Gamma'_2(B)|\leq |\Gamma|$. 

Notice that a vertex $u\in U$ may belong to at most $\Delta$ super-blocks in $\zset(H)$. For each super-block $Z\in \zset(H)$, there is at most one pseudo-block $B\in \lset(Z)$, such that $u$ belongs to $\tilde B$ but is not an endpoint of a fake edge of $\tilde B$. This is because for any pair $B_1,B_2$ of blocks of $\lset(Z)$, the only vertices of $\tB_1$ that may possibly lie in $\tB_2$ are endpoints of fake edges of $\tB_1$. Therefore, $\sum_{B\in \bset'}|\Gamma'_3(B)|\leq \Delta\cdot |U|\leq O(\Delta \cdot |\Gamma|)$, from Observation~\ref{obs: few sep vertices}.
		
In order to bound $\sum_{B\in \bset'}|\Gamma'_1(B)|$,
we partition the set $\bset'$ of pseudo-blocks into two subsets: set $\bset'_1$ contains all pseudo-blocks $B$ such that $v(B)$ has at least two children in the forest $\tset$, and $\bset'_2$ contains all remaining pseudo-blocks.
Let $\bset^*\subseteq \bset(H)$ be the set of all pseudo-blocks whose corresponding vertex $v(B)$ has degree $1$ in $\tset$. Since the original graph $G$ is $3$-connected, and since for each pseudo-block $B\in \bset^*$, $\tilde B$ contains at most one fake edge, for each pseudo-block $B\in \bset^*$, the corresponding graph $\tilde B$ must contain a terminal $t\in \Gamma$ that is not one of its endpoints. 
If $t\not\in U$, then $B$ is the only pseudo-block in $\bset^*$ such that $t\in \tilde B$ and $t$ is not one of the endpoints of $B$. Otherwise, there are at most $\Delta$ such pseudo-blocks in $\bset^*$.
Therefore, $|\bset^*|\leq |\Gamma|+\Delta\cdot |U|\leq O(\Delta\cdot |\Gamma|)$. From the definition of the set $\bset'_1$, it is immediate to see that the total number of fake edges in all pseudo-blocks of $\bset'_1$ is at most $O(|\bset^*|)\leq O(\Delta\cdot|\Gamma|)$. Therefore, $\sum_{B\in \bset'_1}|\Gamma'_1(B)|\leq O(\Delta \cdot |\Gamma|)$.
Consider now a pseudo-block $B\in \bset'_2$. Then $\tilde B$ contains at most two fake edges and at least one vertex of $\Gamma\cup U$, that is not an endpoint of a fake edge. Therefore, for each pseudo-block $B\in \bset'_2$, $|\Gamma'_1(B)|\leq 4$, and $\sum_{B\in \bset'_2}|\Gamma'_1(B)|\leq 4\sum_{B\in \bset'_2}(|\Gamma'_2(B)|+|\Gamma'_3(B)|)\leq O(\Delta \cdot |\Gamma|)$. Altogether, we conclude that $\sum_{B\in \bset'}|\Gamma'(B)|\leq O(\Delta\cdot |\Gamma|)$.
\end{proof}

The final set $E_1$ of edges  that is the outcome of Theorem \ref{thm: getting good pseudoblocks} is the union of the input set $E'$ of edges and four other edge sets, $\tilde E_1,\tilde E_2,\tilde E_3$ and $\tilde E_4$, that we define next.

%\vspace{-4mm}
\paragraph{Sets $\tilde E_1$ and $\tilde E_2$.}
%Set $\tilde E_1$ of edges is defined as follows. 
We let set $\tilde E_1$ contain, for every pseudo-block $B\in \bset'$ and for every fake edge $e=(x,y)$ of $\tilde B$, all edges of $G$ that are incident to $x$ or $y$. From Observation \ref{obs: bound gamma for marked blocks} and the definition of the set $\Gamma'(B)$, it is immediate that $|\tilde E_1|\leq \sum_{B\in \bset'}\Delta\cdot |\Gamma'(B)|\leq  O(\Delta^2\cdot |\Gamma|)$.
We let set $\tilde E_2$ contain all edges incident to vertices of $U$. From Observation \ref{obs: few sep vertices}, $|\tilde E_2|\leq O(\Delta \cdot |\Gamma|)$.

%\vspace{-4mm}
\paragraph{Set $\tilde E_3$.}
We now define the set $\tilde E_3$ of edges and identify a set $\cset_1'$ of connected components of $H\setminus (\tilde E_1\cup \tilde E_2\cup \tilde E_3)$, each of which contains at most $\mu$ vertices that serve as endpoints of edges in $E'\cup \tilde E_1\cup \tilde E_2\cup \tilde E_3$. %The components of $\cset_1'$ will eventually become type-1 acceptable clusters in the final decomposition, though we may create additional type-1 acceptable clusters.

Consider the graph obtained from the forest $\tset$ by deleting all marked vertices in it. It is easy to verify that the resulting graph is a collection of disjoint paths, that we denote by $\qset$. 
Observe that the total number of paths in $\qset$ is bounded by the total number of fake edges in graphs of $\set{\tilde B\mid B\in \bset'}$, so $|\qset|\leq O(\Delta\cdot |\Gamma|)$ from Observation \ref{obs: bound gamma for marked blocks}.
Next, we will process the paths in $\qset$ one-by-one.

Consider now a path $Q\in \qset$. Notice that, from the definition of marked vertices, for every vertex $v(B)\in Q$, graph $\tB$ may contain at most two fake edges and at most four vertices of $\Gamma\cup U$, and all such vertices must be endpoints of the fake edges of $\tB$. 
For any sub-path $Q'\subseteq Q$, we define the graph $H(Q')$ as the union of all graphs $\tilde B'$, for all pseudo-blocks $B\in \bset(H)$ with $v(B)\in V(Q')$ (recall that graph $\tilde B'$ is obtained from graph $\tilde B$ by removing all fake edges from it). The \emph{weight} $w(Q')$ of the path $Q'$ is defined to be the total number of vertices of $H(Q')$ that belong to $\Gamma\cup U$. We need the following simple observation.

\begin{observation}\label{obs: partition paths}
There is an efficient algorithm that computes, for every path $Q\in \qset$, a partition $\Sigma(Q)$ of $Q$ into disjoint sub-paths $Q_1,\ldots,Q_{z}$, such that for all $Q_i\in \Sigma(Q)$, $w(Q_i)\leq \mu/(2\Delta)$, and all but at most one path of $\Sigma(Q)$ have weight at least $\mu/(4\Delta)$. Moreover, every vertex of $Q$ lies on exactly least one path of $\Sigma(Q)$.
\end{observation}
\begin{proof}
We start with $\Sigma(Q)=\emptyset$ and then iterate as  long as $w(Q)>\mu/(2\Delta)$. In an iteration, we let $Q'$ be the shortest sub-path of $Q$ that contains one endpoint of $Q$ and has weight at least $\mu/(4\Delta)$. Since for every pseudo-block $B$ with $v(B)\in Q$, $|(\Gamma\cup U)\cap V(\tilde B)|\leq 4$ and $\mu>16\Delta$, we get that $w(Q')\leq \mu/(2\Delta)$. We add $Q'$ to $\Sigma(Q)$, delete all vertices of $Q'$ from $Q$, and terminate the iteration. Once $w(Q)\leq \mu/(2\Delta)$ holds, we add the current path $Q$ to $\Sigma(Q)$ and terminate the algorithm. %\znote{maybe omit this proof?}
\end{proof}

Consider a sub-path $Q_i\in \Sigma(Q)$ and let $Q_i=(v(B_1),\ldots,v(B_x))$. We assume that $v(B_1)$ is an ancestor of $v(B_x)$ in the forest $\tset$, and we denote by $v(B_{x+1})$ the unique child of $v(B_x)$ in the forest, if it exists. We denote by $\eta(Q_i)$ the set that consists of the endpoints of $B_1$ and the endpoints of $B_{x+1}$ (if $B_{x+1}$ exists).
Note that the only vertices that $H(Q_i)$ may share with other graphs $H(Q_j)$ (for $i\neq j$) are the vertices of $\eta(Q_i)$. Moreover, for any path $Q'\neq Q$ in $\qset$ and any sub-path $Q'_j\in \Sigma(Q')$, the only vertices of $H(Q_i)$ that may possibly belong to $H(Q'_j)$ are the vertices of $\eta(Q_i)$ and $U\cap H(Q_i)$. On the other hand, note that a vertex $u\in U$ may belong to at most $\Delta$ super-blocks of $\zset(H)$. Therefore, for each $u\in U$, the number of graphs in $\set{H(Q_i)\mid Q\in \qset,Q_i\in \Sigma(Q)}$ such that $u\in V(H(Q_i))\setminus \eta(Q_i)$ is at most $\Delta$.
% endpoints of blocks $B_1$ and $B_{x+1}$. Therefore, if $w(Q_i)\geq \mu/(2\Delta)$, then there are at least $\mu/(2\Delta)-4\geq \mu/(4\Delta)$ vertices of $U\cup \Gamma$ that belong to $H(Q_i)$ but do not belong to any other graph $H(Q'_j)\neq H(Q_i)$. Recall that there is at most one path $Q_i\in \Sigma(Q)$ with $w(Q_i)<\mu/(2\Delta)$.
Denote $\Sigma=\bigcup_{Q\in \qset}\Sigma(Q)$ and $\eta=\bigcup_{Q_i\in \Sigma}\eta(Q_i)$. We let $\tilde E_3$ contain all edges of $G$ that are incident to the vertices of $\eta$.
The following observation bounds the cardinality of $\tilde E_3$.
\begin{observation}
\label{obs: third edge set small}
 $|\tilde E_3|\leq O(\Delta^2 \cdot |\Gamma|)$.
\end{observation}
\begin{proof}
Consider any path $Q\in \qset$ and the corresponding subset $\Sigma(Q)$ of its sub-paths. From Observation~\ref{obs: partition paths}, there is at most one path $Q_i\in \Sigma(Q)$ with $w(Q_i)<\mu/(4\Delta)$.
Denote $\Sigma'(Q)=\Sigma(Q)\setminus \set{Q_i}$ and $\Sigma'=\bigcup_{Q\in \qset}\Sigma'(Q)$. Observe that $|\Sigma\setminus \Sigma'|\leq |\qset|\leq O(\Delta \cdot|\Gamma|)$. We claim that $|\Sigma'|\leq O(\Delta\cdot |\Gamma|)$. Note that this implies Observation~\ref{obs: third edge set small}, since every path in $\Sigma$ contributes at most four vertices to set $\eta$, and the maximum vertex degree of a vertex in $\eta$ is at most $\Delta$.
	
It remains to show $|\Sigma'|\leq O(\Delta\cdot |\Gamma|)$.
Consider again some path $Q\in \qset$.	 
Each sub-path $Q_i\in\Sigma'(Q)$ has weight $w(Q_i)\geq \mu/(4\Delta)$. Since $\mu>16\Delta$ and $|\eta(Q_i)|\leq 4$, there are at least $\mu/(8\Delta)$ vertices of $H(Q_i)\setminus \eta(Q_i)$ that belong to $\Gamma\cup U$. Let $S(Q_i)$ be the set of all these vertices. Note that every terminal vertex $t\in \Gamma$ may belong to at most one set $S(Q_i)$ for all paths $Q_i\in \Sigma'$, and every vertex $u\in U$ may belong to at most $\Delta$ such sets. Therefore, $|\Sigma'|\leq \frac{|\Gamma|+\Delta |U|}{\mu/(8\Delta)}\leq O (\Delta^2\cdot |\Gamma|)$, since $\mu=\Theta(\Delta\cdot \log^{1.5}n)$.
\end{proof}

Consider now the graph $H'=H\setminus (\tilde E_1\cup \tilde E_2\cup \tilde E_3)$. 
From the definition of $\tilde E_1, \tilde E_2, \tilde E_3$, it is immediate that, for every path $Q_i\in \Sigma$,
$\out_G(V(H(Q_i)))\subseteq E'\cup \tilde E_1\cup \tilde E_2\cup \tilde E_3$. Therefore, for every connected component $C$ of $H'$, either $V(C)\subseteq V(H(Q_i))$, or $V(C)\cap V(H(Q_i))=\emptyset$. We let $\cset'_1$ contain all connected components $C$ of $H'$, such that $V(C)\subseteq V(H(Q_i))$ for some path $Q_i\in \Sigma$.
\begin{observation}
Each connected component $\cset'_1$ contains at most $\mu$ vertices that are endpoints of edges in $E'\cup \tilde E_1\cup \tilde E_2\cup \tilde E_3$.
\end{observation}
\begin{proof}
Let $C\in \cset'_1$ be any component, and let $Q_i\in \Sigma$ be the path such that $V(C)\subseteq V(H(Q_i))$. 
Let $S(Q_i)$ be the set of all vertices $v\in V(Q_i)$, such that $v\in U\cup \Gamma\cup \eta(Q_i)$. From the construction of the path $Q_i$ and the definition of $w(Q_i)$, $|S(Q_i)|\leq w(Q_i)+4\leq \mu/(2\Delta)+4$. Notice that a vertex $v$ of $H(Q_i)$ may be an endpoint of an edge in $E'\cup \tilde E_1\cup \tilde E_2\cup \tilde E_3$ iff $v\in S(Q_i)$, or $v$ has a neighbor that lies in $S(Q_i)$. Therefore, the total number of vertices of $C$ that may be incident to edges of  $E'\cup \tilde E_1\cup \tilde E_2\cup \tilde E_3$ is at most $(\Delta+1)|S(Q_i)|\leq (\Delta+1)\cdot (\mu/(2\Delta)+4)\leq \mu$.
\end{proof}

\paragraph{Set $\tilde E_4$.}
We now define the set $\tilde E_4$ of edges. 
Let $\bset''\subseteq \bset'$ be the set of pseudo-blocks with $|V(\tilde B)|>3$. %Note that all pseudo-blocks in $\bset'\setminus \bset''$ are good, since for each pseudo-block $B\in \bset'\setminus \bset''$, graph $\tilde B$ is isomorphic to $K_3$.
Recall that, for every pseudo-block $B\in \bset''$, $\tilde B'$ is the graph obtained from $\tilde B$ by deleting all its fake edges. %Note that the unique planar drawing $\rho_{\tilde B}$ of $B$ induces a drawing of $\tilde B'$, that we denote by $\rho'_{\tilde B'}$. 
We will define, for each pseudo-block $B\in \bset''$, 
a set $E^*(B)$ of edges of $\tilde B'$, that have the following useful property: for every bridge $R\in \rset_{G}(\tilde B')$, all vertices of $L(R)$ lie on the boundary of a single face in the drawing of $\tilde B'\setminus E^*(B)$ induced by $\rho_{\tilde B}$.
(Notice that this property already holds for all pseudo-blocks in $\bset'\setminus \bset''$, since for each pseudo-block $B\in \bset'\setminus \bset''$, graph $\tilde B$ is isomorphic to $K_3$).
 %This property will allow us to establish the Bridge Consistency Property in the type-2 acceptable clusters in our final decomposition. 
%\znote{this sentence does not relate to $E^*(B)$}
We will then set $\tilde E_4=\bigcup_{B\in \bset''}E^*(B)$.

In order to be able to bound $|\tilde E_4|$, we start by setting up an accounting scheme.
We use Lemma~\ref{lem:block_endpoints_routing} to compute, for each pseudo-block $B\in \bset''$, an embedding $\pset_{\tilde B}=\set{P_{\tilde B}(e)\mid e\in A_{\tilde B}}$ of the set $A_{\tilde B}$ of fake edges of $\tilde B$ into paths that are internally disjoint from $\tilde B$ and are mutually internally disjoint. Recall that all paths in $\pset=\bigcup_{ B\in \bset''}\left(\pset_{\tilde B}\setminus \set{P_{\tilde B}(e^*_{\tilde B})} \right)$ cause edge-congestion at most $6$ in $G$, where $e^*_{\tilde B}$ is the fake parent-edge for $B$, that connects the endpoints of $B$ (it is possible that $e^*_{\tilde B}$ is undefined). %\mynote{If the lemma statement remains non-constructive we should add "Our algorithm does not compute the embedding $\pset$ explicitly; we only use it for accounting purposes.". But it's better if it can be made constructive}.

Consider now some pseudo-block $B\in \bset''$. 
%We define a new graph $\tilde B^*$ as follows. We start from graph $\tilde B'$, and then add to it all vertices and edges participating in the paths in $\pset_{\tilde B}$. 
Recall that $\phi^*$ is some fixed optimal drawing of $G$. We define a set $\chi(B)$ of crossings to be the union of (i) all crossings $(e,e')$ in $\phi^*$, such that either $e$ or $e'$ (or both) are real edges of $\tilde B$; and (ii) all crossings $(e,e')$ in $\phi^*$, such that $e,e'$ lie on two distinct paths of $\pset_{\tilde B}$ (that is, $e\in \pset_{\tilde B}(e_1)$, $e'\in \pset_{\tilde B}(e_2)$ and $e_1\neq e_2$ are two distinct fake edges of $\tilde B$). 
We need the following simple observation.
\begin{observation}\label{claim: chi is small for each B}
$\sum_{B\in \bset''}|\chi(B)|\leq O(\optcro(G))$.
\end{observation}
\begin{proof}
Consider any crossing $(e,e')$ in the optimal drawing $\phi^*$ of $G$. Recall that $e,e'$ may belong to $\chi(B)$ in one of two cases: either at least one of $e,e'$ is a real edge of $\tilde B$; or $e,e'$ lie on two distinct paths in $\pset_{\tilde B}$. In particular, in the latter case, one of the two edges $e,e'$ must lie on a path $P_{\tilde B}(\hat e)$, where $\hat e\neq e^*_{\tilde B}$ (that is, $\hat e$ is not the fake parent-edge of $B$). Note that there may be at most one pseudo-block $B\in \bset''$ for which $e$ is a real edge, and the same is true for $e'$. Moreover, there are at most $O(1)$ pseudo-blocks $B\in \bset''$, such that edge $e$ lies on a path of $\pset_{\tilde B}\setminus \set{P_{\tilde B}(e^*_{\tilde B})}$, and the same holds for $e'$. Therefore, there are at most $O(1)$ pseudo-blocks $B\in \bset''$ with $(e,e')\in \chi(B)$.
\end{proof}

In order to construct the sets $\set{E^*(B)}_{B\in \bset''}$, we process the pseudo-blocks in $\bset''$ one by one, using the following lemma.
\begin{lemma}\label{lem: process a block}
	There is an efficient algorithm, that, given a pseudo-block $B\in \bset''$, computes a subset $E^*(B)\subseteq E(\tilde B')$ of edges, such that, if $\rho'$ is the drawing of the graph $\tilde B'\setminus E^*(B)$ induced by the unique planar drawing $\rho_{\tilde B}$ of graph $\tilde B$, then for every bridge $R\in \rset_G(\tilde B')$, all vertices of $L(R)$ lie on the boundary of a single face in $\rho'$. Moreover, $|E^*(B)|\leq O((|\chi(B)|+|\Gamma'(B)|)\cdot \poly(\Delta \log n ))$.
\end{lemma}
Notice that Lemma~\ref{lem: process a block} only considers bridges that are defined with respect to graph $\tilde B'$, namely bridges in $\rset_G(\tilde B')$. Graph $\tilde B'\setminus E^*(B)$ may not even be connected, and its bridges in $G$ can be completely different. However, this weaker property turns out to be sufficient for us. % in order to establish the bridge consistency property for the final type-2 acceptable clusters in our decomposition (for each such cluster, all its real edges and the embeddings of all its fake edges will be contained in a single graph $\tilde B$ with $B\in \bset''$).
We defer the proof of Lemma~\ref{lem: process a block} to Section~\ref{sec:proof_of_process_a_block}.
Now we complete the proof of Theorem~\ref{thm: getting good pseudoblocks} using it.

\iffalse
%----------------------
%----------------------
Let $\tilde E$ be a set of edges that contains, for every block $B\in \bset(H)$, the edge connecting the endpoints of $B$, if such an edge is present in $H$. We need the following observation.

\begin{observation}\label{obs: few edges connecting endpoints}
$|\tilde E|\leq O(|E'|)$.	
\end{observation}
\begin{proof}
	Let $C$ be any connected component of $H$, and let $Z$ be any super-block in $\zset(C)$. Consider the block decomposition $\lset=\lset(Z)$ of $Z$, and let $\tau=\tau(\lset)$ be its corresponding block decomposition. 
\end{proof}
%------------------------
%-----------------------
\fi
%
%
We let $\tilde E_4=\bigcup_{B\in \bset''}E^*(B)$. From Observations~\ref{obs: bound gamma for marked blocks} and~\ref{claim: chi is small for each B},
\[
\begin{split}
|\tilde E_4|&=\sum_{B\in \bset''}|E^*(B)|  
\leq \sum_{B\in \bset''} O\left ((|\chi(B)|+|\Gamma'(B)|)\cdot\poly(\Delta\log n)\right )\\ 
&\leq O((|\Gamma|+\optcro(G))\cdot\poly(\Delta\log n))\leq O((|E'|+\optcro(G))\cdot\poly(\Delta\log n)).\end{split}\]

Lastly, we define the edge set $E_1$ to be the union of edge sets $E'$, $\tilde E_1$, $\tilde E_2$, $\tilde E_3$ and $\tilde E_4$. From the above discussion, $|E_1|\leq O((|E'|+\optcro(G))\cdot\poly(\Delta\log n))$. 

Recall that, in graph $G\setminus (E'\cup \tilde E_1\cup \tilde E_2\cup \tilde E_3)$, all vertices in the set $\eta\cup U$ are isolated. From the definition of $\cset'_1$, it is immediate that for any cluster $C\in \cset_1'$, if a vertex of $C$ is incident to an edge in $\tilde E_4$, then this vertex must lie in $\eta\cup U$,
% this is because $\cset'_1$ contains not-marked blocks, while the edge set $\tilde E_4$ only contains edges incident to nodes in marked blocks
and therefore $C$ contains a single vertex. Therefore, every cluster $C\in \cset_1'$ remains a connected component of $G\setminus E_1$, and it contains at most $\mu$ vertices that are endpoints of edges in $E_1$. 
\iffalse
\paragraph{Summary of Stage 1:}
To summarize, in this stage we have defined a set $E_1$ of $O((|E'|+\optcro(G))\poly(\Delta\cdot \log n))$ edges, with $E'\subseteq E_1$. Let $\Gamma_1\subseteq V(G)$ be the set of all vertices that serve as endpoints of the edges of $E_1$. We have also defined a collection $\cset_1'$ of connected components of $G\setminus E_1$, such that for each $C\in \cset_1'$, $|C\cap \Gamma_1|\leq \mu$. The components in $\cset_1'$ will become type-1 acceptable clusters in the final decomposition, though we may introduce additional type-1 acceptable clusters. We have also defined a collection $\bset''\subseteq \bset(H)$ of blocks in the block decomposition of the graph $H=G\setminus E'$. For every block $B\in \bset''$, no vertex of $\tilde B$ lies in the components of $\cset_1$. Additionally, for every fake edge $e\in \tilde B$, all edges of $G$ that are incident to the endpoints of $e$ belong to $E_1$. Lastly, we ensured that for every bridge $R\in \rset_{\tilde B'}(G)$, there is a face $F$ in the unique drawing of $\tilde B'\setminus E_1$ whose boundary contains all vertices of $L(R)$ (we have achieved this property for the set $E^*(B)$ of edges, but augmenting the set $E^*(B)$ of edges to $E_1$ cannot destroy this property).
\fi

Denote $H_1=G\setminus E_1$. 
Consider now a connected component $C$ of $H_1$ with $C\notin\cset'_1$, and a block $B$ in the block decomposition $\bset(C)$ of $C$. Clearly, there is a block $B_0\in\bset'$ such that $\tilde B_0'$ contains $B$ as a subgraph (this is because we have deleted all edges incident to vertices that serve as endoints of fake edges of every graph $\tilde B_a$, for $B_a\in \bset'$). We then let $\hat\psi_{B}$ be the drawing of $B$ induced by $\rho_{\tilde B_0}$, the unique planar drawing of $\tilde B_0$.
It is now enough to prove the following claim.
\begin{claim}\label{claim: every block is good}
For every connected component $C$ of $H_1$ with $C\not\in\cset_1'$, every pseudo-block $B$ in the block decomposition $\bset(C)$ of $C$ is good, with $\hat{\psi}_{B}$ being its associated drawing.
\end{claim}
%\vspace{-2mm}
\begin{proof}
Assume for contradiction that the claim is false. Let $B\in \bset(C)$ be a bad pseudo-block, and
%, so there must be some pseudo-block $B_1\in \bset_{H_1}$ that is bad. 
let $R\in \rset_G(B)$ be a bridge that is a witness for $B$ and drawing $\hat \psi_B$. %Note that $|V(B_1)|>3$.
Recall that there is a block $B_0\in \bset$ with $B\subseteq \tilde B_0'$. For brevity, we denote $\rho=\rho_{\tilde B_0}$, $\rho'=\rho'_{\tilde B'_0}$, and denote by $\psi=\hat\psi_{B}$ the drawing of $B$ induced by $\rho'$.

Let $\fset$ be the set of faces in the drawing $\psi$ of $B$. From Lemma~\ref{lem: process a block}, for every bridge $R'\in \rset_G(\tilde B_0')$, if $L(R')\cap V(B)\ne \emptyset$, then there is some face $F\in \fset$, such that all vertices of $L(R')\cap V(B)$ lie on the boundary of $F$ in the drawing $\psi$. (If $B_0\in \bset'\setminus \bset''$, so $\tilde B_0$ is isomorphic to $K_3$, this property must also hold).
On the other hand,
for every vertex $v\in V(\tilde B_0)\setminus V(B)$, there is a unique face $F(v)\in \fset$, such that the image of vertex $v$ in $\rho$ lies in the interior of the face $F(v)$.

Consider now the witness bridge $R$ for $B$.
Recall that $T_R\subseteq R$ is a tree whose leaves are precisely the vertices of $L(R)$. 
Assume first that $V(T_R)\cap V(\tilde B_0)=L(R)$. In this case, there is some bridge $R'\in \rset_G(\tilde B_0')$ that contains the tree $T_R$, so $L(R)\subseteq L(R')$. However, from Lemma~\ref{lem: process a block}, all vertices of $L(R')$ lie on the boundary of the same face in the drawing $\rho'$, and therefore they also lie on the boundary of the same face in the drawing $\psi$. This leads to a contradiction to $R$ being a witness bridge for $B$ and $\psi$. %Since $B_2$ is a good block, there must be a face in the drawing $\rho'_{\tilde B_2'}$ of $\tilde B'_2$ whose boundary contains all vertices of $L(R)$.

Assume now that there is some vertex $v\in V(T_R)\cap  V(\tilde B_0)$ that does not lie in $L(R)$. %Let $F=F(v)$ be the face of $\fset$ in whose interior the image of $v$ in $\rho_{\tilde B_2}$ lies. 
We will show that all vertices of $L(R)$ must lie on the boundary of $F(v)$, again leading to a contradiction. Let $u$ be an arbitrary vertex of $L(R)$. %It suffices to show that $u$ must lie on the boundary of the face $F(v)$.
Let $P\subseteq T_R$ be the unique path connecting $v$ to $u$ in $T_R$. Since the leaves of tree $T_R$ are precisely the vertices of $L(R)$, except for $v$, every vertex $x$ of $P$ lies outside $V(B)$, and, if $x\in V(\tilde B_0)$, then the image of $x$ in $\rho$ lies in the interior of some face in $\fset$. Let $v=v_1,v_2,\ldots,v_r=u$ be all vertices of $P$ that belong to $V(\tilde B_0)$, and assume that they appear on $P$ in this order. 
It remains to prove the following observation.

\begin{observation}
For all $1\leq i<r-1$, $F(v_i)=F(v_{i+1})$. Moreover, vertex $v_r$ lies on the boundary of face $F(v_{r-1})$.
\end{observation}
\begin{proof}
Fix some index $1\leq i\leq r-1$. Assume for contradiction that the observation is false.
Then there is some face $F'\in \fset$, such that $v_i$ lies in the interior of $F'$, but $v_{i+1}$ does not lie in the interior or on the boundary of $F'$ (the latter case is only relevant for $i=r-1$). Since the boundary of $F'$ separates $v_i$ from $v_{i+1}$, they  cannot lie on the boundary of the same face in the drawing $\rho'$ of $\tilde B_0'$. 
Let $\sigma_i$ be the subpath of $P$ between $v_i$ and $v_{i+1}$. If $\sigma_i$ consists of a single edge $(v_i,v_{i+1})$, then it is either a bridge in $\rset_G(\tilde B_0')$, or it is an edge of $\tilde B_0'$. In the former case, Lemma \ref{lem: process a block} ensures that the endpoints of $\sigma_i$  may not be separated by the boundary of a face of drawing $\psi$. In the latter case, the same holds since the drawing $\psi$ is planar. In either case, we reach a contradiction. Otherwise, there exists a bridge $R'\in \rset_G(\tilde B_2')$ containing $\sigma'_i$, with $v_{i},v_{i+1}\in L(R')$. However, from Lemma \ref{lem: process a block}, it is impossible that the boundary of a face in the drawing $\psi$ separates the two vertices, a contradiction. 
	%	We conclude that for all $1\leq i<r$, there must be a face $F_i$ in the drawing $\rho'_{\tilde B_2'}$ of $\tilde B_2'$ such that $v_i,v_{i+1}$ lie on the boundary of $F_i$. Notice that, since $\tilde B_1'\subseteq \tilde B_2'$, and from the way the drawing $\rho'_{\tilde B_1'}$ was defined, there must be a face $F'_i$ of $\rho'_{\tilde B_1'}$ that contains $F_i$.	
\end{proof}	
\end{proof}

%In order to complete the proof of Theorem \ref{thm: getting good pseudoblocks} it suffices to prove Lemma \ref{lem: process a block}, which we do next.

%-------------------------------------------------------
%-------------------------------------------------------
%-------------------------------------------------------
\subsection{Proof of Lemma \ref{lem: process a block}}
\label{sec:proof_of_process_a_block}
%-------------------------------------------------------
%-------------------------------------------------------
%-------------------------------------------------------
%-------------------------------------------------------
In this section we provide the proof of Lemma~\ref{lem: process a block}. 
We fix a pseudo-block $B\in \bset''$ throughout the proof. For brevity, we denote $\Gamma'=\Gamma'(B)$ and we denote by $\rho=\rho_{\tilde B}$ the unique planar drawing of $\tilde B$.
%, and $\rho'$ is the drawing of $\tilde B'$ that is induced by $\rho$. 
Recall that vertex set $\Gamma'$ contains all terminals of $\Gamma$ (that is, endpoints of edges of $E'$) lying in $\tilde B$, all endpoints of all fake edges of $\tilde B$, and all separator vertices in $U\cap V(\tilde B)$. The set $\Gamma'$ of vertices remains fixed throughout the algorithm. Abusing the notation, we call the vertices of $\Gamma'$ \emph{terminals} throughout the proof.
	
Throughout the algorithm, we maintain a subgraph $J$ of $\tilde B$ and gradually construct the set $E^*(B)$. Initially, we set $E^*(B)=\emptyset$ and $J=\tilde B'$, the graph obtained from $\tilde B$ by deleting all its fake edges. Over the course of the algorithm, we will remove some edges from $J$ and add them to the set $E^*(B)$. We will always use $\rho'$ to denote the drawing of the current graph $J$ induced by the drawing $\rho$ of $\tilde B$. 
%We note that graph $J$ may become disconnected over the course of the algorithm, but its drawing $\rho'$ is always fixed to be the drawing of $J$ induced by $\rho$.

Notice that a terminal of $\Gamma'$ may belong to the boundaries of several faces in the drawing $\rho'$, which is somewhat inconvenient for us. As our first step, we remove all edges that are incident to the terminals in $\Gamma'$ from $J$, and add them to $E^*(B)$. Notice that now each terminal of $\Gamma'$ becomes an isolated vertex
%but its drawing in $\rho'$ is fixed, as $\rho'$ is the unique drawing of the current graph $J$ induced by the drawing $\rho$ of $\tilde B$ (throughout this proof, we only refer to the vertices in the original set $\Gamma'$ as terminals). 
and lies on the (inner) boundary of exactly one face of the current drawing $\rho'$. Clearly, $|E^*(B)|\leq \Delta\cdot |\Gamma'|$.

From now on, we denote by $\fset$ the set of all faces in the drawing $\rho'$ of the current graph $J$.
For every terminal $t\in \Gamma'$, there is a unique face $F(t)\in \fset$, such that $t$ lies on the (inner) boundary of $F(t)$. For a face $F\in \fset$, we denote by $\Gamma(F)\subseteq \Gamma'$ the set of all terminals $t$ with $F(t)=F$.

\paragraph{Bad Faces.} 
We denote by $\rset=\rset_{G}(\tilde B')$ the set of bridges for the graph $\tilde B'$ in $G$.
For each bridge $R\in \rset$, all vertices in $L(R)$ must be terminals. Let $\fset(R)=\set{F\in \fset\mid L(R)\cap \Gamma(F)\neq \emptyset}$ be the set of all faces in the drawing $\rho'$ of the current graph $J$, whose inner boundaries contain terminals of $L(R)$. We say that a bridge $R\in \rset$ is \emph{bad} for $J$ iff $|\fset(R)|>1$, namely, not all vertices of $L(R)$ lie on the boundary of the same face of $\fset$. In such a case, we say that every face in $\fset(R)$ is a \emph{bad face}, and for each face $F\in \fset(R)$, we say that bridge $R$ is \emph{responsible for $F$ being bad}. As the algorithm progresses and the graph $J$ changes, so does the set $\fset$. The set $\rset$ of bridges does not change over the course of the algorithm, and the definitions of the sets $\fset(R)$ of faces for $R\in \rset$ and of bad faces are always with respect to the current graph $J$ and its drawing. 
The main subroutine that we use in our algorithm is summarized in the following lemma.
\begin{lemma}\label{lem: one step for bad faces}
	There is an efficient algorithm, that, given the current graph $J$ and its drawing $\rho'$, computes a subset $\hat E$ of at most $O\left((|\chi(B)|+|\Gamma'|)\cdot \poly(\Delta\log n)\right)$ edges, such that, if $n_1$ is the number of bad faces in the drawing $\rho'$ of $J$, and $n_2$ is the number of bad faces in the drawing of $J\setminus \hat E$ induced by $\rho'$, then $n_2\leq n_1/2$.
\end{lemma}
It is easy to complete the proof of Lemma \ref{lem: process a block} using Lemma \ref{lem: one step for bad faces}. As long as the  drawing $\rho'$ of the current graph $J$ contains bad faces (note that the number of bad faces is always either $0$ or at least $2$), we apply the algorithm from Lemma \ref{lem: one step for bad faces} to graph $J$ to compute a set $\hat E$ of edges, then delete the edges of $\hat E$ from $J$ and add them to $E^*(B)$, and continue to the next iteration. Once the drawing $\rho'$ of the current graph $J$ contains no bad faces, the algorithm terminates. It is easy to see that the number of iterations in the algorithm is $O(\log n)$. Therefore, at the end, $|E^*(B)|\leq O\left ((|\chi(B)|+|\Gamma'|)\cdot \poly(\Delta\log n)\right )$.
Consider now the graph $J$ obtained at the end of the algorithm. Since the drawing $\rho'$ of $J$ contains no bad faces, for every bridge $R\in \rset$, there is a single face of $\rho'$ whose boundary contains all vertices of $L(R)$ (we emphasize that the graph $J$ is not connected, and the vertices of $L(R)$ are isolated since they are terminals; but the drawing of each such vertex and the face to whose boundary it belongs are fixed by the original drawing of $\tilde B'$ induced by $\rho$). In order to complete the proof of Lemma \ref{lem: process a block}, it suffices to prove Lemma \ref{lem: one step for bad faces}.

From now on we focus on the proof of Lemma \ref{lem: one step for bad faces}.
Throughout the proof, we fix the drawing $\rho'$ of the current graph $J$.
Consider a pair $F,F'$ of faces in $\fset$. Let $P$ be the shortest path connecting $F$ to $F'$ in the dual graph of $J$ with respect to $\rho'$. This path defines a curve $\gamma(F,F')$, that starts at the interior of $F$, terminates at the interior of $F'$, and intersects the image of $J$ only at edges. Let $E(\gamma(F,F'))$ be the set of all edges whose image intersects $\gamma(F,F')$. Equivalently, $\gamma(F,F')$ can viewed as the curve that, among all curves $\gamma$ connecting a point in the interior of $F$ to a point in the interior of $F'$ that only intersects the image of $J$ at its edges, minimizes $|E(\gamma)|$. We define the distance between $F$ and $F'$ to be $\dist(F,F')=|E(\gamma(F,F'))|$. Equivalently, $\dist(F,F')$ is the minimum cardinality of a set $\tilde E\subseteq E(J)$ of edges, such that, in the drawing of $J\setminus \tilde E$ induced by $\rho'$, the faces $F$ and $F'$ are merged into a single face.

Let $\fset'\subseteq \fset$ be the set of bad faces. For each $F\in \fset'$, denote $\hat F=\arg\min_{F'\in \fset'}\set{\dist(F, F')}$. We also denote $\Pi=\set{(F,\hat F)\mid F\in \fset'}$. 
%Note that a bad face may participate in several such pairs. 
We define $\hat E=\bigcup_{F\in \fset'}E(\gamma(F,\hat F))$. In other words, set $\hat E$ contains, for every pair $(F,\hat F)\in \Pi$, the set $E(\gamma(F,\hat F))$ of edges.
% -- the minimum-cardinality set of edges whose removal from $J$ results in a merge of these faces. 
Notice that, since a bad face may participate in several pairs in $\Pi$, it is possible that more than two faces may be merged into a single face. We remove the edges of $\hat E$ from $J$. Note that no new bad faces may be created, since the bad faces are only defined with respect to the original set $\rset_{G}(\tilde B')$ of bridges. Therefore,  the number of bad faces decreases by at least a factor of $2$. It now remains to show that $|\hat E|\leq O\left ((|\Gamma'|+|\chi(B)|)\cdot \poly(\Delta\log n)\right )$. 
This is done in the next claim, whose proof completes the proof of Lemma \ref{lem: one step for bad faces}.
\begin{claim}
\label{claim:bound on hatE}
$|\hat E|\leq O\left((|\Gamma'|+|\chi(B)|)\cdot \Delta\log n\right)$. 
\end{claim}
\begin{proof}
For each $F\in \fset'$, we denote $c(F)=\dist(F,\hat F)$. To prove Claim~\ref{claim:bound on hatE}, it is sufficient to show that 
$\sum_{F\in \fset'}c(F)\leq O\left((|\Gamma'|+|\chi(B)|)\cdot \Delta\log n\right)$.
We partition the set $\fset'$ of bad faces into $O(\log n)$ classes $\fset'_1,\ldots,\fset'_z$, for $z\leq O(\log n)$ as follows. For each $1\leq i\leq z$, face $F\in \fset'$ lies in class $\fset'_i$ iff $2^i\leq c(F)<2^{i+1}$. Clearly, there must be an index $i^*$, such that $\sum_{F\in \fset'_{i^*}}c(F)\geq \sum_{F'\in \fset'}c(F')/O(\log n)$. We denote $\fset'_{i^*}=\fset^*$ and $c^*=2^{i^*}$. Therefore, for every face $F\in \fset^*$, $c^*\leq c(F)<2c^*$. 
%Notice that we can assume that $c^*$ is greater than a large enough constant. 
Since each bad face contains at least one terminal of $\Gamma'$ on its inner boundary, $|F^*|\leq |\Gamma'|$. Therefore, if $c^*$ is upper bounded by a constant, then $\sum_{F'\in \fset'}c(F')\leq O(c^*\cdot |\fset^*|\cdot \log n)\leq O(|\Gamma'|\log n)$. We will assume from now on that $c^*$ is greater than some large constant. We use the following claim.
\begin{claim}\label{claim: Fstar is large enough}
 $|\fset^*|\leq O(|\chi(B)|\cdot \Delta/c^*)$. 
\end{claim}
Note that Claim~\ref{claim: Fstar is large enough} completes the proof of Claim~\ref{claim:bound on hatE}, since
\[\sum_{F'\in \fset'}c(F')\leq O(\log n)\cdot \sum_{F\in \fset^*}c(F)\leq O(|\fset^*|\cdot c^*\cdot \log n)\leq O(|\chi(B)|\Delta\log n).
\]
From now on we focus on the proof of Claim \ref{claim: Fstar is large enough}. 
The main idea of the proof is to associate, with each face $F\in \fset^*$, a collection $\chi^F\subseteq \chi(B)$ of $\Omega(c^*/\Delta)$ crossings of $\chi(B)$, such that each crossing in $\chi(B)$ appears in at most $O(1)$ sets of $\set{\chi_F}_{F\in \fset^*}$. Clearly, this implies that $|\fset^*|\leq O(|\chi(B)|\Delta/c^*)$. In order to define the sets $\chi^F$ of crossings, we carefully construct a witness graph $W(F)\subseteq \tilde B$ for each face $F\in \fset^*$, such that, for every pair $F,F'\in \fset^*$ of distinct faces, graphs $W(F)$ and $W(F')$ are disjoint. We define the set $\chi^F$ of crossings for the face $F$ by carefully considering the crossings in which the edges of $W(F)$ participate in the optimal drawing $\phi^*$ of $G$. %The witness graph $W(F)$ are defined in three steps. First, we compute a ``shell'' around the face $F$, which is a collection of disjoint subgraphs of $J$. The shell is used in order to compute a collection $\yset(F)$ of $\Omega(c^*/\Delta)$ disjoint cycles, which are then in turn used to define the graph $W(F)$.
The remainder of the proof consists of three steps. In the first step, we define a ``shell'' around each face $F$. In the second step, we use the shells in order to define the witness graphs $\set{W(F)}_{F\in \fset^*}$. In the third and the last step, we use the witness graphs in order to define the collections $\chi^F\subseteq \chi(B)$ of crossings associated with each face $F\in \fset^*$.
%---------------------------
%---------------------------
%---------------------------
%---------------------------
%---------------------------
%---------------------------
%\vspace{-3mm}
\paragraph{Step 1: Defining the Shells.}
%---------------------------
%---------------------------
%---------------------------
%---------------------------
%---------------------------
%---------------------------
%---------------------------
%---------------------------
%---------------------------
We denote $z=\floor{c^*/(16\Delta)}$. In this step we define, for every face $F\in \fset^*$, a shell $\hset(F)=\set{L_1(F),\ldots,L_z(F)}$, which is a collection  of $z$ disjoint subgraphs $L_1(F),\ldots,L_z(F)$ of $J$, that we refer to as \emph{layers}. 

We now fix a face $F\in \fset^*$ and  define its shell $\hset(F)$ inductively, as follows. We consider the drawing $\rho'$ of the graph $J$, and we view the face $F$ as the outer face of the drawing. We let $L_1(F)$ be the boundary of the face $F$ (note that this boundary may not be connected). 
%If we let $p^*$ be any point in the interior of $F$, then a vertex $v$ belongs to $L_1(F)$ iff there is a curve $\gamma$ connecting $p^*$ to the image of $v$, such that $\gamma$ only intersects $J$ at its endpoint $v$. Similarly, an edge $e$ belongs to $L_1(F)$ iff there is a curve $\gamma$ connecting $p^*$ to a point on the image of $e$,  such that $\gamma$ only intersects $J$ at its endpoint.
Assume now that we have defined layers $L_1(F),\ldots,L_{i-1}(F)$. In order to define $L_i(F)$, we again consider the drawing $\rho'$ of $J$, with $F$ being its outer face, and we delete from this drawing the images of all vertices  of $L_1(F),\ldots,L_{i-1}(F)$, and of all edges that are incident to these vertices. We then let $L_i(F)$ be the boundary of the outer face in the resulting plane graph.
This completes the definition of the shell $\hset(F)=\set{L_1(F),\ldots,L_z(F)}$.
 We denote $\Lambda(F)=\bigcup_{i=1}^zL_i(F)$. 
 
In order to analyze the properties of the shells, we need the following notion of $J$-normal curves.
\begin{definition}
Given a plane graph $\hat J$ and a curve $\gamma$, we say that $\gamma$ is a $\hat J$-normal curve iff it intersects the image of $\hat J$ only at the images of its vertices.
\end{definition}
We state some simple properties of the shells in the next observation and its two corollaries. 
\begin{observation}\label{obs: J-normal curve for shell}
Let	$\hset(F)=\set{L_1(F),\ldots,L_z(F)}$ be a shell for some face $F\in \fset^*$. Then for each $1\leq i\leq z$, for every vertex $v\in L_i(F)$, there is a $J$-normal curve $\gamma$ connecting $v$ to a point in the interior of $F$, such that $\gamma$ intersects exactly $i$ vertices of $J$ -- one vertex from each graph $L_1(F),\ldots,L_i(F)$.
\end{observation}
\begin{proof}
It suffices to show that, for each $1< j\leq z$ and for every vertex $v'\in L_j(F)$, there is a vertex $v''\in L_{j-1}(F)$ and a curve $\gamma_j$ connecting $v'$ to $v''$, intersecting $J$ only at its endpoints. The existence of the curve $\gamma_j$ follows immediately from the definition of $L_j(F)$. 
Indeed, consider the drawing obtained from $\rho'$ after we delete the images all vertices of $L_1(F),\ldots,L_{j-2}(F)$ together with all incident edges from it. Then there must be a face $F'$ in the resulting drawing, that contains $v'$ on its boundary, and also contains, on its boundary, another vertex $v''\in L_{j-1}(F)$. This is because $v'$ does not lie on the boundary of the outer face in the current drawing, but it lies on the boundary of the outer face in the drawing obtained from the current one by deleting all vertices of $L_{j-1}(F)$ and all incident edges from it. Therefore, there must be a curve $\gamma_j$ connecting $v'$ to $v''$, that is contained in $F'$ and intersects $J$ only at its endpoints.
\end{proof}
\begin{corollary}\label{cor: no faces captured by shell}
Let $F\in \fset^*$ and let $F'\neq F$ be any other bad face in $\fset'$. Then graph $\Lambda(F)$ contains no vertex that lies on the boundary of $F'$.
\end{corollary}
\begin{proof}
Assume for contradiction that the claim is false, and let $v\in \Lambda(F)$ be a vertex that lies on the boundary of $F'$. Then from Observation \ref{obs: J-normal curve for shell}, there is a $J$-normal curve $\gamma$ in the drawing $\rho'$ of $J$, connecting $v$ to a point in the interior of $F$, such that $\gamma$ intersects at most $z\leq c^*/(16\Delta)$ vertices of $J$. Let $U\subseteq V(J)$ be the set of these vertices. By slightly adjusting $\gamma$ we can ensure that it originates in the interior of $F'$, terminates in the interior of $F$, does not intersect any vertices of $J$, and only intersects those edges of $J$ that are incident to the vertices of $U$. But then $\dist(F,F')\leq |U|\cdot \Delta<c^*\leq c(F)$, a contradiction to the definition of $c(F)$.
\end{proof}

Notice that, since $|\fset'|\geq 2$, Corollary \ref{cor: no faces captured by shell} implies, that for every face $F\in \fset^*$, $L_z(F)$ is non-empty.
\begin{corollary}\label{cor:every pair of shells disjoint}
	Let $F,F'\in \fset^*$ be two distinct faces. Then $\Lambda(F)\cap \Lambda(F')=\emptyset$.
\end{corollary}
\begin{proof}
	Assume for contradiction that the claim is false, and let $v$ be a vertex that lies in both $\Lambda(F)$ and $\Lambda(F')$.  Then from Observation \ref{obs: J-normal curve for shell}, there is a $J$-normal curve $\gamma_1$ in the drawing $\rho'$ of $J$, connecting $v$ to a point in the interior of $F$, such that $\gamma_1$ contains images of at most $z\leq c^*/(16\Delta)$ vertices of $J$. 
	Similarly, there is a $J$-normal curve $\gamma_2$, connecting $v$ to a point in the interior of $F'$, such that $\gamma_2$ contains images of at most $c^*/(16\Delta)$ vertices of $J$. Let $U$ be the set of all vertices of $J$ whose images lie on either $\gamma_1$ or $\gamma_2$.
	By concatenating the two curves and slightly adjusting the resulting curve, we can obtain a curve that originates in the interior of $F$, terminates in the interior of $F'$, does not intersect any vertices of $J$, and only intersects those edges of $J$ that are incident to the vertices of $U$. But then $\dist(F,F')\leq |U|\cdot \Delta<c^*\leq c(F)$, a contradiction to the definition of $c(F)$.
\end{proof}
%---------------------------
%---------------------------
%---------------------------
%---------------------------
%---------------------------
%---------------------------
%\vspace{-3mm}
\paragraph{Step 2: Computing the Witness Graphs.}
%---------------------------
%---------------------------
%---------------------------
%---------------------------
%---------------------------
%---------------------------
%---------------------------
%---------------------------
%---------------------------
In this step we compute, for every face $F\in \fset^*$, its witness graph $W(F)$. The witness graph $W(F)$ consists of two parts. The first part is a collection $\yset(F)=\set{Y_1(F),\ldots,Y_{z-3}(F)}$ of $z-3$ vertex-disjoint cycles. We will ensure that $Y_i(F)\subseteq L_i(F)$ for all $1\leq i\leq z-3$. The second part is a collection $\qset(F)=\set{Q_1(F),Q_2(F),Q_3(F)}$ of three vertex-disjoint paths in graph $\tilde B$, each of which has a non-empty intersection with each cycle in $\yset(F)$. Graph $W(F)$ is defined to be the union of the cycles in $\yset(F)$ and the paths in $\qset(F)$. The main challenge in this part is to define the path sets $\set{\qset(F)}_{F\in \fset^*}$ so that the resulting witness graphs are disjoint.

We now fix a face $F\in \fset^*$, and we start by defining the collection $\yset(F)=\set{Y_1(F),\ldots,Y_{z-3}(F)}$ of cycles for it. Let $R\in \rset$ be the bad bridge that is responsible for $F$ being a bad face. Then there must be a bad face $\notF\in \fset'$, such that $F\neq \notF$, and at least one vertex of $L(R)$ lies on the (inner) boundary of $\notF$.
%(here we use the fact that all vertices of $L(R)$ are terminals and hence they are isolated vertices in $J$, and each such vertex lies on a boundary of a single face of $\rho'$). 
Let $P^*(F)$ be any path that is contained in $R$ and connects a vertex of $L(R)$ on the boundary of $F$ to a vertex of $L(R)$ on the boundary of $\notF$. Note that path $P^*(F)$ is internally disjoint from $\tilde B'$.

From Corollary \ref{cor: no faces captured by shell}, no vertex of $\notF$ may lie in $\Lambda(F)$. Therefore, for all $1\leq i\leq z$, there is a simple cycle $Y_i(F)\subseteq L_i(F)$ that separates $\notF$ from $F$. In other words, if we denote by $D(Y_i(F))$ the unique disc in the drawing $\rho'$ with $F$ as the outer face of the drawing, whose boundary is $Y_i(F)$, then $\notF\subseteq D(Y_i(F))$, and $F$ is disjoint from the interior of $D(Y_i(F))$. Clearly, the boundary of $\notF$ is contained in $D(Y_z(F))$, and $D(Y_z(F))\subsetneq D(Y_{z-1}(F))\subsetneq \cdots\subsetneq D(Y_1(F))$, while the boundary of $F$ is disjoint from the interior of $D(Y_1(F))$. We let $\yset(F)=\set{Y_1(F),\ldots,Y_{z-3}(F)}$ be the collection of disjoint cycles associated with $F$ (we note that we exclude that last three cycles $Y_{z-2}(F),\ldots,Y_z(F)$ on purpose).

It remains to define a collection $\qset(F)$ of three disjoint paths in graph $\tilde B$, each of which connects a vertex of $Y_1(F)$ to a vertex of $Y_{z-3}(F)$. We emphasize that, while the cycles in $\yset(F)$ are all contained in the current graph $J\subseteq \tilde B'$ that only contains real edges of $\tilde B$ that have not been deleted yet,
the paths in $\qset(F)$ are defined in graph $\tilde B$ and are allowed to contain fake edges. Since graph $\tilde B$ is $3$-connected, it is not hard to see that such a collection $\qset(F)$ of paths must exist. However, we would like to ensure that all paths in the set $\bigcup_{F'\in \fset^*}\qset(F')$ are mutually vertex-disjoint. In order to achieve this, we show in the next claim that there exist a desired set $\qset(F)$ of paths that only uses vertices in $\Lambda(F)$. Since all graphs in $\set{\Lambda(F')}_{F'\in \fset^*}$ are mutually vertex-disjoint, the path sets $\set{\qset(F')}_{F'\in \fset^*}$ are also mutually vertex-disjoint. The proof uses the fact that we have left the ``padding'' of three layers $L_{z-2}(F),L_{z-1}(F),L_z(F)$ between the cycles in $\yset(F)$ and $J\setminus \Lambda$.
\begin{claim}\label{claim: find the paths}
For every face $F\in \fset^*$, there is a collection $\qset(F)$ of three vertex-disjoint paths in graph $\tilde B$, where each path connects a vertex of $Y_1(F)$ to a vertex of $Y_{z-3}(F)$, and only contains vertices of $\Lambda(F)$.
\end{claim}
\begin{proof}
Fix some face $F\in \fset^*$. 
Since $\tilde B$ is $3$-connected, there must be a collection $\qset$ of three vertex-disjoint paths in graph $\tilde B$, each of which connects a vertex of $Y_1(F)$ to a vertex of $Y_{z-3}(F)$.
%(otherwise, we could disconnect $\tilde B$ by removing two vertices from it, contradicting the fact that it is $3$-connected). 
Among all such sets $\qset$ of paths we select the one that minimizes the number of vertices of $V(\tilde B)\setminus V(\Lambda(F))$ that belong to the paths in $\qset$. We now claim that no vertex of $V(\tilde B)\setminus V(\Lambda(F))$ may lie on a path in $\qset$. 
	
Assume that this is false, and let $v\in V(\tilde B)\setminus V(\Lambda(F))$ be any vertex that lies on some path in $\qset$. 
Let graph $K$ be the union of graph $\Lambda(F)$ and the paths in $\qset$. 
From the definition of the paths in $\qset$, graph $K\setminus \set{v}$ does not contain three vertex-disjoint paths that connect vertices of $Y_1(F)$ to vertices of $Y_{z-3}(F)$. 
In particular, there are two vertices $x,y\in V(K)\setminus\set{v}$, such that in graph $K\setminus\set{v,x,y}$, there is no path connecting a vertex of $Y_1(F)$ to a vertex of $Y_{z-3}(F)$. 
We will prove that this is false, reaching a contradiction. 	
Notice that each of the vertices $v,x,y$ must lie on a distinct path in $\qset$. We let $Q\in \qset$ be the path that contains $v$, so $Q$ does not contain $x$ or $y$. Notice that, from the definition of shells, for each $z-3<j\leq z$, graph $L_j(F)$ must contain a simple cycle $X_j$ that separates $v$ from every cycle in $\yset(F)$ in the drawing $\rho'$ of $J$. At least one of these three cycles $X\in \set{X_{z-2},X_{z-1},X_z}$ is disjoint from $x$ and $y$. Notice that path $Q$
% in order to reach vertex $v$, 
must intersect the cycle $X$ (this is since the drawing $\rho'$ of $J$ is the drawing induced by the unique planar drawing $\rho_{\tilde B}$ of $\tilde B$, and so $X$ separates the cycles of $\yset(F)$ from $v$ in $\rho_{\tilde B}$ as well). We view the path $Q$ as originating at some vertex $a\in V(Y_1(F))$ and terminating at some vertex $b\in V(Y_{z-3}(F))$. Let $v_1$ be the first vertex of $Q$ that lies on $X$, and let $v_2$ be the last vertex of $Q$ that lies on $X$. Then we can use the segment of $Q$ from $a$ to $v_1$, the cycle $X$, and the segment of $Q$ from $v_2$ to $b$ to construct a path connecting $a$ to $b$ in graph $K$. Moreover, neither of these three graphs may contain a vertex of $\set{x,y,v}$, and so $K\setminus\set{x,y,v}$ contains a path connecting a vertex of $Y_1(F)$ to a vertex of $Y_{z-3}(F)$, a contradiction.
\end{proof}

The witness graph $W(F)$ is defined to be the union of all cycles in $\yset(F)$ and the three paths in $\qset(F)$. Note that, from Corollary \ref{cor:every pair of shells disjoint}, all witness graphs in $\set{W(F)\mid F\in \fset^*}$ are mutually vertex-disjoint. We emphasize that the cycles of $\yset(F)$ only contain real edges of $\tilde B$ (that belong to $J$), while the paths in $\qset(F)$ may contain fake edges of $\tilde B$.
%---------------------------
%---------------------------
%---------------------------
%\vspace{-3mm}
\paragraph{Step 3: Defining the Sets of Crossings.}
%---------------------------
%---------------------------
%---------------------------
The goal of this step is to define, for each face $F\in \fset^*$, a set $\chi^F\subseteq \chi(B)$ of $\Omega(c^*/\Delta)$ crossings, such that each crossing in $\chi(B)$ appears in at most two sets of $\set{\chi_F}_{F\in \fset^*}$. This will imply that $|\fset^*|\leq O(|\chi(B)|\cdot\Delta/c^*)$, thus concluding the proof of Claim \ref{claim: Fstar is large enough}.

We now fix a face $F\in \fset^*$ and define the set $\chi^F$ of crossings. We will first partition the graph $W(F)$ into $z'=\floor{(z-3)/3}$ disjoint subgraphs $W_1(F),\ldots,W_{z'}(F)$, each of which consists of three consecutive cycles in $\yset(F)$, and a set of three paths connecting them. Each such new graph will contribute exactly one crossing to $\chi^F$. Recall that $z=\Theta(c^*/\Delta)$, so $z'=\Theta(c^*/\Delta)$.

We now fix an index $1\leq i\leq z'$, and define the corresponding graph $W_i(F)$. We start with the set $\yset_i(F)=\set{Y_{3i-2}(F),Y_{3i-1}(F),Y_{3i}(F)}$ of three cycles. Additionally, we define a collection $\qset_i(F)$ of three disjoint paths, connecting vertices of $Y_{3i-2}(F)$ to vertices of $Y_{3i}(F)$, as follows.
Consider any of the three paths $Q\in \qset(F)$. We view $Q$ as originating at a vertex $a\in Y_1(F)$ and terminating at a vertex $b\in Y_{z-3}(F)$. From the definition of the cycles, path $Q$ must intersect every cycle in $\yset(F)$. We let $v$ be the last vertex of $Q$ that lies on $Y_{3i-2}(F)$, and we let $v'$ be the first vertex that appears on $Q$ after $v$ and lies on $Y_{3i}(F)$. We let $Q^i$ be the segment of $Q$ between $v$ and $v'$. Notice that $Q^i$ originates at a vertex of $Y_{3i-2}(F)$, terminates at a vertex of $Y_{3i}(F)$, and the inner vertices of $Q_i$ are disjoint from all cycles in $\yset(F)$ except for $Y_{3i-1}$ (that $Q^i$ must intersect). Moreover, in the drawing $\rho'$ of $J$ where $F$ is viewed as the outer face of the drawing, the interior of the image of $Q^i$ is contained in $D(Y_{3i-2}(F))\setminus D(Y_{3i}(F))$.
We let $\qset_i(F)=\set{Q^i\mid Q\in \qset(F)}$ be the resulting set of three paths, containing one segment from each path in $\qset(F)$.
Initially, we let the graph $W_i(F)$ be the union of the cycles in $\yset_i(F)$ and the paths in $\qset_i(F)$. 
Notice that for all $1\leq i<i'\leq z'$, $W_i(F)\cap W_{i'}(F)=\emptyset$. 
For convenience, we rename the three cycles $Y_{3i-2}(F),Y_{3i-1}(F),Y_{3i}(F)$ in $\yset_i$ by $Y^1_i(F),Y^2_i(F)$ and $Y^3_i(F)$, respectively.
Next, we slightly modify the graph $W_i(F)$, as follows. We let $\qset'_i(F)$ be a set of $3$ vertex-disjoint paths in $W_i(F)$
that connect vertices of $Y^1_i(F)$ to vertices of $Y^3_i(F)$ and are internally disjoint from $V(Y^1_i(F))\cup V(Y^3_i(F))$, and among all such paths, we choose those that contain fewest vertices of $V(W_i(F))\setminus \left (\bigcup_{j=1}^3V(Y^j_i(F))\right )$ in total, and, 
subject to this, contain fewest edges of $E(W_i(F))\setminus \left(\bigcup_{j=1}^3E(Y_i(F))\right )$.
Clearly, set $\qset'_i(F)$ of paths is well defined, since we can use the set $\qset_i(F)$ of paths. We discard from $W_i(F)$ all vertices and edges except for those lying on the cycles in $\yset_i(F)$ and on the paths in $\qset_i(F)$. This finishes the definition of the graph $W_i(F)$.

To recap, graph $W_i(F)$ is the union of (i) three cycles $Y_i^1(F),Y_i^2(F)$ and $Y_i^3(F)$; each of the three cycles is contained in graph $J$ and only contains real edges of graph $\tilde B$, and (ii) a set $\qset'_i(F)$ of three disjoint paths, each of which connects a distinct vertex of $Y_i^1(F)$ to a distinct vertex of $Y_i^3(F)$, and is internally disjoint from $V(Y^1_i(F))\cup V(Y^3_i(F))$. Set $\qset'_i(F)$ of paths is chosen to minimize the number of vertices of $V(W_i(F))\setminus \left (\bigcup_{j=1}^3V(Y^j_i(F))\right )$ that lie on the paths.  The paths in $\qset'_i(F)$ are contained in graph $\tilde B$ and may contain fake edges. All resulting graphs $W_i(F)$ for all $F\in \fset^*$ and $1\leq i\leq z'$ are disjoint from each other. 
Note that each such graph $W_i(F)\subseteq \tilde B$ is a planar graph. 
We need the following claim.

\begin{claim}\label{claim: drawing the witness graph}
	For each $F\in \fset^*$ and $1\leq i\leq z'$, if $\psi$ is any planar drawing of $W_i(F)$ on the sphere, and $D,D'$ are the two discs whose boundary is the image of $Y_i^2(F)$, then the images of $Y^1_i(F),Y^3_i(F)$ cannot lie in the same disc in $\set{D,D'}$ (in other words, the image of $Y_i^2(F)$ separates the images of $Y^1_i(F)$ and $Y^3_i(F)$).
\end{claim}

\begin{proof}
%Observe that, from the definition of the cycles in $\yset(F)$, in the unique planar drawing $\rho_{\tilde B}$ of $\tilde B$ on the sphere, the image of $Y_i^2(F)$ separates the images of $Y^1_i(F)$ and $Y^2_i(F)$. Therefore, there exists a planar drawing of  $W_i(F)\subseteq \tilde B$ on the sphere, such that, if $D,D'$ are the two discs whose boundary is the image of $Y_i^2(F)$, then the images of $Y^1_i(F),Y^3_i(F)$ do not lie in the same disc in $\set{D,D'}$. %In order to prove the claim, it is now enough to show that graph $W_i(F)$ has a unique planar drawing.
Let $W$ be the graph obtained from $W_i(F)$ after all degree-$2$ vertices are suppressed. We denote by $Y^1,Y^2$ and $Y^3$ the cycles corresponding to $Y^1_i(F),Y^2_i(F)$ and $Y^3_i(F)$ in $W$ respectively, and we denote by $Q_1,Q_3$ and $Q_3$ the paths corresponding to the paths in $\qset'_i(F)$ in $W$. Notice that every vertex of $W$ must lie on one of the cycles $Y^1,Y^2,Y^3$, and on one of the paths $Q_1,Q_2,Q_3$. Moreover, graph $W$ may not have parallel edges (due to the minimality of the set $\qset'_i(F)$ of paths). 
%Drawing $\rho_{\tilde B}$ of $\tilde B$ induces a planar drawing of $W$ on the sphere, such that, if $D,D'$ are the two discs whose boundary is the image of $Y^2$, then the images of $Y^1,Y^3$ do not lie in the same disc in $\set{D,D'}$. Conversely, every planar drawing of $W$ on the sphere induces a corresponding planar drawing of $W_i(F)$ on the sphere.

Observe that, from the definition of the cycles in $\yset(F)$, in the unique planar drawing $\rho_{\tilde B}$ of $\tilde B$ on the sphere, the image of $Y^2$ separates the images of $Y^1$ and $Y^3$. Therefore, there exists a planar drawing of $W$ on the sphere, such that, if $D,D'$ are the two discs whose boundary is the image of $Y^2$, then the images of $Y^1$ and $Y^3$ do not lie in the same disc in $\set{D,D'}$.
Therefore, it suffices to show that $W$ is a $3$-connected graph.

Assume for contradiction that this is not the case, and let $\set{x,y}$ be a pair of vertices of $W$, such that there is a partition $(X,X')$ of $V(W)\setminus\set{x,y}$, with $X,X'\neq \emptyset$, and no edge of $W$ connects a vertex of $X$ to a vertex of $X'$. For brevity, we denote $\yset=\set{Y^1,Y^2,Y^3}$, $\qset=\set{Q_1,Q_2,Q_3}$, and we will sometimes say that a cycle $Y\in \yset$ is contained in $X$ (or in $X'$) if $V(Y)\subseteq X$ (or $V(Y)\subseteq X'$, respectively). We will use a similar convention for paths in $\qset$.
	
We first claim that both $x,y$ must belong to the same cycle of $\yset$. Indeed, assume for contradiction that they belong to different cycles. Then there must be a path $Q\in \qset$ that is disjoint from $x,y$, with all vertices of $Q$ lying in one of the two sets $X,X'$ (say $X$). But, since $Q$ intersects every cycle in $\yset$, for each cycle $Y\in \yset$, all vertices of $V(Y)\setminus\set{x,y}$ lie in $X$ (as $Y\setminus\set{x,y}$ is either a cycle or a connected path). Therefore, $X'=\emptyset$, an contradiction.
	
We denote by $Y$ the cycle in $\yset$ that contain vertices $x,y$.  Note that each of the remaining cycles must be contained in $X$ or contained in $X'$. We now consider two cases.
	
The first case is when there is some path $Q\in \qset$ that contains both vertices $x$ and $y$; assume w.l.o.g. that it is $Q_1$. Since path $Q_2$ is disjoint from $x$ and from $y$, it must be contained in one of the two sets $X,X'$; assume w.l.o.g. that it is $X$. Since $Q_2$ intersects every cycle in $\set{Y^1,Y^2,Y^3}$, and two of these cycles are disjoint from $x,y$, we get that both cycles in $\set{Y^1,Y^2,Y^3}\setminus\set{Y}$ lie in $X$. Since path $Q_3$ is disjoint from $x,y$ but intersects each cycle in $\set{Y^1,Y^2,Y^3}$, it must be contained in $X$ as well. Therefore, every vertex of $X'\cup\set{x,y}$ must belong to $Y\cap Q_1$. But then all vertices of $X'$ must have degree $2$, a contradiction.

\iffalse
Since $Q_1$ is a connected path, and exactly two vertices of $Q_1$ lie in the cut $\set{x,y}$, both endpoints of $Q_1$ must lie in the same set: $X$ or $X'$; we assume that it is $X$. Since the two endpoints of $Q_1$ lie on cycles $Y^1$ and $Y^3$ respectively, one of these cycles (say $Y^1$), that is distinct from $Y$, must be also contained in $X$. The paths $Q_2,Q_3$ are disjoint from the vertices $x,y$, and they both contain at least one vertex of $Y_1$, so both these paths are also contained in $X$. As these paths contain vertices from all three cycles, both cycles of $\yset\setminus\set{Y}$ must be contained in $X$.
	Therefore, $X'$ may only contain vertices of the cycle $Y$, and of the path $Q_1$. Since every vertex of $W$ lies on a cycle in $\yset$ and on a path in $\qset$, graph $W[X'\cup \set{x,y}]$ must be a path, that is a sub-path of the cycle $Y$ and of the path $Q_1$. But since we have suppressed all degree-$2$ vertices, $X'=\emptyset$ must hold, a contradiction.
\fi
	
	It remains to consider the case where $x$ and $y$ lie on two different paths of $\set{Q_1,Q_2,Q_3}$, say $Q_1$ and $Q_2$. Then path $Q_3$ is disjoint from $x,y$, and is contained in one of the sets $X,X'$; assume without loss of generality that it is $X$. Notice that path $Q_3$ contains vertices from all three cycles in $\yset$, therefore, each of the two cycles in $\yset\setminus \set{Y}$ is also contained in $X$. Set $X'$ then contains vertices of a single cycle in $\yset$ -- the cycle $Y$ (that contains the vertices $x$ and $y$). Since the paths $Q_1$ and $Q_2$ connect vertices of $Y^1$ to vertices of $Y^3$, 
	and each of them has one endpoint in $X$ and another in $X'$, $Y\neq Y^2$ must hold. We assume without loss of generality that $Y=Y^1$ (the case where $Y=Y^3$ is symmetric). Therefore, every vertex of $X'\cup \set{x,y}$ lies on cycle $Y^1$, and on either path $Q_1$ or path $Q_2$. However, from our construction of set $\qset'_i(F)$, each path of $\qset$ contains exactly one vertex of $Y_1$ (which serves as its endpoint). Since each of the paths $Q_1,Q_2$ contains one vertex of $\set{x,y}$ that lies on $Y_1$, it follows that $X'=\emptyset$, a contradiction.
	
	We conclude that graph $W$ is $3$-connected and therefore has a unique planar drawing -- the drawing induced by the drawing $\rho_{\tilde B}$ of $\tilde B$. In that drawing (on the sphere), the image of cycle $Y^2$ separates the images of cycles $Y^1$ and $Y^3$. Therefore, in every planar drawing of $W$ on the sphere, the image of $Y^2$ separates the images of $Y^1$ and $Y^3$. Since graph $W_i(F)$ is obtained from $W$ by subdividing some of its edges, in every planar drawing of $W_i(F)$ on the sphere, the image of $Y_i^2(F)$ separates the images of $Y^1_i(F)$ and $Y^3_i(F)$.
\end{proof}

Lastly, we use the following claim to associate a crossing of $\chi(B)$ with the graph $W_i(F)$.
\begin{claim}\label{claim: find one crossing}
Consider some face $F\in \fset^*$ and index $1\leq i\leq z'$. Let $\phi^*$ be the fixed optimal drawing of the graph $G$. Then there is a crossing $(e,e')$ in this drawing, such that:
\begin{itemize}
\item either at least one of the edges $e,e'$ is a real edge of $\tilde B$ that lies in $W_i(F)$; or
\item there are two distinct fake edges $e_1,e_2\in E(\tilde B)$, that belong to $W_i(F)$, such that $e\in P(e_1)$ and $e'\in P(e_2)$, where $P(e_1),P(e_2)\in \pset_{\tilde B}$ are the embeddings of the fake edges $e_1,e_2$, respectively.
\end{itemize}
\end{claim}

Assuming that the claim is correct, the crossing $(e,e')$ must lie in $\chi(B)$. We denote by $\chi_i^F$ the crossing $(e,e')$ obtained by applying Claim \ref{claim: find one crossing} to graph $W_i(F)$. We then set $\chi^F=\set{\chi_i^F\mid 1\leq i\leq z'}$. It is easy to verify that $\chi^F\subseteq \chi(B)$, and that it contains $z'=\Omega(c^*/\Delta)$ distinct crossings. Moreover, since all witness graphs in $\set{W(F)\mid F\in \fset^*}$ are disjoint from each other, 
%for all distinct faces $F,F'\in \fset^*$, $\chi^F\cap \chi^{F'}=\emptyset$. 
each crossing in $\chi(B)$ appears in at most two sets of $\set{\chi_F}_{F\in \fset^*}$.
It remains to prove Claim \ref{claim: find one crossing}.

\begin{proofof}{Claim \ref{claim: find one crossing}}
	We fix a face $F\in \fset^*$ and an index $1\leq i\leq z'$. 
	Consider the fixed optimal drawing $\phi^*$ of graph $G$ on the sphere. If this drawing contains a crossing $(e,e')$, where at least one of the edges $e,e'$ is a real edge of $\tilde B$ that lies in $W_i(F)$, then we are done. Therefore, we assume from now on that this is not the case. In particular, we can assume that the edges of the cycles $\yset_i(F)$ do not participate in any crossings in $\phi^*$ (recall that all edges of these cycles are real edges of $\tilde B$.)
	
	Therefore, the image of cycle $Y^2_i(F)$ in $\phi^*$ is a simple closed curve. Let $D,D'$ be the two discs whose boundaries are $Y^2_i(F)$. We claim that the images of both remaining cycles, $Y^1_i(F),Y^3_i(F)$ must lie inside a single disc in $\set{D,D'}$.

	Recall that we have defined a path $P^*(F)$, that is contained in some bridge $R\in \rset$, and connects some vertex $v$ on the boundary of $F$ to some vertex $v'$ on the boundary of $\notF$. Since both $v,v'$ lie in $\tilde B$, and since graph $\tilde B$ is connected, there is a path $P$ in $\tilde B$ that connects $v$ to $v'$; we view path $P$ as originating from $v$ and terminating at $v'$. Since each cycle $Y\in \yset_i(F)$ separates the boundary of $F$ from the boundary of $\notF$ in the drawing $\rho'$ of $J$, and since $J\subseteq \tilde B$ and $\rho'$ is the drawing of $J$ induced by the planar drawing $\rho_{\tilde B}$ of $\tilde B$, path $P$ must intersect every cycle in $\yset_i(F)$. We let $x$ be the first vertex of $P$ that lies on cycle $Y^1_i(F)$, and $y$ the last vertex on $P$ that lies on cycle $Y^3_i(F)$. Let $P'$ be the sub-path of $P$ connecting $v$ to $x$, and let $P''$ be the sub-path of $P$ connecting $y$ to $v'$. Notice that both paths are internally disjoint from the cycles in $\yset_i(F)$. Next, we denote by $\hat P$ the path that is obtained by concatenating the paths $P',P^*$ and $P''$. Therefore, path $\hat P$ connects a vertex $x\in Y^1_i(F)$ to a vertex $y\in Y^3_i(F)$, and it is internally disjoint from the cycles in $\yset_i(F)$. Note however that path $\hat P$ may contain fake edges of $\tilde B$. For every fake edge $e\in A_{\tilde B}$ that lies on $\hat P$, we replace $e$ with its embedding $P(e)\in \pset_{\tilde B}$ given by using Lemma \ref{lem:block_endpoints_routing}. Recall that the lemma guarantees that the path $P(e)$ is internally disjoint from the vertices of $\tilde B$, and that all paths in $\pset_{\tilde B}=\set{P_{\tilde B}(e')\mid e'\in A_{\tilde B}}$ are mutually internally disjoint. Let $\hat P'$ be the path obtained from $\hat P$ after we replace every fake edge on $\hat P$ with its corresponding embedding path. Notice that $\hat P'$ still connects $x$ to $y$ and it is still internally disjoint from all cycles in $\yset_i(F)$ (that are also present in $G$, as they only contain real edges). If the images of the cycles $Y^1_i(F),Y^3_i(F)$ are contained in distinct discs in $\set{D,D'}$, then the endpoints of the path $\hat P'$ lie on opposite sides of the image of $Y^2_i(F)$. Since path $\hat P'$ is disjoint from cycle $Y^2_i(F)$, at least one edge of $Y^2_i(F)$ must participate in a crossing in $\phi^*$, a contradiction. Therefore, we can assume from now on that the images of both cycles $Y^1_i(F),Y^3_i(F)$ lie inside a single disc in $\set{D,D'}$ (say $D$).

Next, we use the drawing $\phi^*$ of $G$ on the sphere in order to define a corresponding drawing $\phi$ of graph $W_i(F)$, as follows. Recall that every vertex and every real edge of $W_i(F)$ belong to $G$, so their images remain unchanged. Consider now some fake edge $e\in E(W_i(F))$. Let $P(e)$ be the path in $G$ into which this edge was embedded, and let $\gamma(e)$	be the image of this path in $\phi^*$ (obtained by concatenating the images of its edges). If curve $\gamma(e)$ crosses itself then we delete loops from it, until it becomes a simple open curve, and we draw the edge $e$ along the resulting curve. 
Recall that all paths that are used to embed the fake edges of $\tilde B$ are internally disjoint from $V(\tilde B)$ and internally disjoint from each other.

Consider now the resulting drawing $\phi$ of $W_i(F)$. As before, the edges of the cycles in $\yset_i(F)$ do not participate in crossings in $\phi$, and, if we define the discs $D,D'$ as before (the discs whose boundary is $Y^2_i(F)$), then the images of $Y^1_i(F),Y^2_i(F)$ lie in the same disc $D$. From Claim \ref{claim: drawing the witness graph}, the drawing $\phi$ of $W_i(F)$ is not planar. Let $(e_1,e_2)$ be any crossing in this drawing. It is impossible that $e_1$ or $e_2$ are real edges of $W_i(F)$, since we have assumed that no real edges of $W_i(F)$ participate in crossings in $\phi^*$. Therefore, $e_1,e_2$ must be two distinct fake edges of $W_i(F)$, such that there are edges $e\in P(e_1),e'\in P(e_2)$ whose images in $\phi^*$ cross.
\end{proofof}	
\end{proof}

\subsection{Stage 2: Obtaining a Decomposition into Acceptable Clusters}

%-------------------------------------------
%-------------------------------------------
%-------------------------------------------
%-------------------------------------------

In this subsection we complete the proof of Theorem \ref{thm: can find edge set w acceptable clusters}.
 We start with a $3$-connected graph $G$ with maximum vertex degree $\Delta$, and a planarizing set $E'$ of edges of $G$. We then use Theorem \ref{thm: getting good pseudoblocks} to compute a subset $E_1$ of edges of $G$, with $E'\subseteq E_1$, such that $|E_1|\leq O\left((|E'|+\optcro(G))\cdot\poly(\Delta\log n)\right )$, and a set $\cset_1'$ of connected components of $G\setminus E_1$, each of which contains at most $\mu$ vertices that are incident to edges of $E_1$.

The remainder of the algorithm is iterative.
We use a parameter $\alpha'=8\Delta\alpha=\frac{1}{16 \alphasc(n)\log_{3/2} n}$. Recall that $\alpha=\frac{1}{128\Delta \alphasc(n)\log_{3/2} n}$ is the well-linkedness parameter from the definition of type-2 acceptable clusters. Throughout the algorithm, we maintain a set $\hat E$ of edges of $G$, starting with $\hat E=E_1$, and then gradually adding edges to $\hat E$. We also maintain a set $A$ of fake edges, initializing $A=\emptyset$. We denote $\hH=G\setminus \hat E$ with respect to the current set $\hat E$, and we let $\hcset$ be the set of connected components of $\hH\cup A$, that we refer to as \emph{clusters}. We call the endpoints of the edges of $\hat E$ \emph{terminals}, and denote by $\hGamma$ the set of terminals. We will ensure that edges of $A$ only connect pairs of terminals. We will also maintain an embedding $\pset=\set{P(e)\mid e\in A}$ of the fake edges, where for each edge $e=(u,v)\in A$, path $P(e)$ is contained in graph $G$, and it connects $u$ to $v$.  We will ensure that all paths in $\pset$ are mutually internally disjoint.
We also maintain a partition of $\hcset$ into two subsets: set $\cset^A$ of \emph{active} clusters, and set $\cset^I$ of \emph{inactive} clusters.
Set $\cset^I$ of inactive clusters is in turn partitioned into two subsets, $\cset_1^I$ and $\cset_2^I$. We will ensure that every cluster $C\in \cset_1^I$ is a type-1 acceptable cluster. In particular, no edges of $A$ are incident to vertices of clusters in $\cset_1^I$. For every cluster $C\in \cset_2^I$, we denote by $A_C\subseteq A\cap E(C)$ the set of fake edges contained in $C$. We will maintain, together with cluster $C$, a planar drawing $\psi_C$ of $C$ on the sphere, such that  $C$ is a type-2 acceptable cluster with respect to $\psi_C$. Additionally, for every fake edge $e=(x,y)\in A_C$, its embedding $P(e)$ is internally disjoint from $C$. Moreover, we will ensure that there is some cluster $C(e)\in \cset^I_1$ containing $P(e)\setminus \set{x,y}$, and for every pair $e,e'\in A$ of distinct edges, $C(e)\neq C(e')$.  Lastly, we ensure that no fake edges are contained in an active cluster of $\cset^A$.

\paragraph{Vertex Budgets.}
For the sake of accounting, we assign a budget $b(v)$ to every vertex $v$ in $G$.  The budgets are defined as follows. 
If $v\notin \hG$, then $b(v)=0$. 
Assume now that $v\in \hG$, and let $C\in \hcset$ be the unique cluster containing $v$. 
If $C\in \cset^I$, then $b(v)=1$. 
Otherwise, $b(v)=8\Delta\cdot \log_{3/2}(|\hGamma\cap V(C)|)$. 
At the beginning of the algorithm, the total budget of all vertices is $\sum_{t\in \hat\Gamma}b(t)\leq O(|\hat \Gamma|\cdot\Delta\log n)\leq O(|E_1|\cdot\Delta \log n)$. 
Note that, as the algorithm progresses, the sets $\cset^I$ and $\cset^A$ evolve, and the budgets may change. 
We will ensure that, over the course of the algorithm, the total budget of all vertices does not increase. 
Since the budget of every terminal in $\hG$ is always at least $1$, the total budget of all vertices is at least $|\hGamma|$ throughout the algorithm, and this will ensure that the total number of terminals at the end of the algorithm is bounded by $O(|E_1|\cdot\Delta\log n)\leq O\left((|E'|+\optcro(G))\cdot\poly(\Delta\log n)\right)$, and therefore $|\hat E|$ is also bounded by the same amount.

%\vspace{-4mm}
\paragraph{Initialization.}
At the beginning, we let $\hat E=E_1$, and we let $\hcset$ be the set of all connected components of the graph $\hH\cup A$, where $\hH=G\setminus \hat E$, and $A=\emptyset$.  The set $\hG$ of terminals contains all endpoints of edges in $\hE$.
Recall that we have identified a subset $\cset_1'$ of clusters of $\hH$, each of which contains at most $\mu$ terminals. We set $\cset^I=\cset^I_1=\cset_1'$, $\cset^I_2=\emptyset$, and $\cset^A=\hcset\setminus \cset^I$. The algorithm proceeds in iterations, as long as $\cset^A\neq \emptyset$. %We now describe the execution of an iteration.

\subsubsection*{Description of an Iteration}
We now describe a single iteration.
Let $C\in \cset^A$ be any active cluster. If $|\hGamma\cap V(C)|\leq \mu$, then we  move $C$ from $\cset^A$ to $\cset^I_1$ (and to $\cset^I$), and continue to the next iteration. Clearly, in this case $C$ is a type-1 acceptable cluster, and the budgets of vertices may only decrease.

We assume from now on that $|\hGamma\cap V(C)|> \mu$ and denote $\tilde\Gamma=\hGamma\cap V(C)$. 
We then apply the algorithm $\algsc$ for computing the (approximate) sparsest cut in the graph $C$, with respect to the set $\tilde\Gamma$ of terminals. 
Let $(X,Y)$ denote the cut that the algorithm returns, and assume w.l.o.g. that $|X\cap \tilde\Gamma|\leq |Y\cap \tilde\Gamma|$. Denote $E^*=E_C(X,Y)$. Assume first that $|E^*|<\alpha'\cdot \alphasc(n) |X\cap \tilde\Gamma|$. Then we delete the edges of $E^*$ from $\hH$ and add them to the set $\hat E$. We then replace the cluster $C$ in $\cset^A$ by all connected components of $C\setminus E^*$ and continue to the next iteration. Note that we may have added new terminals to $\hGamma$ in this iteration: the endpoints of the edges in $E^*$. Denote by $\Gamma^*$ the set of endpoints of these edges.
The changes in the budgets of vertices are as follows. On the one hand, for every new terminal $t\in \Gamma^*$, the budget $b(t)$ may have grown from $0$ to at most $8\Delta\cdot \log_{3/2}n$.
%(this is since the total number of terminals that lie in $X$ at the end of this iteration is at most $|\tilde\Gamma|$ -- the number of terminals that originally lied in $C$, and the same is true for $Y$).
Since $|\Gamma^*|\leq 2|E^*|\leq 2 \alpha'\cdot \alphasc(n) |X\cap \tilde\Gamma|$, the total increase in the budget of new terminals is at most $16 \alpha'\cdot \alphasc(n) |X\cap \tilde\Gamma|\cdot \Delta\log_{3/2} n$.
Note that for terminals in $\tilde\Gamma\cap Y$, their budgets can only decrease.
On the other hand, since $|\tilde\Gamma\cap X|\leq |\tilde\Gamma\cap Y|$ and the cut $(X,Y)$ is sufficiently sparse, the total number of terminals that lie in $X$ at the end of the current iteration is at most $2|\tilde\Gamma|/3$. 
Therefore, the budget of every terminal in $\tilde\Gamma\cap X$ decreases by at least $8\Delta$, and the total decrease in the budget is therefore at least $8\Delta\cdot|X\cap \tilde\Gamma|$. 
Since $\alpha'=\frac{1}{16\alphasc(n)\log_{3/2}n}$, the total budget of all terminals does not increase.

We assume from now on that algorithm $\algsc$ returned a cut of sparsity at least $\alpha'\cdot \alphasc(n)$. Then we are guaranteed that the set $\tilde \Gamma$ of terminals is $\alpha'$-well-linked in $C$. 
%Since $\alpha'>\alpha$, they are also $\alpha$-well-linked in $C$, but we will need to use this stronger $\alpha'$-well-linkedness property. 
%We denote by $U$ the set of all separator vertices of $C$.
%that is, a vertex $v\in U$ iff graph $C\setminus v$ is  . 
%
We use the following standard definition of vertex cuts:
\begin{definition}
Given a graph $\hat G$, a \emph{vertex cut} in $\hat G$ is a partition $(W,X,Y)$ of $V(\hat G)$ into three disjoint subsets, with $W,Y\neq\emptyset$, such that no edge of $\hat G$ connects a vertex of $W$ to a vertex of $Y$. We say that the cut is a \emph{$1$-vertex cut} if $|X|=1$, and we say that it is a $2$-vertex cut if $|X|=2$.
\end{definition}
In the remainder of the proof we consider three cases. The first case is when graph $C$ has a $1$-vertex cut $(W,X,Y)$, with $W$ and $Y$ containing at least two terminals of $\tilde \Gamma$ each. In this case, we delete some edges from $C$, decomposing it into smaller clusters, and continue to the next iteration. The second case is when $C$ has a $2$-vertex cut $(W,X,Y)$, where both $W$ and $Y$ contain at least three terminals of $\tilde \Gamma$. In this case, we also delete some edges from $C$, decomposing it into smaller connected components, and continue to the next iteration. The third case is when neither of the first two cases happens. In this case, we decompose $C$ into a single type-$2$ acceptable cluster, and a collection of type-$1$ acceptable clusters. We now proceed to describe each of the cases in turn.

\paragraph{Case 1.} We say that Case 1 happens if there is a $1$-vertex cut $(W,X,Y)$ of $C$, with $|W\cap \tilde \Gamma|,|Y\cap \tilde \Gamma|\geq 2$. Set $X$ contains a single vertex, that we denote by $v$. Assume w.l.o.g. that $|W\cap \tilde \Gamma|\leq |Y\cap \tilde \Gamma|$. We start with the following simple claim.
\begin{claim}\label{claim: few vertices on one side of 1-sep}
	$|W\cap \tilde \Gamma|<\mu/4$.
\end{claim}
\begin{proof}
Assume for contradiction that the claim is false.
Consider a bi-partition $(W',Y')$ of $V(C)$, where $W'=W\cup \set{v}$ and $Y'=Y$. We denote by $E^*$ the set of all edges of $C$ incident to the separator vertex $v$, so $|E^*|\leq \Delta$. Since $E_C(W',Y')\subseteq E^*$, $|E_C(W',Y')|\leq \Delta$. However, since vertices of $\tilde \Gamma$ are $\alpha'$-well-linked in $C$, we have $|E_C(W',Y')|\geq \alpha'\cdot |W'\cap \tilde \Gamma|\geq \alpha' \cdot \mu/4>\Delta$, (as $\mu=512\Delta \alphasc(n)\log_{3/2} n$, while $\alpha'=\frac{1}{16\alphasc(n)\log_{3/2} n}$), a contradiction. 
\end{proof}

Let $E_0$ be the set of all edges that connect the separator vertex $v$ to the vertices of $W$. 
We add to $\hat E$ the edges of $E_0$. 
Consider now the graph $C\setminus E_0$. Note that for every connected component of $C\setminus E_0$, either $V(C')\subseteq W$, or $V(C')=Y\cup\set{v}$. We let $\sset$ contain all components $C'$ with $V(C')\subseteq W$. Note that every component $C'\in \sset$ is a type-1 acceptable cluster. This is because $|\tilde \Gamma\cap V(C')|\leq |\tilde \Gamma\cap W|\leq \mu/4$, and we have created at most $\Delta$ new terminals in $C'$: the endpoints of the edges in $E_0$. As $\mu>4\Delta$, every component $C'\in \sset$ now contains at most $\mu$ terminals. We add all components in $\sset$ to the set $\cset^I_1$ of inactive components (and also to the set $\cset^I_1$). Additionally, we replace the cluster $C$ in $\cset^A$ by the subgraph of $C$ induced by vertices of $Y\cup\set{v}$.

It remains to prove that the total budget of all vertices does not grow. 
Recall that $|\tilde \Gamma\cap W|\ge 2$, %The budget of each such terminal before the current iteration was $8\Delta \cdot \log_{3/2}(|\tilde \Gamma|)$, and at the end of the current iteration, their budgets became $1$ each. 
and the budget of each terminal $t\in \tilde \Gamma\cap W$ has decreased from $8\Delta \cdot \log_{3/2}(|\tilde \Gamma|)$ to $1$. Therefore, the total budget decrease of these terminals is at least $16\Delta \cdot \log_{3/2}(|\tilde \Gamma|)-2$. We have created at most $\Delta$ new terminals in set $W$. For each new terminal, its new budget is $1$ since at the end of this iteration it belongs to an inactive cluster. We have created at most one new terminal in set $X\cup Y$ -- the vertex $v$. Since $|\tilde \Gamma\cap W|\ge 2$, the total number of terminals in $X\cup Y$ at the end of the iteration is at most $|\tilde \Gamma|-1$. Therefore, the budgets of terminals in $\tilde\Gamma \cap Y$ do not increase, and the budget of $v$ is at most $8\Delta \cdot \log_{3/2}(|\tilde \Gamma|)$. 
Altogether, the total budget increase is at most $8\Delta \cdot \log_{3/2}(|\tilde \Gamma|)+\Delta$, which is less than $16\Delta \cdot \log_{3/2}(|\tilde \Gamma|)-2$, the total budget decrease of vertices of $\tilde \Gamma\cap W$.
We conclude that the total budget of all vertices does not increase.

From now on we assume that Case 1 does not happen. We need the following simple observation.
\begin{observation}\label{obs: no case 1 small cuts}
Assume that Case 1 does not happen. Let $(W,X,Y)$ be any $1$-vertex cut in $C$, then either $|W|=1$, or $|Y|=1$.
\end{observation}
\begin{proof}
Assume for contradiction that $|W|,|Y|>1$. Since Case 1 does not happen, either $|W\cap \tilde \Gamma|\leq 1$, or $|Y\cap \tilde \Gamma|\leq 1$. Assume w.l.o.g. that $|W\cap \tilde \Gamma|\leq 1$. 
Let $t\in W\cap \tilde \Gamma$ be the unique terminal that lies in $W$ (if it exists; otherwise $t$ is undefined), and let $u$ be any vertex that lies in $W\setminus \tilde \Gamma$ (since $|W|>1\ge |W\setminus \tilde \Gamma|$, such a vertex always exists). Note that the removal of $v$ and $t$ (if it exists) from $G$ separates $u$ from vertices of $Y$ in $G$, causing a contradiction to the fact that $G$ is $3$-connected.
\end{proof}

Let $U$ be the set of separator vertices of $C$. From the above observation, for every separator vertex $u\in U$, there is a unique vertex $u'\in V(C)$ that is a neighbor of $u$, such that $u'$ has degree $1$ in $C$. %Moreover, vertex $u$ is not a separator vertex for $C\setminus\set{u'}$. 
Since graph $G$ is $3$-connected, $u'$ must belong to $\tilde \Gamma$. Let $U'$ be the set of all such vertices. It is easy to verify that graph $C\setminus U'$ is $2$-connected.

%Assume otherwise. Consider a partition $(X,Y)$ of $V(C)$, that is defined as follows. Set $X$ contains, for every super-node $v_{Z'}\in V(T')$, all vertices of $Z'$, while set $Y$ contains all remaining vertices of $C$. Since $w(T')\geq \mu/4$, $|X\cap \tilde\Gamma|\geq \mu/4$. Moreover, from the choice of $Z_0$, $w(Y)\geq \mu/16$ must hold. Let $u'$ be the parent vertex of $v_{Z_0}$ in the tree, that must be a separator vertex, and let $E^*$ be the set of all edges of $C$ that are incident to $u'$. Clearly, $E_C(X,Y)\subseteq E^*$, and so $|E_C(X,Y)|\leq \Delta$. On the other hand, since the set $\tilde\Gamma$ of terminals is $\alpha'$-well-linked in $C$, $|E_C(X,Y)|\geq \alpha'\cdot \min\set{|X\cap \tilde\Gamma|,|Y\cap \tilde\Gamma|}\geq \alpha'\mu/16>\Delta$ (as $\mu=512\Delta \alphasc(n)\log_{3/2} n$, while $\alpha'=\frac{1}{16\alphasc(n)\log_{3/2} n}$), a contradiction.
%---------------------------------
%---------------------------------
%---------------------------------
%---------------------------------
\paragraph{Case 2.}
%---------------------------------
%---------------------------------
%---------------------------------
%---------------------------------
We say that Case 2 happens if Case 1 does not happen, and there is a $2$-vertex cut $(W,X,Y)$, with $|W\cap \tilde \Gamma|,|Y\cap \tilde \Gamma|\geq 3$. Set $X$ contains exactly two vertices, that we denote by $x,y$. Assume w.l.o.g. that $|W\cap \tilde \Gamma|\leq |Y\cap \tilde \Gamma|$. 
The algorithm for Case 2 is very similar to the algorithm for Case 1.
We start with the following simple claim, that is similar to Claim \ref{claim: few vertices on one side of 1-sep}, and its proof is almost identical.
\begin{claim}\label{claim: few vertices on one side of 2-sep}
$|W\cap \tilde \Gamma|<\mu/4$.
\end{claim}
\begin{proof}
Assume for contradiction that the claim is false.
Consider a bi-partition $(W',Y')$ of $V(C)$, where $W'=W\cup \set{x,y}$ and $Y'=Y$. Let $E^*$ be the set of all edges of $C$ incident to vertices $x$ or $y$, so $|E^*|\leq 2\Delta$. Since $E_C(W',Y')\subseteq E^*$, $|E_C(W',Y')|\leq 2\Delta$. However, since vertices of $\tilde \Gamma$ are $\alpha'$-well-linked in $C$, we have $|E_C(W',Y')|\geq \alpha'\cdot |W'\cap \tilde \Gamma|\geq \alpha' \cdot \mu/4>2\Delta$, (as $\mu=512\Delta \alphasc(n)\log_{3/2} n$, while $\alpha'=\frac{1}{16\alphasc(n)\log_{3/2} n}$), a contradiction. 
\end{proof}

Let $E_0$ be the set of all edges that connect the vertices $x,y$ to the vertices of $W$. 
We add to $\hat E$ the edges of $E_0$. 
Consider now the graph $C\setminus E_0$ and let $\sset$ be the set of its connected components. Note that, for every component of $C'\in \sset$, either $V(C')\subseteq W$ or $V(C')\subseteq Y\cup\set{x,y}$. We let $\sset_1\subseteq \sset$ contain all components $C'$ with $V(C')\subseteq W$, and we let $\sset_2$ contain all remaining connected components. Note that every component $C'\in \sset_1$ is a type-1 acceptable cluster. This is because $|\tilde \Gamma\cap V(C')|\leq |\tilde \Gamma\cap W|\leq \mu/4$, and we have created at most $2\Delta$ new terminals in $C'$: the endpoints of edges in $E_0$. As $\mu>8\Delta$, every component in $\sset_1$ now contains at most $\mu$ terminals. We add all components in $\sset_1$ to the set $\cset^I_1$ (and also to $\cset^I_1$). Additionally, we replace the cluster $C$ in $\cset^A$ by all components in $\sset_2$.

It remains to prove that the total budget of all vertices does not grow. The proof again is very similar to the proof in Case 1.
Recall that $|\tilde \Gamma\cap W|\ge 3$, and the budget of each terminal in $\tilde \Gamma\cap W$ has decreased from $8\Delta \cdot \log_{3/2}(|\tilde \Gamma|)$ to $1$. Therefore, the decrease of their total budgets is at least $24\Delta \cdot \log_{3/2}(|\tilde \Gamma|)-3$. We have created at most $2\Delta$ new terminals in set $W$ -- the neighbors of $x$ and $y$, and each such new terminal has new budget $1$ since it belongs to an inactive cluster at the end of this iteration. We have created at most two new terminals in set $X\cup Y$ -- terminals $x$ and $y$. Since $|\tilde \Gamma\cap W|\ge 3$, the total number of terminals that belong to set $X\cup Y$ at the end of the current iteration is at most $|\tilde \Gamma|-1$. Therefore, the budgets of terminals in $\tilde{\Gamma}\cap Y$ do not increase, and the budgets of the vertices $x$ and $y$ are at most $8\Delta \cdot \log_{3/2}(|\tilde \Gamma|)$. 
Altogether, the total budget increase is at most $16\Delta \cdot \log_{3/2}(|\tilde \Gamma|)+2\Delta$, which is less than $24\Delta \cdot \log_{3/2}(|\tilde \Gamma|)-3$, the total budget decrease of vertices of $\tilde \Gamma\cap W$.
We conclude that the total budget of all vertices does not increase.
From now on, we can assume that Case 1 and Case 2 did not happen.

\paragraph{Case 3.} The third case happens if neither Case 1 nor Case 2 happen. Recall that we have denoted by $U$ the set of all separator vertices of the cluster $C$. We have defined $U'$ to be the set of all vertices $u'$, such that $u'$ has degree $1$ in $C$, and it has a neighbor in $U$. We have also shown that $U'\subseteq \tilde \Gamma$. Our first step is to add to the edge set $\hat E$ every edge of $E(C)$ that connects a vertex of $U$ to a vertex of $U'$, and let $C'=C\setminus U'$. Every vertex $u'\in U'$ is now an isolated vertex in $G\setminus \hat E$. For each such vertex $u'\in U'$, we add the cluster $\set{u'}$ to the set $\cset^I_1$ of inactivate type-1 acceptable clusters (and also to $\cset^I$). Notice that every vertex $u\in U$ now becomes a terminal. We denote by $\tilde \Gamma'=(\tilde \Gamma\cup U)\setminus U'$ the set of terminals in the current cluster $C'$. From the above discussion, $|\tilde \Gamma'|\leq |\tilde \Gamma|$, and cluster $C'$ is $2$-connected. If $|\tilde \Gamma'|\leq \mu$, then $C'$ is a type-1 acceptable cluster. We then add it to the sets $\cset^I_1$ (and also to $\cset^I$), and terminate the current iteration. Note that the total budget of all vertices does not increase. This is because, before the current iteration, $|\tilde \Gamma|\geq \mu$ held, and every vertex in $\tilde \Gamma$ had budget at least $8\Delta\log \mu$; while at the end of the current iteration, every terminal in $\tilde \Gamma'\cup U'$ has budget $1$, and $|\tilde \Gamma'\cup U'|\leq 2\mu$. Therefore, we assume from now on that $|\tilde \Gamma'|>\mu$, and we refer to the vertices of $\tilde \Gamma'$ as terminals. In the remainder of this iteration, we will split the cluster $C'$ into a single type-2 acceptable cluster, and a collection of type-1 acceptable clusters, and we will prove that the total budget of all vertices does not increase.
Before we proceed further, we first prove the the following observations about the graph $C'$ that we will use later.
\begin{observation}
\label{obs:C'_terminal_wl}
The set $\tGamma'$ of terminals is $\alpha'$-well-linked in $C'$.
\end{observation}
\begin{proof}
Consider any partition $(X,Y)$ of $V(C')$. Then we can augment $(X,Y)$ to a partition $(X',Y')$ of $V(C)$ as follows. Start with $X'=X$ and $Y'=Y$. For every vertex $u\in U$, if $u\in X$, then we add its unique neighbor in $U'$ to $X'$, otherwise we add it to $Y'$. Note that $|\tGamma\cap X'|\geq |\tGamma'\cap X|$. This is because $\tGamma'=(\tGamma\cup U)\setminus U'$, and for every vertex $u\in U\cap X$, while $u$ may or may not belong to $\tGamma$, it always has a neighbor in $U'\cap \tGamma$.
Similarly, $|\tGamma\cap Y'|\geq |\tGamma'\cap Y|$. Since the set $\tGamma$ of terminals is $\alpha'$-well-linked in $C$, we get that $|E_{C'}(X,Y)|=|E_C(X',Y')|\geq \alpha'\cdot\min\set{|\tGamma\cap X'|,|\tGamma\cap Y'|}\geq \alpha'\cdot\min\set{|\tGamma'\cap X|,|\tGamma'\cap Y|}$. Therefore,  the set $\tGamma'$ of terminals is $\alpha'$-well-linked in $C'$.
\end{proof}
\begin{observation}\label{obs: small 2-vertex cuts in smaller cluster}
For every $2$-vertex cut $(W,X,Y)$ of $C'$ with $|W\cap \tilde \Gamma'|\leq |Y\cap \tilde \Gamma'|$, $|W\cap \tilde \Gamma'| \leq 2$.
\end{observation}
\begin{proof}
Let $(W,X,Y)$ be a $2$-vertex cut of $C'$ with $|W\cap \tilde \Gamma'|\leq |Y\cap \tilde \Gamma'|$. We augment it to a $2$-vertex cut $(W',X,Y')$ of $C$ as follows. Start with $W'=W$ and $Y'=Y$. For every vertex $u\in U$, if $u\in W$, then we add its unique neighbor in $U'$ to $W$, otherwise we add it to $Y'$. It is immediate to verify that $(W',X',Y')$ is indeed a $2$-vertex cut in $C$, and that $|W'\cap \tilde \Gamma|\geq |W\cap \tilde \Gamma'|$ and  $|Y'\cap \tilde \Gamma'|\geq |Y\cap \tilde \Gamma'|$. Since we assumed that Case 2 does not happen, $|W'\cap \tilde \Gamma|\leq 2$ must hold, and so $|W\cap \tilde \Gamma'|\leq 2$.
\end{proof}

The next observation gives us a planar drawing of $C'$. %Recall that $\bset'$
%(defined in Section~\ref{sec:stage1}) 
%is the set of marked pseudo-blocks in the block decomposition of $G\setminus E'$, where $E'$ is the initial input planarizing set of edges.
\begin{observation}\label{obs: block containing cluster}
There is a pseudo-block $B_0$ in the block decomposition of $G\setminus E_1$ that is not contained in a component of $\cset'_1$, such that $C'\subseteq B_0$. In particular, the associated drawing $\hat\psi_{B_0}$ of $B_0$ naturally induces a planar drawing $\psi_{C'}$ of the cluster $C'$.
\end{observation}
\begin{proof}
Let $H_1=G\setminus E_1$ and $H_2=G\setminus \hat E$. Clearly $E_1\subseteq \hat{E}$. 
Denote by $\bset(H_1)$ the block decomposition of the graph $H_1$. 
Since $C'$ is a connected graph, there is a pseudo-block in $\bset(H_1)$ that contains $C'$ a subgraph.
%From Claim \ref{claim: every block is good}, $B_0$ is a good pseudo-block.
We denote this block by $B_0$.
Note that it is impossible that $V(B_0)\subseteq V(\hat C)$ for a component $\hat C\in \cset_1'$, since initially all clusters in $\cset_1'$ belong to $\cset^I_1$ and will not be processed. 
%Therefore, $B_0\in \bset'$. 
%Recall that set $E_1\subseteq \hat E$ of edges contains all edges incident to the endpoints of the fake edges of $\tilde B_0$. 
%Lastly, since $\hat B\subseteq C'$, $C'$ is a $2$-connected subgraph of $H_2$, and $H_2\subseteq H_1$, it follows that all edges and vertices of $C'$ lie in $\tilde B_0'$.
\end{proof}

%%%%%%%%%%%%%%%%%%%%%%%%%%

We now proceed to split the cluster $C'$ into one type-2 acceptable cluster and a collection of type-1 clusters.
Recall that $C'$ is $2$-connected. %We denote by $S_2$ the set of all vertices that participate in some $2$-separator of $C'$. Equivalently, a vertex $v\in V(C')$ belongs to $S_2$ iff there is a vertex $v'\in V(C')$, such that $C'\setminus\set{v,v'}$ is not connected.
We use the algorithm from Theorem~\ref{thm: block decomposition}, to obtain a the block decomposition $\lset$ of $C'$,
%Recall that the theorem ensures that for every vertex $v\in S_2$, either $v$ is an endpoint of a block of $\lset$, or $v$ has exactly two  neighbors in $C'$, and there is an edge $(v',v)\in E$, such that $v'$ is an endpoint of a block $B\in \lset$.
and let $\tau=\tau(\lset)$ be the decomposition tree associated with the decomposition $\lset$. We let $B\in \lset$ be a pseudo-block that contains at least $\mu/4$ terminals of $\tilde \Gamma'$, and among all pseudo-blocks with this property, maximizes the distance in $\tau$ between its corresponding vertex $v(B)$ and the root of $\tau$, breaking ties arbitrarily. Notice that such a pseudo-block always exists, since $C'$ belongs to $\lset$ as a pseudo-block and contains at least $\mu$ terminals of $\tGamma'$. Let $v(B_1),\ldots,v(B_q)$ denote the child vertices of $v(B)$ in $\tau$, and let $B^c$ denote the complement block of $B$.
%(that is, $V(B^c)$ has the same endpoints as $B$, and contains, in addition to these endpoints, all vertices that do not lie in $B$; graph $B^c$ is the sub-graph of $C'$ induced by $V(B^c)$. 
We denote the endpoints of $B$ by $(x,y)$, and, for all $1\leq i\leq q$, the endpoints of $B_i$ by $(x_i,y_i)$. Recall that $\nset(B)=\set{B^c, B_1,\ldots,B_q}$. We use the following simple observations and their corollaries.
\begin{observation}
\label{obs:leaf_contain_terminals}
Let $\hat B$ be a pseudo-block in $\lset$ such that $v(\hat B)$ is a leaf of the tree $\tau$. Then $\hat B$ contains terminal of $\tGamma'$ that is not an endpoint of $\hat B$. Moreover, if $B^c$ is defined, then it contains a terminal of $\tGamma'$, that is not one of its endpoints.
\end{observation}
\begin{proof}
Let $x,y$ be the endpoints of $\hat B$. Assume that $\hat B$ does not contain a terminal that does not belong to $\set{x,y}$, then the removal of $\set{x,y}$ separates $V(\hat B)\setminus \set{x,y}$ from $V(G)\setminus V(\hat B)$, contradicting the fact that $G$ is $3$-connected. The proof for $B^c$ is similar.
\end{proof}

We obtain the following immediate corollary.
\begin{corollary}
\label{obs:number_of_branch}
$|\nset(B)|\leq |\tilde \Gamma'|$.
\end{corollary}
%\begin{proof}
%Since graph $G$ is $3$-connected, for all $1\leq i\leq q$, at least one vertex of $V(B_i)\setminus\set{x_i,y_i}$ must belong to $\tGamma'$. Similarly, at least one vertex of $V(B^c)\setminus \set{x,y}$ belongs to $\tGamma'$. Clearly, these vertices are distinct. Therefore, $|\nset(B)|\leq |\tilde \Gamma'|$.
%\end{proof}
\begin{observation}\label{obs: few terminals in every other block}
For all $1\leq i\leq q$, at most two vertices of $V(B_i)\setminus\set{x_i,y_i}$ belong to $\tGamma'$. Similarly, at most two vertices of $V(B^c)\setminus\set{x,y}$ belong to $\tGamma'$.
\end{observation}
\begin{proof}
Fix some $1\leq i\leq q$, and consider a $2$-vertex cut $(W,X,Y)$ of $C'$, where $W=V(B_i)\setminus\set{x_i,y_i}$, $X=\set{x_i,y_i}$, and $Y=V(C')\setminus(X\cup W)$. From the definition of the block $B$, $|W\cap \tilde \Gamma'|<\mu/4$. Since $|\tilde \Gamma'|>\mu$, $|Y\cap \tilde \Gamma'|\geq 3\mu/4-2>\mu/4>|W\cap \tilde \Gamma'|$. From Observation \ref{obs: small 2-vertex cuts in smaller cluster}, $|W\cap \tGamma'|\leq 2$. Therefore, $V(B_i)\setminus\set{x_i,y_i}$ contains at most two terminals of $\tGamma'$.
The proof that $V(B^c)\setminus\set{x,y}$ contains at most two terminals of $\tGamma'$ is similar. 
%We define a $2$-vertex cut $(W,X,Y)$ of $C'$, where $W=V(B^c)\setminus \set{x,y}$, $X=\set{x,y}$, and $W=V(C')\setminus(X\cup Y)$. From the definition of the block $B$, and since $V(B)\setminus\set{x,y}\subseteq Y$, we get that $|Y\cap \tilde \Gamma'|\geq \mu/4-2>4$. But then, from \ref{obs: small 2-vertex cuts in smaller cluster}, $|W\cap \tGamma'|\leq 2$. Therefore, $V(B^c)\setminus\set{x,y}$ contains at most two terminals of $\tGamma'$.
\end{proof}
%The proof follows immediately from the fact that $G$ is a $3$-connected graph. By combining the above observation with Observation \ref{obs: few terminals in every other block}, we obtain the following immediate corollary.

We obtain the following immediate corollary of Observations \ref{obs:leaf_contain_terminals} and \ref{obs: few terminals in every other block}.

\begin{corollary}\label{cor: few leaves in each sub-block}
There are at most two leaves in tree $\tau$ that are not descendants of $v(B)$. Moreover, for all $1\leq i\leq q$, there are at most two leaves in the subtree of $\tau$ rooted at $v(B_i)$.
\end{corollary}

We now describe the next steps for processing cluster $C'$, starting with a high-level intuitive overview. 
From Observation~\ref{obs: block containing cluster}, the associated drawing $\hat\psi_{B_0}$ of $B_0$ naturally induces a planar drawing $\psi_{C'}$ of $C'$. One can show that $C'$ is a type-2 acceptable cluster with respect to the drawing $\psi_{C'}$, except that we cannot ensure that the size of set $S_2(C')$ -- the set of all vertices that participate in $2$-separators in $C'$, is sufficiently small (recall that the requirement in the definition of type-$2$ acceptable clusters is that $|S_2(C')|\leq O(\Delta|\tilde \Gamma'|)$). 
To see this, consider the situation where the sub-tree of $\tau$ rooted at the child vertex $v(B_i)$ of $v(B)$, that we denote by $\tau_i$, is a long path. The cardinality of set $S_2(C')$ may be as large as the length of the path, while there are only at most two terminals in $B_i$.

In order to overcome this difficulty, we need to ``prune'' the sub-tree $\tau_i$.
%Let $v(\hat B)$ be a leaf $\tau_i$. From Observation~\ref{obs:leaf_contain_terminals}, $\hat B$ contains a terminal $t\in \tilde \Gamma'$ that is not one of its endpoints, so $t$ also belongs to $B_i$ and it is not an endpoint of $B_i$. 
From Observation~\ref{obs: few terminals in every other block}, tree $\tau_i$ may contain at most two leaves. 
Assume first for simplicity that $\tau_i$ contains exactly one leaf, so $\tau_i$ is a path. 
%Note that, for every vertex $v(\hat B)$ on this path, the endpoints of $\hat B$ belong to $S_2(C')$. 
If we denote by $v(B^1_i),v(B^2_i),\ldots,v(B^r_i)$ the vertices that appear on path $\tau_i$ in this order, with $B^1_i=B_i$. We simply add to $\hat{E}$ all edges of $B^2_i$ incident to its endpoints, and replace $B^2_i$ with a fake edge connecting its endpoints in $C'$. 
The block $B^2_i$ then decomposes into a number of a type-1 acceptable clusters. Notice that now the total number of vertices of $S_2(C')$ that block $B_i$ contributes is $O(1)$. If tree $\tau_i$ has two leaves, then the pruning process for $\tau_i$ is slightly more complicated but similar. We treat the complement block $B^c$ similarly as blocks in $\set{B_1,\ldots,B_q}$. 

We now provide a formal proof. We start with the cluster $\hat C=C'$ and the set $A_{\hat C}=\emptyset$ of fake edges, and then we iterate. In every iteration, we process a distinct block of $\nset(B)$, and modify the cluster $\hat C$ by deleting some edges and vertices and adding some fake edges to $A_{\hat C}$ and to $\hat C$. Throughout the algorithm, we will maintain the following invariants:
\begin{properties}{I}
\item $\hat C$ is a simple graph; \label{inv: C is simple}
\item each $2$-separator of $\hat C$ is also a $2$-separator of $C'$; and\label{inv: 2-cuts preserved}
\item graph $\hat C\setminus A_{\hat C}$ is $2$-connected. \label{inv: 2-connectivity}
\end{properties}
%Throughout the algorithm, we will maintain a collection $\lset'$ of pseudo-blocks (that we call ``eliminated blocks''), starting from $\lset'=\emptyset$.
We now describe an iteration.
%\paragraph{Processing Blocks $B_i$.}
Consider a child block $B_i$ of block $B$ (the block $B^c$ will be processed similarly), and let $\tau_i$ be the sub-tree of $\tau$ rooted at $v(B_i)$. As our first step, we construct a set $V_i$ of nodes in $\tau_i$ that we will use in order to ``prune'' block $B_i$, using the following observation.
\begin{observation}
\label{obs:prune}
There is an efficient algorithm, that constructs a subset $V_i\subseteq V(\tau_i)$ containing at most two vertices (where possibly $V_i=\emptyset$), that satisfies the following properties:
\begin{itemize}
\item for each vertex $\hat v\in V_i$, $\hat v$ has degree exactly $2$ in $\tau_i$, it is not $v(B_i)$ or one of its children, and the parent of $\hat v$ in $\tau_i$ has degree exactly $2$; 
\item if $|V_i|=2$ then neither vertex is a descendant of the other in $\tau_i$; 
\item if $v(B^*)\in V_i$, then graph $\tilde B^*$ is not isomorphic to $K_3$;
\item for each vertex $\hat v\in V_i$, there is some ancestor vertex $v(B^p)$ of $\hat v$ in $\tau_i$, that is not $v(B_i)$, such that $\tilde B^p$ is not isomorphic to $K_3$, and every vertex on the unique path in tree $\tau$ connecting $\hat v$ to  $v(B^p)$ has degree exactly $2$;
and
\item if we let $\tau_i'$ be the sub-tree obtained from $\tau_i$ after we delete, for every vertex $\hat v\in V_i$, the sub-tree rooted at the child vertex of $\hat v$, then $|V(\tau_i')|\leq 500$.
\end{itemize}
%The algorithm also constructs a subset $V_c$ of vertices of $V(B^c)$ that satisfies the above properties.
\end{observation} 
\begin{proof}
We fix a child block $B_i$ of $B$ and show how to construct vertex set $V_i$. %The sets $\set{V_{j}}_{1\le j\le q, j\ne i}$ are constructed similarly.
We say that a vertex $v(B^*)\in \tau_i$ is bad iff graph $\tilde B^*$ is isomorphic to $K_3$. We say that a path $P\subseteq \tau_i$ is bad iff every vertex of $P$ has degree exactly $2$ in $\tau_i$, and it is a bad vertex. 

\begin{observation}\label{obs: no long bad path}
	If $P$ is a bad path in $\tau_i$, then $P$ contains at most $20$ vertices.
\end{observation}
\begin{proof}
	Let $P$ be a bad path in $\tau_i$, and assume for contradiction that it contains $21$ vertices. Then every vertex $v(B^*)\in V(P)$ has exactly one child in $\tau_i$, and graph $\tilde B^*$ is isomorphic to $K_3$. It is then easy to verify that graph $C'$ must contain a path $P'$ containing at least five vertices, each of which has degree exactly $2$ in $C'$ (every vertex on $P'$ is an endpoint of some block in $\set{B^*\mid v(B^*)\in P}$). Since graph $G$ is $3$-connected, every vertex of $P'$ must lie in $\tilde \Gamma'$. But then there are at least three vertices of $V(B_i)\setminus\set{x_i,y_i}$ that lie in $\tilde \Gamma'$, contradicting Observation \ref{obs: few terminals in every other block}.
\end{proof}

From Corollary \ref{cor: few leaves in each sub-block}, $\tau_i$ contains at most two leaves.
Therefore, there is at most one node in $\tau_i$ that has degree $3$, and all other nodes has degree $1$ or $2$. We consider the following cases:
\begin{enumerate}
	\item if $\tau_i$ contains no degree-$3$ node, and $|V(\tau_i)|\le 50$, then we set $V_i=\emptyset$;
	\item if $\tau_i$ contains no degree-$3$ node, and $|V(\tau_i)|> 50$, then we let $V_i$ consist of a single vertex $v$, that is at distance at least $25$ and at most $50$ from the root, and is not a bad vertex; from Observation \ref{obs: no long bad path}, such a vertex must exist;
	\item if $\tau_i$ contains a degree-$3$ node $v'$, and the distance between $v'$ and $v(B_i)$ is at least $51$ in $\tau_i$, then $V_i$ consists of a single vertex $v$, that is defined exactly like in the previous case;
	\item if $\tau_i$ contains a degree-$3$ node $v'$, and the distance between $v'$ and $v(B_i)$ is at most $51$ in $\tau_i$, then the subtree of $\tau_i$ rooted at $v'$ can be viewed as the union of two paths that share a common endpoint $v'$; we denote by $\tau^1_i$ and $\tau^2_i$ the two paths. We select at most one vertex on each of the two paths to add to $V_i$, exactly like in Cases 1 and 2. %; and
	%\begin{enumerate}
	%	\item if $|V(\tau^1_i)|\le 3$, we set $V^1_i=\emptyset$;
	%	\item if $|V(\tau^1_i)|\ge 4$, let $v'=v'_1, v'_2, v'_3, v'_4$ be the first four nodes of $\tau^1_i$ from the root; then we set $V^1_i=\set{v'_3}$;
	%\end{enumerate}
	%we construct $V^2_i$ similarly, and set $V_i=V^1_i\cup V^2_i$.
\end{enumerate}
It is easy to see that $V_i$ contains most two vertices, and satisfies the properties in Observation \ref{obs:prune}.
\end{proof} 

%(the constant $16$ above is arbitrary, and can be replaced by any other large enough constant). Note that we allow $V_i$ to be empty.

We consider the vertices in $V_i$ one-by-one. Let $v(B^{**})\in V_i$ be a vertex of $V_i$. %, and let $B^{**}$ be its corresponding block, namely $\hat v=v(B^{**})$.
 Let $v(B^{*})$ be the unique child vertex of $v(B^{**})$ in $\tau_i$, and let $(x^{*},y^*)$ be the endpoints of block $B^*$. Note that block $B^{*}$ contains a path $P$ connecting $x^*$ to $y^*$. Let $\tilde E$ be the set of all edges of $B^{*}$ that are incident to $x^*$ or to $y^*$. We add the edges of $\tilde E$ to $\hat E$. Consider now the graph $\hat C\setminus \tilde E$. It is immediate to verify that there is one component $\tilde C$ of this graph, containing all vertices of $(\hat C\setminus B^{*})\cup \set{x^*,y^*}$, and all  remaining components are a type-1 acceptable clusters that are contained in $B^{*}$. We add the components of the latter type to $\cset^I_1$ (and also to $\cset^I$). If the edge $(x^*,y^*)$ does not lie in $\tilde C$, then we add the edge $e'=(x^*,y^*)$ to the set $A_{\hat C}$ of fake edges, and let $P(e')=P$ be its embedding. %For the sake of the analysis, we set $P'(e)=P'$ be the path connecting $x'$ to $y'$ in $\hat B\setminus (\hat B'\setminus\set{x',y'})$. 
It is immediate to verify that there is some type-1 acceptable cluster in $\hat C\setminus \tilde E$ that contains the path $P(e')\setminus\set{x^*,y^*}$. 
%We let $C(e)$ be that connected component. 
Finally, we update the cluster $\hat C$ by first removing all vertices and edges of $B^*\setminus\set{x^*,y^*}$ form it, and then adding the edge  $(x^*,y^*)$ if it does not belong to $\hat C$. % does not lie in $\hat C$, then we add this edge to the new cluster $\hat C$. In other words, the new cluster $\hat C$ is obtained from the original cluster $\hat C$ by deleting all vertices of $B^{**}\setminus\set{x',y'}$ from it, and adding the fake edge $(x',y')$. 
We say that the block $B^{*}$ is \emph{eliminated} when processing $B_i$.

The processing of the block $B^c$ is very similar to the processing of the blocks $B_1,\ldots,B_q$, though the details are somewhat more tedious and are omitted here.

%We add $B^{**}$ to the set $\lset'$ of eliminated blocks.
\begin{claim}
After each vertex $ v(B^{**})\in V_i$ is processed, the invariants continue to hold.
\end{claim}
\begin{proof}
We denote by $C_1$ and $C_2$ the cluster $\hat C$ before and after vertex $v(B^{**})$ was processed, respectively. Similarly, we denote by $A_1$ and $A_2$ the set $A_{\hat C}$ of vertices before and after vertex $v(B^{**})$ was processed, respectively. We assume that all invariants held before vertex $v(B^{**})$ was processed. It is immediate to verify that $C_2$ is a simple graph, since we add a fake edge to $\hat C$ only if $\hat C$ did not contain it. Therefore, Invariant \ref{inv: C is simple} continues to hold.
	
Next, we show that $C_2\setminus A_2$ is $2$-connected. Assume for contradiction that this is not the case, and let $(W,X,Y)$ be a $1$-vertex cut of $C_2\setminus A_2$. 
Let $B^*$ be the child block of $B^{**}$, and let $x^*,y^*$ be its endpoints.
 We claim that either (i) $x^*\in W$, $y^*\in Y$, or (ii) $x^*\in Y$, $y^*\in W$ must hold. Indeed, if neither of these holds, then can assume w.l.o.g. that $x^*,y^*\in W\cup X$. Then, by adding to $W$ all vertices of $B^{*}\setminus\set{x^*,y^*}$, we obtain a $1$-vertex cut in $C_1\setminus A_1$, contradicting the assumption that Invariant \ref{inv: 2-connectivity} held for $C_1$.
We now assume without loss of generality that $x^*\in W$ and $y^*\in Y$. 

Since graph $\tilde B^{**}$ is not isomorphic to $K_3$, it must be $3$-connected. Therefore, graph $\tilde B^{**}$ contains three internally disjoint paths connecting $x^*$ to $y^*$. Two of these paths, that we denote by $P_1,P_2$, do not contain the fake edge $(x^*,y^*)$. If we denote by $(x^{**},y^{**})$ the endpoints of the block $B^{**}$, then at least one of these two paths (say $P_1$) is disjoint from the fake parent-edge $(x^{**},y^{**})$ of block $B^{**}$.  Path $P_2$ may contain the fake edge $(x^{**},y^{**})$, but all other edges of $P_2$ must be real edges of $\tilde B^{**}$, as vertex $v(B^{**})$ only has one child in $\tau_i$. Note that there must be a path $P'\subseteq C_2$ that connects $x^{**}$ to $y^{**}$, and is internally disjoint from $B^{**}$ (if we denote by $v(B^p)$ the parent vertex of $v(B^{**})$, then, if $\tilde B^p$ is not isomorphic to $K_3$, such a path exists in $\tilde B^p$, as graph $\tilde B^p$ is $3$-connected, and only has one child block. Otherwise, we let $v(B^p)$ be an ancestor of $v(B^{**})$ that is closest to $v(B^{**})$, such that $\tilde B^p$ is not isomorphic to $K_3$, and $v(B^p)$ has only one child in $\tau_i$. We can define the path $P'$ using the graph $\tilde B^p$).
%\mynote{I can't understand the argument from here on. The paths $P_1$ and $P_2$ exist in which graph? why do they exist? what are $x$ and $y$?}
%Note that there exists two paths $P_1$ and $P_2$ connecting $x'$ to $y'$, such that $P_1$ consists of only edges in $\hat B$, and $P_2$ may contain the fake edge $(x,y)$ 
%(while all other edges of $P_2$ are real edges in $B^{*}$).
%Since $v(B^*)\neq v(B_i)$ and it is not a child vertex of $v(B_i)$, no fake edges 
%Let $v(B^p)$ be the parent vertex of $v(B^{*})$ in $\tau$. 
%Since $v(B^p)$ has degree exactly $2$ in $\tau_i$, there are at most two fake edges in $\tilde B^p$. Since $\tilde B^p$ is $3$-connected, there is a path $P$ connecting $x$ to $y$ in $\tilde B^p$, that does not contain fake edges. 
By combining path $P_2$ with $P'$ (namely, replacing the fake edge $(x^{**},y^{**})$ of $P_2$ with the path $P'$, if $P_2$ contains such an edge), we obtain a path $P_3$ connecting $x^*$ to $y^*$, that is disjoint from $P_1$, and contains no fake edges. Note that both paths $P_1,P_3$ are contained in $C_2$. But then both paths $P_1,P_3$ must contain the separator vertex that lies in $X$, a contradiction. Therefore, Invariant~\ref{inv: 2-connectivity} continues to hold.
	
In order to show that Invariant \ref{inv: 2-cuts preserved} continues to hold, it suffices to show that every $2$-separator in $C_2$ is also a $2$-separator in $C_1$. Consider any $2$-vertex cut $(W,X,Y)$ in $C_2$, and assume w.l.o.g. that $x^*,y^*\in W\cup X$ (since the fake edge $(x^*,y^*)$ belongs to $C_2$, the two vertices cannot lie in sets $W$ and $Y$ respectively). By adding all vertices of 
$B^{*}\setminus\set{x^*,y^*}$ to set $W$, we obtain a $2$-vertex cut in $C_1$. Therefore, $X$ is also a $2$-separator in $C_1$.
\end{proof}
%\paragraph{Processing the Block $B^c$.} The algorithm for processing the block $B^c$ is very similar to the algorithm for processing the blocks $B_i$. The details are omitted.
%\mynote{If you have time and energy please feel free to fill them in but this is not a priority right now, unless there is a doubt in correctness}

%\mynote{next paragraph is wrong, it does not reflect the new Observation  \ref{obs: block containing cluster}. It should be using block $B_0 $ and its associated drawing.}There is a pseudo-block $B_0$ in graph $G\setminus E_1$ that is not contained in a component of $\cset'_1$, such that $C'\subseteq B_0$. In particular, the associated drawing $\hat\psi_{B_0}$ of $B_0$ naturally induces a planar drawing $\psi_{C'}$ of the cluster $C'$.

Let $\hat C$ be the cluster we obtain after all blocks in $\nset(B)$ are processed. 
Recall that from Observation \ref{obs: block containing cluster}, there is a pseudo-block $B_0$ in the block decomposition of graph $G\setminus E_1$, such that $B_0$ is not contained in a cluster of $\cset_1'$, and $C'\subseteq B_0$. The associated drawing $\hat\psi_{B_0}$ of $B_0$ induces a planar drawing $\psi_{\hat C}$ of $\hat C$, where for each fake edge $e\in A_{\hat C}$, the edge $e$ is drawn along the drawing of the path $P(e)$ in $\hat\psi_{B_0}$. We prove the following Lemma in Section~\ref{subsec:type-2-acceptable}.
\begin{lemma}\label{lemma: acceptable}
Cluster $\hat C$ is a type-2 acceptable cluster with respect to the drawing $\psi_{\hat C}$.
\end{lemma}

In order to complete the proof of Theorem \ref{thm: can find edge set w acceptable clusters}, it remains to show that if Case 3 happens, then the total budget of all vertices does not increase after the cluster $C$ is processed.
Recall that $\tilde \Gamma$ denotes the set of terminals in $C$ before $C$ was processed, and $|\tilde \Gamma|\geq \mu$. Let $\tGamma^{\neww}$ be the set of all new terminals that were added to $\tGamma$ over the course of processing $C$. 
Notice that every vertex in $\tilde \Gamma$ had a budget of at least $8\Delta$ before $C$ was processed. After $C$ was processed, the budget of every terminal in $\tilde \Gamma\cup\tGamma^{\neww}$ became $1$. Therefore, in order to show that the total budget of all vertices did not grow, it is sufficient to show that $|\tGamma^{\neww}|\leq 7\Delta |\tGamma|$. Recall that we have created a set $U$ of at most $|\tilde \Gamma|$ terminals, and, whenever a block $ B^*$ was eliminated, we created at most $2\Delta$ new terminals (neighbors of endpoints of $ B^*$). Since each such block $ B^*$ contained at least one terminal of $\tGamma'=(\tilde \Gamma\cup U)\setminus U'$, the total number of new terminals that we have created is bounded by $|\tilde \Gamma|\cdot 4\Delta$. Therefore, the total budget of all vertices does not grow.

In order to complete the proof of Theorem \ref{thm: can find edge set w acceptable clusters}, it remains to prove Lemma \ref{lemma: acceptable}, which we do next.

\subsection{Proof of Lemma \ref{lemma: acceptable}}
\label{subsec:type-2-acceptable}
Recall that $\tilde \Gamma$ denotes the set of all terminals that belonged to cluster $C$ before it was processed. We denote by $\Gamma^*$ the final set of terminals that lie in $\hat C$ after $C$ is processed.

Recall that we have already established that graph $\hat C$ is planar, and defined its planar drawing $\psi_{\hat C}$.
From the invariants, $\hat C$ is a simple graph, and $\hat C\setminus A_{\hat C}$ is $2$-connected. 

Next, we bound the cardinality of the set $S_2(\hat C)$. Recall that the invariants ensure that a $2$-separator in $\hat C$ is also a $2$-separator in $C'$, so $S_2(\hat C)\subseteq S_2(C')$. Moreover, Theorem \ref{thm: block decomposition} ensures that for every vertex $x\in S_2(C')$, either $x$ is an endpoint of a block of $\lset$, or there is a vertex $x'$ that is a neighbor of $x$ in $C'$, and it is an endpoint of a block of $\lset$. In the former case, we denote by $B(x)$ the largest (with respect to $|V(B(x))|$) block of $\lset$ such that $x$ is an endpoint of $B(x)$, and in the latter case, we denote by $B(x)$ the largest block of $\lset$, such that $x'$ is an endpoint of $B(x)$. Notice that, if $B^*$ is a block that was eliminated, and $x$ lies in $B^*$ but is not an endpoint of $B^*$, then $x$ is not a vertex of $\hat C$, and it does not belong to $S_2(\hat C)$. Our algorithm ensures that, for every child block $B_i$ of $B$, there are at most $O(1)$ vertices $x$, such that $x\in V(\hat C)$, and $x$ serves as an endpoint of a block that is a descendant of $B_i$ in $\tau_i$. Therefore, at most $O(\Delta)$ vertices of $S_2(\hat C)$ may lie in $B_i\cap \hat C$. On the other hand, at least one terminal of $\Gamma^*$ lies in $B_i\cap \hat C$. We \emph{charge} these separator vertices to that terminal.  Similarly, if $B^c$ is defined, then at most $O(\Delta)$ vertices of $S_2(\hat C)$ may lie in $B^c\cap \hat C$, and at least one terminal of $\Gamma^*$ lies in $\hat C\cap B^c$. 
Therefore, altogether, $|S_2(\hat C)|\leq O(\Delta |\Gamma^*|)$, as required.

%From Corollary~\ref{obs:number_of_branch}, $|\nset(B)|\leq |\tilde \Gamma'|$. It follows that $|S_2(\hat C)|\leq O(\Delta\cdot |\tilde \Gamma'|)$.
%By definition, $\tilde \Gamma'\subseteq \hat \Gamma\cap V(\hat C)$. Therefore, $|S_2(\hat C)|\leq O(\Delta\cdot |\hat \Gamma\cap V(\hat C)|)$.

Next, we show that cluster $\hat C$ has the well-linkedness property.
\begin{claim}\label{claim: well linkedness}
The set $\Gamma^*$ of terminals is $\alpha$-well-linked in $\hat C\setminus A_{\hat C}$.
\end{claim}
\begin{proof}
%\mynote{needs to be completed. The idea. First, prove that $\hat \Gamma\cap V(C^*)$ is $4\alpha$-well-linked in $C^*$. In order to see this, take any cut $(X,Y)$ of $C^*$. Then augment it to a cut $(X',Y')$ of $C'$ by inserting back the eliminated blocks, like we did in the previous proof. Once you show it for $C^*$, we take the fake edges out. But for each fake edge $e=(x',y')$, we have defined a path $P'(e)$ that stays in $C^*\setminus A_{C^*}$. So the deletion of the fake edges only decreases the number of edges going across the cut by factor at most $2$.}
%Note that $\tilde \Gamma'=\hat \Gamma\cap V(\hat C)$.
%
We start by showing that the set $\Gamma^*$ of terminals is $2\alpha$-well-linked in $\hat C$. Let  $(X,Y)$ be any partition of $V(\hat C)$. Denote $\Gamma_X=\Gamma^*\cap X$ and $\Gamma_Y=\Gamma^*\cap Y$. It suffices to show that $|E_{\hat C}(X,Y)|\ge 2\alpha\cdot \min\set{|\Gamma_X|,|\Gamma_Y|}$.

Let $\mset$ be the set of all blocks $B^*$ that our algorithm eliminated, and let $V^*$ be the set of all vertices that serve as endpoints of the blocks in $\mset$.

Recall that, in addition to the terminals of $\tilde\Gamma'=\tilde \Gamma\cup U$, the sets $\Gamma_X,\Gamma_Y$ of terminals may also contain vertices of $V^*$. We call such terminals \emph{new terminals}.
%(a new terminal cannot belong to $\Gamma'$)

We further partition sets $\Gamma_X$ and $\Gamma_Y$ as follows.
Let $\Gamma^1_X\subseteq \Gamma_X$ contain all terminals $t$, such that there is some block $B^*\in \mset$ with $t$ being one of its endpoints, and the other endpoint $t'$ of $B^*$ lies in $Y$. We let $\Gamma^2_X=\Gamma_X\setminus \Gamma^1_X$.
The partition $(\Gamma^1_Y,\Gamma^2_Y)$ of $\Gamma_Y$ is defined similarly.  Since the endpoints of every block in $\mset$ are connected by a fake edge in $\hat C$, it is immediate to verify that
\begin{equation}
|E_{\hat C}(X,Y)|\geq |\Gamma^1_X|\mbox{  and  } |E_{\hat C}(X,Y)|\geq |\Gamma^1_Y|. 
\label{eq: cut comparable to type 1 terminals}
\end{equation}

Next, we construct a cut $(X',Y')$ in graph $C'$ based on the cut $(X,Y)$ of $\hat C$, and then use the well-linkedness of the terminals in $\tilde\Gamma'$ in graph $C'$ (from Observation~\ref{obs:C'_terminal_wl}) to bound $|E_{\hat C}(X,Y)|$. We start with $X'=X$ and $Y'=Y$, and then consider the blocks $B^*\in \mset$ one-by-one. Denote the endpoints of $B^*$ by $(x^*,y^*)$.

%We define the set $S(B')$ of vertices to be the union of (i) $V(B')\setminus \set{x',y'}$; and (ii) all sets $X_i$ whose corresponding separator vertex $u_i\in U^*$ lies in $V(B')\setminus\set{x',y'}$. 
Note that $x^*,y^*$ may not belong to the terminal set $\tilde\Gamma'=\tilde \Gamma\cup U$. However, since we have assumed that graph $G$ is $3$-connected, vertex set $V(B^*)\setminus \set{x^*,y^*}$ must contain a terminal of $\tilde\Gamma'$. We denote this terminal by $t_{B^*}$, and we will view this terminal as ``paying'' for $x^*$ and $y^*$ (if $x^*,y^*\in \Gamma^2_X\cup \Gamma^2_Y$).
If both $x^*,y^*\in X$, then we add all vertices of $V(B^*)\setminus \set{x^*,y^*}$ to $X$, and otherwise we add them to $Y$. Notice that, if both $x^*,y^*$ lie in the same set in $\set{X,Y}$, then we do not increase the number of edges in the cut $(X,Y)$. 
%Moreover, we can think of replacing the new terminals $x',y'$ with a terminal $t_{B'}\in \Gamma'\setminus \Gamma''$. 
Assume now that $x^*\in X$ and $y^*\in Y$ (the other case is symmetric). 
In this case, we have increased the number of edges in the cut $(X,Y)$ by at most $\Delta$, by adding all edges that are incident to $x^*$ to this cut. Note however that the edge $(x^*,y^*)$ already belonged to this cut (possibly as a fake edge), so we charge this increase in the cut size to this edge. 
Once all blocks $B^*\in \mset$ are processed in this way, we obtain a cut $(X',Y')$ in graph $C'$. From the above discussion:
\begin{equation}
|E_{C'}(X',Y')|\leq \Delta\cdot |E_{\hat C}(X,Y)|.\label{eq: few new edges in cut}
\end{equation}
Consider now the terminals of $\Gamma_X$. Clearly $\Gamma_X\cap \tilde\Gamma'\subseteq X'$. For each terminal in $\Gamma^2_X$, we have added a terminal of $\tilde\Gamma'$ to $X'$ that pays for it, while each newly added terminal of $X'\cap \tilde\Gamma'$ pays for at most two terminals in $\Gamma^2_X$. Therefore, $|\Gamma^2_X|\leq 2|\tilde\Gamma'\cap X'|$, and similarly $|\Gamma^2_Y|\leq 2|\tilde\Gamma'\cap Y'|$. Since the set $\tilde\Gamma'$ of terminals was $\alpha'$-well-linked in $C'$,
$|E_{C'}(X',Y')|\geq \alpha'\cdot\min\set{|\tilde\Gamma'\cap X'|,|\tilde\Gamma'\cap Y'|}\geq \frac{\alpha'}{2}\cdot \min\set{|\Gamma^2_X|,|\Gamma^2_Y|}$.
Combining this with Equation \ref{eq: few new edges in cut}, we get that $|E_{\hat C}(X,Y)| \geq \frac{\alpha'}{2\Delta}\cdot \min\set{|\Gamma^2_X|,|\Gamma^2_Y|}$.
Combining this latter equation with Equation \ref{eq: cut comparable to type 1 terminals}, and using $\alpha'=8\Delta\alpha$, we conclude that  $|E_{\hat C}(X,Y)| \geq \frac{\alpha'}{4\Delta}\cdot\min\set{|\Gamma_X|,|\Gamma_Y|}\geq 2\alpha \cdot\min\set{|\Gamma_X|,|\Gamma_Y|}$.
%This completes the proof of Claim~\ref{claim: well linkedness}.
Therefore, the set $\Gamma^*$ of terminals is $2\alpha$-well-linked in $\hat C$.

We are now ready to complete the proof of Claim \ref{claim: well linkedness}. Consider any fake edge $e=(x^*,y^*)\in A_{\hat C}$. Then $x^*,y^*$ are endpoints of some block $B^*$ that we have eliminated. Recall that, if vertex $v(B^{**})$ is the parent of $v(B^*)$ in the corresponding decomposition tree $\tau_i$ (or the tree associated with $B^c$), then $v(B^{**})$ has degree $2$ in the tree, and moreover, $\tilde B^{**}$ is not isomorphic to $K_3$. Therefore, graph $\tilde B^{**}$ is $3$-connected, and contains a path connecting $x^*$ to $y^*$, that does not contain any fake edges. We denote this path by $P'(e)$. It is immediate to verify that the paths in $\set{P'(e)\mid e\in A_{\hat C}}$ are mutually disjoint. 
Consider now some cut $(X,Y)$ in $\hat C\setminus A_{\hat C}$. Recall that $|E_{\hat C}(X,Y)|\geq 2\alpha \cdot \min \set{|\Gamma^*\cap X|,|\Gamma^*\cap Y|}$. Let $A'\subseteq E_{\hat C}(X,Y)$ be the set of all fake edges in $E_{\hat C}(X,Y)$. Since, for each fake edge $e\in A'$, path $P'(e)$ must contain a real edge in $E_{\hat C}(X,Y)$, we get that $|E_{\hat C\setminus A_{\hat C}}(X,Y)| =|E_{\hat C}(X,Y)\setminus A'|\geq  |E_{\hat C}(X,Y)|/2\geq \alpha  \cdot \min \set{|\Gamma^*\cap X|,|\Gamma^*\cap Y|}$. We conclude that the set $\Gamma^*$ of terminals is $\alpha$-well-linked in $\hat C\setminus A_{\hat C}$.
%
%   the block $B^{**}$ that is the parent of $B^*$ in the corresponding decomposition tree has exactly one child-
%Now we consider the graph $\hat C\setminus A_{\hat{C}}$. Recall that, for each fake edge $e=(x',y')$, we have associated a path $P'(e)$ that is contained in $\hat C\setminus A_{\hat C}$, such that all associated paths for all fake edges are mutually edge-disjoint. Therefore, for each partition $(X,Y)$ of $V(\hat{C})$, the number of fake edges in $E_{\hat C\setminus A_{\hat C}}(X,Y)$ is at most $|E_{\hat C}(X,Y)|/2$. It follows that the terminals in $\hat{\Gamma}\cap V(\hat C)$ is $\alpha$-well-linked in $\hat C\setminus A_{\hat C}$.
\end{proof}

Lastly, the next claim will complete the proof of  Lemma \ref{lemma: acceptable}.
\begin{claim}
Cluster $\hat C$ has the bridge consistency property.
\end{claim}
\begin{proof}
%\mynote{this needs to be completed. But because we already use a drawing of $C^*\setminus A_{C^*}$ that is induced by the drawing $\rho_{\tilde B_0}$ of $\tilde B_0$, we only need to use the second part of the original proof which is included here. The notation needs to be updated.}
%\mynote{the drawing was already defined before, we claim that the cluster is type-2 acceptable with respect to this drawing. It should not be re-defined here}
%Recall that, from Observation \ref{obs: block containing cluster}, $C'$ is contained in some block $B_0$ in $G\setminus E_1$ that is not contained in a component of $\cset'_1$. Therefore, associated drawing $\hat\psi_{B_0}$ induces a planar drawing of $C'$, and in turn induces a planar drawing of $\hat C\setminus A_{\hat C}$, that we denote by $\rho$. We now show that, for each bridge $R\in \rset_{G}(\hat C\setminus A_{\hat C})$, there is a face in $\rho$, whose boundary contains all vertices of $L(R)$.
We denote by $\rho$ the drawing of $\hat C\setminus A_{\hat C}$ induced by the drawing $\psi_{\hat C}$. We now show that, for each bridge $R\in \rset_{G}(\hat C\setminus A_{\hat C})$, there is a face in the drawing $\rho$, whose boundary contains all vertices of $L(R)$. 
Recall that, from Observation \ref{obs: block containing cluster}, all vertices of $V(\hat C)$ belong to some block $B_0$ in the block decomposition of $G\setminus E_1$ that is not contained in a component of $\cset'_1$, and the drawing $\psi_{\hat C}$ of $\hat C$ is induced by the associated drawing $\hat \psi_{B_0}$ of $B_0$. In particular, block $B_0$ is a good block.

Assume for contradiction that $\hat C$ does not have the bridge property, and let $R\in \rset_{G}(\hat C\setminus A_{\hat C})$ be a witness bridge for $\hat C$ and $\rho$, so no face of the drawing $\rho$ contains all vertices of $L(R)$ on its boundary. %lie on the boundary of a single face of the drawing $\rho$.
Let $\fset$ be the set of all faces of the drawing $\rho$. 
For every vertex $v\in V(B_0)\setminus V(\hat C)$, there is a unique face $F(v)\in \fset$, such that the image of vertex $v$ in $\hat\psi_{B_0}$ lies in the interior of the face $F$.
%\mynote{should we define $T_R$ here? Do we need it to have smallest number of edges, or is it enough that it is some tree spanning $L(R)$? (note that $R$ is no longer a bridge of a block, it's a bridge of $C^*$).}
Recall that $T_R$ is a tree whose leaves are the vertices of $L(R)$, and $T_R\setminus L(R)\subseteq R$.
Assume first that $V(T_R)\cap V(B_0)=L(R)$. In other words, no vertex of $T_R\setminus L(R)$ is a vertex of $B_0$. In this case, there is some bridge $R'\in \rset_G(B_0)$ of $B_0$ that contains $T_R\setminus L(R)$, and so $L(R)\subseteq L(R')$. Since $B_0$ is a good pseudo-block, there must be a face in the drawing $\hat\psi_{B_0}$, the associated drawing of $B_0$, whose boundary contains all vertices of $L(R)$. Therefore, for some face $F\in \fset$, all vertices of $L(R)$ must lie on the boundary of $F$. 
Assume now that there is some vertex $v\in V(T_R)\cap  V(B_0)$ that does not lie in $L(R)$. Let $F(v)$ be the face of $\fset$ whose interior contain the image of $v$ in $\hat\psi_{B_0}$ (recall that $v\not\in V(\hat C)$, so it may not lie on a boundary of a face in $\fset$). We will show that all vertices of $L(R)$ must lie on the boundary of $F(v)$, leading to a contradiction. Let $u$ be an arbitrary vertex of $L(R)$. It suffices to show that $u$ must lie on the boundary of the face $F(v)$.
Let $P\subseteq T_R$ be the unique path connecting $v$ to $u$ in $T_R$. 
%Since the leaves of tree $T_R$ are vertices of $L(R)$, except for $v$, every vertex $x$ of $P$ lies outside $V(C^*)$, and the image of $x$ in $\rho_{\tilde B_0}$ lies in the interior of some face in $\fset$. 
Let $v=v_1,v_2,\ldots,v_r=u$ be all vertices of $P$ that belong to $V(B_0)$, and assume that they appear on $P$ in this order. 
It remains to prove the following observation.
\begin{observation}
$F(v_1)=F(v_2)=\cdots=F(v_{r-1})$. Moreover, $v_r$ lies on the boundary of $F(v_{r-1})$.
\end{observation}
\begin{proof}
Fix some $1\leq i\leq r-1$. Assume that the observation is false for some $v_i$ and $v_{i+1}$.
Then there is some face $F'\in \fset$, such that $v_i$ lies in the interior of $F'$, but $v_{i+1}$ does not lie in the interior or on the boundary of $F'$ (the latter case is only relevant for $i=r-1$). Since the boundary of $F'$ separates $v_i$ from $v_{i+1}$, these two vertices cannot lie on the boundary of the same face in the drawing $\hat\psi_{B_0}$ of $B_0$. 
		
Let $\sigma_i$ be the subpath of $P$ between $v_i$ and $v_{i+1}$. If $\sigma_i$ consists of a single edge connecting $v_i$ and $v_{i+1}$, then either it is a bridge in $\rset_G(B_0)$, or it is an edge of $B_0$. In either case, the endpoints of $\sigma_i$ must lie on the boundary of a single face of the drawing $\hat\psi_{B_0}$, a contradiction (we have used the fact that $B_0$ is a good pseudo-block and is therefore a planar graph). Otherwise, let $\sigma'_i$ be obtained from $\sigma_i$ by deleting the two endpoints from it. Then there must be a bridge $R'\in \rset_G(B_0)$ containing $\sigma'_i$, with $v_{i},v_{i+1}\in L(R')$. But then, since $B_0$ is a good pseudo-block, there must be a face in the drawing $\hat\psi_{B_0}$ of $B_0$, whose boundary contains $v_i$ and $v_{i+1}$, leading to a contradiction. 
%	We conclude that for all $1\leq i<r$, there must be a face $F_i$ in the drawing $\rho'_{\tilde B_0'}$ of $\tilde B_0'$ such that $v_i,v_{i+1}$ lie on the boundary of $F_i$. Notice that, since $\tilde B_0_z\subseteq \tilde B_0'$, and from the way the drawing $\rho'_{\tilde B_0_z}$ was defined, there must be a face $F'_i$ of $\rho'_{\tilde B_0_z}$ that contains $F_i$.	
\end{proof}
\end{proof}

\section{Computing Drawings of Type-1 Acceptable Clusters -- Proof of Theorem \ref{thm: drawing of cluster extension-type 1}}\label{sec: fixing drawing of type 1 clusters}

In this section we prove Theorem \ref{thm: drawing of cluster extension-type 1}.
We fix a cluster $C_i\in \cset_1$. For convenience of notation, we omit the subscript $i$ in the remainder of this section, so in particular $\chi_i$ will be denoted by $\chi$.
Recall from  Property \ref{prop: degrees of terminals for type 1}  that we have assumed, every  terminal $t\in \Gamma(C)$ has degree $1$ in 
$C$, and degree $2$ in $G$. In particular, for each terminal $t\in \Gamma(C)$, there is exactly one bridge $R\in \rset_G(C)$ with $t\in L(R)$. We start by defining a new graph $C^+$, that is obtained from graph $C$, as follows. For every bridge $R\in \rset_C(G)$, we consider an arbitrary ordering $\set{t_1(R),t_2(R),\ldots,t_{|L(R)|}(R)}$ of the vertices of $L(R)$, and we add a set $E'(R)$ of $|L(R)|$ edges, connecting these vertices into a cycle according to this ordering. In other words, $E'(R)=\set{(t_i(R),t_{i+1}(R))\mid 1\leq i\leq |L(R)|}$, where the indexing is modulo $|L(R)|$. We denote the cycle defined by the edges of $E'(R)$ and vertices of $L(R)$ by $J_R$. Let $E'=\bigcup_{R\in \rset_G(C)}E'(R)$. Then $C^+=C\cup E'$. We start with the following two useful observations regarding the new graph $C^+$.

\begin{observation}\label{obs: extended graph 3conn}
	Graph $C^+$ is $3$-connected.
\end{observation}

\begin{proof}
	Assume otherwise, and let $\set{x,y}$ be a $2$-separator for graph $C^+$. Recall that graph $C$ is connected, and, since every vertex $t\in \Gamma(C)$ has degree $1$ in $C$, it cannot be the case that both $x$ and $y$ are terminals in $\Gamma(C)$. Therefore, at least one of the two vertices $x,y$ lies in $V(C)\setminus \Gamma(C)$; we assume w.l.o.g. that it is $x$. Note that, since $\set{x,y}$ is a $2$-separator for $C^+$, there must be at least two connected components in $C^+\setminus\set{x,y}$. We let $X$ be the set of vertices of one such connected component, and we let $Y=V(C^+)\setminus(X\cup \set{x,y})$. Consider now any bridge $R\in \rset_G(C)$. If $y\not\in L(R)$, then, due to the edges of $E'(R)$ that connect all vertices of $L(R)$ to each other via a cycle, either all vertices of $R$ lie in $X$, or all vertices of $R$ lie in $Y$. In the former case, we say that bridge $R$ belongs to $X$, and in the latter we say that it belongs to $Y$. If $y\in L(R)$, then, since $J_R\setminus\set{y}$ remains a connected graph, either all vertices of $L(R)\setminus \set{y}$ lie in $X$ (in which case we say that bridge $R$ belongs to $X$), or they all lie in $Y$ (in which case we say that bridge $R$ belongs to $Y$). Consider now a partition $(X',Y')$ of $V(G)\setminus\set{x,y}$, that is defined as follows. For every vertex $v\in V(C)$, if $v\in X$, then we add $v$ to $X'$, and otherwise we add it to $Y'$. For every bridge $R\in\rset_G(C)$, if $R$ belongs to $X$, then we add all vertices of $V(R)$ to $X'$, and otherwise we add them to $Y'$. It is easy to verify that $(X',Y')$ is indeed a partition of $V(G)\setminus\set{x,y}$, and moreover, no edge connects a vertex of $X'$ to a vertex of $Y'$ in $G$. Therefore, $\set{x,y}$ is a $2$-separator in $G$, a contradiction to the fact that $G$ is $3$-connected.
\end{proof}

\begin{observation}\label{obs: cheap solution for extended graph}
	$\optcro(C^+)\leq O((|\chi|+1)\cdot \poly(\Delta\log n))$.
\end{observation}

\begin{proof}
Since every terminal $t\in \Gamma(C)$ belongs to exactly one set in $\set{L(R)\mid R\in \rset_G(C)}$, we get that	$|\rset_G(R)|\leq |\Gamma(C)|\leq \Delta\mu\leq O(\poly(\Delta\log n))$. 
Recall that we have defined the extension $X_{G}(C)$ of cluster $C$ (in Section \ref{sec:prelim}), that  
is a collection of trees, that contains, for every bridge $R\in \rset_{G}(C)$, a tree $T_R$, whose leaves are the vertices of $L(R)$, and whose inner vertices lie in $R$. For each such tree $T_R$, let $T'_R$ be the tree obtained from $T_R$ by suppressing all degree-$2$ vertices. Since $|L(R)|\leq |\Gamma(C)|\leq O(\poly(\Delta\log n))$, $|E(T'_R)|\leq O(\poly(\Delta\log n))$.
Let $C^1$ be the graph obtained from the union of the cluster $C$, and all trees in $\set{T'_R\mid R\in \rset_G(C)}$. Let $\tilde E=\bigcup_{T\in T'_R}E(T'_R)$. From the above discussion, $|\tilde E|\leq O(\poly(\Delta\log n))$.

Consider now the optimal drawing $\phi$ of $G$. We delete from this drawing the images of all edges and vertices, except those lying in $C$ and in $\bigcup_{R\in \rset_G(C)}T_R$. By suppressing all degree-$2$ vertices lying in trees in $\set{T_R\mid R\in \rset_G(C)}$ (or, equivalently, by concatenating the images of the pair of edges incident to each such vertex), we obtain a drawing $\phi^1$ of graph $C^1$. In this drawing, the total number of crossings in which edges of $C$ participate is bounded by $|\chi|$, as in the drawing $\phi$. However, images of edges in $\tilde E$ may cross each other many times. For each edge $e\in \tilde E$, we first modify its image so it does not cross itself, by removing self-loops as necessary. Additionally, as long as there is a pair $e,e'\in \tilde E$ of edges whose images cross more than once, we can modify the images of $e$ and $e'$ to reduce the number of crossings between them, by using a standard uncrossing operation (see Figure \ref{fig:uncross-curves}); this operation does not create any new additional crossings, and does not increase the number of crossings in which the edges of $E(C)$ participate. Let $\phi^1$ be the final drawing of graph $C^1$ that we obtain at the end of this procedure. Then, since $|\tilde E|\leq O(\poly(\Delta\log n))$, the number of crossings in $\phi^1$ is bounded by $O(|\chi|+\poly(\Delta\log n))$.

\begin{figure}[h]
	\centering
	\subfigure[Before: Image of edge $e$ (red) and image of edge $e'$ (blue) cross twice at $p$ and $q$.]{\scalebox{0.54}{\includegraphics{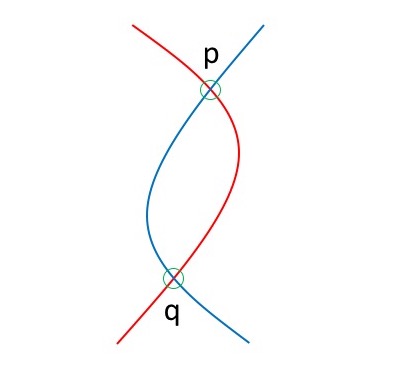}}%\label{fig:non_crossing_representation}
	}
	\hspace{1.5cm}
	\subfigure[After: New image of edge $e$ (red) and new image of edge $e'$ (blue) no longer cross at $p$ or $q$.]{
		\scalebox{0.54}{\includegraphics{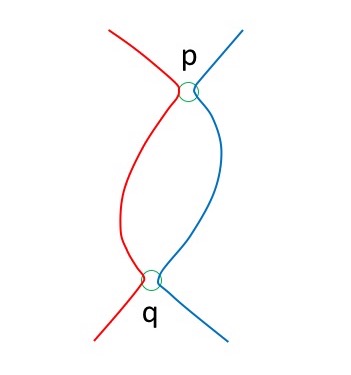}}}
	\caption{An illustration of uncrossing of the images of a pair of edges that cross more than once.}\label{fig:uncross-curves}
\end{figure}

We perform one additional modification to graph $C^1$, to obtain a multigraph $C^2$ as follows. Consider any bridge $R\in \rset_G(C)$, and the corresponding tree $T'_R$. %We root tree $T'_R$ at any vertex $v_R\not\in L(R)$. Let $e=(x,y)\in E(T'_R)$ be any edge of the tree, where $y$ is a child vertex of $x$. We denote by $n_e$ the total number of vertices of $L_R$ in the subtree of $T'_R$ rooted at $y$. 
For every edge $e\in E(T'_R)$, we then create $2\Delta\mu$ parallel copies of the edge $e$. Let $T''_R$ denote the resulting multi-graph that we obtain from tree $T'_R$. Once every bridge $R\in \rset_G(C)$ is processed in this manner, we obtain the final graph $C^2$. We can obtain a drawing $\phi^2$ of the graph $C^2$ from drawing $\phi^1$ of $C^1$ in a natural way: for every tree $T'_R$, for every edge $e\in E(T'_R)$, we draw the $2\Delta\mu$ copies of the edge $e$ in parallel to the drawing of $e$, very close to it. Since $\mu\leq \poly(\Delta\log n)$, every crossing in drawing $\phi^1$ may give rise to at most $\poly(\Delta\log n)$ crossings in $\phi^2$, so the total number of crossings in $\phi^2$ is bounded by $O((|\chi|+1)\poly(\Delta\log n))$. 

We are now ready to define the drawing of the graph $C^+$. We start with the drawing $\phi^2$ of graph $C^2$, which already contains the images of the edges and the vertices of $C$. It now remains to add the images of the edges of $\tilde E$ to this drawing. In order to do so, we consider each bridge $R\in \rset_G(C)$ one by one. Fix any such bridge $R\in \rset_G(C)$, and consider the corresponding vertex set $L(R)=\set{t_1(R),t_2(R),\ldots,t_{|L(R)|}(R)}$. For all $1\leq i\leq |L(R)|$, we let $Q_i(R)$ be a simple path in graph $T''_R$ connecting $t_i(R)$ to $t_{i+1}(R)$, that corresponds to the unique path connecting $t_i(R)$ to $t_{i+1}(R)$ in the tree $T'_R$. Since graph $C^2$ contains $2\Delta\mu$ copies of every edge of $T'_R$, while $|L(R)|\leq \Delta\mu$, we can ensure that the resulting paths $Q_1(R),\ldots,Q_{|L(R)|}(R)$ are mutually edge-disjoint. For all $1\leq i\leq |L(R)|$, we let $\gamma_i(R)$ denote the curve obtained by concatenating the images of all edges lying on path $Q_i(R)$ in drawing $\phi^2$. Intuitively, we would like to map, for all $1\leq i\leq |L(R)|$, the edge $(t_i(R),t_{i+1}(R))$ to the curve $\gamma_i(R)$. One difficulty with this approach is that several curves in the set $\set{\gamma_i(R)}_{i=1}^{|L(R)|}$ may cross in a single point. This is because several paths in $\set{Q_i(R)}_{i=1}^{|L(R)|}$ may pass through a single vertex $v$. Obseve however that $|V(T'(R))|\leq |L(R)|\leq \poly(\Delta\mu)$. For each non-leaf vertex $v\in V(T'(R))$, consider a small disc $\eta(v)$ containing $v$ in its interior, in the current drawing. We slightly modify all curves in $\set{\gamma_i(R)}_{i=1}^{|L(R)|}$ that contain the image of $v$ inside the disc $\eta(v)$ to ensure that every pair of such curves cross at most once inside $\eta(v)$, and no point of $\eta(v)$ is contained in more than two curves. Since the total number of vertices in all graphs in $\set{T'(R)}_{R\in \rset_G(C)}$ is bounded by $O(\poly(\Delta\log n))$, and since $|\tilde E|\leq   O(\poly(\Delta\log n))$, this modification introduces at most  $O(\poly(\Delta\log n))$ additional crossings. We then obtain a valid drawing of graph $C^+$ with at most $O\left ((|\chi|+1)\poly(\Delta \log n)\right )$ crossings.
\end{proof}

We now use the following theorem from~\cite{chuzhoy2011graph}:

\begin{theorem}[Theorem 8 in full version of \cite{chuzhoy2011graph}]\label{thm: old crossing number alg}
There is an efficient algorithm, that, given a $3$-connected graph $G$ with maximum vertex degree $\Delta$ and a planarizing set $E'$ of its edges, computes a drawing $\psi$ of $G$ in the plane with $O\!\left((|E'|^2+|E'|\cdot \optcro(G)) \poly(\Delta)\right)$ crossings. Moreover, the drawing of $G\setminus E'$ induced by $\psi$ is planar.
\end{theorem}

We note that the statement of Theorem 8 in \cite{chuzhoy2011graph} does not include the claim that  the drawing of $G\setminus E'$ induced by $\psi$ is planar. However, it is easy to verify that their algorithm ensures this property, as the algorithm first selects a planar drawing of $G\setminus E'$ and then adds edges of $E'$ to this drawing.

Observe that the set $E'$ of edges is a planarizing set for $C^+$, as $C^+\setminus E'=C$ is a planar graph. Therefore, we can use the algorithm from Theorem \ref{thm: old crossing number alg} to compute a drawing $\psi$ of graph $C^+$, such that the drawing of $C$ induced by $\psi$ is planar. The number of crossings in $\psi$ is bounded by:
\[O\left((|E'|^2+|E'|\cdot \optcro(G)) \poly(\Delta)\right )\leq O\left ((|\chi|+1)\poly(\Delta \log n)\right ).\]
Note that the drawing $\psi$ of graph $C^+$ naturally induces a planar drawing $\psi'$ of graph $C$. 
Since every terminal $t\in \Gamma(C)$ has degree $1$ in $C$, for every terminal $t\in \Gamma(C)$, there is a unique face $F(t)$ in the drawing $\psi'$, such that $t$ lies on the (inner) boundary of $F(t)$. Unfortunately, it is possible that two terminals $t,t'$ lie in the set $L(R)$ of legs for some bridge $R\in \rset_G(C)$, but $F(t)\neq F(t')$. This situation is undesirable as it precludes us from defining the discs $D(R)$ for bridges $R\in \rset_G(C)$ with the required properties. In order to overcome this difficulty, we start with $E^*=\emptyset$, and then gradually add edges of $E(C)$ to $E^*$. We will ensure that, throughout, graph $C\setminus E^*$ remains connected. Eventually, our goal is to ensure that, in the drawing of $C\setminus E^*$ induced by $\psi'$, for every bridge $R\in \rset_G(C)$, all vertices of $L(R)$ lie on the boundary of the single face. Once we achieve this, we will slightly modify the images of the edges incident to the terminals of $C$ in a way that will allow us to define the desired set of discs $\set{D(R)}_{R\in \rset_G(C)}$.

Before we proceed, we define the notion of distances between faces of the current drawing $\psi'$. Intuitively, the distance between a pair $F,F'$ of faces in drawing $\psi'$ is the smallest number of edges that need to be deleted from the current drawing, so that faces $F$ and $F'$ merge into a single face. Equivalently, it is the distance between $F$ and $F'$ in the dual graph. A third equivalent definition is the following: let $\gamma(F,F')$ be a curve originating at a point of $F$ and terminating at a point of $F'$ that intersects the current drawing $\psi'$ at images of edges only (and avoids images of vertices). Among all such curves, choose the one minimizing the number of edges whose images it intersects. Denote by $E(F,F')$ the set of all edges that $\gamma(F,F')$ intersects. Then the distance between $F$ and $F'$ is $|E(F,F')|$. We need the following simple claim.

\begin{claim}\label{claim: small distances between faces}
	For every bridge $R\in \rset_G(C)$, and every index $1\leq i\leq |L(R)|$, the distance between faces $F(t_i(R))$ and $F(t_{i+1}(R))$ is at most $O\left ((|\chi|+1)\poly(\Delta \log n)\right )$.
\end{claim}
\begin{proof}
	Recall that edge $(t_i(R),t_{i+1}(R))$ lies in graph $C^+$, and its image in $\psi$ is a curve connecting a point in $F(t_i(R))$ to a point in  $F(t_{i+1}(R))$, that intersects the image of $C^+$ at edges only. Since the total number of crossings in drawing $\psi$ is at most $O\left ((|\chi|+1)\poly(\Delta \log n)\right )$, we get that the distance between the two faces is also bounded by $O\left ((|\chi|+1)\poly(\Delta \log n)\right )$.
\end{proof}

We now gradually modify the set $E^*$ of edges, by processing the bridges $R\in \rset_G(C)$ one by one. When a bridge $R\in \rset_G(C)$ is processed, we consider each index $1\leq i\leq |L(R)|$ in turn. We now describe an iteration where index $i$ is processed. Let $F$ and $F'$ be the faces in the drawing of graph $C\setminus E^*$ induced by $\psi'$, containing the images of terminals $t_i(R)$ and $t_{i+1}(R)$, respectively. If $F=F'$, then we do nothing. Otherwise, we are guaranteed that the distance between the two faces is at most $O\left ((|\chi|+1)\poly(\Delta \log n)\right )$. Then there is a set $E^*_i(R)$ of at most $O\left ((|\chi|+1)\poly(\Delta \log n)\right )$ edges of $C\setminus E^*$, such that, in the drawing of the graph $C\setminus(E^*\cup E^*_i(R))$ induced by $\psi$, the two faces are merged, and so both terminals $t_i(R)$ and $t_{i+1}(R)$ lie on the boundary of a single face. Moreover, it is easy to verify that, if graph $C\setminus E^*$ is connected, and $E^*_i(R)$ is a minimum-cardinality set of edges with the above properties, then graph $C\setminus (E^*\cup E^*_i(R))$ is also connected. We add the edges of $E^*_i(R)$ to $E^*$, and continue to the next iteration. Once every bridge $R$ is processed, we are guaranteed that, in the drawing of $C\setminus E^*$ induced by $\psi$, for every bridge $R\in \rset$, there is a single face $F(R)$, whose boundary contains images of all vertices in $L(R)$. We are also guaranteed that graph $C\setminus E^*$ is connected. Since we add at most $O\left ((|\chi|+1)\poly(\Delta \log n)\right )$ edges to $E^*$ in each iteration, and the number of iterations is bounded by $|\Gamma(C)|\leq \poly(\Delta\log n)$, we get that at the end of the algorithm $|E^*|\leq O\left ((|\chi|+1)\poly(\Delta \log n)\right )$.

Consider now the drawing of graph $C\setminus E^*$ induced by $\psi$. Recall that for every bridge $R\in \rset_G(C)$, we have denoted by $F(R)$ the face in this drawing whose boundary contains all vertices of $L(R)$. We denote by $\fset=\set{F(R)\mid R\in \rset_G(C)}$. Next, we consider each face $F\in \fset$ one by one. Let $\rset(F)\subseteq \rset_G(C)$ be the set of all bridges $R$ with $F(R)=F$. We select one arbitrary disc $D$ in the interior of the face $F$, and set, for every bridge $R\in \rset(F)$, $D(F)=D$. For every terminal $t\in \bigcup_{R\in \rset(F)}L(R)$, we consider the unique edge $e_t$ that is incident to $t$ in $C$. We extend the image of this edge, so that its endpoint $t$ lies on disc $D$, but the interior of the edge remains disjoint from $D$. This is done by appending, to the current drawing of the edge $e_t$, a curve $\gamma_t$, connecting the image of $t$ to a point on the disc $D$. The set $\Gamma(F)=\set{\gamma_t\mid t\in \bigcup_{R\in \rset(F)}L(R)}$ of curves is defined as follows. Consider a terminal $t\in\bigcup_{R\in \rset(F)}L(R)$. We start by letting $\gamma_t$ be a curve connecting the image of $t$ to a point on the boundary of the disc $D$, so that $\gamma_t$ is disjoint from the images of the edges in $E(C)\setminus E^*$, and crosses the image of each edge in $E^*$ at most once. It is easy to verify that such a curve can be constructed, for example, by following the curves $\gamma(F,F')$ that we used in order to merge pairs of faces by adding edges to set $E^*$. Next, we use standard uncrossing procedure to ensure that the resulting curves in $\Gamma(F)=\set{\gamma_t\mid t\in \bigcup_{R\in \rset(F)}L(R)}$ do not cross each other. This step only modifies the curves in $\Gamma(F)$ and does not introduce any new crossings.

Once we process every face $F\in \fset$, we obtain the final drawing $\psi^*$ of the cluster $C$. Notice that the only difference between $\psi^*$ and the planar drawing of $C$ induced by $\psi$ is that we have modified the images of the edges in set $\set{e_t\mid t\in \Gamma(C)}$, by appending a curve $\gamma_t$ to the image of each such edge $e_t$. The total number of crossings in $\psi^*$ is bounded by $|E^*|\cdot |\Gamma(C)|\leq  O\left ((|\chi|+1)\poly(\Delta \log n)\right )$. From our construction, we also guarantee that graph $C\setminus E^*$ is connected, and that the drawing of $C\setminus E^*$ induced by $\psi^*$ is planar. Additionally, for every bridge $R\in \rset_G(C)$, we have defined a disc $D(R)$,  such that the images of all vertices in $L(R)$ are drawn on the boundary of $D(R)$, the interior of $D(R)$ is disjoint from the drawing $\psi^*$ of $C$, and the image of every edge of $C$ is disjoint from $D(R)$, except possibly for its endpoint that lies on $D(R)$. For every pair $R\neq R'$ of bridges, either $D(R)=D(R')$, or $D(R)\cap D(R')=\emptyset$ must hold. This completes the proof of Theorem \ref{thm: drawing of cluster extension-type 1}.
\section{Obtaining a Canonical Drawing: Proof of Theorem \ref{thm: canonical drawing}}
\label{sec:canonical drawing}

In this section we provide the proof of Theorem~\ref{thm: canonical drawing}. For brevity, we will refer to type-1 and type-2 acceptable clusters of $\cset_1\cup \cset_2$ as type-1 and type-2 clusters, respectively.
Throughout this section, we assume that we are given a  $n$-vertex graph $G$ 
with maximum vertex degree at most $\Delta$, and a decomposition $\dset=\left ( E'',A,\cset_1,\cset_2,\{\psi_C\}_{C\in\cset_2},\pset(A)\right )$  of $G$ into acceptable clusters. Recall that $E''$ is a planarizing set of edges for $G$; the endpoints of the edges in $E''$ are called terminals, and we denote the set of all terminals by $\Gamma$. Set $A$ contains fake edges, whose endpoints must be in $\Gamma$.
The set of all connected components of $(G\setminus E'')\cup A$ is $\cset_1\cup \cset_2$, and we refer to the elements of $\cset_1\cup\cset_2$ as clusters.
Additionally, we are given, for each cluster $C\in \cset_1\cup \cset_2$, a drawing $\psi_C$ of $C$ on the sphere, and, for every bridge $R\in \rset_G(C\setminus A_C)$, a disc $D(R)$ on the sphere. We are guaranteed that every cluster in $\cset_1$ is a type-1  cluster, and every cluster in $\cset_2$ is a type-2 cluster with respect to the drawing $\psi_C$. Lastly, the set $\pset(A)$ of paths defines a legal embedding of the fake edges. 
 We also assume that we are given some drawing $\phi$ of $G$, and our goal is to transform this drawing, so that it becomes canonical with respect to all clusters in $\cset_1\cup \cset_2$, such that the total number of crossings only increases slightly. At a high level, the algorithm processes all clusters in $\cset_1\cup \cset_2$ one-by-one. In each iteration, we will modify the current drawing of $G$ so that it will become canonical with respect to the cluster $C$ that is processed in that iteration. % We will also add the drawing of the fake edge $e_i$ associated with $C_i$ (if such an edge exists) to the drawing. Once every type-1 cluster is processed, we turn to processing type-2 clusters, where again in every iteration we process a distinct cluster $C$ and modify the current drawing of $G$ so that it becomes canonical with respect to $C$. This step exploits the drawings of the fake edges of $A_C$ that were added when their corresponding type-1 clusters were processed. Once cluster $C$ is processed, we delete the drawings of the edges in $A_C$ from the resulting drawing.
The main tool that we use in order to iteratively modify the graph is procedure \procdraw that is described in Section \ref{subsec: main procedure}. Before we describe the procedure, we need two additional tools: the notion of non-interfering paths, and the notion of irregular vertices and edges that was introduced in \cite{chuzhoy2011graph}. We describe these two tools in Sections \ref{sec: non-interfering} and \ref{subsec: irregular} respectively.

\subsection{Non-Interfering Paths}
%--------------------------------------------
%--------------------------------------------
%--------------------------------------------
\label{sec: non-interfering}

In this subsection we define the notion of non-interfering paths and prove a lemma that allows us to find such paths. The notion of the non-interfering paths is defined with respect to any given graph $H'$, and will eventually be applied to various subgraphs of $G\cup A$.

We assume that we are given any graph $H'$ and a drawing $\psi$ of $H'$ on the sphere.
For every vertex $v\in V(H')$, we let $\eta(v)$ be a small closed disc that contains $v$ in its interior. In particular, no image of the vertices of $V(H')\setminus\set{v}$ appears in the disc; if $\psi(e)\cap \eta(v)\neq\emptyset$ for any edge $e$, then $e$ must be incident to the vertex $v$, and $\psi(e)\cap \eta(v)$ must be a simple curve (a curve that does not cross itself) that intersects the boundary of $\eta(v)$ at exactly one point. The discs $\eta(v)$ that correspond to distinct vertices must be disjoint.

\begin{figure}[h]
	\centering
	\scalebox{0.5}{\includegraphics{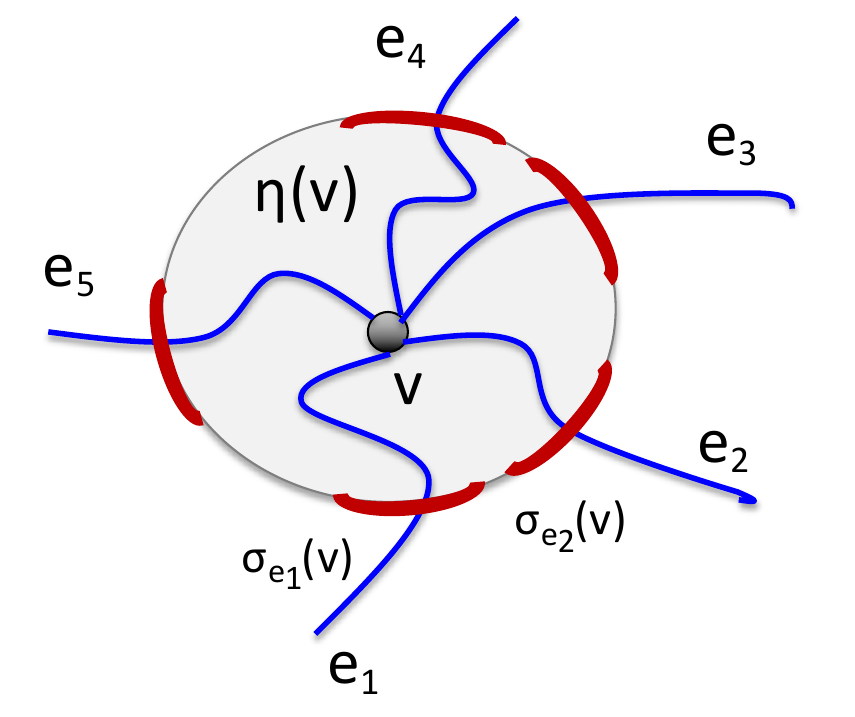}}
	\caption{Disc $\eta(v)$ and segments $\sigma_{e_i}(v)$.\label{fig: disc and segments}}
	\end{figure}

We now fix some vertex $v\in V(H')$. Let $\delta(v)=\set{e_1,\ldots,e_r}$ be the set of all edges that are incident to vertex $v$ in $H'$. For each such edge $e_i$, let $p_i$ be the unique point on the image of edge $e_i$ that lies on the boundary of the disc $\eta(v)$. Notice that the circular ordering of the points $p_1,\ldots,p_r$ on the boundary of $\eta(v)$ defines a circular ordering $\tilde \oset(v)$ of the edges in $\delta(v)$. 
We call this ordering \emph{the ordering of the edges of $\delta(v)$ in $\psi$, as they enter vertex $v$}.
 For each edge $e_i\in \delta(v)$, we let $\sigma_{e_i}(v)$ be a small closed segment of the boundary of the disc $\eta(v)$, that contains the point $p_i$ in its interior, such that all segments in $\set{\sigma_{e_i}(v)}_{i=1}^r$ are mutually disjoint (see Figure \ref{fig: disc and segments}). Next, we define the notion of a thin strip around a path in $H'$.

\paragraph{Thin strip around a path.}
 Let $P$ be any path in $H'$. We denote the endpoints of $P$ by $u$ and $v$, we denote by $e_v$ the unique edge on path $P$ that is incident to vertex $v$, and we denote by $e_u$ the unique edge of $P$ that is incident to vertex $u$. Recall that the image $\psi(P)$ of path $P$ in $\psi$ is a curve obtained by concatenating the images of its edges in $\psi$. 
We define a \emph{thin strip} $S_P$ around the image of $P$ in $\psi$, by adding two curves, $\gamma_L$ and $\gamma_R$, immediately to the left and to the right of the image of $P$ respectively, that follow the image of $P$. The two curves originate at the two endpoints of the segment $\sigma_{e_v}(v)$, and terminate at the two endpoints of the segment $\sigma_{e_u}(u)$. They do not cross each other except when the image of $P$ crosses itself, and do not intersect the interiors of the discs $\eta(v)$ and $\eta(u)$. The two curves are extremely close to the image of $P$ in $\psi$. The region of the sphere, whose boundary is the concatenation of $\sigma_{e_v}(v),\gamma_L,\sigma_{e_u}(u)$ and $\gamma_R$, and that contains the image of $P$ (except for $\psi(P)\cap \eta(u)$ and $\psi(P)\cap \eta(v)$), defines the thin strip $S_P$ around $P$.
We draw the curves $\gamma_L$ and $\gamma_R$ so that the resulting strip $S_P$ contains images of vertices of $P\setminus\set{u,v}$, and no other vertices of $H'$. Additionally, the only edges of $G$ whose images have a non-empty intersection with $S_P$ are (i) edges that are incident to vertices of $P\setminus\set{u,v}$; and (ii) edges whose images cross the edges of $P$; 
see Figure~\ref{fig:strip} for an illustration.
For each such edge $e$, $\psi(e)\cap S_P$ is a collection of disjoint open curves, where each curve contains a point that belongs to the image of $P$ (the corresponding point is either an image of a vertex of $P$, or a crossing point of $e$ with an edge of $P$).
We can similarly define a thin strip $S_{e'}$ around the image of an edge $e'$ in $\psi$, by considering a path that only consists of the edge $e'$. %\snote{should we mention that non-interfering paths are similar to confluent paths used previously in the literature?}

\begin{figure}[h]
	\centering
	\scalebox{0.30}{\includegraphics{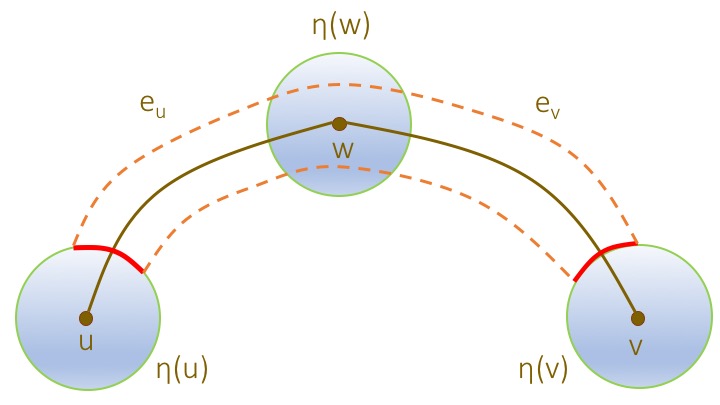}}
\caption{A thin strip $S_P$ around path $P=(u,w,v)$. The segments $\sigma_{e_v}(v)$ and $\sigma_{e_u}(u)$ are shown in red and the curves $\gamma_L$ and $\gamma_R$ are dashed orange lines. \label{fig:strip}}
\end{figure}

\iffalse
\begin{figure}[h]
	\centering
	\subfigure[The thin strip $S_P$ around path $P=(u,w,v)$. The segments $\sigma_{e_v}(v)$ and $\sigma_{e_u}(u)$ are shown in red and the curves $\gamma_L$ and $\gamma_R$ are dashed orange lines.]{\scalebox{0.30}{\includegraphics{strip.jpg}}\label{fig:strip}}
	\hspace{1cm}
	\subfigure[The ordering $\tilde O(e)$ of the paths in $\qset'_e$ in $S_e$.]{
		\scalebox{0.30}{\includegraphics{edge-ordering.jpg}}\label{fig:order_strip}}
	\caption{An illustration of a thin strip and the ordering of paths in the strip.}\label{fig:strips} %\mynote{this figure needs to be fixed. First, the discs are $\eta(v)$, not $\eta_v$. The quality seems quite bad, the $\sigma$ segments and the orange dashed lines are almost impossible to see. In the second figure, the curves are so close to each other it's impossible to see with a naked eye what's going on. Please make the strip wider, the lines wider and more visible, and bigger spacing between the lines. }
\end{figure}
\fi

\begin{figure}[h]
\centering
\subfigure[In this figure we consider a collection of paths connecting every leaf of the tree to its root. These paths are non-interfering, and the curves that are shown in red are their non-interfering representation with respect to this drawing of the tree.]{\scalebox{0.33}{\includegraphics{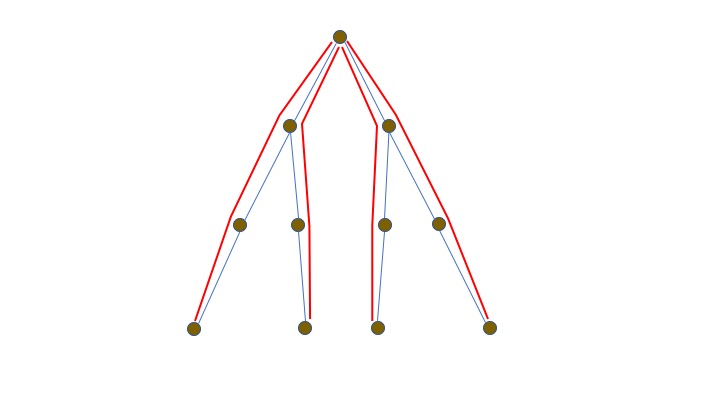}}\label{fig:non_crossing_representation}}
\hspace{1cm}
\subfigure[The red path and the green path in this figure are not non-interfering with respect to this drawing.]{
		\scalebox{0.28}{\includegraphics{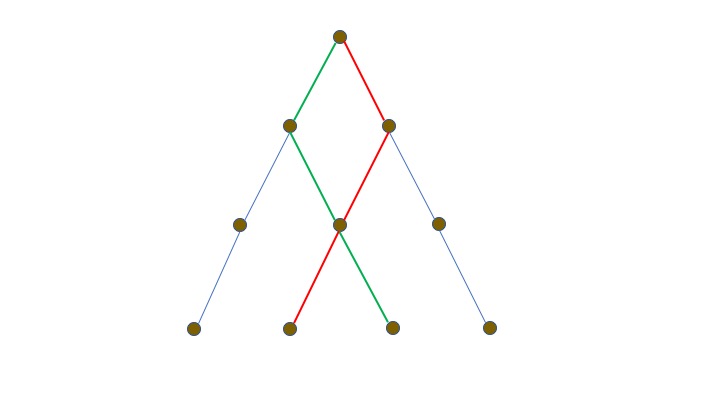}}}
	\caption{Non-interfering paths and non-interfering representations.%Path $P$ is shown in red and path $P'$ is shown in green. (a) also shows a non-interfering representation of paths $P$ and $P'$.%\mynote{this figure and the caption need to be fixed. It is not clear what are the paths and what are the curves. The red and green are not paths but curves representing the paths. I would give a tree as an example and say that the paths are all paths connecting the leaves to the root.}
	}\label{fig:non_crossing}
\end{figure}

\begin{definition}[Non-interfering Paths]
	Let $H'$ be a planar graph and let $\psi$ be a planar drawing of $H'$. Let $\pset$ be a set of paths in $H'$, where for each path $P\in \pset$, we denote the endpoints of $P$ by $u_P$ and $v_P$. We say the paths of $\pset$ are \emph{non-interfering with respect to $\psi$} (see Figure~\ref{fig:non_crossing}), iff there exists a collection $\set{\gamma_P}_{P\in\pset}$ of curves, such that:
	\begin{enumerate}
		\item for each path $P\in \pset$, the curve $\gamma_P$ connects $\psi(u_P)$ to $\psi(v_P)$, and is contained in $\eta(u_P)\cup \eta(v_p)\cup S_P$, where $S_P$ is the thin strip around $P$ in $\psi$;
		and
		\item for every pair $P,P'\in \pset$ of distinct paths, the curves $\gamma_P$ and $\gamma_{P'}$ are disjoint.
	\end{enumerate}
	The set $\set{\gamma_P}_{P\in\pset}$ of curves with the above properties is called a \emph{non-interfering representation} of $\pset$ with respect to $\psi$.
\end{definition}
Note that the curve $\gamma_P$ in the above definition may cross $\psi(P)$ (the image of the path $P$ in drawing $\psi$) arbitrarily many times.
Also note that, as shown in Figure~\ref{fig:non_crossing}, non-interfering paths may share vertices and edges. 
We  also use the following two definitions.

%\mynote{I think it would be much more useful to draw a figure of a tree with say 4-5 leaves and show how these curves look like. With 2 paths it's easy to see that they exist but with more paths it's confusing.}

\begin{definition}
Given a graph $H'$, a set $\Gamma$ of its vertices, together with another vertex $u$ (that may belong to $\Gamma$), a \emph{routing of the vertices of $\Gamma$ to $u$} is a set $\qset=\set{Q_v\mid v\in \Gamma}$ of paths, where for each vertex $v\in \Gamma$, path $Q_v$ connects $v$ to $u$. We sometimes say that set $\qset$ of paths \emph{routes vertices of $\Gamma$ to $u$}. 
\end{definition}

\begin{definition}
Given a set $\qset$ of paths in a graph $H'$, for each edge $e\in E(H')$, we denote by $\cong_{\qset}(e)$ the \emph{congestion} of the paths in $\qset$ on edge $e$ -- the number of paths in $\qset$ that contain $e$. We denote by $\cong_{H'}(\qset)=\max_{e\in E(H')}\set{\cong_{\qset}(e)}$ the total congestion caused by the set $\qset$ of paths in $H'$. 
%\znote{In prelim we do have a notation  $c(\qset)$ for this}\mynote{if we don't use it anywhere else it can be defined here. If we do I think $\cong$ is a better notation because $c$ is used all the time for constants and will be hard to remember what it is}
\end{definition}

%		. We say that a set $\qset$ of paths in $H'$ \emph{routes the vertices of $\Gamma$ to $u$} if $|\qset|=|\Gamma|$ and, for every vertex $v\in \Gamma$, there is a path $Q_v\in \qset$, connecting $v$ to $u$.
Assume now that we are given a set $\Gamma$ of vertices of $H'$, a vertex $u\in V(H')$, and a routing $\qset$ of the vertices of $\Gamma$ to $u$ in $H'$. Let $\oset$ be any ordering of the vertices of $H'$. We say that $\oset$ is \emph{consistent} with the set $\qset$ of paths if, for every path $Q_v\in \qset$, for every pair $x,y$ of distinct vertices of $Q$, where $y$ lies closer to $u$ than $x$ on $Q$, vertex $x$ appears before vertex $y$ in $\oset$.
The following lemma allows us to transform any routing of a set $\Gamma$ of vertices to a given vertex $u$ into a collection of non-interfering paths, and to compute an ordering of vertices of $H'$ that is consistent with the resulting set of paths. The sets of paths produced by this lemma will be used as guiding paths by procedure \procdraw in order to modify the drawing of $G$. The proof of the lemma is deferred to Appendix~\ref{apd:non_interfering}.

\begin{lemma}
\label{lem:uncrossing}
There is an efficient algorithm, that, given a planar graph $H'$ with a planar drawing $\psi$ of $H'$, a collection $\Gamma$ of vertices of $H'$ and another vertex $u\in V(H')$ (where possibly $u\in \Gamma$), together with a set $\qset$ of paths routing $\Gamma$ to $u$ in $H'$, computes another set $\qset'$ of paths routing $\Gamma$ to $u$, such that the set $\qset'$ of paths is non-interfering with respect to $\psi$, and for every edge $e\in E(H')$, $\cong_{\qset'}(e)\le \cong_{\qset}(e)$.
Additionally, the algorithm computes a non-interfering representation $\set{\gamma_Q\mid Q\in \qset'}$ of $\qset'$ and an ordering $\oset$ of the vertices of $H'$ that is consistent with the set $\qset'$ of paths.
\end{lemma}

Consider the set $\qset'$ of paths given by Lemma \ref{lem:uncrossing}. Even though the paths in $\qset'$ are undirected, it may be convenient to think of them as being directed towards $u$. Let $e=(x,y)\in E(H')$ be an edge, and assume that $x$ appears before $y$ in the ordering $\oset$. Let $\pset(e)\subseteq\qset'$ be the subset of all paths $Q$ with $e\in E(Q)$. Notice that all paths in $\pset(e)$ must traverse the edge $e$ in the direction from $x$ to $y$, since the ordering $\oset$ of $V(H')$ is consistent with respect to $\qset'$. Moreover, if we consider the intersections of the curves $\set{\gamma_Q\mid Q\in \pset(e)}$ with the thin strip $S_e$ around the image of the edge $e=(u,v)$ in $\psi$, then the order in which these curves traverse $S_e$ (e.g. the order in which they intersect the segment $\sigma_e(u)$) naturally defines an ordering of the paths in $\pset(e)$. We denote this ordering by $\tilde\oset(e)$, and we refer to it as the \emph{ordering of the paths in $\pset(e)$ in the strip $S_e$}; see Figure~\ref{fig:order_strip} for an illustration.
%\mynote{it would be good to add a figure}

\begin{figure}[h]
	\centering
	\scalebox{0.30}{\includegraphics{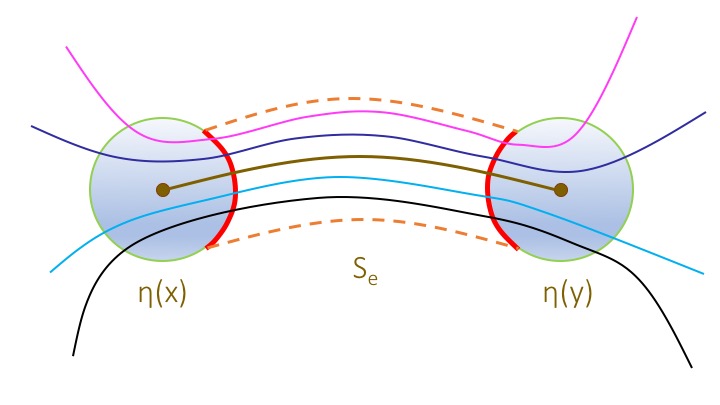}}
	\caption{The ordering $\tilde O(e)$ of the paths in $\qset'_e$ in $S_e$. \label{fig:order_strip}}
\end{figure}

\subsection{Irregular Vertices and Edges}\label{subsec: irregular}
In this subsection, we provide the definitions of irregular vertices and edges from~\cite{chuzhoy2011graph}, and then state a lemma from~\cite{chuzhoy2011graph} about them. Let $H'=(V,E)$ be a connected graph and let $\phi$ and $\psi$ be a pair of drawings of $H'$ in the plane.

As before, we denote by $S_2(H')$ the set of all vertices that participate in $2$-separators in $H'$, that is, vertex $v\in S_2(H')$ iff there is another vertex $u\in V$, such that graph $H'\setminus\set{u,v}$ is not connected. We denote by $E_2(H')$ the set of all edges that have both endpoints in set $S_2(H')$. 
%We also denote by $S_1(H')$ the set of all vertices that are separator vertices in $H'$, and we denote by $E_1(H')$ the set of all edges that are incident to vertices of $S_1(H')$.

\begin{definition}[Irregular Vertices]
	\label{def:irregular_vertex}
	We say that a vertex $v$ of $H'$ is \emph{irregular} (with respect to $\phi$ and $\psi$) iff (i) its degree in $H'$ is greater than $2$;
	and (ii) the circular ordering of the edges incident on it, as their images enter $v$, is different in $\phi$ and $\psi$ (ignoring the orientation). 
\end{definition}
We denote the set of all vertices that are irregular with respect to $\phi$ and $\psi$ by $\ir_V(\phi,\psi)$, and we call all other vertices \emph{regular}.
\begin{definition}[Irregular Paths and Edges]
	\label{def:irregular_edge}
	
	We say that a path $P$ with endpoints $x$ and $y$ in $H'$ is \emph{irregular} iff $x$ and
	$y$ both have degree at least $3$ in $H'$, all other vertices of $P$ have degree $2$ in $H'$, vertices $x$ and $y$ are regular, but their
	orientations differ in $\phi$ and $\psi$. In other words, the orderings of the edges adjacent to $x$ and to $y$ are identical in
	both drawings, but the pairwise orientations are different: for one of the two vertices, the orientations are
	identical in both drawings (say clock-wise), while for the other vertex, the orientations are opposite (one is
	clock-wise, and the other is counter-clock-wise). An edge $e$ is an \emph{irregular edge} with respect to $\phi$ and $\psi$ iff it is the first or the last
	edge on an irregular path. In particular, if the irregular path only consists of a single edge $e$, then $e$ is an irregular
	edge.
	%
	%
	\iffalse
	We say that an edge $(v,v')$ of $H'$ is \emph{irregular} (with respect to $\phi$ and $\psi$), iff $v$ and $v'$ have degree at least $3$ in $H'$, vertices $v$ and $v'$ are {\bf regular}, but their
	orientations differ in $\phi$ and $\psi$. That is, the orderings of the edges adjacent to $v$ and to $v'$, as their images enter these vertices, are identical in both drawings, but the pairwise orientations are different: for one of the two vertices, the orientations are identical in both drawings (say clock-wise), while for the other vertex, the orientations are opposite (one is clock-wise, and the other is counter-clock-wise). 
	\fi
\end{definition}
We denote the set of all edges that are irregular with respect to $\phi$ and $\psi$ by $\ir_E(\phi,\psi)$, and we call all other edges regular.
%The following lemma bounds the number of irregular vertices and irregular edges between a (non-planar) drawing and a planar drawing of a $2$-connected planar graph. 
The following lemma is a re-statement of Lemma 2 from Section B from the arxiv version of \cite{chuzhoy2011graph} for the special case where the graph $H'$ is $2$-connected.

\begin{lemma} (\cite{chuzhoy2011graph})
	\label{lem:irregular_bounded_by_crossing}
	Let $H'$ be a $2$-connected planar graph, let $\phi$ be an arbitrary drawing of $H'$ in the plane, and let $\rho$ be a planar drawing of $H'$. %Moreover, let $S_2$ be the set of vertices of $H'$ that participate in some $2$-separator of $H'$, %(i.e., $S_2 = \{u\in V(H')\colon \exists v\in V(H')$ s.t. $(u,v)$ is a $2$-separator in $H'\}$) 
%and let $E_2$ be the set of edges of $H'$ with both endpoints in $S_2$. 
Then
	\[
	|\ir_V(\phi,\rho)\setminus S_2(H')| + |\ir_E(\phi,\rho)\setminus E_2(H')| \leq O(\cro(\phi)).
	\]
\end{lemma}

%--------------------------------------------
%--------------------------------------------
%--------------------------------------------
%--------------------------------------------
\subsection{Main Subroutine: Procedure \procdraw}\label{subsec: main procedure}

%--------------------------------------------
%--------------------------------------------
%--------------------------------------------
%--------------------------------------------

In this subsection we describe and analyze procedure \procdraw, that is central to the proof of Theorem \ref{thm: canonical drawing}. 
We note that a similar procedure was introduced in \cite{chuzhoy2011algorithm} (see Section D of the full version).
The procedure will be applied repeatedly to every cluster $C\in \cset_1\cup \cset_2$ (and more precisely, to several faces in the drawing $\psi_C$ of the cluster $C$), with the goal of transforming the current drawing of the graph $G$ into a drawing that is canonical with respect to $C$.

\begin{figure}[h]
\centering
\subfigure[Schematic view of graph $G$.]{\scalebox{0.6}{\includegraphics{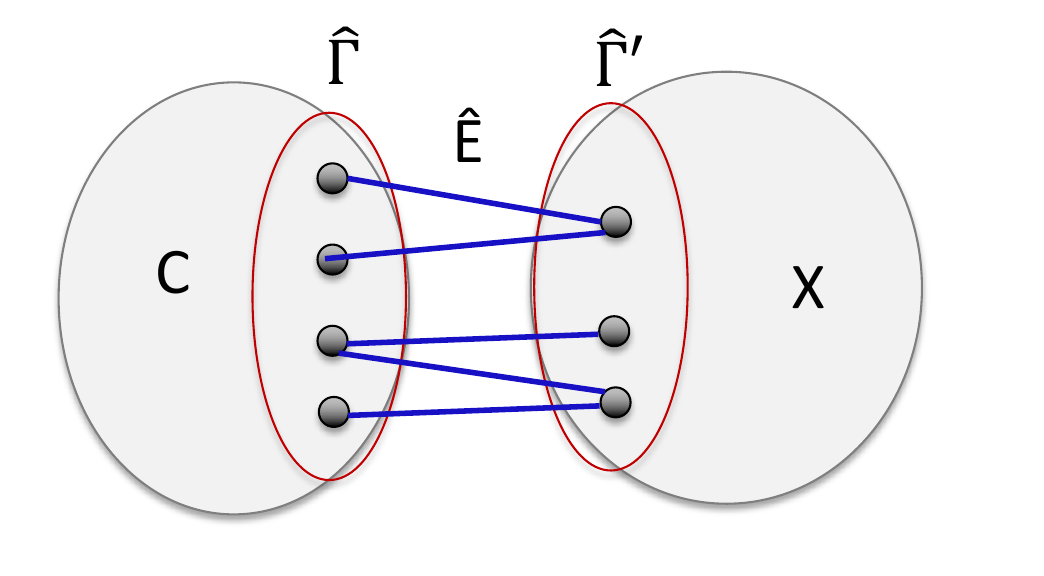}}\label{fig: graph G}}
\hspace{1cm}
\subfigure[Drawing $\psi$ of $C$ and the disc $D$. The vertices of $\Gamma$ are shown in red.]{
\scalebox{0.6}{\includegraphics{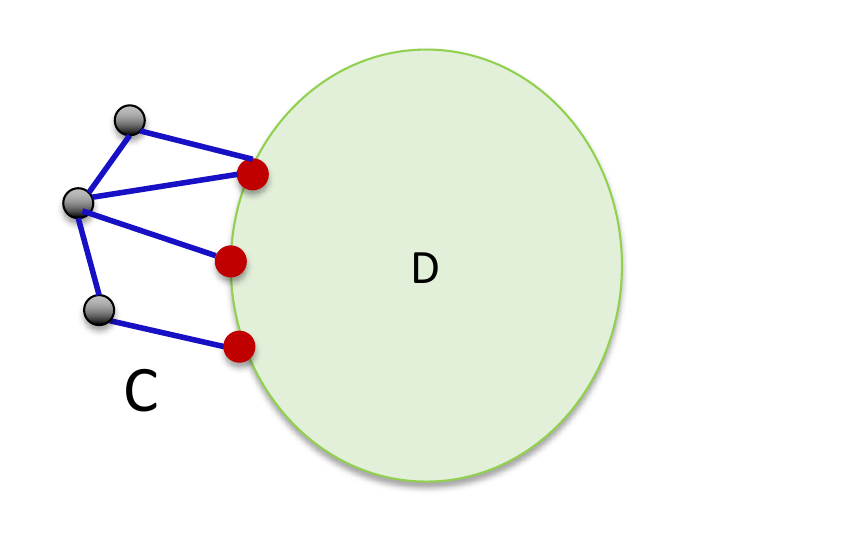}}\label{fig: C and disc}}
%\hspace{1cm}
%\subfigure[caption subfigure 3]{\scalebox{0.2}{\includegraphics{file3.pdf}}\label{fig: label3}}
\caption{Input to \procdraw\label{fig: input to procdraw}}
\end{figure}

Intuitively, the input to the procedure consists of two disjoint graphs: graph $C$ (that we can think of as a cluster of $\cset_1\cup \cset_2)$, and graph $X$ (that we can think of, somewhat imprecisely, as the rest of the graph $G$, or as some bridge in $\rset_G(C)$). Additionally, we are given a set $\hat E$ of edges that connect some vertices of $X$ to some vertices of $C$. We denote by $\hat \Gamma$ and by $\hat \Gamma'$ the sets of endpoints of the edges in $\hat E$ lying in $C$ and $X$, respectively. Abusing the notation, in this subsection, we denote by $G$ the graph that is the union of $X,C$, and the set $\hat E$ of edges (see Figure \ref{fig: graph G}).

We assume that we are given some drawing $\phi$ of $G$ on the sphere (which represents, intuitively, the current drawing of the whole graph $G$), and another drawing $\psi$ of $C$ on the sphere (which will eventually be the canonical drawing $\psi_C$ of $C$). Furthermore, we are given a closed disc $D$ on the sphere, such that, in the drawing $\psi$, the images of all edges of $C$ are internally disjoint from $D$, and the images of all vertices of $C$ are disjoint from $D$, except that the images of vertices of $\hat \Gamma$ lie on the boundary of $D$ (see Figure \ref{fig: C and disc}). This disc $D$ will correspond to the discs $D(R)$ that we have defined for the various bridges $R\in \rset_G(C)$.

Lastly, we are given some vertex $u^*\in V(C)$, and a set $\qset=\set{Q_v\mid v\in \hat \Gamma}$ of paths in $C$ that route $\hat \Gamma$ to $u^*$. %, such that the paths in $\qset$ are non-interfering with respect to $\psi$, together with a non-interfering representation $\set{\gamma_Q\mid Q\in \qset}$ of paths in $\qset$ with respect to $\psi$. %, and an ordering $\oset$ of the vertices of $C$ that is consistent with the set $\qset$ of paths. 
Intuitively, we will use the images of the paths in $\qset$ (after slightly modifying them) in the drawing $\phi$ as guiding lines in order to modify the drawing $\phi$. %\mynote{to define earlier}   
We let $J\subseteq G$ be the graph containing all vertices and edges that participate in the paths in $\qset$, and we assume that the drawing of $J$ induced by $\psi$ is planar. 

The goal of \procdraw is to compute a new drawing $\phi'$ of $G$, such that the drawing of $C$ induced by $\phi'$ is identical to $\psi$ (but the orientation may be arbitrary), and all vertices and edges of $X$ are drawn in the interior of the disc $D$, that is defined with respect to $\psi$. We would also like to ensure that the number of crossings in $\phi'$ is not much larger than the number of crossings in $\phi$. We now formally summarize the input and the output of the procedure $\procdraw$.

\paragraph{Input.} The input to procedure $\procdraw$ consists of:

\begin{itemize}
	\item Two disjoint graphs $C,X$, subsets $\hat \Gamma\subseteq V(C)$, $\hat \Gamma'\subseteq V(X)$ of vertices, and a set $\hat E$ of edges that connect vertices of $\hat \Gamma$ to vertices of $\hat \Gamma'$, such that every vertex in $\hat \Gamma\cup \hat \Gamma'$ is an endpoint of at least one edge of $\hat E$ (see Figure \ref{fig: graph G}). We denote $G=C\cup X\cup \hat E$, and we denote the maximum vertex degree in $G$ by $\Delta$;

	\item A vertex $u^*\in V(C)$, and a set $\qset=\set{Q_v\mid v\in \hat \Gamma}$ of paths in $C$ that route $\hat \Gamma$ to $u^*$. We refer to the paths in $\qset$ as the \emph{guiding paths} for the procedure, and 	
we let $J\subseteq G$ be the graph containing all vertices and edges that participate in the paths in $\qset$;

	\item An arbitrary drawing $\phi$ of graph $G$ on the sphere; and %, such that for every pair of edges of $G$, their images cross at most once in $\phi$;

	\item A drawing $\psi$ of graph $C$ on the sphere, such that the drawing of $J$ induced by $\psi$ is planar, and additionally a closed disc $D$ on the sphere, such that, in drawing $\psi$, the images of vertices of $V(C)$ are disjoint from the disc $D$, except for the vertices of $\hat \Gamma$ whose images lie on the boundary of $D$, and  the images of the edges of $E(C)$ are disjoint from the disc $D$, except for their endpoints that belong to $\hat \Gamma$.

 %, such that the paths in $\qset$ are non-interfering with respect to $\psi$, %and for every path $Q_v\in \qset$, all vertices of $Q_v\setminus\set{v}$ lie in $C$,
%	together with a non-interfering representation $\set{\gamma_Q\mid Q\in \qset}$ of paths in $\qset$ with respect to $\psi$. %, and an ordering $\oset$ of the vertices of $C$ that is consistent with the set $\qset$ of paths. 
\end{itemize}

\paragraph{Output.}
The output of the procedure \procdraw is a drawing $\phi'$ of $G$ on the sphere, that has the following properties:
\begin{itemize}
	\item The drawing of $C$ induced by $\phi'$ is identical to $\psi$ (but the orientation may be different); 
	\item All vertices and edges of $X$ are drawn in the interior of the disc $D$ in $\phi'$ (the disc $D$ is defined with respect to $\psi$); and
	\item The edges of $\hat E$ are drawn inside the disc $D$, and they only intersect the boundary of $D $ at their endpoints that belong to $\hat \Gamma$. 
\end{itemize}

We now describe the execution of the procedure \procdraw.
We start from the drawing $\phi$ of the graph $G$, and then modify it to obtain the desired drawing $\phi'$.
For simplicity of exposition, in the remainder of this subsection, we use the following notation.
For any drawing $\hat \phi$ of any graph $\hat G$, and for any subgraph $\hat H\subseteq \hat G$, we denote by $\hat \phi_{\hat H}$ the unique drawing of $\hat H$ induced by the drawing $\hat \phi$ of $\hat G$.
%It would be convenient to think about the drawing $\phi$ of $G$ as being on the sphere instead of on the plane. 
The procedure consists of two steps.

\begin{figure}[h]
	\centering
	\subfigure[Discs $\eta(u^*)$ and $\eta'(u^*)$, and their boundaries $\lambda$ and $\lambda'$, respectively.]{\scalebox{0.4}{\includegraphics{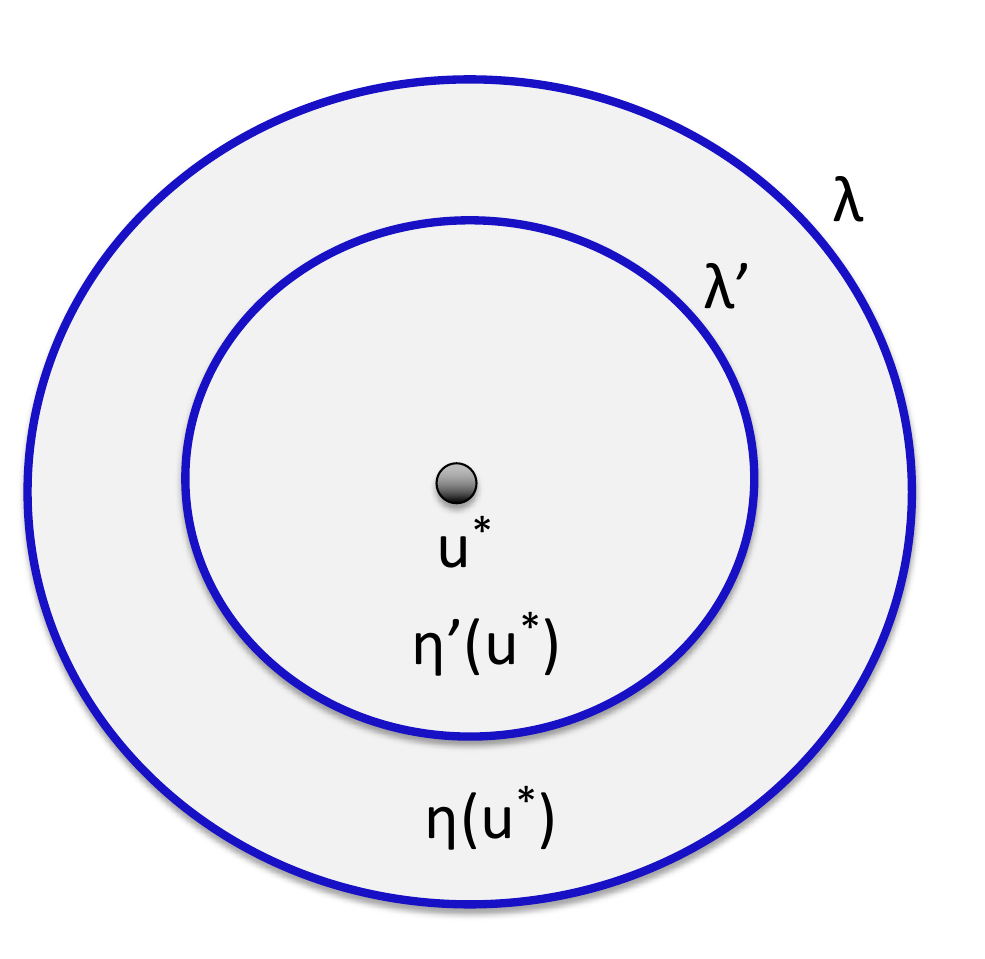}}\label{fig: two discs}}
	\hspace{1cm}
	\subfigure[After Step 1. Graph $C$ is now drawn inside disc $\eta'(u^*)$, using the drawing $\psi$, with the vertices of $\hat \Gamma$ (shown in red) appearing on $\lambda'$. Graph $X$ is drawn outside $\eta(u^*)$, preserving the original drawing $\phi$. The vertices of $\hat \Gamma'$ are shown in red, and the images of the edges of $\hat E$ in $\phi$ are shown in green. We also show, in dashed green curves, the images of the paths in $\qset$ in the original drawing $\phi$.]{
		\scalebox{0.4}{\includegraphics{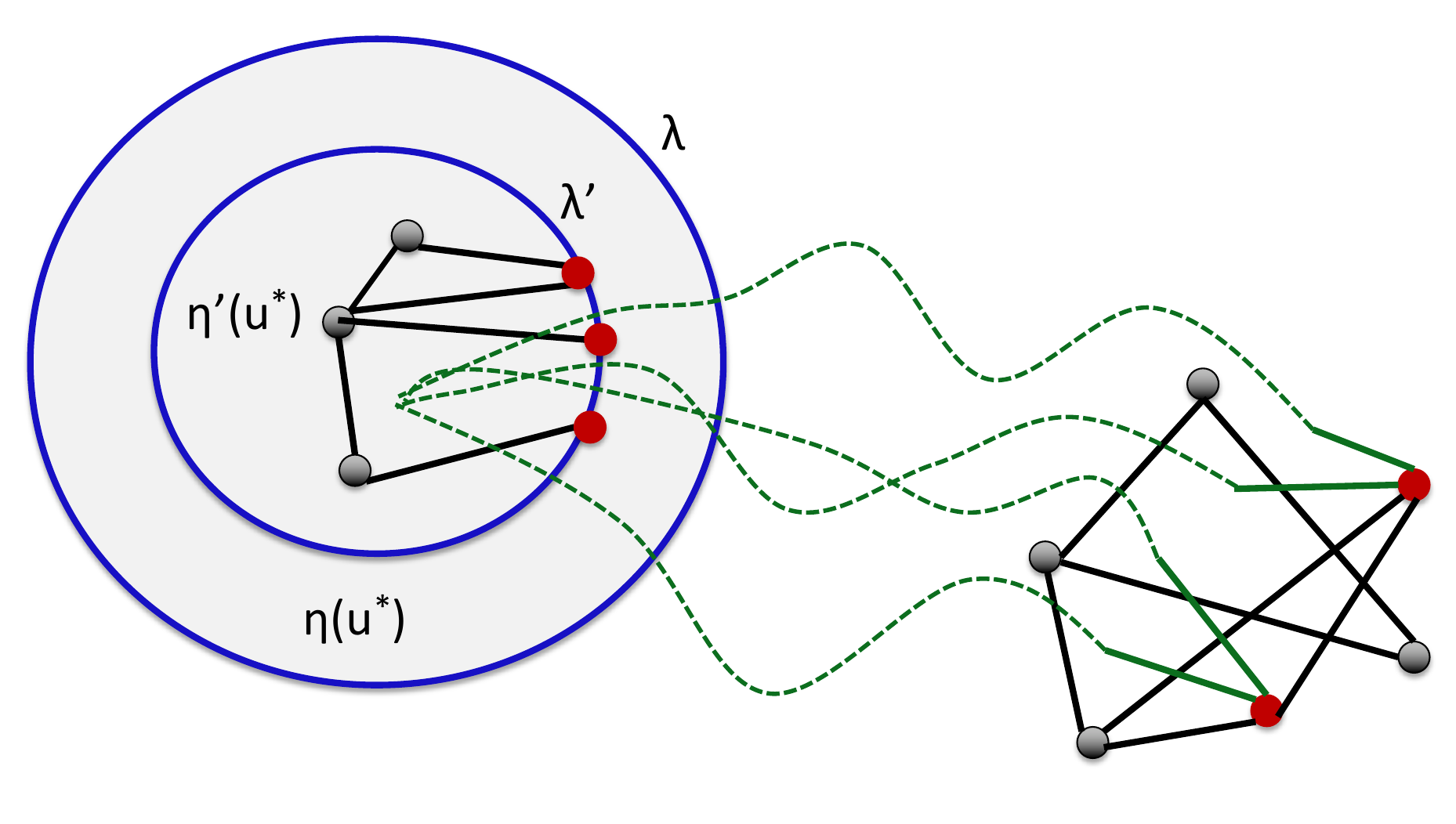}}\label{fig: after step 1}}
	%\hspace{1cm}
	%\subfigure[caption subfigure 3]{\scalebox{0.2}{\includegraphics{file3.pdf}}\label{fig: label3}}
	\caption{Illustration for Step 1 of \procdraw.\label{fig: step 1}}
\end{figure}

\paragraph{Step 1.}
In this step, we consider the drawing $\phi$ of $G$ on the sphere, and the disc $\eta(u^*)$ around the vertex $u^*$. We denote the boundary of this disc by $\lambda$, and we let $D'$ be the disc whose boundary is $\lambda$, that is disjoint from $\eta(u^*)$ except for sharing the boundary with it.  By shrinking the disc $\eta(u^*)$ a little, we obtain another disc $\eta'(u^*)\subseteq \eta(u^*)$, whose boundary is denoted by $\lambda'$, such that $\lambda'$ is disjoint from $\lambda$ (see Figure \ref{fig: two discs}).
%This curve partitions the sphere into two discs, that we denote by $D'$ and $D''$. We draw the curve $\eta$ so that the only vertex whose image in $\phi$ lies in $D'$ is $u$, and the only edges whose image in $\phi$ intersects $D'$ are the edges that are incident to $u$, and for each such edge, the intersection of its image with $D'$ is a contiguous curve.

We then erase, from $\phi$, the images of the vertices and the edges of $C$ (but we keep the images of the edges of $\hat E$), and instead place the drawing $\psi$ of $C$ inside the disc $\eta'(u^*)$.
Recall that we are given a disc $D$ in the drawing $\psi$ of $C$, such that the images of the vertices of $C$ are disjoint from the disc $D$, except that the images of the vertices in $\hat \Gamma$, that lie on the boundary of $D$, and the images of the edges of $C$ are disjoint from the disc $D$, except for their endpoints that belong to $\hat \Gamma$ and lie on the boundary of $D$.
We plant the drawing $\psi$ inside the disc $\eta'(u^*)$ in such a way that the boundary of the disc $D$ coincides with $\lambda'$ (see Figure \ref{fig: after step 1}). Therefore, all vertices of $\hat \Gamma$ are now drawn on the curve $\lambda'$, and the image of $C$ now appears inside the disc $\eta'(u^*)$. Note that the drawing of $X\cup \hat E$ remains unchanged and its image still lies in the interior of the disc $D'$. We denote the drawing that we obtained after the first step by $\hat \phi$.
In order to obtain the final drawing of the graph $G$, we need to extend the drawings (in $\hat \phi$) of the edges in $\hat E$, so that they connect the original images of the vertices in $\hat \Gamma'$ to the new images of the vertices in $\hat \Gamma$.

\paragraph{Step 2.}
The goal of this step is to extend the images of the edges $e\in \hat E$ in the current drawing, so that they terminate at the new images of the vertices of $\hat \Gamma$. We do so by exploiting the images of the paths in $\qset$ in the original drawing $\phi$ of $G$, after slightly modifying them. Specifically, consider the set $\qset=\set{Q_v\mid v\in \hat \Gamma}$ of paths.
Recall that $J\subseteq G$ is the graph containing all vertices and edges that participate in the paths in $\qset$. %Note that the drawing $\phi$ of $G$ and the drawing $\psi$ of $C$ induce drawings $\phi_J$ and $\psi_J$ of the graph $J$, respectively. 
We say that a vertex $v\in V(J)$ is \emph{irregular} if it is irregular with respect to the drawings $\phi_J$ and $\psi_J$ of $J$ (that are induced by the drawings $\phi$ of $G$ and $\psi$ of $C$, respectively, where $\phi$ is the original drawing of the graph $G$). We define irregular edges and paths in graph $J$ similarly.

Recall that for every edge $e'\in E(C)$, we have denoted by $\cong_{\qset}(e')$ the congestion of the set $\qset$ of edges on $e'$ -- that is, the total number of paths in $\qset$ that contain the edge $e'$. Consider now some edge $e=(u,v)\in \hat E$, with $u\in \hat \Gamma$ and $v\in \hat \Gamma'$. We subdivide the edge $e$ with a new vertex $t_e$, and we denote by $\Gamma^*=\set{t_e\mid e\in \hat E}$ this new set of vertices. We let set $\hat E'$ of edges contain, for each edge $e=(u,v)\in \hat E$ with $u\in \hat \Gamma$, the edge $(u,t_e)$. Consider the current drawing $\hat \phi$. Once we subdivide each edge $e\in \hat E$ with the vertex $t_e$, this new drawing (that we still denote by $\hat \phi$) now contains the images of the edges in $\hat E'$. Similarly, we add the vertices of $\Gamma^*$ to the original drawing $\phi$ of $G$, and we still denote this new drawing by $\phi$.
We denote by $\phi(C)$ the drawing of $C$ induced by $\phi$, and we similarly denote by $\phi(C\cup \hat E')$ the drawing of $C\cup \hat E'$ induced by $\phi$. Lastly, we denote by $J'$ the graph $J\cup \hat E'$.

Notice that graph $J'$ is planar, since $J$ is planar.
 We let $\psi'_{J'}$ be a drawing of $J'$, obtained from the drawing $\psi_J$ of $J$ induced by $\psi$, by adding the drawings of the edges in $\hat E'$ to it, without introducing any new crossings (recall that each such edge connects a vertex of $\hat \Gamma$ to a vertex of $\Gamma^*$, and that the latter has degree $1$ in $J'$). Consider any edge $(u,t_e)\in \hat E'$. If vertex $u$ is a regular vertex (recall that this is defined with respect to the drawings $\phi_J$ and $\psi_J$ of $J$), then we add the drawing of the edge $(u,t_e)$ to $\psi$ so that $u$ remains a regular vertex with respect to the drawings $\phi_{J'}$ and $\psi'_{J'}$ of $J'$. In other words, the drawing of the edges in $\hat E'$ is added to $\psi_J$ in such a way that every vertex $v\in \hat \Gamma$ that was regular with respect to $\psi_J$ and $\phi_J$, remains regular with respect to $\psi'_{J'}$ and $\phi_{J'}$. Similarly, we can ensure that every edge of $E(J)$ that was regular with respect to $\psi_J$ and $\phi_J$ remains regular with respect to $\psi'_{J'}$ and $\phi_{J'}$.

Consider now some edge $e=(u,v)\in \hat E$, with $u\in \hat \Gamma$. We denote the corresponding new edge $(u,t_e)\in \hat E'$ by $\hat e$. 
By concatenating the path $Q_u\in \qset$ with the edge $(u,t_e)$, we obtain a new path, that we denote by $Q'_e$, connecting $t_e$ to $u^*$, such that every vertex of $Q'_e$ except for $t_e$ lies in $J$. Let $\qset'=\set{Q'_e\mid e\in \hat E}$ be the resulting set of paths. It is easy to verify that, for every edge $\hat e\in \hat E'$, $\cong_{\qset'}(\hat e)=1$, and for every edge $e'\in E(J)$, $\cong_{\qset'}(e')\leq \Delta\cdot \cong_{\qset}(e')$.

Next, we apply Lemma \ref{lem:uncrossing} to graph $J'$, its planar drawing $\psi'_{J'}$, and the set $\qset'$ of paths, to obtain a new set $\qset''$ of paths, routing $\Gamma^*$ to $u^*$ in $J'$, that are non-interfering with respect to $\psi'_{J'}$. The lemma ensures that,  for every edge $\hat e\in \hat E'$, $\cong_{\qset''}(\hat e)=1$, and for every edge $e'\in E(J)$, $\cong_{\qset''}(e')\leq \cong_{\qset'}(e')\leq \Delta\cdot \cong_{\qset}(e')$. We denote, for each edge $e\in \hat E$, by $Q''_e\in \qset''$ the unique path originating at the vertex $t_e$. 
Additionally, the lemma provides a non-interfering representation $\set{\gamma_{Q''_e}\mid e\in \hat E}$ of the paths in $\qset''$, and an ordering $\oset$ of the vertices of $V(J)\cup \Gamma^*$ that is consistent with the paths in $\qset''$. Notice that every vertex $t_e\in \Gamma^*$ has degree $1$ in $C\cup \hat E'$, so the first edge on path $Q''_e$ must be the edge $\hat e\in \hat E'$.

Consider some edge $e=(u,u')\in \hat E$, with $u\in \hat \Gamma$. We will define a curve $\zeta_e$ that connects the image
of $u$ in the current drawing $\hat \phi$ to the image of $u'$, and is contained in the thin strip $S_{Q''_e}$ around the drawing of path $Q''_e$ in $\phi$, as follows. 
Denote $Q''_e=(t_e=v_1,v_2,\ldots,v_r=u^*)$. For all $1\leq i<r$, we denote by $e_i=(v_i,v_{i+1})$ the $i$th edge on this path. In order to define the curve $\zeta_e$, we will define, for every edge $e_i\in E(Q''_e)$ with $i>1$, a curve $\zeta_e(e_i)$, that is contained in the thin strip $S_{e_i}$ around the image of $e_i$ in the original drawing $\phi$ of $G$, and connects some point $p'_e(v_i)$ %on the segment $\sigma_{v_i}(e_i)$
 on the boundary of the disc $\eta(v_i)$ to some point $p_e(v_{i+1})$ %on the segment $\sigma_{v_{i+1}}(e_{i})$ 
 on the boundary of the disc $\eta(v_{i+1})$.
 We also define a curve $\zeta_e(e_1)$, connecting the image of vertex $v_1=t_e$ to some point $p_e(v_2)$ on the boundary of the disc $\eta(v_2)$.
  Additionally, for all $2\leq i<r$, we define a curve $\zeta_e(v_i)$, that is contained in $\eta(v_i)$, and connects the point $p_e(v_i)$ to the point $p_e'(v_i)$. Lastly, we define a curve $\zeta_e(v_r)$, that is contained in $\eta(u^*)\setminus \eta'(u^*)$, and connects the point $p_e(v_r)$ that lies on $\lambda$ to the image of the vertex $u\in \hat \Gamma$, that lies on $\lambda'$. 
The final drawing of the edge $e=(u,u')$ is obtained by concatenating  the image of the edge $(u',t_e)$ in the current drawing $\hat \phi$, and the curves $\zeta_e(e_1),\zeta_e(v_2),\zeta_e(e_2),\ldots,\zeta_e(e_r),\zeta_e(v_r)$. The resulting curve connects the image of the vertex $u'$ to the image of the vertex $u$, as required. It now remains to define each of these curves.

\paragraph{Drawing around the vertices.}
Consider some vertex $v\in V(J)$. Let $\pset(v)\subseteq \qset''$ be the set of all paths $Q''_e\in \qset''$ that contain the vertex $v$. 
We assume first that $v\neq u^*$.
For each path $Q''_e\in \pset(v)$, consider the corresponding curve $\gamma_{Q''_e}$ that was defined as part of the non-interfering representation of the paths in $\qset''$ in  the drawing $\psi'_{J'}$. We think of this curve as being directed towards the vertex $u^*$. Note that the curve $\gamma_{Q''_e}$ intersects the boundary of $\eta(v)$ in $\psi$ in exactly two points; we denote the first point by $q_e(v)$ and the second point by $q_{e}'(v)$. If we denote by $e_i,e_{i+1}$ the edges of $Q''_e$ that appear immediately before and immediately after $v$ on path $Q''_e$, then point $q_e(v)$ must lie on the segment $\sigma_{e_i}(v)$, and point $q'_e(v)$ must lie on the segment $\sigma_{e_{i+1}}(v)$ of the boundary of $\eta(v)$ in the drawing $\psi$ (see Figure \ref{fig: disc and segments}). 

Assume first that vertex $v$ is a regular vertex with respect to the drawings $\psi'_{J'}$ and $\phi_{J'}$. Then the set $\set{q_e(v),q_{e}'(v)}_{Q_{e}''\in \pset(v)}$ of points on the boundary of $\eta(v)$ in the drawing $\psi$ naturally defines the  corresponding set $\set{p_e(v),p_{e}'(v)}_{Q_{e}''\in \pset(v)}$ of points on the boundary of $\eta(v)$ in the drawing $\phi$ (if the orientation of the vertex $v$ is different in the two drawings, then we flip the sets of points accordingly). Moreover, for every path $Q_{e}''\in \pset(v)$, the intersection of the curve $\gamma_{Q''_e}$ with the disc $\eta(v)$ in the drawing $\psi'_{J'}$ naturally defines a curve $\zeta_e(v)$ in the drawing $\phi$, that is contained in the disc $\eta(v)$, and connects point $p_e(v)$ to point $p'_e(v)$. Notice that the resulting curves in $\set{\zeta_e(v)}_{Q_e''\in \pset(v)}$ are all mutually disjoint.

Assume now that vertex $v$ is irregular with respect to $\psi'_{J'}$ and $\phi_{J'}$. Consider any path $Q''_e\in \pset(v)$, and let $e_i,e_{i+1}$ be the edges of $Q''_e$ that appear immediately before and immediately after $v$ on path $Q''_e$. In this case, we let $p_e(v)$ be a point on the segment $\sigma_{e_i}(v)$ of the boundary of the disc $\eta(v)$ in $\phi$, and similarly we let $p'_e(v)$ be a point on the segment $\sigma_{e_{i+1}}(v)$ of the boundary of the disc $\eta(v)$ in $\phi$. We ensure that all points that are added to each segment $\sigma_{e'}(v)$, for all $e'\in \delta(v)$ are distinct, and their ordering within each segment $\sigma_{e'}(v)$ is the same as the ordering of the corresponding points of  $\set{q_e(v),q_{e}'(v)}_{Q_{e}''\in \pset(v)}\cap \sigma_{e'}(v)$ in $\psi'_{J'}$.
For every path $Q''_e\in \pset(v)$, we let $\zeta_e(v)$ be an arbitrary curve in $\phi$, that is contained in the disc $\eta(v)$, and connects point $p_e(v)$ to point $p'_e(v)$; we ensure that every pair of curves in $\set{\zeta_e(v)}_{Q_e''\in \pset(v)}$ intersect at most once.

Lastly, we consider the case where $v=u^*$. In this case, for each path $Q''_e\in \qset''$, the intersection of the curve $\gamma_{Q''_e}$ with the boundary of the disc $\eta(u^*)$ in $\psi$ is exactly one point, that is denoted by $q_e(u^*)$. 
We use the set $\set{q_e(v)}_{Q_{e}''\in \qset''}$ of points on the boundary of $\eta(u^*)$ in the drawing $\psi'_{J'}$ to define the  corresponding set $\set{p_e(v)}_{Q_{e}''\in \qset''}$ of points on the boundary $\lambda$ of the disc $\eta(u^*)$ 
exactly as before (where we again consider the cases where $u^*$ is regular or irregular separately). It now remains to define the curves $\zeta_e(u^*)$ for all paths $Q''_e\in \qset''$.

Assume first that vertex $u^*$ is irregular with respect to the drawings $\psi'_{J'}$ and $\phi_{J'}$. Then for every edge $e=(u,u')\in \hat E$ with $u\in \hat \Gamma$, we let $\zeta_e(u^*)$ be any curve that is contained in $\eta(u^*)\setminus\eta'(u^*)$, that connects point $p_e(u^*)$ to the image of the vertex $u$ (that lies on the boundary $\lambda'$ of $\eta(u^*)$), such that each pair of such curves cross at most once.

 Lastly, we assume that vertex $u^*$ is regular with respect to $\psi'_{J'}$ and $\phi_{J'}$.
  For every vertex $u\in \hat \Gamma$, let $\sset(u)\subseteq \qset''$ be the set of paths whose second vertex is $u$. In other words, a path $Q_e''\in \sset(u)$ iff $u$ is an endpoint of the edge $e\in \hat E$. 
 Since the curves in $\set{\gamma_{Q''_e}}_{Q''_e\in \qset''}$ are a non-interfering representation of the paths in $\qset''$ in the drawing $\psi$, for all $u\in \hat \Gamma$, there is a contiguous segment $\sigma'(u)$ of the boundary $\lambda$ of $\eta(u^*)$ in the current drawing $\hat \phi$, such that for every path $Q''_e\in \sset(u)$, the point $p_e(u^*)$ lies on segment $\sigma'(u)$. Moreover, we can ensure that the segments $\set{\sigma'(u)\mid u\in \hat \Gamma}$ are disjoint from each other. Since the curves in  $\set{\gamma_{Q''_e}}_{Q''_e\in \qset''}$ are non-interfering, the circular ordering of the segments 
 $\set{\sigma'(u)\mid u\in \hat \Gamma}$ along $\lambda$ is identical to the circular ordering of the images of the vertices in $\hat \Gamma$ on $\lambda'$. 
 If the orientations of the two orderings are different, then we flip the current drawing of $C$, by replacing the current drawing contained in disc $\eta'(u^*)$ with its mirror image.
 Therefore, we can define, for every vertex $u\in \hat \Gamma$, for every path $Q''_e\in \sset(u)$, a curve $\zeta_e(u^*)$, that is contained in $\eta(u^*)\setminus \eta'(u^*)$ in the drawing $\hat\phi$, and connects point $p_e(u^*)$ to the image of the vertex $u$, while ensuring that all resulting curves in $\set{\zeta_e(u^*)\mid e\in \hat E}$ are mutually disjoint from each other.

\paragraph{Drawing along the first edge on each path.} Consider again an edge $e=(u,u')\in \hat E$, with $u\in \hat \Gamma$, and denote by $e_1=(v_1,v_2)$ the first edge on path $Q''_e$, where $v_1=t_e$. Recall that the current drawing $\hat \phi$ contains the drawing of the edge $(t_e,v_2)=(v_1,v_2)$. We slightly shorten the corresponding curve, so it still originates at the image of $t_e$, but now it terminates at the point $p_e(v_2)$ on the boundary of $\eta(v_2)$. This defines the curve $\zeta_e(e_1)$.

\paragraph{Drawing along the edges.} Lastly, we define, for every edge $e\in \hat E$, the curves $\zeta_e(e_i)$, where $e_i$ is an edge on the path $Q''_e$, that is not the first edge on the path. In order to do so, we consider any edge $e'\in E(C)$, denoting $e'=(x,y)$. Recall that we are given an ordering $\oset$ of the vertices of $J$ that is consistent with the paths in $\qset''$. We assume that $x$ appears before $y$ in this ordering, so every path in $\qset''$ that contains the edge $e'$, traverses it in the direction from $x$ to $y$. We denote by $\pset(e')\subseteq\qset''$ the set of paths that contain the edge $e'$. 
Recall that for each such path $Q''_e\in \pset(e')$, we have defined a point $p'_e(x)$ on the segment $\sigma_e(x)$ of the boundary of the disc $\eta(x)$ in $\phi$, and a point $p_e(y)$ on the segment $\sigma_e(y)$ on the boundary of the disc $\eta(y)$ in $\phi$.
We now consider two cases.

\begin{figure}[h]
	\centering
	\subfigure[Drawing along a regular edge $e'$.]{\scalebox{0.4}{\includegraphics{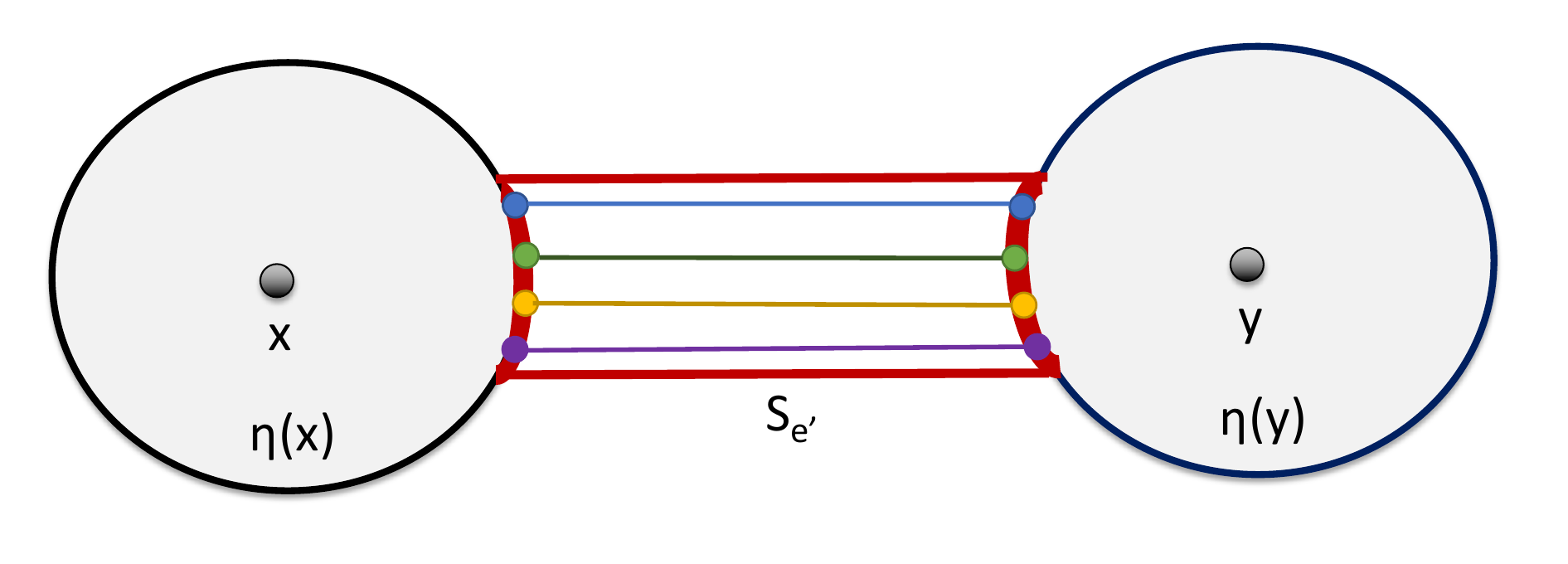}}\label{fig: regular edge}}
	\hspace{1cm}
	\subfigure[Drawing along an edge $e'$ that is the last irregular edge of an irregular path.]{
		\scalebox{0.4}{\includegraphics{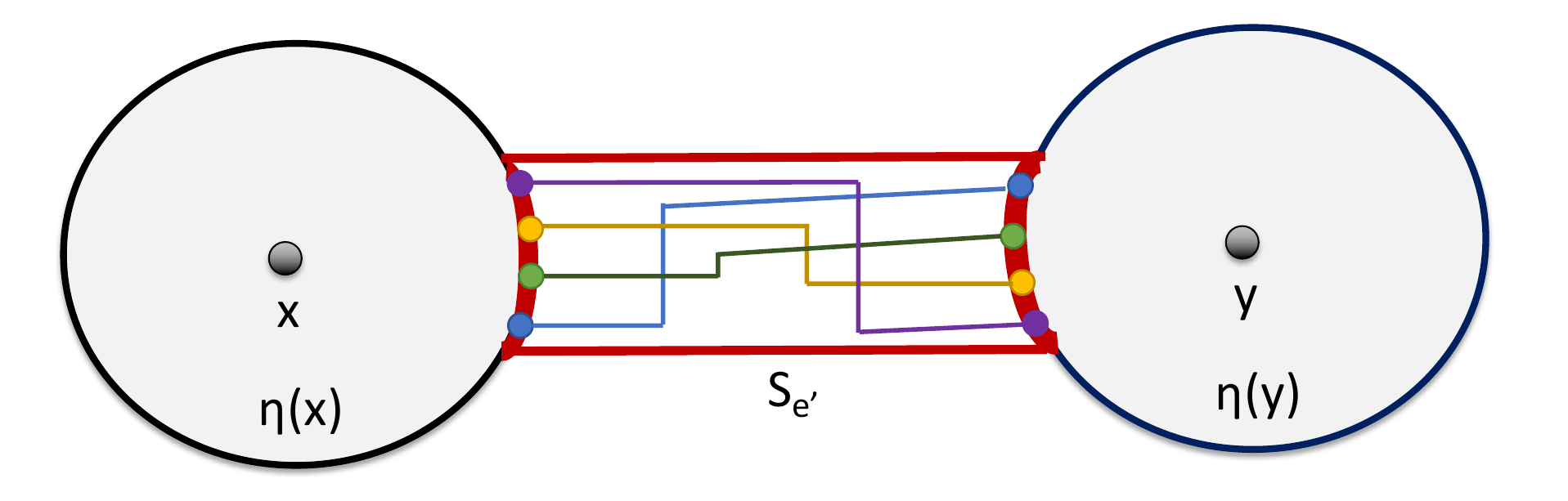}}\label{fig: irregular-edge}}
	%\hspace{1cm}
	%\subfigure[caption subfigure 3]{\scalebox{0.2}{\includegraphics{file3.pdf}}\label{fig: label3}}
	\caption{Drawing along edges of $E(C)$.\label{fig: drawing along edges}}
\end{figure}

The first case is when either (i) edge $e'$ is a regular edge (with respect to $\phi_{J'}$ and $\psi'_{J'}$), or (ii) $e'$ is an irregular edge, but it is not the last edge on the corresponding irregular path (since we can view the paths in $\qset''$ as directed towards $u^*$, and since we are given an ordering $\oset$ of the vertices of $V(J)$ that is consistent with the paths in $\qset''$, the notion of the last edge on a path is well defined). 
In this case, the ordering of the points in  $\set{p'_e(x)\mid Q''_e\in \pset(e')}$ on segment $\sigma_e(x)$ is identical to the ordering of the points in $\set{p_e(y)\mid Q''_e\in \pset(e')}$ on segment $\sigma_e(y)$. We can then define, for each path $Q''_e\in \pset(e')$, a curve $\zeta_e(e')$ connecting points $p'_e(x)$ and $p_e(y)$ that is contained in the thin strip $S_{e'}$ around $e'$ in $\phi$, such that all resulting curves are mutually disjoint, in a straightforward way (see Figure \ref{fig: regular edge}).

%Consider the curves in $\set{\gamma_Q}_{Q\in \pset(e')}$ that are given by the non-interfering representation of the set $\qset''$ of paths. For each path $Q_e''\in \pset(e')$, let $\gamma_{Q''_e}(e')$ be the intersection of the curve $\gamma_Q$ with the thin strip $S_{e'}$ around the edge $e'$ in the drawing $\psi$ of $C$. We then let $\eta_{e}(e')$ be a curve that is contained in the thin strip $S_{e'}$ around the edge $e'$ in the current drawing $\phi$, that is identical to $\gamma_{Q''_e}(e')$. In other words, the drawing of the paths in $\pset(e')$ inside the strip $S_{e'}$ in $\phi$ is exactly the same as the drawing of their corresponding curves in the thin strip $S_{e'}$ around $e'$ in $\psi$. Note that there are two ways to define this drawing (by ``flipping'' the ordering of the curves in the strip $S_{e'}$). 
%If both endpoints of $e'$ are regular vertices, then the ``flip'' is determined in a way that ensures that the curves intersect the boundaries of $\eta(x),\eta(y)$ in the same order as the curves $\set{\gamma_Q}_{Q\in \pset(e')}$ intersect them in the drawing $\psi$. If one of the vertices is irregular, then the flip is determined to maintain consistency with the regular vertex. If both vertices are irregular, then the flip is arbitrary.

The second case is when edge $e'$ is an irregular edge with respect to $\phi_{J'}$ and $\psi'_{J'}$, and it is the last edge on an irregular path. In this case, 
the ordering of the points in  $\set{p'_e(x)\mid Q''_e\in \pset(e')}$ on segment $\sigma_e(x)$ and the ordering of the points in $\set{p_e(y)\mid Q''_e\in \pset(e')}$ on segment $\sigma_e(y)$ are reversed. We can then define, for each path $Q''_e\in \pset(e')$, a curve $\zeta_e(e')$ connecting points $p'_e(x)$ and $p_e(y)$ that is contained in the thin strip $S_{e'}$ around $e'$ in $\phi$, such that every pair of the resulting curves intersect exactly once (see Figure \ref{fig: irregular-edge}).

This completes the definition of the images $\zeta_e$ of the edges $e\in \hat E$, and completes the definition of the new drawing $\phi'$ of $G$. It is immediate to verify that drawing $\phi'$ has all required properties. It now remains to analyze the number of crossings in this drawing. This analysis will be used later in order to bound the number of crossings in the modified drawing of the input graph $G$ that our algorithm constructs.

\paragraph{Analysis of the Number of Crossings.}
%We denote by $\phi'$ the resulting drawing obtained from the procedure \procdraw. 
We now analyze the number of crossings in the final drawing $\phi'$ of the graph $G$. First, the images of the edges of $E(C)$ may cross each other in the new drawing $\phi'$, and the number of such crossings is bounded by $\cro(\psi)$.
The edges of $E(X)$ may also cross each other, and the number of such crossings is bounded by $\cro(\phi)$.
The crossings caused by pairs of edges in $E(C)$ or pairs of edges in $E(X)$ are called \emph{old crossings}. From the above discussion, the total number of old crossings is bounded by $\cro(\phi)+\cro(\psi)$. We now bound the number of additional crossings, that we call \emph{new crossings}.
Note that, the edges of $E(C)$ may only cross the edges of $E(C)$, as the edges of $E(C)$ are drawn inside the disc $\eta'(u^*)$, while all other edges are drawn outside this disc.
Therefore, all new crossings are those in which at least one edge of $\hat E$ participates, and it remains to bound (i) the number of crossings of the edges of $\hat E$ with each other, and (ii) the number of crossings between the edges of $\hat E$ and the edges of $E(X)$.

We assume without loss of generality that, in both $\phi$ and $\psi$, and every pair of edges cross at most once. This assumption is only made for the ease of notation; the analysis below works even if the images of pairs of edges are allowed to cross multiple times.
We denote by $(e_1,e_2)$ the crossing caused by the pair $e_1,e_2$ of edges. Consider the original drawing $\phi$ of $G$.
We denote by $\hat \chi_1$ the set of all crossings $(e_1,e_2)$ in $\phi$, where $e_1,e_2\in E(J)$. We denote by $\hat \chi_2$ the set of all crossings $(e_1,e_2)$ in $\phi$, where $e_1\in E(J)$ and $e_2\not\in E(C)$.

Consider some crossing $(e_1,e_2)\in \hat \chi_1$. For every pair $e,e'\in \hat E$ of edges such that $e_1\in E(Q''_e)$ and $e_2\in E(Q''_{e'})$, the new images of the edges $e$ and $e'$ in $\phi'$ must cross. Therefore, each crossing $(e_1,e_2)\in \hat \chi_1$ contributes $\cong_{\qset''}(e_1)\cdot \cong_{\qset''}(e_2)=O(\cong_{\qset}(e_1)\cdot \cong_{\qset}(e_2)\cdot \Delta^2)$ crossings to $\cro_{\phi'}(G)$. 
%We call such crossings \emph{internal}.
%because each such new crossing is between an edge of $C$ and an edge of $\hat E$.
Consider some crossing $(e_1,e_2)\in \hat \chi_2$, where $e_1\in E(J)$ and $e_2\not\in E(C)$. For every edge $e\in \hat E$ such that the path $Q''_e$ contains $e_1$, the new drawing of $e$ must intersect the new drawing of $e_2$ in $\phi'$. Therefore, each crossing $(e_1,e_2)\in \hat \chi_2$ contributes $\cong_{\qset''}(e_1)=O(\cong_{\qset}(e_1)\cdot \Delta)$ crossings to $\cro_{\phi'}(G)$. 
%(We note that it is possible that $e_2\in \hat E$, but in this case the analysis is similar, as $e_2$ may belong to at most one path of $\qset''$).

The only additional crossings in $\phi'$ are crossings between the images of the edges of $\hat E$ due to the re-ordering of the corresponding curves along irregular vertices and irregular edges.  %\mynote{maybe define the re-ordering procedure formally?}
%We call such crossings \emph{external}, because each such crossing is between an edge of $\hat E$, and an edge of $X$.
If an edge $e'\in E(J)$ is irregular with respect to $\psi'_{J'}$ and $\phi_{J'}$, then the paths  in set $\pset(e')\subseteq \qset''$ may incur up to  $(\cong_{\qset''}(e'))^2\leq (\cong_{\qset}(e')\cdot \Delta)^2$ crossings as they are drawn along the edge $e'$. 
Assume now that a vertex $v\in V(J)$ is irregular with respect to $\psi'_{J'}$ and $\phi_{J'}$, and let $n_v=|\pset(v)|$ be the total number of all paths in $\qset''$ that contain the vertex $v$. Then the number of crossings due to the drawing of these paths in the disc $\eta(v)$ is at most $n_v^2$. If $e'\in \delta(v)$ is the edge with maximum congestion $\cong_{\qset''}(e')$, then this number of crossings is bounded by $(\Delta\cdot \cong_{\qset''}(e'))^2\leq \Delta^4(\cong_{\qset}(e))^2$. 
Recall that we have ensured that, if a vertex $v\in V(J)$ is irregular with respect to $\psi'_{J'}$ and $\phi_{J'}$, then it is irregular with respect to $\psi_J$ and $\phi_J$. Similarly, if an edge $e\in E(J)$ is irregular with respect to $\psi'_{J'}$ and $\phi_{J'}$, then it is irregular with respect to $\psi_J$ and $\phi_J$.
%Notice that, if vertex $v$ is irregular with respect to $\psi'$ and $\phi$, then either $v$ is irregular with respect to $\psi$ and $\phi$, or $v\in \hat \Gamma$ holds (this is since drawing $\psi'$ contains the edges of $\tilde E$ which do not belong to the drawing $\phi$.) 

% or its first endpoint is in $\ir_V(\phi(C),\psi)$, then reordering the images of the edges whose paths traverse $e$ may add up to $(c_{\qset'}(e))^2=\Theta((c_\qset(e))^2)$ new crossings to $\phi'$. 
%\mynote{this needs to be proved formally} 
%We call each resulting crossing \emph{neutral}.
Denote by $E^*$ the set of all edges $e\in E(J)$, such that either $e$ is an irregular edge with respect to $\psi_J$ and $\phi_J$, or at least one endpoint of $e$ is an irregular vertex with respect to $\psi_J$ and $\phi_J$. Then the total number of new crossings in $\phi'$ is bounded by:
\[ O\left(\sum_{(e_1,e_2)\in \hat \chi_1}\Delta^2\cdot \cong_{\qset}(e_1)\cdot \cong_{\qset}(e_2)+\sum_{(e_1,e_2)\in \hat \chi_2}\Delta\cdot\cong_{\qset}(e_1)+\sum_{e\in  E^*} \Delta^4(\cong_{\qset}(e))^2\right).\]

Finally, the total number of crossings in $\phi'$ is bounded by:
\[\cro(\phi)+\cro(\psi)+O\left(\sum_{(e_1,e_2)\in \hat \chi_1}\Delta^2\cdot\cong_{\qset}(e_1)\cdot \cong_{\qset}(e_2)+\sum_{(e_1,e_2)\in \hat \chi_2}\Delta\cdot\cong_{\qset}(e_1)+\sum_{e\in  E^*}\Delta^4 (\cong_\qset(e))^2\right).\]
%Moreover, the total number of new external crossings is bounded by: $\sum_{(e_1,e_2)\in \hat \chi_2}c_{\qset}(e_1)$.

The following observation is immediate from the analysis above and will be useful in bounding the total number of crossings in our algorithm.

\begin{observation}\label{obs: classifying new crossings}
	Let $(e_1,e_2)$ be a new crossing in $\phi'$ (that is, edges $e_1$ and $e_2$ cross in $\phi'$ but do not cross in $\phi$ or in $\psi$). Then either $e_1,e_2\in \hat E$, or one of these edges belongs to $\hat E$ and the other to $X$. Moreover, if $e\in E(X)$, and $K_e$ is the set of all edges of $J$ whose image in $\phi$ crosses the image of $e$, then the total number of new crossings in $\phi'$ in which edge $e$ participates is at most $\sum_{e'\in K_e}\Delta\cdot\cong_{\qset}(e')$.
\end{observation}

\subsection{Completing the Proof of Theorem~\ref{thm: canonical drawing}}

We now provide the proof of  Theorem~\ref{thm: canonical drawing}, by showing an algorithm that produces the desired drawing $\phi'$ of the graph $G$. %By using standard uncrossing we can assume w.l.o.g. that for every pair of edges in $G$, their images cross at most once in $\phi$. 
The algorithm processes all clusters in $\cset_1\cup \cset_2$ one-by-one, with clusters in $\cset_1$ processed before clusters in $\cset_2$. When a cluster $C$ is processed, we modify the current drawing of the graph $G$, so that it becomes canonical with respect to $C$. %Additionally, if the fake edge $e(C)$ exists, then we add this edge to the drawing. In the second part, we process all type-2 clusters one-by-one. When cluster $C\in \cset_2$ is processed, we modify the current drawing $\phi'$ of $G$, so that it becomes canonical with respect to $C$, and we delete the edges of $A_C$ from the current drawing $\phi'$. We now describe each of these parts in turn.
For convenience, we denote $\cset=\cset_1\cup \cset_2$, and we denote the clusters in $\cset$ by $C_1,\ldots,C_r$. We assume that the type-1 clusters appear before the type-2 clusters in this ordering. %For all $1\leq i\leq r$, we denote by $\chi_i$ the set of all crossings $(e,e')$ in the original drawing $\phi$ of $G$, in which either $e\in E(C_i)$, or $e'\in E(C_i)$ (or both). Therefore, $\sum_{i=1}^r|\chi_i|\leq O(\cro(\phi))$. 

Our algorithm proceeds by repeatedly invoking procedure \procdraw. As part of input, the procedure requires a collection $\qset$ of guiding paths. We start by defining, for every cluster $C_i\in \cset$, a collection $\qset_i$ of paths that are contained in $C_i$, and connect every vertex of $\Gamma(C_i)=\Gamma\cap V(C_i)$ to some fixed vertex $u^*_i$ of $C_i$.

\subsubsection{Defining the Guiding Paths}

In this subsection, we define, for every cluster $C_i\in \cset$,  a collection $\qset_i$ of paths that are contained in $C_i$, and connect every vertex of $\Gamma(C_i)=\Gamma\cap V(C_i)$ to some fixed vertex $u^*_i$ of $C_i$. These paths will eventually be used by procedure \procdraw as guiding paths. The definition of the set $\qset_i$ of paths is different depending on whether $C_i$ is a type-1 or a type-2 cluster. %It will be convenient for us to define, for every cluster $C_i\in \cset$, for every edge $e\in E(C)$, its weight $w(e)$, that we do next.

%Next, we show how to compute the sets $\qset_i$ of guiding paths for type-1 and type-2 clusters $C_i$.

\paragraph{Type-1 clusters.} Let $C_i\in \cset_1$ be a type-1 cluster. Recall that the number of terminals in $\Gamma(C_i)$ is at most $\mu\Delta$. We let $u^*_i$ be an arbitrary vertex of $C_i$. 
Recall that we have defined, in Theorem \ref{thm: drawing of cluster extension-type 1}, a set $E^*(C_i)$ of edges, such that graph $C_i\setminus E^*(C_i)$ is connected, and the drawing of $C_i\setminus E^*(C_i)$ induced by $\psi_{C_i}$ is planar. 
Consider now any spanning tree of $C_i\setminus E^*(C_i)$, rooted at the vertex $u^*_i$. For every terminal $t\in \Gamma(C_i)$, let $Q_t$ be the unique path connecting $t$ to $u^*_i$ in this tree. We then set $\qset_i=\set{Q_t\mid t\in \Gamma(C_i)}$ be the set of the guiding paths for the cluster $C_i$. Since $|\qset_i|\leq \mu\Delta$, for every edge $e\in E(C_i)$, $\cong_{\qset_i}(e)\leq \mu\Delta$.

Let $J_i\subseteq C_i$ be the graph obtained from  the union of the paths in $\qset_i$. Then $J_i$ is a tree with at most $O(\mu\Delta)$ leaves, and so it has $O(\mu\Delta)$ vertices of degree greater than $2$. 
We denote by $\psi_{J_i}$ the planar drawing of $J_i$ induced by the drawing $\psi_{C_i}$ of $C_i$, and we denote by $\phi_{J_i}$ the drawing of $J_i$ induced by the original drawing $\phi$ of $G$. 
Clearly, drawing $\psi_{J_i}$ of $J_i$ is planar, and the number of vertices and of edges of $J_i$ that are irregular with respect to $\psi_{J_i}$ and $\phi_{J_i}$ is at most $O(\mu\Delta)$. Let $\ir_i\subseteq E(J_i)$ be the set of all edges $e\in E(J_i)$, such that either $e$ is irregular with respect to the drawings $\psi_{J_i}$ and $\phi_{J_i}$ of $J_i$, or at least one endpoint of $e$ is irregular with respect to these drawings. The following observation is immediate from the above discussion.

\begin{observation}\label{obs: few irreg vertices type 1}
	If $C_i\in \cset$ is a type-1 cluster then $|\ir_i|\leq O(\mu \Delta)$.
\end{observation}

%Consider a cluster $C_i\in \cset$. Let $C'_i=C_i\setminus A$.
For every edge $e\in E(C_i)$, we define its weight $w(e)$ as follows.  We start with $w(e)$ being the number of crossings in the drawing $\phi$ of $G$, in which edge $e$ participates. Additionally, if $e\in \ir_i$, then we increase $w(e)$ by $1$. %Also, for every endpoint of $e$ that is irregular with respect to these two drawings, we increase $w(e)$ by $1$. 
From the above discussion, we get that the following inequality that will be useful for us later:
\begin{equation}\label{eq: bound for type 1}
\sum_{e\in E(C_i)}w(e)\cdot (\cong_{\qset_i}(e))^2\leq O(\mu^2\Delta^2)\cdot \sum_{e\in E(C_i)}w(e).
\end{equation}

\paragraph{Type-2 clusters.} 
Let $C_i\in \cset_2$ be a type-2 cluster. For convenience, we denote $C'_i=C_i\setminus A_{C_i}$.
In order to define the set $\qset_i$ of guiding paths for $C_i$, we use the following lemma, that generalizes Lemma D.10 from~\cite{chuzhoy2011algorithm}. For completeness, we provide the proof of the lemma in Appendix~\ref{apd:find_confluent_paths}. %\mynote{Isn't this lemma implicit from my crossing number paper? If it is we should say so and put the proof in Appendix}
\begin{lemma}
	\label{lem:find_confluent_paths}
	There is an efficient algorithm, that, given 
	an $n$-vertex planar graph $H$, non-negative  weights $\set{w(e)}_{e\in E(H)}$ on its edges, and a subset $S\subseteq V(H)$ of vertices of $H$ that is $\alpha'$-well-linked in $H$, for any parameter $0<\alpha'<1$, computes a vertex $u^*\in V(H)$ together with a set $\qset$ of $|S|$ %confluent
	paths in $H$ routing the vertices of $S$ to $u^*$, such that:
	$$\sum_{e\in E(H)}w(e)\cdot (\cong_{\qset}(e))^2= O\left(\frac{\log n}{(\alpha')^4}\cdot\sum_{e\in E(H)}w(e)\right).$$
\end{lemma}
%\znote{remove $\Delta$ and later use of it}

For every edge $e\in E(C_i')$, we define its weight $w'(e)$ as follows.  We start with $w'(e)$ being the number of crossings in the drawing $\phi$ of $G$, in which edge $e$ participates. Additionally, if $e$ is an irregular vertex with respect to the drawing $\psi_{C'_i}$ of $C'_i$, and the drawing $\phi_{C'_i}$ of $C'_i$ induced by $\phi$, then we increase $w'(e)$ by $1$. Also, for every endpoint of $e$ that is irregular with respect to these two drawings, we increase $w'(e)$ by $1$. 

 We  apply Lemma \ref{lem:find_confluent_paths} to graph $C_i'$ with the edge weights $w'(e)$ that we have defined, and the set $S=\Gamma(C_i)$ of its vertices; recall that from the well-linkedness property of type-2 clusters, set $\Gamma(C_i)$ of vertices is $\alpha$-well-linked in $C_i'$.
The algorithm from the lemma then returns a vertex $u^*_i\in V(C_i)$, and a set $\qset_i=\set{Q_t\mid t\in \Gamma(C_i)}$ of paths in graph $C_i'$, where each path $Q_t$ connects $t$ to $u^*_i$, and moreover:
\begin{equation}
\sum_{e\in E(C_i')}w'(e)\cdot (\cong_{\qset_i}(e))^2\leq  O\left(\frac{\log n}{\alpha^4}\cdot\left (\sum_{e\in E(C'_i)}w'(e)+1 \right )\right).
\end{equation}
As before, we let $J_i$ be the graph obtained from the union of the paths in $\qset_i$.  
Note that $J_i$ may no longer be a tree. As before,  
we denote by $\psi_{J_i}$ the drawing of $J_i$ induced by the drawing $\psi_{C_i'}$ of $C'_i$ (which must be planar), and we denote by $\phi_{J_i}$ the drawing of $J_i$ induced by the drawing $\phi$ of $G$.

Let $\ir_i\subseteq E(J_i)$ be the set of all edges $e\in E(J_i)$, such that either $e$ is irregular with respect to the drawings $\psi_{J_i}$ and $\phi_{J_i}$ of $J_i$, or at least one endpoint of $e$ is irregular with respect to these drawings. For every edge $e\in E(C'_i)$, we let its new weight $w(e)$ be defined as follows. Initially, we let $w(e)$ be the number of crossings in the drawing $\phi$ of $G$ in which edge $e$ participates. If $e\in \ir_i$, then we increase $w(e)$ by $1$. Clearly, $w(e)\leq w'(e)$, and so:
\begin{equation}\label{eq: bound weights type 2}
\sum_{e\in E(C'_i)}w(e)\cdot (\cong_{\qset}(e))^2\leq  O\left(\frac{\log n}{\alpha^4}\cdot\left (\sum_{e\in E(C'_i)}w'(e)+1\right )\right).
\end{equation}

%----------------------------------
%----------------------------------
%----------------------------------
In the remainder of the algorithm, we perform $r$ iterations. The input to the $i$th iteration is a drawing $\phi_{i-1}$ of the graph $G$, that is canonical with respect to the clusters $C_1,\ldots,C_{i-1}$. The output of the $i$th iteration is a drawing $\phi_i$ of $G$, that is canonical with respect to clusters $C_1,\ldots,C_i$. Initially, we set $\phi_0=\phi$.
We now focus on the description of the $i$th iteration, when cluster $C_i$ is processed.

\subsubsection{Processing a Cluster}

We now describe an iteration when cluster $C_i\in \cset$ is processed. 
Recall that, if $C_i$ is a type-2 cluster, then we have denoted $C'_i=C_i\setminus A_{C_i}$. In order to simplify the notation, if $C_i$ is a type-1 cluster, we will denote $C'_i=C_i$.
Let $\discset_i$ be the set of all discs $D(R)$ for the bridges $R\in \rset_G(C'_i)$. For every disc $D\in \discset_i$, we let $\rset_i^D\subseteq \rset_G(C'_i)$ be the set of all the bridges $R\in \rset_G(C'_i)$ with $D(R)=D$. We also let $X_i^D$ be the graph obtained by the union of all bridges in $\rset_i^D$, and $\hat E_i^D$ be the set of all edges of $G$ connecting vertices of $C'_i$ to vertices of $X_i^D$. Lastly, we let $\hat \Gamma_i^D$ be the set of all terminals in $\bigcup_{R\in \rset_i^D}L(R)$, and we let $\qset_i^D\subseteq \qset_i$ be the set of all paths originating at the vertices of $\hat \Gamma_i^D$. Observe that, in the drawing $\psi_{C'_i}$ of $C'_i$, all vertices of $\hat \Gamma_i^D$ lie on the boundary of the disc $D$. Let $J_i^D$ be the graph obtained from the union of the paths in $\qset_i^D$. Note that every path $Q_t\in \qset_i$ may participate in at most $\Delta$ different path sets in $\set{\qset_i^D}_{D\in \discset_i}$.
 
Let $G_i^D\subseteq G$ be the sub-graph of $G$ consisting of the union of the graphs $C'_i,X^D_i$, and the edges of $\hat E_i^D$. Let $\phi^D$ be the drawing of $G_i^D$ that is induced by the current drawing $\phi_{i-1}$ of $G$. We apply \procdraw to graph $G_i^D$, with the subgraphs 
$C=C'_i$, $X=X_i^D$, together with the vertex $u^*_i$, and the set $\qset_i^D$ of paths routing the vertices of $\hat \Gamma_i^D$ to $u^*_i$ in $C'_i$. Recall that the corresponding graph $J_i^D$ (which is obtained from the union of the paths in $\qset_i^D$) is guaranteed to be planar, and its drawing induced by $\psi_{C_i'}$ is also planar. We denote the drawing of the graph $G_i^D$ produced by \procdraw by $\hat \phi^D_i$. Recall that the drawing of $C'_i$ induced by $\hat \phi_i^D$ is identical to $\psi_{C'_i}$, and that all vertices of $X_i^D$ are drawn inside the disc $D$, with the vertices of $\hat \Gamma_i^D$ drawn on the boundary of $D$.

Once every disc $D\in \discset_i$ is processed, we combine the resulting drawings $\hat \phi^D$  together, in order to obtain the final drawing $\phi_{i}$ of the graph $G$. In order to do so, we start by placing the drawing $\psi_{C'_i}$ on the sphere. Next, for every disc $D\in \discset_i$, we copy the drawing of graph $X_i^D\cup \hat E_i^D$ in $\hat \phi_i^D$ to this new drawing, so that the two copies of the disc $D$ coincide with each other, and the images of the vertices of $\hat \Gamma_i^D$ in both drawings coincide. It is immediate to verify that the resulting drawing $\phi_i$ of $G$ is canonical with respect to $C_i$.
We next claim that, if drawing $\phi_{i-1}$ was canonical with respect to some cluster $C\in \cset$, then so is drawing $\phi_i$.

\begin{claim}\label{claim: remain canonical}
	Let $C_j\in \cset$ be a cluster, such that drawing $\phi_{i-1}$ was canonical with respect to $C_j$. Then drawing $\phi_i$ remains canonical with respect to $C_j$.
\end{claim}
\begin{proof}
	Observe that there must be some bridge $R\in \rset_G(C'_j)$ that contains the graph $C'_i$. Consider the corresponding disc $D(R)$ in the drawing $\psi_{C'_j}$ of $C'_j$, and the corresponding disc, that we also denote by $D(R)$, in the drawing $\phi_{i-1}$ of $G$. Recall that in the drawing $\phi_{i-1}$, all vertices and edges of $R$ are drawn in the disc $D(R)$. Let $D^*$ be the disc on the sphere that is the complement of $D(R)$, so the two discs share their boundaries but are otherwise disjoint. 
	
	Note that similarly, there must be some bridge  $R'\in \rset_G(C'_i)$ that contains the graph $C'_j$. We let $D(R')$ be the corresponding disc in the drawing $\psi_{C'_i}$ of $C'_i$. Note that the cluster $C'_j$ was unaffected when discs $D\in \discset_i\setminus \set{D(R')}$ were processed, as $C'_j$ is disjoint from the corresponding graphs $G_i^D$. When disc $D(R')$ was processed, we have deleted the images of vertices and edges of $C'_i$, and modified the images of the edges of $\hat E_i^{D(R')}$ accordingly. However, since graph $C'_i$ is drawn outside disc $D^*$ in $\phi_{i-1}$, this did not change the part of the drawing that lies in $D^*$. When we computed the final drawing $\phi_i$ of $G$, we have copied the drawing inside the disc $D(R')$ in $\hat \phi_i^D$ to the same disc in $\phi_i$. Since $D^*\subseteq D(R')$, this again did not affect the drawing in $D^*$. Therefore, the part of the drawing $\phi_{i-1}$ of the graph $G$ that appeared in disc $D^*$ remains unchanged in the drawing $\phi_i$. It is then easy to verify that drawing $\phi_i$ remains canonical with respect to $C_j$.	
\end{proof}

We let $\phi'=\phi_r$ be the drawing of $G$ that we obtain after all clusters of $\cset$ are processed. It now remains to analyze the number of crossings in $\phi'$.

\subsubsection{Analyzing the Number of Crossings}

Consider some cluster $C_i\in \cset$. Our goal is to bound the increase in the number of crossings due to iteration $i$, that is, $\cro(\phi_i)-\cro(\phi_{i-1})$. Let $\chi_i$ be the set of all crossings $(e_1,e_2)$ in the original drawing $\phi$ of $G$, with $e_1,e_2\in E(J_i)$. Notice that the drawings of $C'_i$ in $\phi$ and $\phi_{i-1}$ are identical. Let $\chi'_i$ be the set of all crossings $(e_1,e_2)$ in the drawing $\phi_{i-1}$ of $G$ with $e_1\in E(J_i)$ and $e_2\not\in E(C'_i)$. Recall that we have denoted by $\ir_i\subseteq E(J_i)$ the set of all edges $e\in E(J_i)$, such that either $e$ is irregular with respect to the drawings $\psi_{J_i}$ and $\phi_{J_i}$ of $J_i$, or at least one endpoint of $e$ is irregular with respect to these drawings. 

Consider now some disc $D\in \discset_i$. Let $\chi_i(D)$ be the set of all crossings $(e_1,e_2)$ in the original drawing $\phi$ of $G$, with $e_1,e_2\in E(J^D_i)$, and let $\chi'_i(D)$ be the set of all crossings $(e_1,e_2)$ in the drawing $\phi_{i-1}$ of $G$ with $e_1\in E(J^D_i)$ and $e_2\not\in E(C'_i)$. We also denote by $\ir_i(D)\subseteq E(J^D_i)$ the set of all edges $e\in E(J^D_i)$, such that either $e$ is irregular with respect to the drawings $\psi_{J^D_i}$ and $\phi_{J^D_i}$ of $J^D_i$ induced by the drawing $\psi_{C_i}$ of $C_i$ and $\phi$ of $G$, respectively, or at least one endpoint of $e$ is irregular with respect to these drawings. It is easy to verify that, if $e\in \ir_i(D)$, then $e\in \ir_i$ must hold.

From the analysis of \procdraw, we get that the number of new crossings in the drawing $\hat \phi_i^D$ is at most:
\[
%\begin{split}
z_i(D) \leq  O\left(\sum_{(e_1,e_2)\in \chi_i(D)}\Delta^2\cong_{\qset^D_i}(e_1)\cdot \cong_{\qset^D_i}(e_2)+\sum_{(e_1,e_2)\in \chi_i'(D) }\Delta\cdot\cong_{\qset_i^D}(e_1)+\sum_{e\in  \ir^D_i} \Delta^4(\cong_{\qset_i^D}(e))^2 \right).
%\end{split}
\]
Consider some crossing $(e_1,e_2)\in \chi_i(D)$. We can view this crossing as contributing $\cong_{\qset^D_i}(e_1)\cdot \cong_{\qset^D_i}(e_2)\cdot\Delta^2$ crossings to the first term of $z_i(D)$. If $\cong_{\qset^D_i}(e_1)\geq \cong_{\qset^D_i}(e_2)$, then we let the edge  $e_1$ ``pay'' $\left(\cong_{\qset^D_i}(e_1)\cdot\Delta\right )^2\geq \cong_{\qset^D_i}(e_1)\cdot \cong_{\qset^D_i}(e_2)\cdot \Delta^2$ units for these crossings, and otherwise we let the edge $e_2$ pay $\left(\cong_{\qset^D_i}(e_2)\cdot\Delta\right )^2\geq \cong_{\qset^D_i}(e_1)\cdot \cong_{\qset^D_i}(e_2)\cdot \Delta^2$ units for these crossings. Therefore, we obtain the following bound:
\[
%\begin{split}
z_i(D)\leq  O\left(\sum_{e_1\in E(J_i^D)}\sum_{e_2: (e_1,e_2)\in \chi_i(D)}\left (\cong_{\qset^D_i}(e_1)\cdot\Delta\right )^2 \\
+\sum_{e\in  \ir^D_i} \Delta^4(\cong_{\qset_i^D}(e))^2 +\sum_{(e_1,e_2)\in \chi_i'(D) }\cong_{\qset_i^D}(e_1)\cdot\Delta\right).
%\end{split}
\]
Summing up over all discs $D\in \discset_i$, and noting that, for every edge $e\in E(J_i)$, $\sum_{D\in \discset_i}\cong_{\qset_i^D}(e)\leq O(\Delta \cong_{\qset_i}(e))$, we get that the total increase $\cro(\phi_i)-\cro(\phi_{i-1})$ in the number of crossings is bounded by:
\[O\left(\sum_{e_1\in E(J_i)}\sum_{e_2: (e_1,e_2)\in \chi_i}\Delta^4\cdot\left (\cong_{\qset_i}(e_1)\right )^2+\sum_{e\in  \ir_i} \Delta^6(\cong_{\qset_i}(e))^2
+\sum_{(e_1,e_2)\in \chi_i'}\Delta^2\cdot \cong_{\qset_i}(e_1)\right).\]
Recall that for every edge $e\in E(C'_i)$, we defined its weight $w(e)$ as follows. First, we let $w(e)$ be the number of crossings in $\phi$ in which edge $e$ participates. Then, if $e\in \ir_i$, we increased $w(e)$ by $1$.  Therefore, we get that:
\[\cro(\phi_i)-\cro(\phi_{i-1}) \leq O\left(\sum_{e\in E(J_i)}\Delta^6 \cdot w(e)\cdot\left (\cong_{\qset_i}(e)\right )^2 \right ) +  O\left (\sum_{(e_1,e_2)\in \chi_i'}\Delta^2\cdot \cong_{\qset_i}(e_1)\right).\]
We denote $\Upsilon_i=O\left(\sum_{e\in E(J_i)}\Delta^6 \cdot w(e)\cdot\left (\cong_{\qset_i}(e)\right )^2\right )$
and $\Upsilon'_i=O\left (\sum_{(e_1,e_2)\in \chi_i'}\Delta^2\cdot \cong_{\qset_i}(e_1)\right),$
 and we analyze these terms separately.

\paragraph{Bounding $\sum_i\Upsilon_i'$.}
Consider a cluster $C_i\in \cset$, and let $e \in E(G\setminus (C'_i\cup \delta_G(C'_i)))$ be an edge of $G$ that does not lie in $C'_i\cup \delta_G(C'_i)$. Denote by $K_e$ the set of all edges of $J_i$ whose image in $\phi_{i-1}$ crosses the image of $e$. From Observation \ref{obs: classifying new crossings}, the total number of crossings $(e',e)$ in $\phi_i$ that do not belong to $\phi_{i-1}$, over all edges $e'\in E(G)$ is at most:
\[ \sum_{e'\in K_e}\sum_{D\in \discset_i} \Delta\cdot\cong_{\qset_i^D}(e')\leq \sum_{e'\in K_e}\Delta^2 \cdot \cong_{\qset_i}(e').  \]
Observe that for each such new crossing $(e'',e)$, edge $e''$ must lie in $E''$. For every edge $e'\in K_e$, we say that the crossing $(e',e)$ in $\phi_{i-1}$ is responsible for $\Delta^2 \cong_{\qset_i}(e')$ new crossings in $\phi_i$. We also say that crossing $(e',e)$ contributes $\Delta^2 \cong_{\qset_i}(e')$ crossings to $\Upsilon_i'$. 
If $(e',e)$ is a crossing of $\phi_{i-1}$ with $e\in \delta_G(C'_i)$ and $e'\in J_i$, then we say that it contriburtes $\Delta^2\cong_{\qset_i}(e')$ crossings to $\Upsilon'_i$, but it is not responsible for any new crossings.
Observe that the sum of the contributions of all crossings $(e',e)$ of $\phi_{i-1}$ with $e\in E(G\setminus C'_i)$ and $e'\in J_i$ is at least $\Omega(\Upsilon_i')$.

Consider now some crossing $(e_1,e_2)$ in the original drawing $\phi$ of $G$, and assume that $e_1\in C_i$ and $e_2\in C_j$, where $i<j$. The crossing $(e_1,e_2)$ contributes $\cong_{\qset_i}(e_1)\cdot\Delta^2$ crossings to $\Upsilon_i'$. It is also responsible for  $\cong_{\qset_i}(e_1)\cdot\Delta^2$ new crossings of the edge $e_2$. When cluster $C_j$ is processed, each one of these new crossings contributes $\cong_{\qset_j}(e_2)\cdot\Delta^2$ crossings to $\Upsilon_j'$. Therefore, altogether, crossing $(e_1,e_2)$ is responsible for $\cong_{\qset_i}(e_1)\cdot \cong_{\qset_j}(e_2)\cdot\Delta^4$ crossings in $\sum_{i'=1}^r\Upsilon'_{i'}$. If $\cong_{\qset_i}(e_1)\leq  \cong_{\qset_j}(e_2)$, then we make edge $e_2$ responsible for all these crossings and charge it $( \cong_{\qset_j}(e_2))^2\cdot\Delta^4\geq \cong_{\qset_i}(e_1)\cdot  \cong_{\qset_j}(e_2)\cdot\Delta^4$ %\snote{changed $\leq$ to $\geq$} 
for them, and otherwise, we make edge $e_1$ responsible for these crossings, and charge it $(\cong_{\qset_j}(e_1))^2\cdot \Delta^4\geq \cong_{\qset_i}(e_1)\cdot  \cong_{\qset_j}(e_2)\cdot \Delta^4$ %\snote{same here} 
for them.

If $(e_1,e_2)$ is a crossing in $\phi$ where exactly one of the two edges $e_1,e_2$ lies in some cluster $C_i$ and the other edge lies in $E''$, then the analysis is similar except that this crossing only contributes to $\Upsilon'_i$ and is charged to the corresponding edge. If both $e_1,e_2$ lie in the same cluster $C_i$, then crossing $(e_1,e_2)$ does not contribute to $\sum_{i'=1}^r\Upsilon'_{i'}$.
Recall that for every cluster $C_i\in \cset$, for every edge $e\in E(C_i')$, we have defined weight $w(e)$, which is at least the number of crossings in the drawing $\phi$ of $G$ in which edge $e$ participates.
To summarize, from the above discussion, we get that:
\[\sum_{i=1}^r\Upsilon'_{i} \leq O\left(\sum_{i=1}^r\sum_{e\in C'_i}\Delta^4\cdot w(e)\left(\cong_{\qset_i}(e)\right )^2\right ).  \]
Altogether, we then get that:
\[\cro(\phi')-\cro(\phi)\leq  \sum_{i=1}^r O\left (  \Delta^6\cdot \sum_{e\in C'_i}w(e)\left(\cong_{\qset_i}(e)\right )^2  \right ).\]

\paragraph{Final Accounting.}
Recall that we have denoted, for every cluster $C_i\in \cset$, by
$\ir_i\subseteq E(J_i)$ the set of all edges $e\in E(J_i)$, such that either $e$ is irregular with respect to the drawings $\psi_{J_i}$ and $\phi_{J_i}$ of $J_i$, or at least one endpoint of $e$ is irregular with respect to these drawings. We also denote by $x_i$ the total number of crossings in the drawing $\phi$ of $G$ in which the edges of $C'_i$ participate. It is then easy to verify that for every cluster $C_i\in \cset$, $\sum_{e\in E(C'_i)}w(e)\leq x_i+|\ir_i|$.

Consider now some cluster $C_i\in \cset$, and assume first that $C_i\in \cset_1$.  From Equation \ref{eq: bound for type 1}:
\[\begin{split}
\sum_{e\in E(C_i)}w(e)\cdot (\cong_{\qset_i}(e))^2&\leq O(\mu^2\Delta^2)\cdot \sum_{e\in E(C_i)}w(e)\\
&\leq O(\mu^2\Delta^2)(x_i+|\ir_i|)\\
&\leq O(\mu^3\Delta^3 (x_i+1))\\
&\leq  O(\poly(\Delta\log n)(x_i+1)).
\end{split}
\]
(We have used the fact that, if $C_i$ is a type-1 cluster then $|\ir_i|\leq O(\mu\Delta)$, and that $\mu=O(\poly(\Delta\log n)$.)

Assume now that $C_i$ is a type-2 cluster.  From Equation \ref{eq: bound weights type 2}, we get that:
\[
\sum_{e\in E(C'_i)}w(e)\cdot (\cong_{\qset}(e))^2\leq  O\left(\frac{\log n}{\alpha^4}\cdot\left (\sum_{e\in E(C_i)}w'(e)+1\right)\right).\]
Let $\ir_i'\subseteq E(C_i')$ denote the set of all edges $e\in E(C_i')$, such that either $e$ is an irregular edge with respect to the drawing $\psi_{C'_i}$ of $C'_i$, and the drawing $\phi_{C'_i}$ of $C'_i$ induced by $\phi$, or at least one endpoint of $e$ is irregular with respect to these drawings. Recall that $\sum_{e\in E(C_i')}w'(e)=O(x_i+|\ir_i'|)$. Therefore, we get that:
\[\sum_{e\in E(C'_i)}w(e)\cdot (\cong_{\qset}(e))^2\leq  O\left(\frac{\log n}{\alpha^4}\cdot (x_i+|\ir'_i|+1)\right )\leq O\left(\poly(\Delta\log n)\cdot (x_i+|\ir'_i|+1)\right ),\]
since $\alpha=\Theta(1/\poly(\Delta\log n))$.
Altogether, the number of crossings in the new drawing $\phi'$ of $\phi$ can now be bounded as:
\[
\begin{split}
\cro(\phi')&\leq \cro(\phi)+ O\left(\poly(\Delta\log n) \right )\cdot \left (\sum_{i=1}^rx_i+\sum_{C_i\in \cset_2}|\ir_i'|+|\cset|\right )\\
&\leq O\left (\poly(\Delta \log n)  (\cro(\phi)+|E''|)\right )+O\left (\poly(\Delta \log n) \right )\cdot \sum_{C_i\in \cset_2}|\ir_i'|.
\end{split} 
\]
The next claim will then finish the proof of Theorem \ref{thm: canonical drawing}.

\begin{claim}\label{claim: few irregular}
	$\sum_{C_i\in \cset_2}|\ir_i'|\leq O\left (\Delta^2(|E''|+\cro(\phi))\right)$.
\end{claim}
\begin{proof}
Consider some cluster $C_i\in \cset_2$. Recall that set $\ir_i'\subseteq E(C_i')$ contains all edges $e\in E(C_i')$, such that either $e$ is an irregular edge with respect to the drawing $\psi_{C'_i}$ of $C'_i$, and the drawing $\phi_{C'_i}$ of $C'_i$ induced by $\phi$, or at least one endpoint of $e$ is irregular with respect to these drawings. In other words, $|\ir_i'|\leq \Delta \cdot |\ir_V(\phi_{C'_i},\psi_{C'_i})| + |\ir_E(\phi_{C'_i},\psi_{C'_i})|$. Lemma \ref{lem:irregular_bounded_by_crossing} guarantees that:
\[
|\ir_V(\phi_{C'_i},\psi_{C'_i})\setminus S_2(C'_i)| + |\ir_E(\phi_{C'_i},\psi_{C'_i})\setminus E_2(C'_i)| \leq O(\cro(\phi_{C'_i}))\leq O(x_i),
\]
where $S_2(C'_i)$ is the set of all vertices that participate in $2$-separators in $C'_i$, and $E_2(C'_i)$ is the set of all edges of $C'_i$ that have both endpoints in $S_2(C'_i)$. Unfortunately, the definition of type-2 acceptable clusters does not provide any bound on the cardinality of the set $S_2(C'_i)$. It does, however, ensure that $|S_2(C_i)|\leq O(\Delta|\Gamma(C_i)|)$, where $C_i$ is the original cluster, that may contain artificial edges. Unfortunately, the original drawing $\phi$ of $G$ does not include the drawings of the artificial edges. However, using the embeddings of these edges, we can easily transform drawing $\phi$ of $G$ into a drawing $\tilde \phi$ of $\bigcup_{C_i\in \cset_2} C_i$, without increasing the number of crossings. Applying  Lemma \ref{lem:irregular_bounded_by_crossing} to the resulting drawings of graphs $C_i\in \cset_2$ will then finish the proof. We now turn to provide a more detailed proof.
	
	Consider the original drawing $\phi$ of graph $G$. We transform it into a drawing $\tilde \phi$ of $\bigcup_{C_i\in \cset_2} C_i$, as follows. Recall that the decomposition $\dset$ of $G$ into acceptable clusters contains an embedding $\pset=\set{P(e)\mid e\in A}$ of all artificial edges via paths that are internally disjoint. Moreover, for every edge $e=(x,y)\in A$, there is a type-1 cluster $C(e)\in \cset_1$, such that $P(e)\setminus\set{x,y}$ is contained in $C(e)$, and the clusters $C(e)$ are distinct for all edges $e\in A$. We delete from $\phi$ all vertices and edges except those participating in graphs $C'_i$ for $C_i\in \cset_2$, and in paths in $\pset$. By suppressing all inner vertices on the paths in $\pset$, we obtain a drawing $\tilde \phi$ of $\bigcup_{C_i\in \cset_2} C_i$, that contains at most $\cro(\phi)$ crossings. Consider now some cluster $C_i\in \cset_2$. Let $\tilde \phi_i$ be the drawing of $C_i$ that is induced by $\tilde \phi$. Observe that, if a vertex $v\in V(C_i)$ is irregular with respect to 
	$\phi_{C'_i}$, $\psi_{C'_i}$, then it must be irregular with respect to $\tilde \phi_i$ and $\psi'_{C_i}$. Similarly, if an edge $e\in E(C'_i)$ is  irregular with respect to 
	$\phi_{C'_i}$, $\psi_{C'_i}$, then it must be irregular with respect to $\tilde \phi_i$ and $\psi'_{C_i}$. Therefore, if we denote by $\ir_i''$ the set of all edges $e\in C_i$, such that either $e$ is irregular with respect to $\tilde \phi_i$ and $\psi'_{C_i}$, or at least one endpoint of $e$ is irregular with respect to these two drawings, then $|\ir_i''|\geq |\ir_i'|$. Let $x'_i$ be the total number of crossings in $\tilde \phi_i$. Let $E_2^i\subseteq E(C_i)$ be the set of all edges that are incident to vertices of $S_2(C_i)$ (vertices that participate in $2$-separators in $C_i$). Then, from  Lemma \ref{lem:irregular_bounded_by_crossing}:
\[|\ir_i''\setminus E_2^i|\leq O(\Delta \cdot x'_i). \]
Moreover, from the definition of type-2 acceptable clusters, $|S_2(C_i)|\leq O(\Delta|\Gamma(C_i)|)$, and so $|E_2^i|\leq O(\Delta^2|\Gamma(C_i)|)$. Overall, we conclude that:
\[|\ir_i'|\leq |\ir_i''| \leq O(\Delta^2(x'_i+|\Gamma(C_i)|)). \]
Summing up over all clusters $C_i\in \cset_2$, we get that:
\[\begin{split}
		\sum_{C_i\in \cset_2}|\ir_i'|& \leq O(\Delta^2)\sum_{C_i\in \cset_2}x'_i+O(\Delta^2)\sum_{C_i\in \cset_2}|\Gamma(C_i)|\\
		&\leq O\left(\Delta^2\cdot(\cro(\tilde \phi)+|E''|)\right)\\
		&\leq O\left(\Delta^2\cdot(\cro(\phi)+|E''|)\right).
	\end{split}\]	
\end{proof}

\section{Handling Non 3-Connected Graphs}
\label{sec:non_3}

So far we have provided the proofs of Theorem \ref{thm: main} and Theorem \ref{thm: reduction} for the special case where the input graph $G$ is $3$-connected.
In this section we extend the proofs to arbitrary graphs, and also provide the proof of Theorem \ref{thm:main_non_3}. We start by extending the proof of Theorem \ref{thm: main} to arbitrary graphs. 

It is sufficient to prove Theorem \ref{thm: main}  for the special case where the input graph $G$ is $2$-connected. Indeed, given any graph $G$, let $\zset(G)$ be the set of all super-blocks of $G$ (maximal $2$-connected components). Given a planarizing set $E'$ of edges for graph $G$, for each graph $Z\in \zset(G)$, we let $E'(Z)=E'\cap E(G)$. We then apply the algorithm from Theorem \ref{thm: main} to each graph $Z\in \zset$, together with the planarizing set $E'(Z)$ of its edges separately, obtaining a new planarizing edge set $E''(Z)$ with $E'(Z)\subseteq E''(Z)$. Moreover, from  Theorem \ref{thm: main}, we are guaranteed that:
\[ \sum_{Z\in \zset(G)}|E''(Z)|\leq \sum_{Z\in \zset(G)}O\left((|E'(Z)|+\optcro(Z))\cdot\poly(\Delta\log n)\right)\leq  O\left((|E'|+\optcro(G))\cdot\poly(\Delta\log n)\right). \]
We are also guaranteed that, for each graph $Z\in \zset$, there is a drawing $\phi(Z)$ of $Z$, such that the number of crossings in $\phi(Z)$ is bounded by $O\left((|E'(Z)|+\optcro(Z))\cdot\poly(\Delta\log n)\right)$, and the edges of $G\setminus E''(Z)$ do not participate in any crossings in $\phi(Z)$. We set $E''=\bigcup_{Z\in \zset}E''(Z)$, so that $|E''|\leq O\left((|E'|+\optcro(G))\cdot\poly(\Delta\log n)\right)$. It is also easy to combine the drawings $\set{\phi(Z)}_{Z\in \zset}$ into a drawing $\phi$ of $G$, without introducing any new crossings, so that the total number of crossings in $\phi$ is at most:
\[\sum_{Z\in \zset(G)}O\left((|E'(Z)|+\optcro(Z))\cdot\poly(\Delta\log n)\right)\leq O\left((|E'|+\optcro(G))\cdot\poly(\Delta\log n)\right),  \]
and the edges of $E''$ do not participate in crossings in $\phi$. 

Therefore, from now on we assume that the input graph $G$ is $2$-connected. We let $\bset=\bset(G)$ be the block decomposition of $G$ given by Theorem \ref{thm: block decomposition}. Let $\tilde\bset=\set{\tilde B\mid B\in \bset}$, and let $\tilde \bset^*\subseteq \tilde\bset$ contain all graphs $\tilde B$ that are not isomorphic to $K_3$. We use the algorithm from Lemma \ref{lem:block_endpoints_routing} to compute, for each graph $\tilde{B}\in \tilde\bset^*$, a collection $\pset_{\tilde B}=\set{P_{\tilde B}(e)\mid e\in A_{\tilde B}}$ of paths in $G$, such that, for each fake edge $e=(u,v)\in \aset_{\tilde B}$,  path $P_{\tilde B}(e)$ connects $u$ to $v$ in $G$ and it is internally disjoint from $\tilde B$. Recall that the lemma guarantees that for every graph $\tilde B\in \tilde \bset$, all paths in $\pset_{\tilde B}$ are mutually internally disjoint, and moreover, if we denote $\pset=\bigcup_{\tilde B\in \tilde\bset^*(G)}\left(\pset_{\tilde B}\setminus \set{P_{\tilde B}(e^*_{\tilde B})} \right)$, then every edge of $G$ participates in at most $6$ paths in $\pset$. For simplicity of notation, for each graph $\tilde B\in \tbset$, we refer to the fake parent-edge $e^*_{\tilde B}$ as the \emph{bad fake edge} of $\tilde B$; all other fake edges of $\tB$ are called good fake edges.

Assume that we are given a planarizing set $E'$ of edges for the input graph $G$. The next lemma allows us to compute, for each graph $\tB\in \tbset$, a planarizing set $E'(\tB)$ of edges for $\tB$, so that $\sum_{B\in \bset}|E'(\tB)|$ is sufficiently small.

\begin{lemma}\label{lem: computing small planarizing edge sets}
	There is an efficient algorithm, that, given a planarizing set $E'$ of edges for graph $G$, computes, for each graph $\tB\in \tbset$, a planarizing edge set $E'(\tB)$ for $\tB$, such that $\sum_{\tB\in \tbset}|E'(\tB)|\leq O(|E'|+\optcro(G))$.
\end{lemma}
\begin{proof}
	Consider a graph $\tB\in \tbset$. In order to construct the planarizing set $E'(\tB)$ of edges, we start by adding every real edge of $\tB$ that lies in $E'$ to set $E'(\tB)$. Additionally, for every fake edge $e\in \tB$ that is not a bad fake edge, if the path $P_{\tB}(e)\in \pset_{\tilde B}$ contains an edge of $E'$, then we add $e$ to $E'(\tB)$. Since every edge of $G$ may belong to at most $6$ paths in $\pset$, this ensures that $\sum_{\tB\in \tbset}|E'(\tB)|\leq O(|E'|)$. If $\tB\setminus E'(\tB)$ is a planar graph, then we say that $B$ is a \emph{good block}, and otherwise we say that it is a bad block. We need the following claim.
	
	\begin{claim}\label{claim: planarizing a block} If $B$ is a bad block, then graph $\tB\setminus (E'(\tB)\cup e^*_{\tB})$ is planar.
	\end{claim}
\begin{proof}
	 Notice that every real edge of $\tB\setminus (E'(\tB)\cup e^*_{\tB})$ belongs to $G\setminus E'$, and for every fake edge $e\in E(\tB)\setminus (E'(\tB)\cup e^*_{\tB})$, the corresponding path $P_{\tB}(e)$ is contained in $G\setminus E'$. Recall that graph $G\setminus E'$ is planar. It is now immediate to verify that a planar drawing of $G\setminus E'$ defines a planar drawing of $\tB\setminus (E'(\tB)\cup e^*_{\tB})$.
\end{proof}
	
The next claim will then finish the proof of Lemma \ref{lem: computing small planarizing edge sets}.

\begin{claim}\label{claim: few bad blocks}
	The total number of bad blocks in $\bset$ is bounded by $O(\optcro(G))$.
\end{claim}
If $B$ is a bad block, then we add the bad fake edge $e^*_{\tB}$ to set $E'(\tB)$. From Claim \ref{claim: planarizing a block}, set $E'(\tB)$ of edges is a planarizing set for graph $\tB$. Moreover, from the above analysis, $\sum_{B\in \bset}|E'(\tB)|\leq O(|E'|+\optcro(G))$, as required.
Therefore, in order to complete the proof of Lemma \ref{lem: computing small planarizing edge sets}, it is	enough to prove Claim \ref{claim: few bad blocks}. 

\begin{proofof}{Claim \ref{claim: few bad blocks}}
Consider the optimal drawing $\phi^*$ of graph $G$. We erase from this drawing all edges and vertices, except for the vertices and the real edges of $\tilde B$, and all vertices and edges participating in the paths $\pset_{\tB}$. The resulting drawing can be viewed as a drawing of the graph $\tB$, after we suppress all inner vertices on the paths in $\pset_{\tB}$. Notice that, since $B$ is a bad block, there is some pair $e_1,e_2$ of edges that crosses in the resulting drawing, and moreover, $e_1,e_2$ do not both belong to the path $P_{\tB}(e^*_{\tB})\in \pset_{\tilde B}$. Therefore, at least one of the two edges $e_1,e_2$, is either a real edge of $\tB$, or it lies on some path $P_{\tB}(e)$, where $e$ is a good fake edge  for $\tB$. We say that the crossing $(e_1,e_2)$ of $\phi^*$ is \emph{responsible} for the bad block $B$. Since every edge of $G$ may participate in at most one graph $\tB\in \tbset$ as a real edge, and since every edge of $G$ may belong to at most $6$ paths in $\pset$, we get that every crossing in $\phi^*$ may only be responsible for a constant number of bad blocks, and so the total number of bad blocks is bounded by $O(\optcro(G))$.
\end{proofof}
\end{proof}

We also need the following lemma:

\begin{lemma}\label{lem: few crossings in all blocks}
	$\sum_{B\in \bset}\optcro(\tilde B)\leq O(\optcro(G))$.
\end{lemma}

\begin{proof}	
Let $\phi^*$ be the optimal drawing of $G$, so $\cro(\phi^*)=\optcro(G)$. We assume w.l.o.g. that every pair of edges crosses at most once in $\phi^*$, and for every edge, its image does not cross itself in $\phi^*$. We now define, for each graph $\tilde B\in \tbset$, a drawing $\psi_{\tilde B}$ in the plane, using the drawing $\phi^*$. Consider any graph $\tilde B\in \tbset$. Note that, if $\tilde B$ is isomorphic to $K_3$, then it has a planar drawing, so we let $\psi_{\tilde B}$ be that planar drawing. Assume now that $\tilde B$ is not isomorphic to $K_3$, so $\tilde B\in \tbset^*$.

In order to obtain the drawing  $\psi_{\tilde B}$ of $\tilde B$, we start from the drawing $\phi^*$ of $G$, and delete from it all edges and vertices, except for the vertices and the real edges of $\tilde B$, and all vertices and edges participating in the paths in  $\pset_{\tilde B}$.
We partition all crossings of the resulting drawing into five sets. Set $\chi_1(\tB)$ contains all crossings $(e_1,e_2)$, where both $e_1$ and $e_2$ are real edges of $\tilde B$. Set $\chi_2(\tB)$ contains all crossings $(e_1,e_2)$, where  $e_1$ is a real edge of $\tilde B$, and $e_2$ lies on some path $P_{\tilde B}(e)$, where $e$ is a fake edge of $\tilde B$. Set $\chi_3(\tB)$ contains all crossings $(e_1,e_2)$, where $e_1\in P_{\tilde B}(e)$, $e_2\in P_{\tilde B}(e')$; both $e$ and $e'$ are fake edges of $\tB$ (where possibly $e=e'$); and neither of these edges is the parent-edge $e^*_{\tB}$. Set $\chi_4(\tB)$ contains all crossings $(e_1,e_2)$, where $e_1\in P_{\tilde B}(e)$, $e_2\in P_{\tilde B}(e')$, such that $e,e'$ are both fake edges of $\tB$, and exactly one of these edges is the parent-edge $e^*_{\tB}$. Lastly, set $\chi_5(\tB)$ contains all crossings $(e_1,e_2)$, where both $e_1,e_2\in P_{\tB}(e^*_{\tB})$.
We obtain the final drawing $\psi_{\tilde B}$ of $\tilde B$ from the current drawing, by suppressing all inner vertices on the paths of $\tilde \pset_{\tB}$. Additionally, if the image of the edge $e^*_{\tB}$ crosses itself, then we remove all loops to ensure that this does not happen. Clearly, $\cro(\psi_{\tilde B})\leq \sum_{i=1}^4|\chi_i(\tB)|$.

We denote, for all $1\leq i\leq 4$, $\chi_i=\bigcup_{\tB\in \tbset^*}\chi_i(\tB)$ (we view $\chi_i$ as a multiset, so a crossing $(e_1,e_2)$ that belongs to several sets $\chi_i(\tB)$ is added several times to $\chi_i$). It is now enough to show that $\sum_{i=1}^4|\chi_i|\leq O(\cro(\phi^*))$.

Consider some crossing $(e_1,e_2)$ in $\phi^*$. Observe that this crossing may lie in set $\chi_1(\tB)$ for a graph $\tB\in \tbset^*$ only if both $e_1$ and $e_2$ are real edges of $\tB$. Since each edge of $G$ may belong to at most one graph in $\tbset$ as a real edge, crossing $(e_1,e_2)$ appears at most once in $\chi_1$. Therefore, $|\chi_1|\leq \cro(\phi^*)$. Notice that crossing $(e_1,e_2)$ may lie in set $\chi_2(\tB)$ for a graph $\tB\in \tbset^*$ only if either $e_1$ or $e_2$ are real edges of $\tB$. Therefore, using the same reasoning as before, crossing $(e_1,e_2)$ may appear at most twice in $\chi_2$, and  so $|\chi_2|\leq 2\cro(\phi^*)$. Assume now that $(e_1,e_2)\in \chi_3(\tB)$, for some graph $\tB\in \tbset^*$. 
Then there are fake edges $e,e'$ in $\tB$, neither of which is the parent-edge $e^*_{\tB}$, such that $e_1\in P_{\tB}(e)$ and $e_2\in  P_{\tB}(e')$ (where it is possible that $e=e'$). Since each edge of $G$ may belong to at most $6$ paths in $\pset$, we get that crossing $(e_1,e_2)$ may appear at most $6$ times in $\chi_3$, so $|\chi_3|\leq 6\cro(\phi^*)$. Lastly, assume that $(e_1,e_2)\in \chi_4(\tB)$, for some graph $\tB\in \tbset^*$. 
Then there are fake edges $e,e'$ in $\tB$,  exactly one of which is the parent-edge $e^*_{\tB}$, such that $e_1\in P_{\tB}(e)$ and $e_2\in  P_{\tB}(e')$. Since each edge  of $G$ may belong to at most $6$ paths in $\pset$, we get that crossing $(e_1,e_2)$ may appear at most $12$ times in $\chi_4$, so $|\chi_4|\leq 12\cro(\phi^*)$. 

Overall, we get that $\sum_{\tB\in \tbset}\optcro(\tilde B)\leq \sum_{\tB\in \tbset}\cro(\psi_{\tilde B})\leq \sum_{i=1}^4|\chi_i|\leq O(\optcro(G))$.
\end{proof}

We are now ready to complete the proof of  Theorem \ref{thm: main}.
Assume that we are given a planarizing set $E'$ of edges for graph $G$. Using the algorithm from Lemma \ref{lem: computing small planarizing edge sets}, we compute, for each graph $\tB\in \tbset$, a planarizing edge set $E'(\tB)$, such that $\sum_{B\in \bset}|E'(\tB)|\leq O(|E'|+\optcro(G))$. Then we use our algorithm for Theorem \ref{thm: main} in $3$-connected graphs to compute, for each graph $\tB\in \tbset$, another planarizing set $E''(\tB)$ with $E'(\tB)\subseteq E''(\tB)$, such that $|E''(\tB)|\leq O\left((|E'(\tB)|+\optcro(\tB))\poly(\Delta\log n)\right)$, and there is a drawing $\phi_{\tB}$ of graph $\tB$ with at most $O\left((|E'(\tB)|+\optcro(\tB))\poly(\Delta\log n)\right)$ crossings, such that the edges of $E''(\tB)$ do not participate in the crossings of $\phi_{\tB}$.  From Lemma \ref{lem: few crossings in all blocks}, we get that:
\[\sum_{B\in \bset}|E''(\tB)|\leq  O\left((|E'|+\optcro(G))\poly(\Delta\log n)\right), \]
and similarly:
\[\sum_{B\in \bset}\cro(\phi_{\tB})\leq  O\left((|E'|+\optcro(G))\poly(\Delta\log n)\right). \]

Next, we define the final set $E''$ of edges for the graph $G$. We start with $E''=\emptyset$. For every graph $\tB\in \tbset$, for every edge $e=(u,v)\in E''(\tB)$, if $e$ is a real edge, then we add $e$, and every edge of $G$ that is incident to either $u$ or $v$ to $E''$. Otherwise, $e$ is a fake edge, and then we add to $E''$ every edge of $G$ that is incident to either $u$ or $v$. This completes the definition of the set $E''$ of edges. It is immediate to verify that $E'\subseteq E''$ (in fact if this does not hold we can simply add the edges of $E'$ to the set $E''$). Moreover, it is easy to see that:
\[|E''|\leq O(\Delta)\cdot \sum_{B\in \bset}|E''(\tB)|\leq  O\left((|E'|+\optcro(G))\poly(\Delta\log n)\right),\]
as required. In order to complete the proof of Theorem \ref{thm: main}, it is now enough to show that there is a drawing $\phi$ of $G$ with at most 
$ O\left((|E'|+\optcro(G))\poly(\Delta\log n)\right)$ crossings, such that the edges of $G\setminus E''$ do not participate in any crossings in $\phi$. We obtain the drawing $\phi$ by ``gluing'' together the drawings $\set{\phi_{\tB}}_{\tB\in \bset}$ (this part is similar to the algorithm of \cite{chuzhoy2011graph}). Before we prove the existence of the desired drawing $\phi$ of $G$, we set up some notation.

For each pseudo-block $B_0\in \bset$, we denote by $\desc(B_0)$ the collection of all pseudo-blocks $B_1\in \bset$, such that vertex $v(B_1)$ is the descendant of vertex $v(B_0)$ in the decomposition tree $\tau$. We note that $B_0\in \desc(B_0)$. Consider now some pseudo-block $B_0\in \bset$ and the drawing $\phi_{\tB_0}$ of graph $\tB_0$ that we defined above. For simplicity of notation, we denote by $N(\tB_0)$ the total number of crossings in the drawing $\phi_{\tB_0}$. We also denote by $N_0(\tB_0)$ the total number of crossings in the drawing $\phi_{\tB_0}$ in which the fake parent-edge $e^*_{\tB_0}$ participates, and we denote by $N_1(\tB_0)=N(\tB_0)-N_0(\tB_0)$ the total number of all other crossings in $\phi_{\tB_0}$.
The following lemma allows us to ``glue'' the drawings $\set{\phi_{\tB}}_{\tB\in \tbset}$ together.

\begin{lemma}\label{lem: gluing the drawings}
	There is an efficient algorithm, that, given a pseudo-block $B_0\in \bset$, and drawings $\set{\phi_{\tB}}_{B \in \desc(B_0)}$, computes a drawing $\hat \phi_{B_0}$ of graph $B_0\cup\set{e^*_{\tB_0}}$ (if $B_0=G$ and edge $e^*_{\tB_0}$ is undefined, then $\hat \phi_{B_0}$ is a drawing of $B_0$), such that the following hold:
	
	\begin{itemize}
		\item edges of $E''\cap E(B_0)$ do not participate in any crossings in $\hat \phi_{B_0}$;
		\item the total number of crossings in $\hat \phi_{B_0}$ in which the fake edge $e^*_{\tB_0}$ participates is at most $4\Delta N_0(\tB_0)$; and
		
		\item the total number of all other crossings in $\hat \phi_{B_0}$ is at most $16\Delta^2 \left (\sum_{B\in \desc(B_0)}N(\tB)-N_0(\tB_0)\right )$.  
	\end{itemize}
	\end{lemma}

We note that Lemma \ref{lem: gluing the drawings} provides somewhat stronger guarantees than what is needed for the proof of Theorem \ref{thm: main}: it would be sufficient to provide an existential version of the lemma, and the efficient algorithm for constructing the drawing $\hat \phi_{B_0}$ from the drawings $\set{\phi_{\tB_1}}_{B_1\in \desc(B_0)}$ is not needed. But we will also exploit this lemma in the proofs of Theorem \ref{thm:main_non_3} and Theorem \ref{thm: reduction}, for which the constructive version is necessary.

It is now immediate to obtain the proof of Theorem \ref{thm: main}, by applying Lemma \ref{lem: gluing the drawings}  to the pseudo-block $B_0=G$. Let $\hat\phi=\hat\phi_{B_0}$ denote the resulting drawing of $G$. We are then guaranteed that no edges of $E''$ participate in crossings in $\hat\phi$, and moreover:
\[ \cro(\hat\phi)\leq O\left (\Delta^2 \cdot \sum_{B_1\in \desc(B_0)}N(\tB_1)\right )\leq O\left (\Delta^2 \cdot \sum_{B\in \bset}\cro(\phi_{\tB}) \right )\leq  O\left((|E'|+\optcro(G))\poly(\Delta\log n)\right).\]
In order to complete the proof of Theorem \ref{thm: main}, it is now enough to prove Lemma \ref{lem: gluing the drawings}.

\begin{proofof}{Lemma \ref{lem: gluing the drawings}}
The proof is by induction on the length of the longest path from vertex $v(B_0)$ to a leaf vertex of $\tau$ that is a descendant of $v(B_0)$ in $\tau$. The base case is when $v(B_0)$ is a leaf vertex of $\tau$. In this case, graph $\tB_0$ is exactly $B_0\cup \set{e^*_{\tB_0}}$, and we set $\hat \phi_{B_0}=\phi_{\tB_0}$. From the definition of the set $E''$ of edges, no edges of $E''\cap E(B_0)$ participate in crossings in the resulting drawing; the number of crossings in which edge $e^*_{\tB_0}$ participates is $N_0(\tB_0)$, and the total number of all other crossings in $\hat \phi_{B_0}$ is $N(\tB_0)-N_0(\tB_0)$. Therefore, drawing $\hat \phi_{B_0}$ has all required properties.
		
Next, we consider an arbitrary pseudo-block $B_0\in \bset$, where $v(B_0)$ is not a leaf of $\tau$. Let $B_1,\ldots,B_r$ be pseudo-blocks whose corresponding vertices $v(B_i)$ are the children of $v(B_0)$ in tree $\tau$. 
We denote, for all $1\leq i\leq r$, the endpoints of the block $B_i$ by $(x_i,y_i)$, so that the edge $e^*_{\tB_i}$ (the fake parent-edge of $\tB_i$) connects $x_i$ to $y_i$. Note that graph $\tB_0$ must also contain an edge $e_i=(x_i,y_i)$. Since parallel edges are not allowed in graph $\tB_0$, edge $e_i$ may be a real or a fake edge of $\tB_0$, and moreover, it is possible that for $1\leq i<j\leq r$, $e_i=e_j$.
We use the induction hypothesis in order to compute, for all $1\leq i\leq r$, a drawing $\hat \phi_{B_i}$ of $B_i\cup\set{ e^*_{\tB_i}}$ with the required properties.
	
	We denote, for all $1\leq i\leq r$, by $G_i$ the multi-graph $\tB_0\cup \left (\bigcup_{i'=1}^{i-1}B_{i'}\right )$.
	We start with the drawing $\phi_0=\phi_{\tB_0}$ of $\tB_0$, and then perform $r$ iterations. The input to the $i$th iteration is a drawing $\phi_{i-1}$ of graph $G_{i-1}$, and the output is a drawing $\phi_i$ of $G_i$.
	
	We now describe the execution of the $i$th iteration. 
	Our starting point is the input drawing $\phi_{i-1}$ of graph $G_{i-1}$ on the sphere, and the drawing $\hat \phi_{B_i}$ of $B_i\cup e^*_{\tB_i}$. Recall that $x_i,y_i$ are the endpoints of the block $B_i$, and graph $G_{i-1}$ contains the edge $e_i=(x_i,y_i)$. For convenience of notation, we denote the fake parent-edge of $\tB_i$, whose endpoints are also $x_i,y_i$ by $e'_i$. Note that there must be some point $t'_i$ on the image of the edge $e'_i$ in $\hat \phi_{B_i}$, such that the segment $\sigma'$ of the image of $e'_i$ between $x_i$ and $t'_i$ lies on the boundary of a single face in the drawing $\hat \phi_{B_i}$. Let $F'$ denote this face. We view $\hat \phi_{B_i}$ as a drawing on the plane, whose outer face is $F'$. Similarly, there is a point $t_i$ on the image of the edge $e_i$ in the drawing $\phi_{i-1}$ such that the segment $\sigma$ of the image of $e_i$ between $x_i$ and $t_i$ lies on the boundary of a single face in $\phi_{i-1}$; we denote that face by $F$. We view $\phi_{i-1}$ as a drawing in the plane, where face $F$ is the outer face. Next, we superimpose the drawings $\phi_{i-1}$ and  $\hat \phi_{B_i}$ in the plane, such that the two resulting drawings are disjoint, except that the image of the vertex $x_i$ is unified in both drawings, and the faces $F$ and $F'$ of the two drawings correspond to the outer face of this new drawing, that we denote by $F^*$. We add a curve $\gamma$ to this new drawing, connecting the images of $t_i$ and $t'_i$, such that $\gamma$ does not intersects any parts of the current drawing, except for its endpoints. The image of the vertex $y_i$ in the new drawing becomes the image of $y_i$ in $\phi_{i-1}$. Let $E_i$ be the set of all edges of $E(B_i)$ that are incident to $y_i$. In order to complete the drawing $\phi_i$ of $G_i$, we need to define the drawings of the edges in $E_i$. Consider any such edge $e=(a,y_i)$. In order to obtain a new drawing of $e$, we start with the drawing of $e$ in $\hat \phi_{B_i}$, that connects the image of $a$ to the original image of $y_i$ in $\hat \phi_{B_i}$. Next, we follow along the image of the edge $e'_i$ in $\hat \phi_{B_i}$, until the point $t'_i$. Next, we follow the curve $\gamma$, connecting point $t'_i$ to point $t_i$, and lastly, we follow the image of the edge $e_i$ in the drawing $\phi_{i-1}$, from point $t_i$ to the image of the vertex $y_i$. We can do so in a way that ensures that the images of the edges in $E_i$ do not cross each other. This defines the final drawing $\phi_i$ of the graph $G_i$. We now analyze its crossings. 
	
	Consider any crossing $(e,e')$ in $\phi_i$. We say that it is an \emph{old} crossing iff: (i) crossing $(e,e')$ is present in the drawing $\phi_{i-1}$; or (ii) crossing $(e,e')$ is present in the drawing $\hat \phi_{B_i}$, and neither of the edges $e,e'$ is $e'_i$. All other crossings in $\phi_i$ are called \emph{new crossings}. Note that each such new crossing must involve exactly one edge from $E_i$. Specifically, for every crossing $(e'_i,\hat e)$ in $\hat\phi_{B_i}$, in which the edge $e'_i$ participates, we introduce $|E_i|$ new crossings of $\hat e$ with edges of $E_i$. Notice that, if $e'_i$ participates in any crossings in $\hat \phi_{B_i}$, then $E_i\subseteq E''$. The number of such new crossings is then bounded, from the induction hypothesis, by $4\Delta N_0(\tB_i)\cdot |E_i|\leq 4\Delta^2N_0(\tB_i)$. Additionally, for every crossing $(e_i,\hat e)$ in $\phi_{i-1}$, in which the edge $e_i$ participates, we introduce $|E_i|$ new crossings of $\hat e$. We say that crossing $(e_i,\hat e)$ is \emph{responsible} for these new crossings. Our algorithm ensures that crossing $(e_i,\hat e)$ may only be present in the drawing $\phi_{i-1}$ if edge $e_i$ participated in crossings in $\phi_{\tilde B_0}$. In this case, we are guaranteed that $E_i\subseteq E''$. Therefore, we ensure that all real edges that participate in the new crossings belong to $E''$.
	This completes the description of the $i$th iteration.
	
	Let $\hat \phi=\phi_r$ be the final drawing of the graph $G_r$ that we obtain. We now bound the number of crossings in $\hat \phi$. We partition the crossings of $\hat \phi$ into three sets: set $\chi_1$ contains all crossings $(e,e')$, where, for some $1\leq i\leq r$, $e,e'\in E(B_i)$. Set $\chi_2$ contains all crossings $(e,e')$, where $e=e^*_{\tB_0}$; and set $\chi_3$ contains all other crossings. 
	
	For all $1\leq i\leq r$, let $\chi^i_1\subseteq \chi_1$ be the set of all crossings $(e,e')$, where $e,e'\in E(B_i)$. From the above discussion, if $(e,e')$ is a crossing in $\chi^i_1$, then either it was present in $\hat \phi_{B_i}$, or it is one of the new crossings. The number of crossings of the former type is bounded, from the induction hypothesis, by $16\Delta^2 \left (\sum_{B\in \desc(B_i)}N(\tB)-N_0(\tB_i)\right )$, while the number of crossings of the latter type is bounded by $4\Delta^2N_0(\tB_i)$. Therefore, altogether, $|\chi^i_1|\leq 16\Delta^2 \left (\sum_{B\in \desc(B_i)}N(\tB)\right )$.
	
	We now proceed to bound the number of crossings in $\chi_2$. Consider any crossing $(e,e')$ in the drawing $ \phi_{\tB_0}$ of $\tB_0$, where $e=e^*_{\tB_0}$. If $e'$ is a real edge of $\tB_0$, then this crossing is present in $\chi_2$. Otherwise, $e'=e_i$ for some $1\leq i\leq r$. Then crossing $(e,e')$ is responsible for $|E_i|$ new crossings in $\chi_2$. It is then easy to verify that every crossing $(e,e')$ of $ \phi_{\tB_0}$ with $e=e^*_{\tB_0}$ may be responsible for at most $\Delta$ crossings in $\chi_2$, and so $|\chi_2|\leq \Delta N_0(\tB_0)$.
	
	Lastly, we need to bound $|\chi_3|$. We partition set $\chi_3$ of crossings into three subsets: set $\chi_3'$ contains all crossings $(e,e')$, where both $e,e'$ are real edges of $\tB_0$. It is easy to verify that $|\chi_3'|\leq N_1(\tB_0)$. Set $\chi_3''$ contains all crossings $(e,e')$, where $e$ is a real edge of $\tB_0$, and $e'\in E(B_i)$, for some $1\leq i\leq r$. In this case, drawing $\phi_{i-1}$ of $G_{i-1}$ contained a crossing $(e,e_i)$, that was charged for this new crossing, and the total charge to each such crossing $(e,e_i)$ was at most $|E_i|$. It is then easy to verify that $|\chi_3''|\leq \Delta N_1(\tB_0)$. Lastly, set $\chi_3'''$ contains all remaining crossings $(e,e')$, where $e\in B_i,e'\in B_j$, for some $1\leq i\neq j\leq r$. In this case, it is easy to verify that crossing $(e_i,e_j)$ must have been present in drawing $\hat \phi_{\tB_0}$ of $\tB_0$. Moreover, each such crossing $(e_i,e_j)$ may be responsible for at most $\Delta^2$  crossings in $\chi_3'''$, and so, overall $|\chi_3'''|\leq \Delta^2 N_1(\tB_0)$. We conclude that $|\chi_3|\leq 2 \Delta^2 N_1(\tB_0)$.
	
	We obtain the final drawing $\hat\phi_{B_0}$ of $B_0\cup \set{e^*_{\tB_0}}$ by deleting, from the drawing $\hat \phi$, the images of all fake edges of $\tB_0$, except for the edge $e^*_{\tB_0}$. From the above discussion, the total number of crossings in $\hat \phi_{B_0}$ in which the fake edge $e^*_{\tB_0}$ participates is at most $4\Delta N_0(\tB_0)$; and
	 the total number of all other crossings in $\hat \phi_{B_0}$ is at most $16\Delta^2 \left (\sum_{B \in \desc(B_0)}N(\tB)-N_0(\tB_0)\right )$. Moreover, from the discussion above, if edge $e\in E(B_0)$ participates in a crossing in $\hat \phi_{B_0}$, then $e\in E''$ must hold.
\end{proofof}

\subsection{Proof of Theorem \ref{thm:main_non_3}}
The proof of Theorem \ref{thm:main_non_3} follows the algorithm outlined above. We can assume that the input graph $G$ is $2$-connected, for the same reason as before, and compute its block decomposition $\bset=\bset(G)$. Given an input planarizing edge set $E'$ for $G$, for each pseudo-block $B\in \bset$, we compute a planarizing edge set $E'(\tB)$ for $\tB$ exactly as before, so that: 
$\sum_{B\in \bset}|E'(\tB)|\leq O(|E'|+\optcro(G))$. Since each graph $\tB$ is $3$-connected, we can compute a drawing $\phi_{\tB}$ of $\tB$ with at most   $f(n,\Delta)\cdot(\optcro(\tB)+|E'(\tB)|)$ crossings. Applying the algorithm from Lemma \ref{lem: gluing the drawings} to the pseudo-block $B=G$, we obtain a drawing of $G$ with total number of crossings bounded by:
\[ 
\begin{split}
O(\Delta^2)\cdot \sum_{B\in \bset}\cro(\phi_{\tB})& \leq O(\Delta^2)\cdot\sum_{B\in \bset}f(n,\Delta)\cdot(\optcro(\tB)+|E'(\tB)|)\\
&\leq O\left (\Delta^2 \cdot f(n,\Delta) (\optcro(G)+|E'|)\right ),
\end{split}
   \]
where we have used Lemma \ref{lem: few crossings in all blocks} in the last inequality.

\subsection{Proof of Theorem \ref{thm: reduction}}
Lastly, we extend our proof of Theorem  \ref{thm: reduction} from Section \ref{subsec: reduction} to graphs that may not be $3$-connected.

We first compute the block decomposition $\bset=\bset(G)$ of the input graph $G$, and denote $\tbset=\set{\tB\mid B\in \bset}$. Since each graph $\tB\in \tbset$ is $3$-connected, we can use the algorithm from Section \ref{subsec: reduction} to compute, for each graph $\tB\in \tbset$, an instance $(G_{\tB},\Sigma_{\tB})$ of the  \CNwRS problem, such that the number of edges in $G_{\tB}$ is at most $O\left(\optcro(\tB)\cdot \poly(\Delta\log n)\right)$, and   $\optcrors(G_{\tB},\Sigma_{\tB})\leq O\left(\optcro(\tB)\cdot \poly(\Delta\log n)\right )$.
From Lemma \ref{lem: few crossings in all blocks}, the total number of edges in all graphs in $\set{G_{\tB}}_{\tB\in \tbset}$ is at most $O\left(\optcro(G)\cdot \poly(\Delta\log n)\right )$, and $\sum_{\tB\in \tbset}\optcrors(G_{\tB},\Sigma_{\tB})\leq O\left(\optcro(G)\cdot \poly(\Delta\log n)\right )$. We obtain a final instance $(G',\Sigma)$ of the \CNwRS problem by letting $G'$ be the disjoint union of all graphs in 
$\set{G_{\tB}}_{\tB\in \tbset}$, and letting $\Sigma=\bigcup_{\tB\in \tbset}\Sigma_{\tB}$. 
From the above discussion, $|E(G')|\leq O\left(\optcro(G)\cdot \poly(\Delta\log n)\right)$, and, since solutions to all instances $(G_{\tB},\Sigma_{\tB})$ can be combined together to obtain a solution to instance $(G',\Sigma)$, we get that $\optcrors(G',\Sigma)\leq O\left(\optcro(G)\cdot \poly(\Delta\log n)\right )$. Assume now that we are given a solution to instance $(G',\Sigma)$ of \CNwRS of value $X$.
This solution immediately provides solutions $\phi_{\tB}$ to all instances in $\set{(G_{\tB},\Sigma_{\tB})}_{\tB\in \tbset}$, such that, if we 
denote by $X_{\tB}$ the value of the solution $\phi_{\tB}$, then $\sum_{\tB\in \tbset}X_{\tB}\leq X$. Using the algorithm from Section \ref{subsec: reduction}, we can compute, for each graph $\tB\in \tbset$, a drawing $\phi_{\tB}$ of $\tB$ with at most $O\left (X_{\tB}+\optcro(\tB)\right )\cdot \poly(\Delta\log n)$ crossings. By combining these drawings using Lemma \ref{lem: gluing the drawings}, we obtain a drawing of $G$, whose number of crossings is bounded by:
\[O(\Delta^2)\cdot \sum_{\tB\in \tbset}\left (X_{\tB}+\optcro(\tB)\right ) \cdot \poly(\Delta\log n)\leq O\left ((X+\optcro(G)\cdot \poly(\Delta\log n))\right ).  \]

\section{Crossing Number with Rotation System}
\label{sec:cr_rotation_system}

In this section we provide the proof of Theorem~\ref{thm: main_rot}.
Throughout this section, we allow graphs to have parallel edges, but no self-loops. A graph with no parallel edges will be explicitly referred to as a simple graph.
Recall that we are given as input an instance $(G,\Sigma)$ of the \CNwRS problem, where $G$ is a graph, and $\Sigma=\set{\oset_v}_{v\in V(G)}$ is a rotation system for $G$. 
Our goal is to compute a drawing of $G$ that respects the rotation system $\Sigma$, while minimizing the number of its crossings.
We let $\phi^*$ be some fixed optimal solution to input instance $(G,\Sigma)$.
%We are guaranteed that $|E(\tG)|\le o(\cro(\phi^*)^{1+\eps/2})$. 
%We will compute a drawing of $\tG$ respecting $\Sigma$ with relatively few crossings. 
Given a rotation system $\Sigma$ for graph $G$, we denote by $\Phi(G,\Sigma)$ the set of drawings of $G$ that respect the rotation system $\Sigma$. 
Throughout this section, we denote $\eac=\cro(\phi^*)+|E(G)|$, and fix the value of the constant $\eps$ from the statement of Theorem \ref{thm: main_rot} to be $\eps=1/20$.
We will design a randomized algorithm that, with high probability, computes a solution to instance $(G,\Sigma)$ of \CNwRS, of cost at most $O(\eac^{2-\eps}\cdot\poly(\log n))$. Over the course of the algorithm, we consider subinstances $(G',\Sigma')$ of instance $(G,\Sigma)$, but the parameter $\eac$ is always defined with respect to the original input instance $(G,\Sigma)$.

\subsection{High-Level Overview}

We start with a high-level overview of the algorithm. 
Our algorithm is recursive. Over the course of the algorithm, we consider subinstances $(G',\Sigma')$ of the input instance $(G,\Sigma)$ of \CNwRS.  Given such a subinstance $(G',\Sigma')$,  if maximum vertex degree in graph $G'$ is bounded by $\eac^{1-\eps}$, then we can use known techniques in order to compute a solution for instance $(G',\Sigma')$ of small enough cost. Otherwise, we compute two subinstances $(G_1,\Sigma_1)$ and $(G_2,\Sigma_2)$ of $(G',\Sigma')$ that we solve recursively; graphs $G_1,G_2$ are obtained from $G'$ by deleting some edges (and also possibly contracting some other edges). We then carefully combine the two resulting solutions together, in order to obtain a solution to instance $(G',\Sigma')$. The decomposition of graph $G'$ into graphs $G_1$ and $G_2$ is computed as follows. Let $v$ be any vertex of $G'$ whose degree in $G'$ is at least $\eac^{1-\eps}$.
%Now we assume the maximum vertex degree of $G$ is greater than $\eac^{1-\eps}$, and our goal is to decompose $G$ into subgraphs with maximum degree bounded by $\eac^{1-\eps}$, and then use the algorithm in~\cite{shahrokhi1994book} to compute a drawing of each of the subgraphs.
%Let $v\in V(G)$ be a vertex of $G$ with degree at least 
Let $\delta_{G'}(v)=\set{e_1,\ldots,e_r}$ be the set of edges incident to $v$ in $G'$, indexed according to their ordering in $\oset_v\in \Sigma'$.
Denote $E_1=\set{e_1,\ldots,e_{\floor{r/2}}}$ and $E_2=\set{e_{\floor{r/2}+1},\ldots,e_{r}}$. 
We first split the vertex $v$ by replacing it with two vertices, $v',v''$, where $v'$ is incident to all edges in $E_1$, and $v''$ is incident to all edges in $E_2$. Then we compute the minimum-cardinality subset $E'$ of edges, whose removal separates $v'$ from $v''$ in the resulting graph, and consider two cases. The first case is when $|E'|$ is sufficiently small. In this case, we let $G_1$ and $G_2$ be the two subgraphs that are obtained by removing the edges of $E'$ from $G'$. Their corresponding rotation systems $\Sigma_1$ and $\Sigma_2$ are defined in a natural way using the rotation system $\Sigma'$. We solve instances $(G_1,\Sigma_1)$ and $(G_2,\Sigma_2)$ separately, obtaining solutions $\phi_1$ and $\phi_2$, respectively. We ``glue'' these solution by unifying the images of vertices $v'$ and $v''$, and then carefully insert the images of the edges in $E'$, to ensure that we do not create too many crossings.

The second, and the more challenging case, is when $|E'|$ is large. In this case, from Menger's Theorem, there is a large set of edge-disjoint paths connecting $v'$ to $v''$.
We show a randomized efficient algorithm to compute a set $\qset^*$ containing $3$ paths connecting $v'$ to $v''$, that we call a \emph{skeleton}, such that the following hold. First, with high probability, the images of the paths  in $\qset^*$ in the optimal solution do not cross each other. Together with the ordering $\oset_v$ of the edges of $\delta_{G'}(v)$ around $v$, this gives us almost complete information about how these paths are drawn with respect to each other in the optimal drawing (the information is ``almost'' complete since  the images of some of these paths may cross themselves in the optimal drawing). Second, we can ensure that with high probability, the edges that belong to each of these paths participate in few crossings in the optimal drawing. Lastly, we can ensure that, for each path $P\in \qset^*$, there are few edges that are incident to the vertices of $P$ (this is an over-simplification and is only given here for the sake of intuition). 
We then use the skeleton to compute a decomposition of instance $(G',\Sigma')$ into the subinstances $(G_1,\Sigma_1),(G_2,\Sigma_2)$ that are solved recursively. We now proceed to describe the algorithm more formally.

\subsection{The Algorithm}
Recall that we are given as input a graph $G$ and a rotation system $\Sigma$ for $G$. We fix some optimal drawing $\phi^*\in \Phi(G,\Sigma)$, so $\cro(\phi^*)=\optcrors(G,\Sigma)$. We assume w.l.o.g. that every pair of edges crosses at most once in $\phi^*$. Our goal is to design a randomized algorithm that, with high probability, computes a drawing in $\Phi(G,\Sigma)$ with $O(\eac^{2-\eps}\cdot\poly(\log n))$ crossings, where $\eac=\optcrors(G,\Sigma)+|E(G)|$.
%In this section, for a graph $G$ and an integer $k>0$, we denote by $V_{k}(G)$ the set of all vertices of $G$ with degree at least $k$ in $G$.

The algorithm is recursive, and we ensure that the depth of the recursion will be bounded by $O(\eac^{2\eps})$. 
%Throughout the algorithm, we maintain a set $E^*$ of \emph{discarded edges}, that is initialized to be $\emptyset$. As the algorithm proceeds, we will gradually add edges to $E^*$. Let $\hat{E}^*$ be the final status of $E^*$. We will compute a drawing of $\tG'\setminus \hat E^*$, and add the edges in $\hat E^*$ back to the drawing, to eventually obtain a drawing of $\tG'$.
Over the course of the algorithm, we will consider instances $(G',\Sigma')$ of \CNwRS. For each such instance, graph $G'$ can be obtained from graph $G$ by first deleting some of its vertices and edges, and then contracting some edges. Therefore, every edge of $G'$ corresponds to some edge of $G$, and we treat them as the same edge. %Instances that lie in the same recursive level will be disjoint in their edges. 
We say that every such instance $(G',\Sigma')$ that our algorithm considers is a \emph{subinstance} of the input instance $(G,\Sigma)$.
Given a subinstance $(G',\Sigma')$, our algorithm will either solve it directly (this is the recursion base) or decompose it into two new subinstances $(G_1,\Sigma_1)$, $(G_2,\Sigma_2)$, such that $E(G_1)\cap E(G_2)=\emptyset$.

\paragraph{Difficult edges and special solutions.}
For every subinstance $(G',\Sigma')$, we will define a set $\hat E(G')\subseteq E(G')$ of edges of $G$, that we call \emph{difficult edges}. We will eventually ensure that the total number of difficult edges, over all instances that the algorithm considers, is small. Next, we define \emph{special solutions} for the instances $(G',\Sigma')$ that we consider.
%We ensure that, for each instance $G'$ that we construct, $|\hat{E}(G')|=\tilde O(\eac^{7\eps})$.
%If $G'$ is in the $i$th recursive level, then we will ensure that $|\hat E(G')|\leq i\eac^{1/3+\eps}$.

\begin{definition}[Special solutions]
Given a subinstance $(G',\Sigma')$ of $(G,\Sigma)$, and a set $\hat E(G')\subseteq E(G')$ of difficult edges for $G'$, we say that a solution  $\phi_{G'}\in \Phi(G',\Sigma)$ to instance $(G',\Sigma')$ of \CNwRS is a \emph{special solution} (or special drawing), if, for every crossing $(e,e')$ in $\phi_{G'}$, either the crossing $(e,e')$ is present in $\phi^*$, or at least one of the edges $e,e'$ belongs to the set $\hat E(G')$ of difficult edges for $G'$.
\end{definition}

Throughout the algorithm, we ensure that the following invariants hold for every instance $(G',\Sigma')$ that we consider, and its corresponding set $\hat E(G')$ of difficult edges:

\begin{properties}{I}
	\item $|E(G')|\geq \chi^{1-2\eps}/4$ \label{inv: many edges}
	(we will later show that this implies that the number of the recursive levels is bounded by $4\eac^{2\eps}$); 
	\item If $(G',\Sigma')$ lies in the $i$th level of the recursion, then $|\hat E(G')|\leq i\cdot \chi^{1-7\eps}$; in particular, since the number of the recursive levels is bounded by $4\chi^{2\eps}$,
	%\znote{needs to claim that in each recursion we separates the current graph into two subgraphs that are disjoint in edges}\mynote{I added this above, this is not a property of a subinstance so adding this as invariant is probably not the right place.}, 
	$|\hat E(G')|\leq 4\chi^{1-5\eps}$ always holds; and \label{inv: few difficult edges}
	\item There is a special solution to instance $(G',\Sigma')$. \label{inv: special solution}
\end{properties}

We emphasize that  the set  $\hat E(G')$ of difficult edges is defined based on the optimal solution $\phi^*$ to the input instance $(G,\Sigma)$. Therefore, neither the set $\hat E(G')$ nor the special solution $\phi_{G'}$ is known to the algorithm; they are only used for the sake of analysis.

The following two lemmas will be used by the algorithm; the proofs are deferred to Section \ref{subsec: proof of base of recursion} and \ref{subsec: proof of adding back edges} of Appendix, respectively.

\begin{lemma}
	\label{lem:base}
	There is an efficient algorithm, that, given an instance $(G,\Sigma)$ of \CNwRS, such that the maximum vertex degree in $G$ is at most $\Delta$, computes a solution to this instance of cost at most $O\left((\optcrors(G,\Sigma)+\Delta\cdot |E(G)|)\cdot\poly(\log n)\right)$.
\end{lemma}

\begin{lemma}
	\label{lem:add_back_discarded_edges new}
	There is an efficient algorithm, that, given an instance $(H,\Sigma)$ of the \CNwRS problem, a subset $E'\subseteq E(H)$ of edges of $H$, and a drawing $\phi$ of graph $H\setminus E'$ that respects $\Sigma$, computes a drawing $\phi'$ of $H$ that respects $\Sigma$, such that $\cro(\phi')\le \cro(\phi)+|E'|\cdot|E(H)|$.
\end{lemma}

Given a subinstance $(G',\Sigma')$ of $(G,\Sigma)$, we denote by $N(G')$ the total number of crossings $(e,e')$ in the drawing $\phi^*$ of $G$, in which at least one of the edges $e,e'$ lies in $E(G')$.
We obtain the following immediate corollary of Lemma \ref{lem:add_back_discarded_edges new}, that follows from the definition of special solutions.

\begin{corollary}\label{obs: few crossings special solutions}
Let $(G',\Sigma')$ be a subinstance of $(G,\Sigma)$ with a set $\hat E(G')$ of difficult edges, for which Invariant \ref{inv: special solution} holds. Then there is a solution to instance $(G',\Sigma')$ of \CNwRS of cost at most $N(G')+|\hat E(G')|\cdot |E(G')|$.
\end{corollary}

We now proceed to describe the algorithm. For the initial instance $(G,\Sigma)$ of \CNwRS, we let the set $\hat E(G)$ of difficult edges be $\emptyset$. Clearly, Invariants \ref{inv: many edges} -- \ref{inv: special solution} hold for this choice of $\hat E(G)$. 

We now assume that we are given some subinstance $(G',\Sigma')$ of $(G,\Sigma)$, with a set $\hat E(G')$ of difficult edges (we emphasize that the edge set $\hat E(G')$ is not know to the algorithm and is only used in the analysis), such that Invariants \ref{inv: many edges} -- \ref{inv: special solution} hold for this subinstance. If the maximum vertex degree in $G'$ is at most $\eac^{1-\eps}$, then we say that instance 
$(G',\Sigma')$ is a \emph{base instance}. Base instances serve as the base of the recursion, that we discuss next.
\subsection{Recursion Base}
We assume that we are given some subinstance $(G',\Sigma')$  of $(G,\Sigma)$, with associated set $\hat E(G')$ of difficult edges (that is not known to the algorithm), such that Invariants \ref{inv: many edges} -- \ref{inv: special solution} hold for this subinstance, and moreover, the maximum vertex degree in $G'$ is at most $\eac^{1-\eps}$.
Recall that, from Corollary \ref{obs: few crossings special solutions}, $\optcrors(G',\Sigma')\leq N(G')+|\hat E(G')|\cdot |E(G')|$.
We use the algorithm from Lemma \ref{lem:base} to compute a solution $\phi_{G'}$ to instance $(G',\Sigma')$ of \CNwRS, with the number of crossings bounded by:
\[
\begin{split}
&O\left((\optcrors(G',\Sigma')+\chi^{1-\eps}\cdot |E(G')|)\cdot\poly(\log n)\right)\\
&\quad\quad\quad\quad\quad\quad \leq O\left ((N(G')+ |\hat E(G')|\cdot |E(G')| + \chi^{1-\eps}\cdot |E(G')|)\cdot\poly(\log n)\right) \\
&\quad\quad\quad\quad\quad\quad \leq O\left((N(G')+ \chi^{1-\eps}\cdot |E(G')|)\cdot\poly(\log n)\right).
\end{split}
\]
Here we have used Invariant \ref{inv: few difficult edges} in order to bound $|\hat E(G')|$.

\subsection{Recursion Step}
%Let $\alpha=\cro^{1-10\eps}$. We apply Lemma~\ref{lem:deco_wrt_high-deg_nodes} to $G'$ with parameter $\alpha$, and let $G''_1,\dots,G''_k$ be the graphs that we obtain. We now describe how to process one graph of $\set{G''_1,\dots,G''_k}$, and will process all other graphs in the same way. Consider the graph $G''_1$ and for brevity we rename it by $G''$. Note that if $G''$ does not contain a vertex of degree at least $\cro^{1-\eps}$, then we are in the recursion base case.

\begin{figure}[h]
\centering
\subfigure[Images of edges in $\delta_{G'}(v)$ in the special solution for instance $(G',\Sigma')$.]{\scalebox{0.31}{\includegraphics{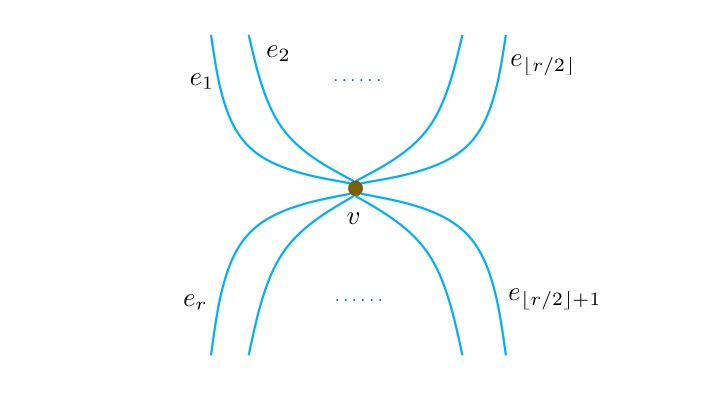}}}
\hspace{0.5cm}
\subfigure[Images of edges in $\delta_{G''}(v')$ and $\delta_{G''}(v'')$ in drawing $\hat \phi$ of $G''$, and the curve $\gamma^*$.]{
\scalebox{0.31}{\includegraphics{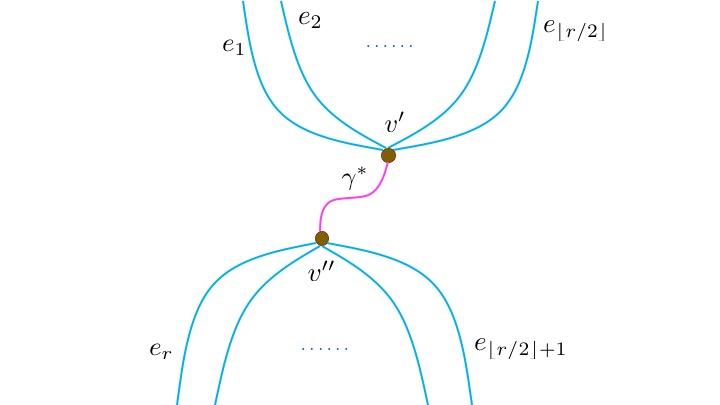}}}
\caption{Splitting the vertex $v$ in $G'$ into $v'$ and $v''$. \label{fig: split_v}}
\end{figure}

We now assume that we are given a sub-instance $(G',\Sigma')$ of the input instance $(G,\Sigma)$, together with associated set $\hat E(G')$ of difficult edges in $G'$ (that is not known to the algorithm), such that Invariants \ref{inv: many edges} -- \ref{inv: special solution} hold for this sub-instance, and, moreover, there is at least one vertex in $G'$ of degree at least $\eac^{1-\eps}$.
%We assume without loss of generality that graph $G'$ is connected, as otherwise we can solve the problem on each of its connected components separately.
%We apply Lemma~\ref{lem:determine_flip_status} to $G''$, and let $\tilde\Sigma$ be the rotation system on $G''$ that we get.
Let $v\in V(G')$ be any vertex of degree at least $\eac^{1-\eps}$, and let $\delta_{G'}(v)=\set{e_1,\ldots,e_r}$ be the set of all edges that are incident to $v$ in $G'$, indexed according to the ordering $\oset_v\in \Sigma'$. 
Denote $E_1=\set{e_1,\ldots,e_{\floor{r/2}}}$ and $E_2=\set{e_{\floor{r/2}+1},\ldots,e_{r}}$. 
Let $G''$ be a graph that is obtained from $G'$, by replacing the vertex $v$ with two new vertices, $v'$ and $v''$, such that $v'$ is incident to all edges in $E_1$, and $v''$ is incident to all edges in $E_2$. Notice that rotation system $\Sigma'$ for $G'$ naturally defines a rotation system $\Sigma''$ for $G''$. %Moreover, it is easy to verify that we can efficiently transform any feasible solution to instance $(G'',\Sigma'')$ of \CNwRS into a solution to instance $(G',\Sigma')$, without increasing the solution cost.  
 For convenience, we set $\hat E(G'')=\hat E(G')$. While instance $(G'',\Sigma'')$ is not a sub-instance of $G$ (that is, we may not be able to obtain graph $G''$ from graph $G'$ by only edge-deletion, vertex-deletion and edge-contraction operations), each edge of $G''$ naturally corresponds to an edge of $G$ and Invariants \ref{inv: many edges} -- \ref{inv: special solution} continue to hold for $(G'',\Sigma'')$ and $\hat E(G'')$.
Consider the special drawing of graph $G'$. We can obtain from it a special drawing of graph $G''$, by simply replacing the image of vertex $v$ with images of $v'$ and $v''$ in a natural way. We also define a simple curve $\gamma^*$,
%\znote{It seems that this curve $\gamma^*$ is never used?}\mynote{it is used in Step 3 to show that $R^*$ together with $\gamma^*$ decompose $G''$ into two sub-graphs}
that connects the images of $v'$ and $v''$ in this new drawing and, other than its endpoints, does not intersect the drawing of $G''$. Curve $\gamma^*$ enters vertex $v'$ between the edges $e_1$ and $e_{\floor {r/2}}$, and it enters the vertex $v''$ between the edges $e_{\floor{r/2}+1}$ and $e_r$; see Figure~\ref{fig: split_v} for an illustration. We denote the resulting drawing of $G''$ by $\hat \phi$. For every crossing $(e,e')$ in this drawing, we say that $(e,e')$ is a \emph{special crossing} iff at least one of $e,e'$ lies in the set $\hat E(G'')$ of difficult edges; otherwise we say that it is a \emph{regular crossing}.

Let $E'\subseteq E(G'')$ be a minimum-cardinality set of edges in $G''$, such that in $G''\setminus E'$, no path connects $v'$ to $v''$. We now consider two cases, depending on whether the cardinality of $E'$ is sufficiently small.

% Next, we compute the minimum edge-cut in $G'$ separating $v'$ from $v''$. Denote by $E'$ the set of edges in the cut. We distinguish between two cases, based on whether $|E'|$ is large or small.

\subsubsection{Case 1. $|E'|< \eac^{1-2\eps}$}

Consider the graph $G''\setminus E'$. Let $G_1$ be the unique connected component of this graph containing the vertex $v'$, and let $G_2=(G''\setminus E')\setminus G_1$, so $v''\in V(G_2)$. Notice that the rotation system $\Sigma''$ for $G''$ naturally induces rotation systems $\Sigma_1$ for $G_1$ and $\Sigma_2$ for $G_2$. 
We define the sets $\hat E(G_1),\hat E(G_2)$ of difficult edges in the natural way: $\hat E(G_1)=E(G_1)\cap \hat E(G')$ and $\hat E(G_2)=E(G_2)\cap \hat E(G')$. Since the special solution to instance $(G',\Sigma')$ naturally induces special solutions to instances $(G_1,\Sigma_1)$ and $(G_2,\Sigma_2)$, Invariant \ref{inv: special solution} holds for both of these instances. It is immediate to verify that Invariant \ref{inv: few difficult edges} also holds for both of these instances. In order to establish Invariant \ref{inv: many edges},
we use the following observation, that provides a slightly stronger bound, that will be useful for us later:

\begin{observation}
	\label{RS_obs:case_1}
	If Case 1 happens, then $|E(G_1)|, |E(G_2)|\ge \eac^{1-\eps}/3$.
\end{observation}
\begin{proof}
 Observe that $|E_1|,|E_2|\geq \eac^{1-\eps}/2-1$, while $|E'|\leq \eac^{1-2\eps}$. Since $|E(G_1)|\geq |E_1\setminus E'|$ and $|E(G_2)|\geq |E_2\setminus E'|$, the observation follows.
\end{proof}

We solve each of the two instances $(G_1,\Sigma_1)$ and $(G_2,\Sigma_2)$ recursively, obtaining solutions $\phi_{G_1}$ for the first instance and $\phi_{G_2}$ for the second instance.
We now show how to combine these solutions together in order to obtain a solution to instance $(G',\Sigma')$. Consider the drawing $\phi_{G_1}$ of graph $G_1$. Since this drawing is a feasible solution to the instance $(G_1,\Sigma_1)$, there must be a face $F$ in this drawing, whose boundary contains a segment of the image of edge $e_1$ that includes its endpoint $v'$, and a segment of the image of edge $e_{\floor{r/2}}$, that includes its endpoint $v'$. We denote this face by $F_1$, and we view it as the outer face of drawing $\phi_{G_1}$. Similarly, drawing $\phi_{G_2}$ contains a face, that we denote by $F_2$, whose boundary contains a segment  of the image of edge $e_{\floor{r/2}+1}$ that includes its endpoint $v''$, and a segment of the image of edge $e_{r}$ that includes its endpoint $v''$. We view $F_2$ as the outer face of the drawing $\phi_{G_2}$. We now superimpose the two drawings on the plane, such that (i) the drawing of $G_2$ is contained in face $F_1$ of the drawing of $G_1$; (ii) the drawing of $G_1$ is contained in face $F_2$ of the drawing of $G_2$; (iii) the images of vertices $v',v''$ coincide, and (iv) the circular order of the edges of $E_1\cup E_2$ entering the image of $v'$ (and also the image of $v''$) is consistent with the ordering $\oset_v\in \Sigma'$. Lastly, in order to complete the drawing of the graph $G'$, we need to insert the edges of $E'$ into the current drawing. We do so using the algorithm from Lemma \ref{lem:add_back_discarded_edges new}. As a result, we obtain a feasible solution $\phi_{G'}$ to instance $(G',\Sigma')$ of \CNwRS, such that the number of crossings in $\phi_{G'}$ is bounded by:
\[ \cro(\phi_{G_1})+ \cro(\phi_{G_2})+ |E'|\cdot |E(G')|\leq \cro(\phi_{G_1})+ \cro(\phi_{G_2})+ \eac^{1-2\eps}\cdot |E(G')| .\]
This finishes the algorithm and the analysis for Case 1. 

We now proceed to discuss the second and the more difficult case, where $|E'|\geq \eac^{1-2\eps}$.

\subsubsection{Case 2. $|E'|\geq \eac^{1-2\eps}$}
\label{subsec:case2}
We now focus on the case where $|E'|\geq \eac^{1-2\eps}$. From Menger's theorem, graph $G''$ contains a collection of at least $\eac^{1-2\eps}$ edge-disjoint paths connecting $v'$ to $v''$. The algorithm for Case 2 consists of three steps. In the first step, we define a set $\pset$ of at least $\eac^{1-2\eps}$ edge-disjoint paths connecting $v'$ to $v''$ in graph $G''$, that have some other additional useful properties. In the second step, we construct a skeleton for graph $G''$ by sub-sampling some of the paths in $\pset$. In the third step, we define two new sub-instances $(G_1,\Sigma_1)$ and $(G_2,\Sigma_2)$, together with the corresponding sets of difficult edges, that we solve recursively. We then combine the resulting solutions in order to obtain a solution to instance $(G',\Sigma')$.

\subsubsection*{Step 1. Computing the Path Set}
%\znote{here we use $\pset$, while at the end of this step we used $\pset'$ to denote the outcome. On the other hand, in the high-level overview we use $\pset$ to denote a set of $O(1)$ paths}\mynote{deleted $\pset$ from the title, and fixed overview}

We start with some definitions. 

\begin{figure}[h]
\centering
\subfigure[The curves $\gamma_1$ and $\gamma_2$ cross at point $p$.]{\scalebox{0.32}{\includegraphics{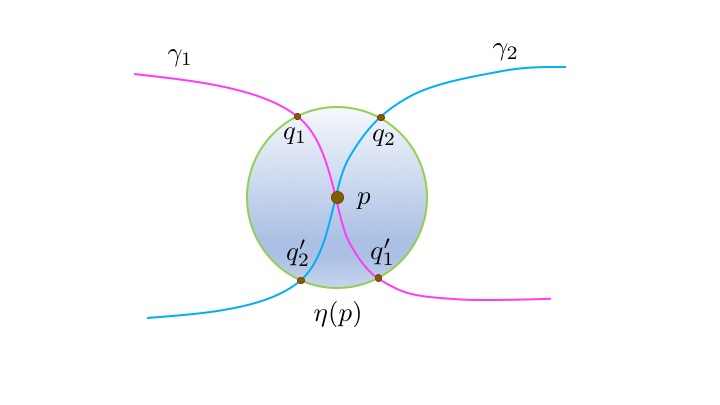}}}
\hspace{0.1cm}
\subfigure[The curves $\gamma_1$ and $\gamma_2$ touch at point $p$.]{
\scalebox{0.32}{\includegraphics{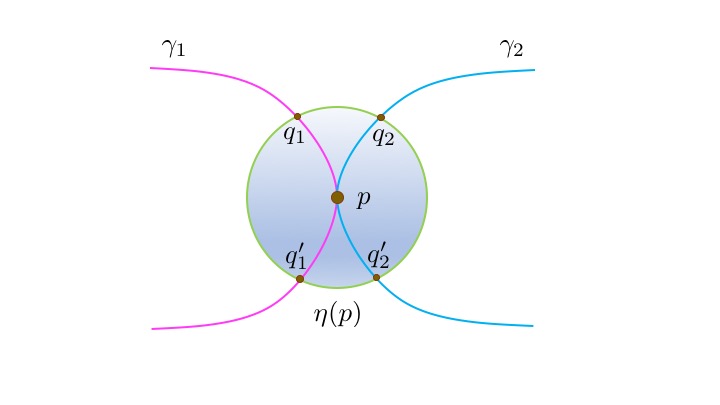}}}
\caption{An illustration of a pair of curves that touch or cross. \label{fig: curve_cross_touch}}
\end{figure}

Suppose we are given two curves $\gamma_1,\gamma_2$ in the plane, that intersect at a finite number of points, and a point $p$ that is an inner point for both $\gamma_1$ and $\gamma_2$. Consider a very small disc $\eta(p)$ around $p$, so that the intersection of each of the two curves with $\eta(p)$ is a simple contiguous curve. Let $q_1,q_1'$ be the two points of $\gamma_1$ that lie on the boundary of $\eta(p)$, and let $q_2,q_2'$ be defined similarly for $\gamma_2$. We say that $\gamma_1$ and $\gamma_2$ \emph{cross at point $p$} if the circular ordering of the points $q_1,q_1',q_2,q_2'$ along the boundary of $\eta(p)$ is either $(q_1,q_2,q_1',q_2')$ or $(q_1',q_2,q_1,q_2')$ (equivalently, if $\sigma$ is any curve connecting $q_1$ to $q_1'$, $\sigma'$ is any curve connecting $q_2$ to $q_2'$, and both curves are contained in $\eta(p)$, then the two curves must share a point). Otherwise, we say that $\gamma_1$ and $\gamma_2$ \emph{touch at point $p$}; see Figure~\ref{fig: curve_cross_touch} for an illustration.
We say that two curves $\gamma_1,\gamma_2$ \emph{cross} iff there is at least one point $p$ that is an inner point for both curves, such that $\gamma_1$ and $\gamma_2$ cross at $p$.

Assume now that we are given two edge-disjoint paths $P_1,P_2$ in graph $G''$, and some vertex $u$ that is an inner vertex of both $P_1$ and $P_2$. Denote by $e_1,e_1'$ the two edges of $P_1$ that are incident to $u$, and similarly denote by $e_2,e_2'$ the two edges of $P_2$ that are incident to $u$. We say that the intersection of the paths $P_1,P_2$ at vertex $u$ is \emph{non-transversal} iff the edges $e_1,e_1',e_2,e_2'$ appear in the ordering $\oset_u\in \Sigma''$ in one of the following two orders: either  (i) $(e_1,e_1',e_2,e_2')$ or (ii) $(e_1',e_1,e_2,e_2')$ (equivalently, in any drawing that respects the rotation system $\Sigma''$, the images of the two paths touch at $u$). We say that a set $\pset$ of edge-disjoint paths is \emph{non-transversal with respect to $\Sigma''$},
%\znote{maybe simply say $\pset$ is non-transversal w.r.t. $\Sigma''$? it seems to me that the word ``consistent'' is not very informative here}\mynote{sure, switch it to non-transversal, but it needs to be fixed in other places as well then} 
iff for every pair $P_1,P_2$ of distinct paths in $\pset$, for every vertex $u$ that is an inner vertex on both paths, the intersection of $P_1$ and $P_2$ at $u$ is non-transversal. The first step is summarized in the following lemma, whose proof appears in Appendix~\ref{apd:proof_of_initial path set}.

\begin{lemma}\label{lem: initial path set}
There is an efficient algorithm to compute a set $\pset$ of $\ceil{\chi^{1-2\eps}}$ paths in $G''$, such that all paths in $\pset$ are mutually edge-disjoint, each path connects $v'$ to $v''$, and the set $\pset$ of paths is non-transversal with respect to $\Sigma''$.
\end{lemma}
%\mynote{needs proof. Should be similar to proof of Lemma 7.1 where the process is formally defined}

%------------------------
%------------------------
%------------------------
%------------------------
%------------------------
\iffalse
Let $\pset$ be a set of paths. We denote $E(\pset)=\bigcup_{P\in \pset}E(P)$.
Let $G$ be a graph and let $\phi$ be a drawing of $G$. We say a pair $G_1,G_2$ of disjoint subgraphs \emph{cross} in $\phi$, iff there is an edge $e_1\in G_1$ and another edge $e_2\in G_2$, such that the image of $e_1$ intersects the image of $e_2$ in $\phi$.
We use the following lemma, whose proof is deferred to Appendix~\ref{apd:proof_of_finding_local_non_inf}.

\begin{lemma}
\label{cor:finding_local_non_inf}
Given a graph $G$, a rotation system $\Sigma$ on $G$ and a set $\tilde\pset$ of edge-disjoint paths connecting the same pair $v,v'$ of distinct vertices of $G$, we can efficiently find (i) a set $\pset$ of $|\pset|=|\tilde\pset|$ edge-disjoint paths in $G$ connecting $v$ to $v'$ such that $E(\pset)\subseteq E(\tilde\pset)$; and (ii) a cyclic order $\oset$ on the paths in $\pset$, such that:
\begin{enumerate}
\item for any drawing $\phi\in \Phi(G,\Sigma)$ and any two paths $P_1,P_2\in \pset$, if $P_1$ and $P_2$ cross in $\phi$, then either every path in $\oset[P_1,P_2]$ crosses $P_1\cup P_2$ in $\phi$, or every path in $\oset[P_2,P_1]$ crosses $P_1\cup P_2$ in $\phi$; and
\item if the image of some path $P\in \pset$ is a blocking curve, then $P$ crosses every other.
\end{enumerate}
\end{lemma}

\fi

 We assume without loss of generality that each path in $\pset$ is a simple path. We say that a path $P\in \pset$ is \emph{long} if it contains at least $10\eac^{2\eps}$ edges.
Since $|E(G'')|\leq \eac$, the number of long paths in $\pset$ is at most $\eac^{1-2\eps}/10$. We let $\pset'\subseteq \pset$ be the set of all short paths. Additionally, we discard from $\pset'$ all paths whose first edge $e_i$ (that is incident to $v'$) has index $1\leq i\leq \chi^{1-2\eps}/3$ or $\floor {r/2}-\chi^{1-2\eps}/3\leq i\leq \floor{r/2}$. Clearly, $|\pset'|\geq |\pset|/4\geq \eac^{1-2\eps}/4$ (recall that edges $e_1,\ldots,e_{\floor{r/2}}$ are the edges that are incident to vertex $v'$, and that $r\geq \chi^{1-2\eps}$). 

\subsubsection*{Step 2. Constructing the Skeleton}
We use the following two parameters: $h=200\ceil{\eac^{5\eps}}$ 
%\znote{$r$ is used to denote the number of edges incident to $v$: $\set{e_1,\ldots,e_r}$, probably we need to change the letter~}\mynote{can you replace $r$ with $h$ throughout? (either here or in the indexing of the edges)} 
and $q=\floor{\eac^{1-7\eps}/1000}$. Note that $|\pset'|\geq  \eac^{1-2\eps}/4\geq h q \geq \eac^{1-2\eps}/8$. 
We discard arbitrarily all but $hq$ paths from $\pset'$, obtaining a subset $\pset''\subseteq \pset'$ containing exactly $hq$ paths. Notice that the ordering $\oset_{v'}\in \Sigma''$ of the edges of $\delta_{G''}(v')$ naturally defines an ordering of the paths in $\pset''$, where the paths are ordered according to the ordering of the first edge on each path in $\oset_{v'}$. We denote this ordering by $\tilde \oset$.

We choose a path $P\in \pset''$ uniformly at random (this is the only randomized step in the algorithm). We then denote $\pset''=\set{P_1,P_2,\ldots,P_{hq}}$, where $P_1=P$, and the paths are indexed according to their ordering in $\tilde \oset$. Next, we partition the paths in $\pset''$ into $q$ groups $\uset_1,\ldots,\uset_q$, each of which contains exactly $h$ consecutive paths of $\pset''$, so for $1\leq i\leq q$, $\uset_i=\set{P_j\mid h(i-1)<j\leq hi}$. 
We denote $\gset=\set{\uset_1,\ldots,\uset_q}$ the collection of these groups.

%Before we define the skeleton, we analyze some properties of these groups.
For each $1\leq i\leq q$, we define three special paths for group $\uset_i$: 
we let $Q_L^i$ and $Q^i_R$ be the first and the last paths in group $\uset_i$, and we let $Q^i$ be the ``middle'' path in that group (that is, $Q^i$ is the $(h/2)$th path in group $\uset_i$). We denote $\qset^i=\set{Q^i_L,Q^i_R,Q^i}$, and 
we let $\qset^*=\qset^1$. We let the skeleton $K$ be the union of the paths in $\qset^*$.

Next, we define several bad events and show that, with a high enough probability, none of them happens. 
Recall that we denoted by $\hat \phi$ the  special drawing of $G''$, and that, for every crossing $(e,e')$ in this drawing, we say that $(e,e')$ is a \emph{special crossing} iff at least one of $e,e'$ lies in the set $\hat E(G'')$ of difficult edges, and otherwise we say that it is a \emph{regular crossing}. Clearly, the number of regular drawings in $\hat \phi$ is bounded by $\cro(\phi^*)\leq \eac$.

\paragraph{Type-1 Bad Groups.}
\begin{definition}
We say that a group $\uset_i$ is a \emph{type-$1$ bad group} iff there is some path $P\in \qset^i$, such that:
\begin{itemize}
	\item either there are at least $\chi^{5\eps}$ regular crossings in $\hat \phi$ in which the edges of $P$ participate; or
	\item some edge of $P$ belongs to the set $\hat E(G'')$ of difficult edges.
\end{itemize}
\end{definition}

\begin{claim}\label{claim: few type 1 bad groups}
For a fixed index $1\leq i\leq q$, the probability that group $\uset_i$ is a type-1 bad group is $O(\chi^{-3\eps})$.
\end{claim}
\begin{proof}
We say that a path $P\in \pset''$ is \emph{bad} if one of the following two happen: either (i) there are at least  $\chi^{5\eps}$ regular crossings in $\hat \phi$ in which the edges of $P$ participate, or (ii) some edge of $P$ belongs to the set $\hat E(G'')$ of difficult edges. 
Recall that, from Invariant \ref{inv: few difficult edges}, $|\hat E(G'')|\leq O(\chi^{1-5\eps})$, so at most $O(\chi^{1-5\eps})$ paths in $\pset''$ may contain a difficult edge.
Since the total number of regular crossings in $\hat\phi$ is at most $\chi$, there are at most $O(\chi^{1-5\eps})$ paths in $\pset''$ whose edges participate in at least $\chi^{5\eps}$ regular crossings in $\hat \phi$. Therefore, the total number of bad paths in $\pset''$ is bounded by $O(\chi^{1-5\eps})$.  The probability of any path to be chosen into $\qset^i$ is at most $O(1/(hq))\leq O(1/\eac^{1-2\eps})$. Therefore, the probability that a bad path lies in $\qset^i$ is at most $O(\chi^{1-5\eps}/\eac^{1-2\eps})\leq O(\chi^{-3\eps})$.
\end{proof}

%We say that bad event $\event_1$ happens if there is a path $P\in \qset^*$, such that there are at least $\chi^{5\eps}$ regular crossings in $\hat \phi$ in which the edges of $P$ participate.

%\paragraph{Bad Event $\event_2$.}
%We say that bad event $\event_2$ happens if at least one path in $\qset^*$ contains a difficult edge in $\hat E(G'')$. Since, from Invariant \ref{inv: few difficult edges} there are at most $\chi^{1-7\eps}$ difficult edges in $\hat E(G'')$, at most $\chi^{1-5\eps}$ paths in $\pset''$ may contain a difficult edge. Since the probability of a path to be chosen into $\qset$ is at most $24/\eac^{1-2\eps}$, we get that the probability that a path containing a difficult edge is chosen into $\qset^*$ is at most $O(1/\chi^{3\eps})$. 

\paragraph{Type-2 Bad Groups.}
\begin{definition}
We say that a group $\uset_i$ is a \emph{type-2 bad group} iff the paths in $\uset_i$ contain in total at least $50\chi^{5\eps}$ difficult edges.
\end{definition}

\begin{claim}\label{claim: few type 2 bad groups}
For a fixed index $1\leq i\leq q$, the probability that group $\uset_i$ is a type-2 bad group is $O(\chi^{-3\eps})$.
\end{claim}
\begin{proof}
From Invariant \ref{inv: few difficult edges}, there are at most $4\chi^{1-5\eps}$ difficult edges in $\hat E(G'')$. Therefore, the total number of bad groups in $\gset$ is bounded by $\chi^{1-10\eps}/10$. Since $\uset_i$ is essentially a random group from $\gset$, and since the total number of groups $q=\Omega(\chi^{1-7\eps})$, the probability that $\uset_i$ is a type-2 bad group is at most $O(\chi^{1-10\eps}/\chi^{1-7\eps})\leq O(\chi^{-3\eps})$.
 \end{proof}

We let $\gset'\subseteq \gset$ contain all groups $\uset_i$ that is not a type-1 bad group or a type-2 bad group. 
From Claim \ref{claim: few type 1 bad groups} and Claim \ref{claim: few type 2 bad groups}, there is some constant $c$, such that $\expect{|\gset'|}\geq (1-\frac{c}{\chi^{3\eps}})\cdot q$.

\paragraph{Bad Event $\event_1$.}
We let $\event_1$ be the bad event that $|\gset'|<\frac{qc}{2\chi^{3\eps}}$.

\begin{claim}\label{claim: first bad event}
$\prob{\event_1}\leq \frac {2c}{\chi^{3\eps}}$.
\end{claim}
\begin{proof}
Assume otherwise. Then:

\[\expect{|\gset'|}\leq \prob{\event_1}\cdot\frac{cq}{2\chi^{3\eps}}+\left (1-\frac {2c}{\chi^{3\eps}}\right )\cdot q \le\frac{cq}{2\chi^{3\eps}}+\left (1-\frac {2c}{\chi^{3\eps}}\right )\cdot q<\left (1-\frac{c}{ \chi^{3\eps}}\right )\cdot q,\]
a contradiction.
\end{proof}

We need the following claim.
\begin{claim}\label{claim: no crossings in skeleton}
Assume that $|\gset'|>1$. Then for each group $\uset^i\in \gset'$ and for every pair $P,P'$ of paths in $\qset^i$, the images of the paths $P$ and $P'$ in drawing $\hat \phi$ of $G''$ do not cross (but they may touch at their shared vertices). Moreover, for every pair $\uset_i,\uset_j$ of distinct and non-consecutive groups in $\gset'$, and for every pair $P\in \qset^i$, $P'\in \qset^j$ of paths, the images of $P$ and $P'$ in drawing $\hat \phi$ of $G''$ do not cross.
\end{claim}
%We notice that the above claim ensure that images of edges in $E(P)$ and in $E(P')$ do not cross each other in $\hat \phi$. However, paths $P$ and $P'$ may still share vertices, so their images are not entirely disjoint. However, since we have ensured that all paths in $\pset$ are non-transversal with respect to $\Sigma''$, the images of the paths in $\qset^*$ may ``touch'' each other but they cannot cross each other.
\begin{proof}
We start with the first claim. Let $\uset^i\in \gset'$ be a group, and let $P,P'\in \uset^i$ be a pair of its paths. Assume first that $P=Q_L^i$ and $P'=Q^i$. Since the paths in $\pset$ are non-transversal with respect to $\Sigma''$, and since the drawing $\hat \phi$ respects the rotation system $\Sigma''$, if $u$ is a vertex that is shared by $P$ and $P'$, then their images $\hat{\phi}(P)$ and $\hat{\phi}(P')$ do not cross at $u$. Therefore, if $\hat{\phi}(P)$ and $\hat{\phi}(P')$ cross, there must be a pair of edges $e\in E(P)$, $e'\in E(P')$, whose images cross in $\hat \phi$. Consider the curve $\hat{\phi}(P)$, that we view as being directed from $v'$ to $v''$. Let $p$ be the first point on this curve that belongs to $\hat{\phi}(P')$ and is not the image of a vertex. Let $\gamma$ be the segment of $\hat{\phi}(P)$ from $v'$ to $p$; we delete all loops from $\gamma$ so that it becomes a simple curve connecting $v'$ to $p$. We let $\gamma'$ be the segment of $\hat \phi(P')$ from $v'$ to $p$, and we delete all loops from $\gamma'$ similarly. Therefore, curves $\gamma$ and $\gamma'$ both originate at $\hat\phi (v')$ and terminate at $p$. They may contain other common points, but they cannot cross at those points. For every vertex $u\neq v'$ whose image is contained in both $\gamma$ and $\gamma'$, we consider a small disc $\eta(u)$ around $u$, and we modify the curves $\gamma$ and $\gamma'$ so that they are disjoint from the interior of $\eta(u)$, and instead follow its boundary on either side (see Figure~\ref{fig: curve_separation}). 
Let $\hat \gamma$ and $\hat \gamma'$ denote the resulting two curves. Then the union of $\hat \gamma$ and $\hat \gamma'$ defines a closed simple curve that we denote by $\lambda$. Curve $\lambda$ partitions the plane into two faces, $F$ and $F'$. We assume w.l.o.g. that vertex $v''$ lies on the interior of face $F'$. Note that, from our definition of the paths in $\qset^i$, there are at least $h/2-2\geq 99\chi^{5\eps}$ paths in $\uset^i$ that lie between $P$ and $P'$ in the circular ordering $\tilde \oset$, and there are at least $99\chi^{5\eps}$ other paths that lie between $P'$ and $P$ (we think of the orientation of the ordering as being fixed). Therefore, there is a set $\hat \pset\subseteq \pset''$ of at least $99\chi^{5\eps}$ paths $\hat P$, such that $\hat \phi(\hat P)$ intersects the interior of the face $F$. Since vertex $v''$ lies in the interior of face $F'$, the image of every path in $\hat P$ needs to intersect the curve $\lambda$. Since $\uset_i$ is not a type-2 bad group, there is a subset $\hat \pset'\subseteq \hat \pset$ of at least $49\chi^{5\eps}$ paths that contain no difficult edges.  Note also that, since the edges of $E(P)\cup E(P')$ participate in at most $2\eac^{5\eps}$ regular crossings in $\hat \phi$ in total (this is since $\uset_i$ is not a type-1 bad group), at most $2\chi^{5\eps}$ paths in $\hat \pset'$ may contain edges whose images cross images of edges of $E(P)\cup E(P')$. Therefore, there is at least one path $P''\in \hat \pset$, such that $\hat{\phi}(P'')$ crosses $\lambda$ and $P''$ does not contain an edge whose image crosses the images of edges in $E(P)\cup E(P')$. Therefore, the only way for the image of $P''$ to cross the curve $\lambda$ is through some disc $\eta(u)$ of some vertex $u\in V(P)\cap V(P')$. In particular, $u\in V(P'')$ must hold as well. However, since we have defined the paths in $\pset$ to be non-transversal with respect to the rotation system $\Sigma''$, and since the drawing $\hat \phi$ of $G''$ must respect the rotation system $\Sigma''$, this is impossible.

\begin{figure}[h]
\centering
\subfigure[The curves $\gamma$ and $\gamma'$ touch at the image of $u$ in $\hat \phi$.]{\scalebox{0.32}{\includegraphics{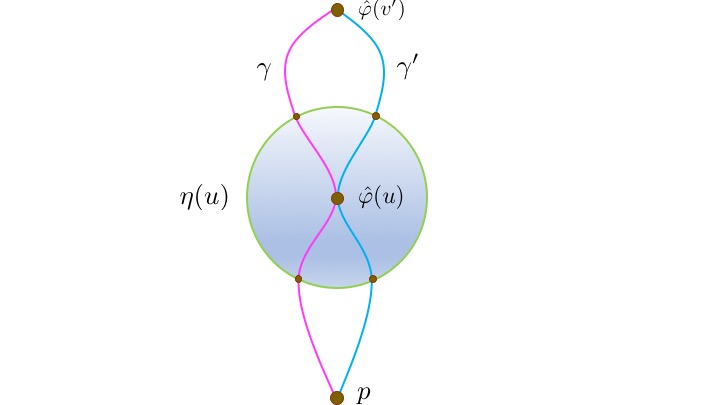}}}
\hspace{0.1cm}
\subfigure[The curves $\hat \gamma$ and $\hat \gamma'$ share only endpoints $p$ and $\hat \phi(v')$.]{
\scalebox{0.32}{\includegraphics{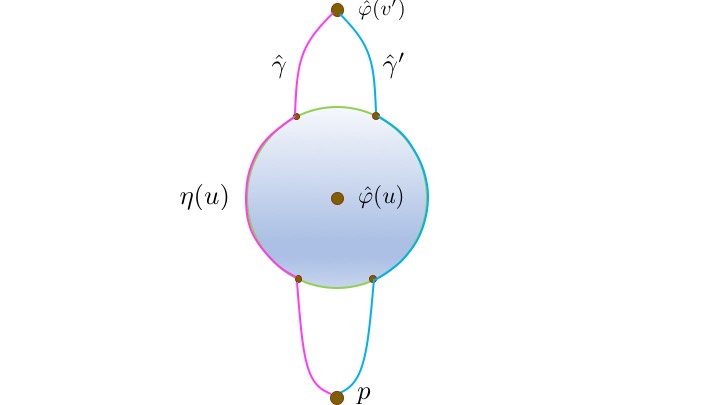}}}
\caption{Illustration for the proof of Claim \ref{claim: no crossings in skeleton}. \label{fig: curve_separation}}
\end{figure}

The proof for the case where $\set{P,P'}=\set{Q^i,Q^i_R}$ is symmetric, and the proof for the case where $\set{P,P'}=\set{Q^i_L,Q^i_R}$ follows from these two cases.
The proof for the case where $P\in \qset^i$, $P'\in \qset^j$ for some pair $\uset_i,\uset_j\in \gset'$ of distinct and non-consecutive groups of $\gset'$ is almost identical (we may need to use a group of $\gset'$ that lies between these two groups in order to define the set $\hat \pset$ of paths). %Assuming that $|\gset'|>1$, we can use the same proof for any other pair $P,P'\in \qset^i$ of paths, for any group $\uset^i\in \gset'$.
\end{proof}

Let $\gset''\subseteq \gset'$ be the set of groups that is obtained by discarding every other group from $\gset'$ (in other words, no two groups of $\gset''$ are consecutive in $\gset'$). Consider the drawing $\hat \phi$ of $G''$ and some group $\uset^i\in \gset''$. We define the \emph{region $R^i$ of the sphere associated with $\uset^i$} as follows. We let $\lambda$ be the union of two curves: curve $\sigma$ whose endpoints are $v'$ and $v''$, that is contained in the image of path $Q^i_L$ in $\hat \phi$,  and curve $\sigma'$ whose endpoints are $v'$ and $v''$, that is contained in the image of path $Q^i_R$ in $\hat \phi$. We choose the two curves $\sigma$ and $\sigma'$ in such a way that a disc whose boundary is $\lambda$ contains the drawing of $Q^i_R\cup Q^i_L$ in $\hat \phi$. %\znote{probably the definition of the disc needs to be modified, please see email. maybe change it to "the disc that contains the drawing of $Q^i$" in $\hat{\phi}$?}\mynote{I don't see what's wrong with the definition. It correctly says that $\lambda$ is chosen in such a way that the disc contains the drawing of $Q^i$, and so this is what we do.} 
Notice that the boundary of this disc is not necessarily a simple curve, since $\sigma$ and $\sigma'$ may touch, and curves $\sigma$ and $\sigma'$ may be non-simple. We let $R^i$ denote the resulting discs.

Let $\rset=\set{R^i\mid \uset_i\in \gset''}$. Note that, from Claim \ref{claim: no crossings in skeleton}, every pair $R^i,R^j$ of distinct discs in $\rset$ are disjoint except for possibly sharing points on their boundaries. We let $R^*=R^1$ be the region associated with group $\uset^1$. We now define several additional bad events, and show that each of them only happens with low probability. 

\paragraph{Bad Event $\event_2$.}
We say that bad event $\event_2$ happens iff $\uset_1$ is either a type-1 or a type-2 bad group. From Claims \ref{claim: few type 1 bad groups} and \ref{claim: few type 2 bad groups}, the probability of event $\event_2$ is bounded by $O(\eac^{-3\eps})$. Notice that, if $\event_2$ does not happen, then $\uset^1\in \gset'$. We can then also assume that $\uset^1\in \gset''$.

\paragraph{Bad Event $\event_3$.}
We say that an edge $e\in E(G'')$ is \emph{interfering} if the intersection of the image of $e$ in $\hat \phi$ with the interior of the region $R^*$ is non-empty. We denote the set of all interfering edges by $\eint$. We say that bad event $\event_3$ happens if  $|\eint\setminus \hat E(G'')|\geq \chi^{1-7\eps}$.

\begin{claim}\label{claim: bound third bad event}
	$\prob{\event_3}\leq O(\eac^{-3\eps})$.
\end{claim}
\begin{proof}
Note that $\prob{\event_3}\leq \prob{\event_3\mid \neg(\event_1\vee \event_2)}+\prob{\event_1\vee \event_2}$. 
Recall that we have shown that $\prob{\event_1\vee \event_2}\leq O(\chi^{-3\eps})$ as proved above. 
Therefore, it suffices to show that $\prob{\event_3\mid \neg(\event_1\vee \event_2)}\leq O(\chi^{-3\eps})$.
	
Assume now that neither of the events $\event_1$, $\event_2$ happen. Notice that an edge $e\in E(G'')$ may be interfering in one of the two cases: either the image of $e$ crosses the image of some edge of $Q^1_L\cup Q^1_R$; or the image of $e$ is contained in the interior of region $R^*$. Since we have assumed that event $\event_1$ did not happen, group $\uset_1$ is not a type-1 bad group. Therefore, the total number of crossings $(e,e')$ with $e\not\in \hat E(G'')$ and $e'\in E(Q^1_L\cup Q^1_R)$ is bounded by $O(\chi^{5\eps})<\chi^{1-7\eps}/2$, as $\eps= 1/20$. It now remains to show that, if events $\event_1$, $\event_2$  do not happen, then, with probability at least $1-O(\chi^{-3\eps})$, the total number of edges $e$ whose image in $\hat \phi$ is contained in region $R^*$ is at most $\chi^{1-7\eps}/2$. Since we have assumed that the event $\event_2$ did not happen, $|\gset''|\geq \Omega(q/\chi^{3\eps})=\Omega(\chi^{1-10\eps})$. 
We say that a group $\uset^i\subseteq \gset''$ is suspicious if there are at least $\chi^{1-7\eps}/2$ edges $e\in E(G'')$, such that the image of $e$ in $\hat \phi$ is contained in the region $R^i$. Since $|E(G'')|\leq \chi$, the number of suspicious regions in $\gset''$ is at most $2\chi^{7\eps}$. Since $R^*$ is essentially a random region from $\set{R^i\mid \uset_i\in \gset''}$, the probability that $\uset^1$ is suspicious is at most $O(\chi^{7\eps}/\chi^{1-10\eps})\leq O(\chi^{-3\eps})$, as $\eps= 1/20$.
\end{proof}

%From now on we will only focus on group $\uset_1$. We denote $Q^1_L,Q^1_R,Q^1$ by $Q_L,Q_R,Q$, respectively, and we denote $\qset^1$ by $\qset^*$.

\paragraph{Bad Event $\event_4$.}
For all $1\leq i\leq q$, let $Z^i=\left (V(Q^i_L)\cup V(Q^i_R)\right )\setminus\set{v',v''}$. We partition the set $Z^i$ of vertices into two subsets: set $Z_1^i$ contains all vertices that appear on both $Q^i_L$ and $Q^i_R$; we refer to vertices of $Z^i_1$ as vertices that are \emph{common} for group $\uset_i$; and set $Z^i_2$ that contains all other vertices, that we refer to as vertices that are \emph{uncommon} for group $\uset_i$. 

\begin{claim}\label{claim: uncommon vertices}
For every vertex $u\in V(G'')$, there are at most two groups $\uset_i,\uset_j\in \gset''$ such that $u\in Z_2^i$ and $u\in Z_2^j$.
\end{claim}
\begin{proof}
It is sufficient to show that (i) there is at most one group $\uset_i$ in $\gset''$, such that $u\in V(Q^i_L)$ but $u\not\in V(Q^i_R)$; and (ii) there is at most one group $\uset_j$ in $\gset''$, such that $u\in V(Q^j_R)$ but $u\not\in V(Q^j_L)$. We prove the former; the proof of the latter is symmetric.
	
Assume for contradiction that there are two distinct groups $\uset_i,\uset_j\in \gset''$, such that $u\in V(Q^i_L)\setminus V(Q^i_R)$ and $u\in V(Q^j_L)\setminus V(Q^j_R)$. Consider the drawing $\hat \phi$ of $G''$, and delete all edges and vertices from it, except for the edges and vertices of $Q^i_R\cup Q^j_R$. Let $\fset$ be the set of faces of the resulting drawing. Then there is some face $F$ that contains the image of $u$ in its interior. Observe however that the images of every pair of the four paths $Q^j_L,Q^j_R,Q^i_L,Q^i_R$ cannot cross, and can only touch. Moreover, they appear in the ordering $\tilde \oset$ in this order (assuming that $j<i$). Therefore, the removal of the images of $Q^i_R$ and $Q^j_R$ from the drawing $\hat \phi$ separates the images of $Q^i_L$ and $Q^j_L$. It is then impossible that both these paths contain the vertex $u$.
\end{proof}

We say that bad event $\event_4$ happens if the total number of edges of $G''$ that are incident to vertices of $Z_2^1$ (the uncommon vertices for group $\uset^1$) is at least $\chi^{1-4\eps}$.

\begin{claim}\label{claim: event 4}
	$\prob{\event_4}\leq O(\chi^{-3\eps})$.
\end{claim}
\begin{proof}
Assume that events $\event_1,\event_2$ did not happen. 
We say that a group $\uset^i\in \gset''$ is problematic if the number of edges of $G''$ that are incident to vertices of $Z^i_2$ is at least $\chi^{1-4\eps}$. Notice that, from Claim \ref{claim: uncommon vertices}, each edge may be incident to a vertex of $Z^i_2$ for at most four groups $\uset_i\in \gset''$. Since $|E(G'')|\leq \chi$, at most $O(\chi^{4\eps})$ groups of $\gset''$ may be problematic. 
Since we assumed that event $\event_1$ did not happen, $|\gset''|\geq \Omega(q/\chi^{3\eps})\geq \Omega(\chi^{1-10\eps})$. 
Since group $\uset^1$ is a random group of $\gset''$, the probability that $\uset^1$ is problematic is at most $O(\chi^{4\eps}/\chi^{1-10\eps})\le O(\chi^{-3\eps})$, as $\eps= 1/20$. 
We conclude that $\prob{\event_4\mid \neg(\event_1\vee \event_2)}\leq O(\chi^{-3\eps})$. Altogether,
$\prob{\event_4}\leq \prob{\event_4\mid \neg(\event_1\vee \event_2)}+\prob{\event_1\vee \event_2} \leq O(\chi^{-3\eps})$.
\end{proof}

We let $\event$ be the bad event that any of the bad events  $\event_1,\event_2,\event_3$ or $\event_4$ happened. From the above discussion, $\prob{\event}\leq O(\chi^{-3\eps})$.

From now on we will only focus on group $\uset_1$. We denote $Q^1_L,Q^1_R,Q^1$ by $Q_L,Q_R,Q$, respectively, and we denote $\qset^1$ by $\qset^*$.
We also denote the sets $Z^1_1,Z^1_2$ of common and uncommon vertices for group $\uset^1$ by $Z_1$ and $Z_2$, respectively. Recall that the region $R^1$ associated with $\uset_1$ is denoted by $R^*$. We now summarize the facts that we need to use in Step 3. Assuming that event $\event$ does not happen:

\begin{properties}{F}
	\item at most $50\chi^{5\eps}$ paths in $\uset_1$ may contain difficult edges; \label{fact: few difficult edges in U1}
	\item paths in $\qset^*$ do not contain difficult edges;\label{fact: no difficult edges in skeleton}
	\item there are at most $3\chi^{5\eps}$ regular crossings in $\hat \phi$ in which the edges of the paths in $\qset^*$ participate; \label{fact: few crossings in skeleton}
	\item the total number of interfering edges (edges whose image intersects the interior of the region $R^*$ associated with group $\uset^1$) that are non-difficult edges is $|\eint\setminus \hat E(G'')|\leq \chi^{1-7\eps}$;  \label{fact: few interfering edges}
	\item the total number of edges that are incident to the vertices of $Z_2$ (the uncommon vertices for $\uset_1$) is bounded by $\chi^{1-4\eps}$; and \label{fact: few edges incident to non-common vertices}
	\item each path in $\qset^*$ contains at most $10\chi^{2\eps}$ edges. \label{fact: short paths}
\end{properties}

\subsubsection*{Step 3. Completing the Algorithm}
\label{subsec:step_3}

Recall that we have denoted the set of edges that are incident to vertex $v'$ by $\set{e_1,\ldots,e_{\floor{r/2}}}$, where the edges are indexed according to their ordering in $\oset_{v'}\in \Sigma''$. Recall that we have ensured that no path in $\pset'$ contains any edge $e_i$ with index $1\leq i\leq \chi^{1-2\eps}/3$ or $\floor {r/2}-\chi^{1-2\eps}/3\leq i\leq \floor{r/2}$. We view the paths in $\qset^*$ as directed from $v'$ to $v''$.
Let $e_{i_L}$ be the first edge on path $Q_L$, and define $e_{i_R}$ similarly for $Q_R$. We denote $E_L=\set{e_1,\ldots,e_{i_L-1}}$ and $E_R=\set{e_{i_R+1},\ldots,e_{\floor{r/2}}}$. Intuitively, sets $E_L$ and $E_R$ contain edges that are incident to $v'$, and lie on the left and on the right side of the region $R^*$ in the drawing $\hat \phi$ of $G''$, respectively. From the above discussion, $|E_L|,|E_R|\geq  \chi^{1-2\eps}/3$.
Let $e_{j_L}$ be the last edge on path $Q_L$, and define $e_{j_R}$ similarly for $Q_R$. We denote $E'_L=\set{e_{j_L},\ldots,e_{r}}$ and $E'_R=\set{e_{\floor{r/2}+1},\ldots,e_{j_R-1}}$. As before, sets $E'_L$ and $E'_R$ contain edges that are incident to $v''$, and lie on the left and the right side of the region $R^*$ in the  drawing $\hat \phi$ of $G''$, respectively.

It will be convenient for us to slightly modify the graph $G''$, by splitting the vertex $v'$ into three vertices: vertex $v'_L$, that is incident to the edges in $E_L$, vertex $v'_R$ that is incident to all edges in $E_R$, and vertex $v'_M$ that is incident to all remaining edges. We split vertex $v''$ into three vertices $v''_L,v''_R$ and $v''_M$ similarly. We denote the resulting graph by $G^*$. We modify the drawing of $\hat \phi$ in a natural way to obtain a drawing of $G^*$, and we keep the curve $\gamma^*$, that now connects vertices $v'_M$ and $v''_M$, is disjoint from the current drawing except for its endpoints, enters vertex $v'_M$ between $e_{i_L}$ and $e_{i_R}$, and enters vertex $v''_M$ between $e_{j_L}$ and $e_{j_R}$. %\snote{should the previous sentence be changed so that $v'_M$ is used instead of $v'_L$ and  $v''_M$ be used instead of $v''_L$?}.
		
%We also denote by $\tilde E$ the set of all edges that interfere with the region $R^*$ (edges whose image in $\hat \phi$ intersects the interior of $R^*$), that are not difficult edges. 

In order to provide some intuition, we first consider a simpler special case, where the set $Z_1$ of common vertices for group $\uset_1$ is empty. In other words, the only vertices that paths $Q_L$ and $Q_R$ share are $v'_M$ and $v''_M$.

\subsubsection*{Special Case: $Z_1=\emptyset$.}

Consider the drawing $\hat \phi$ of $G^*$. Delete from this drawing the images of all edges in $\eint$, and the images of the edges of $E(Q_L)\cup E(Q_R)$. Since we have assumed that paths $Q_L,Q_R$ do not share vertices except for $v'_M$ and $v''_M$, using the definition of the curve $\gamma^*$, it is easy to verify that no path connects a vertex of  $\set{v'_L,v''_L}$ to a vertex of $\set{v'_R,v''_R}$  in the resulting graph. 

Intuitively, we would like to define $G_1$ to be the union of the connected components of the resulting graph that contain $v'_L$ and $v''_L$, and $G_2$ to contain the remainder of the resulting graph. We could then solve the problems on $G_1$ and $G_2$ recursively, and combine their solutions as in Case 1. A problem with this approach is that the set $\eint$ of interfering edges is unknown to us. However, from the above discussion, we do know that there is a relatively small cut separating vertices of $\set{v'_L,v''_L}$ from vertices of $\set{v'_R,v''_R}$ in $G^*$ if $\event$ did not happen. Our algorithm computes such a cut, and then recursively solves the resulting two instances, like in Case 1.

Let $E'_1$ be the set of all edges incident to vertices $v'_M$ and $v''_M$. Observe that all such edges lie in set $\eint$, and from Fact \ref{fact: few interfering edges} and Invariant \ref{inv: few difficult edges}, $|E'_1|\leq |\eint|\leq |\eint\setminus\hat E(G'')|+|\hat E(G'')|\leq 5\chi^{1-5\eps}$.
We let $E'_2$ be a minimum-cardinality set of edges, whose removal from $G^*\setminus E'_1$ separates vertices of $\set{v'_L,v''_L}$ from vertices of $\set{v'_R,v''_R}$ in $G^*$. Finally, let $E'=E'_1\cup E'_2$. The following observation is immediate from the above discussion:

\begin{observation}\label{obs: small cut}
If bad event $\event$ did not happen, then $|E'|\leq 15\chi^{1-5\eps}$.
\end{observation}

\begin{proof}
From the above discussion, the edge set $\eint\cup E(Q_L)\cup E(Q_R)$ separates $v'_L$ from $v'_R$ in $G^*$, and $E'_1\subseteq \eint$. From Invariant \ref{inv: few difficult edges}, $|\hat E(G'')|\leq 4\chi^{1-5\eps}$. If $\event$ does not happen, then, from Fact \ref{fact: few interfering edges}, $|\eint\setminus \hat E(G'')|\leq \chi^{1-7\eps}$, and, since paths $Q_L$, $Q_R$ are short, $|E(Q_L)|,|E(Q_R)|\leq 10\chi^{2\eps}$. Overall, we get that $|E'_2|\leq 10\chi^{1-5\eps}$, so $|E'|\leq |E'_1|+|E'_2|\leq 15\chi^{1-5\eps}$.
\end{proof}

Next, we proceed like in Case 1. We define two new instances $(G_1,\Sigma_1)$, $(G_2,\Sigma_2)$ as follows. We let $G_1$ be a sub-graph of $G^*$ that contains the connected components of $G^*\setminus E'$ containing $v'_L$ and $v''_L$, and let $G_2$ be obtained from $(G^*\setminus E')\setminus G_1$ by discarding isolated vertices from it. Notice that $v'_M,v''_M$ are isolated vertices in $G^*\setminus E'$, since every edge incident to either vertex must lie in the set $\eint$ of interfering edges. Therefore, it is easy to verify that $G_1,G_2\subseteq G''$, and both graphs are disjoint in their edges. We further modify $G_1$ by unifying vertices $v'_L$ and $v''_L$ (we refer to the new vertex as $v_L$), and we 
modify $G_2$ by unifying vertices $v'_R$ and $v''_R$, calling the new vertex $v_R$. Therefore, $G_1,G_2\subseteq G'$ now holds. The rotation systems $\Sigma_1$, $\Sigma_2$ of the graphs $G_1$ and $G_2$ are defined in a natural way using the rotation system $\Sigma''$. We also define the sets of difficult edges in $G_1$ and $G_2$ as $\hat E(G_1)=\hat E(G'')\cap E(G_1)$ and $\hat E(G_2)=\hat E(G'')\cap E(G_2)$.

We now verify that all invariants hold for the resulting two instances. 
First, recall that $|E_L|,|E_R|\geq  \chi^{1-2\eps}/3$, while, from Observation \ref{obs: small cut}, if bad event $\event$ did not happen, then $|E'|\leq 15\chi^{1-5\eps}$. Therefore, if bad event $\event$ did not happen, then $|E(G_1)|\geq |E_L\setminus E'|\geq \chi^{1-2\eps}/4$, and similarly $|E(G_2)|\geq \chi^{1-2\eps}/4$. Therefore, Invariant \ref{inv: many edges} holds for both instances. 
It is also immediate to verify that Invariant \ref{inv: few difficult edges} holds for both instances. Lastly, it remains to show that both resulting instances have special solutions. We show this for instance $(G_1,\Sigma_1)$, the proof for the other instance is symmetric.  
Since $G_1\subseteq G'$, and $\hat E(G_1)=\hat E(G'')\cap E(G_1)$, we can obtain a special solution for instance $(G_1,\Sigma_1)$, by starting from a special solution for instance $(G',\Sigma')$, and erasing from it all vertices and edges of $G'$ that do not lie in $G_1$. We conclude that  Invariant \ref{inv: special solution} holds for both instances.

We solve each of the two instances $(G_1,\Sigma_1)$ and $(G_2,\Sigma_2)$ recursively, obtaining solutions $\phi_{G_1}$ for the first instance and $\phi_{G_2}$ for the second instance.
We now show how to combine these solutions together in order to obtain a solution to instance $(G',\Sigma')$.
The algorithm for combining the two instances is almost identical to the one used in Case 1.
 Consider the drawing $\phi_{G_1}$ of graph $G_1$. Since this drawing is a feasible solution to the instance $(G_1,\Sigma_1)$, there must be a face $F$ in this drawing, whose boundary contains a segment of the image of edge $e_{i_L-1}$ that includes its endpoint $v_L$, and a segment of the image of edge $e_{j_L}$, that includes its endpoint $v_L$. We denote this face by $F_1$, and we view it as the outer face of drawing $\phi_{G_1}$. Similarly, drawing $\phi_{G_2}$ contains a face, that we denote by $F_2$, whose boundary contains a segment  of the image of edge $e_{i_R+1}$ that includes its endpoint $v_R$, and a segment of the image of edge $e_{j_R-1}$ that includes its endpoint $v_R$.  We view $F_2$ as the outer face of the drawing $\phi_{G_2}$. We now superimpose the two drawings on the plane, such that (i) the drawing of $G_2$ is contained in face $F_1$ of the drawing of $G_1$; (ii) the drawing of $G_1$ is contained in face $F_2$ of the drawing of $G_2$; (iii) the images of vertices $v_L,v_R$ coincide, and (iv) the circular order of the edges of $E_L\cup E'_L\cup E_R\cup E'_R$ entering the image of $v_L$ (and also the image of $v_R$) is consistent with the ordering $\oset_v\in \Sigma'$. Lastly, in order to complete the drawing of the graph $G'$, we need to insert the edges of $E'$ into the current drawing. We do so using the algorithm from Lemma \ref{lem:add_back_discarded_edges new}. As a result, we obtain a feasible solution $\phi_{G'}$ to instance $(G',\Sigma')$ of \CNwRS, such that the number of crossings in $\phi_{G'}$ is bounded by:
\[ \cro(\phi_{G_1})+ \cro(\phi_{G_2})+ |E'|\cdot |E(G')|.\]

If event $\event$ did not happen, then $|E'|\leq 15\eac^{1-5\eps}$, and the number of crossings is bounded by:

\[ \cro(\phi_{G_1})+ \cro(\phi_{G_2})+ 15\eac^{1-5\eps}\cdot |E(G')| .\]

This completes the algorithm for the special case where $Z_1=\emptyset$. We now proceed to discuss the general case.

\subsubsection*{The General Case.}

We start with some intuition for the general case. Suppose that the set $Z_1$ of common vertices is non-empty. Notice that now deleting the set $\eint\cup \hat E(G'')\cup E(Q_L)\cup E(Q_R)$ of edges from graph $G^*$ may no longer separate vertices in set $\set{v'_L,v''_L}$ from vertices in set $\set{v'_R,v''_R}$ and so it is not immediately clear how to define the two new sub-instances $(G_1,\Sigma_1),(G_2,\Sigma_2)$. In order to overcome this difficulty, we carefully \emph{split} every vertex $u\in Z_1$, by creating two copies of this vertex. The set of edges incident to $u$ is split between the two copies, so each edge is incident to at most one copy. Intuitively, we would like one copy of $u$ to be incident to all edges of $\delta_{G''}(u)$, whose images lie ``to the left'' of the region $R^*$, and the other copy to be incident to edges whose images lie ``to the right'' of the region $R^*$ in $\hat \phi$. We define these notions formally below. Once we split the vertices of $Z_1$, we set up a minimum-cut problem in the resulting graph (somewhat similar to the special case we discussed above), and use this cut to define the two initial graphs $G_1',G_2'$. Notice, however, that, if we obtain solutions $\phi_1,\phi_2$ to instances associated with graphs $G_1',G_2'$, then we need to be able to ``glue'' the copies of all vertices in $Z_1$ to each other in order to obtain a drawing of $G''$. In order to do this, we need to ensure that the copies of the vertices in $Z_1$ appear on the boundary of a single face in both drawings, and moreover, the order in which they appear on the boundaries of these faces is identical in both drawings. In order to ensure this, we modify the graphs $G_1',G_2'$, by unifying the copies of all vertices in $Z_1\cup \set{v',v''}$ in both graphs, and carefully define orderings of edges incident to this new vertex in each of the two instances. This is done in a way that ensures that any feasible solutions to the resulting two instances can be composed together in order to obtain a feasible solution to instance $(G',\Sigma')$. We now provide a formal description of the algorithm. The main hurdle in the algorithm for the general case is handling the common vertices in $Z_1$. We start with some definitions and observations that allow us to do so.

\paragraph{Dealing with Common Vertices.}
Consider the drawing $\hat \phi$ of the graph $G''$. We denote by $\zeta$ the image of the path $Q\in \qset^*$ in $\hat \phi$. Notice that curve $\zeta$ may cross itself numerous times. Let $\zeta'$ be a simple curve, that is obtained from $\zeta$ by deleting all loops. The following claim is crucial for dealing with common vertices. 

\begin{claim}\label{claim: common vertex not on loop}
If event $\event$ does not happen, then the image of every common vertex of $Z_1$ appears on the curve $\zeta'$ in the drawing $\hat \phi$.
\end{claim}

\begin{proof}
Assume that this is not the case. Then there is some loop $\sigma$ on the curve $\zeta$ that contains the image of some vertex $u\in Z_1$. If we denote by $p$ the unique point of $\sigma$ that lies on $\zeta'$, %\snote{$\zeta'$},
then $u$ is not mapped to $p$ in $\hat \phi$. By further decomposing $\sigma$ into smaller loops, we can find a simple closed curve $\sigma'\subseteq \sigma$, such that $u$ lies on the boundary of $\sigma'$. Let $D$ be a disc, whose boundary is $\sigma'$, such that the image of $v'$ does not lie in $D$ in $\hat \phi$. 
	
	 Recall that vertex $u$ lies on both paths $Q_L$ and $Q_R$. Recall also that in $\uset_1$, there is a set $\sset_1\subseteq \uset_1$ of $99\chi^{5\eps}$ paths that lie between $Q_L$ and $Q$ (excluding $Q$ and including $Q_L$), and similarly, there is a set $\sset_2\subseteq \uset_2$ of $99\chi^{5\eps}$ paths that lie between $Q$ and $Q_R$ (excluding $Q$ and including $Q_R$). If $\event$ does not happen, then from Properties \ref{fact: few difficult edges in U1}, \ref{fact: no difficult edges in skeleton} and \ref{fact: few crossings in skeleton}, at most $50\chi^{5\eps}$ paths in $\uset_1$ may contain difficult edges, paths in $\qset^*$ do not contain difficult edges, and there are at most $3\chi^{5\eps}$ regular crossings in $\hat \phi$ in which the edges of the paths in $\qset^*$ participate. Therefore, there are at most $53\chi^{5\eps}$ paths $P\in \uset_1$, such that the image of $P$ may cross the images of any of the paths $Q,Q_L,Q_R$ in $\hat \phi$ (note that, since the paths in $\pset'$ are non-transversal with respect to $\Sigma''$, and drawing $\hat \phi$ respects $\Sigma''$, each such crossing of images of two paths must involve a crossing of images of two edges). Therefore, there is a subset $\sset'_1\subseteq \sset_1$ of at least $25\chi^{5\eps}$ paths, and a subset  $\sset'_2\subseteq \sset_2$ of at least $25\chi^{5\eps}$ paths, such that for every path $P\in \sset'_1\cup \sset'_2$, the image of $P$ does not cross the images of any of the paths $Q,Q_L,Q_R$ in $\hat \phi$. It is then easy to see that the image of $P$ must be contained in the region $R^*$, and moreover, since $u\in Q_L\cap Q_R$, vertex $u$ must lie on $P$.

	 Let $e,e'$ be the two edges on path $Q$ that are incident to $u$. Consider the ordering $\oset_u\in \Sigma''$ of the edges that are incident to $u$. Removing the edges $e,e'$ from this ordering splits the set $\delta_{G''}(u)\setminus \set{e,e'}$ of edges into two subsets: set $\delta^1$ of edges lying between $e$ and $e'$, and set $\delta^2$ of edges lying between $e'$ and $e$ in $\oset_u$. 
	 
	 Let $\delta_L\subseteq \delta(u)$ be the set of all edges that are incident to $u$ and lie on the paths in $\sset'_1$. Define $\delta_R$ similarly for path set $\sset'_2$. Since the paths in $\pset'$ %\snote{$\pset'$?} 
	 are non-transversal with respect to $\oset_u$, and since the images of the paths in $\sset_1'\cup \sset_2'$ may not cross the image of $Q$, we are guaranteed that either $\delta_L\subseteq \delta^1$ and $\delta_R\subseteq \delta^2$, or the other way around. We assume without loss of generality that it is the former.
	 
	 Depending on the orientation of the vertex $u$ in $\hat \phi$, it then must be the case that for at least one of the sets $\sset_1',\sset_2'$ of paths, the image of every path in the set must have a non-empty intersection with the interior of the disc $D$. We assume without loss of generality that it is $\sset'_1$. Therefore, for each path $P\in \sset'_1$, its image contains the image of the vertex $v'$, that lies outside disc $D$, and it contains a point that lies strictly inside disc $D$. However, the image of path $P$ may not cross the image of path $Q$, a contradiction.	
\end{proof}

Denote the vertices in $Z_1$ by $u_1,u_2,\ldots,u_{|Z_1|}$, where the vertices are indexed in the order of their appearance on path $Q$. It is then immediate to see that the images of the vertices $u_1,u_2,\ldots,u_{|Z_1|}$ appear in this order on the curve $\zeta'$.

\begin{figure}[h]
\centering
\subfigure[Paths $Q_L$, $Q_R$ and edge sets at vertices $u,u'\in Z_1$.]{\scalebox{0.32}{\includegraphics{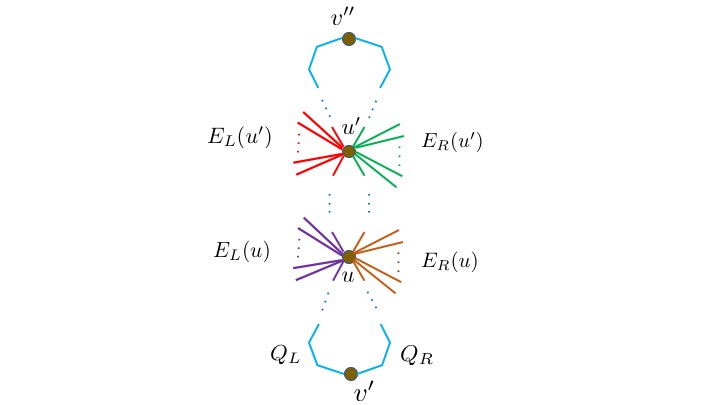}}}
%\hspace{0.1cm}
\subfigure[Zoom in at vertex $u$ in $(a)$.]{
\scalebox{0.32}{\includegraphics{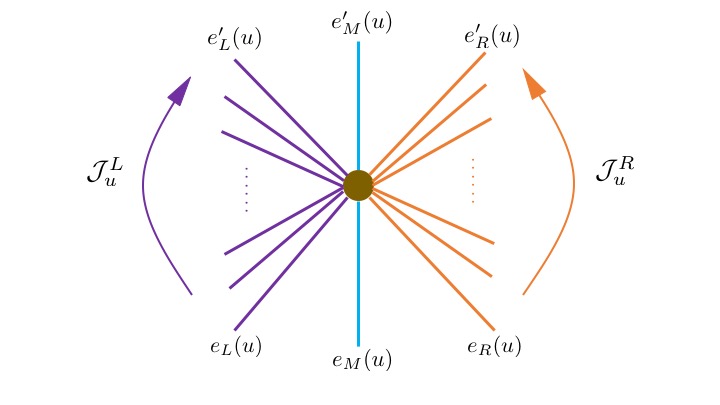}}}
\caption{Edge subsets for vertices in $Z_1$ and their orderings.  %\snote{in the image on the right, $\jset_u^L$ and $\jset_u^R$ should be reversed.}.
\label{fig: LR_path}}
\end{figure}

Consider now some common vertex $u\in Z_1$. We view the paths $Q_L$, $Q$, and $Q_R$ as directed from $v'$ towards $v''$. We denote by $e_L(u)$ the first edge of $Q_L$ that is incident to $u$, and by $e_L'(u)$ the second such edge. We define the edges $e_R(u),e'_R(u)$ for path $Q_R$, and $e_M(u),e'_M(u)$ for path $Q$ similarly. Since all paths in $\pset$ are non-transversal with respect to the ordering $\oset_u\in \Sigma''$, the circular ordering of these edges in $\oset_u$ is: $(e_M(u),e_L(u),e'_L(u),e'_M(u),e'_R(u),e_R(u))$. We denote by $E_L(u)$ the set of all edges that lie between $e_L(u)$ and $e'_L(u)$ (including these two edges). We define an ordering $\jset^L_u$ of the edges in $E_L(u)$ to be consistent with $\oset_u$, with edge $e_L(u)$ appearing first, and edge $e'_L(u)$ appearing last in this ordering (we note that $\jset^L_u$ is no longer a circular ordering). We denote by $E_R(u)$ the set of all edges in $\delta_{G''}(u)$ that lie between $e'_R(u)$ and $e_R(u)$ in $\oset_u$ (including these two edges). We define an ordering $\jset^R_u$ of the edges in $E_R(u)$ to be consistent with $\oset_u$, with edge $e_R(u)$ appearing first, and edge $e'_R(u)$ appearing last in this ordering;
 see Figure~\ref{fig: LR_path} for an illustration. %\mynote{figure would be helpful}

Recall that we have defined sets $E_L,E_R,E'_L,E'_R$ of edges that are incident to vertices $v'_L,v'_R,v''_L$ and $v''_R$,  respectively. We let $\jset_L$ be the ordering of the edges of $E_L$, that is consistent with the ordering $\oset_{v'}\in \Sigma'$, such that edge $e_1$ appears first in this ordering. We define ordering $\jset'_L$ for edges in $E'_L$ similarly, where edge $e_r$ appears last in the ordering. We similarly define orderings $\jset_R$ of edges in $E_R$, where edge $e_{\floor{r/2}}$ appears last in the ordering, and $\jset'_R$ of edges in $E'_R$, where edge $e_{\floor{r/2}+1}$ appears first in the ordering.

Lastly, we set $E^*_L=E_L\cup E'_L\cup \left(\bigcup_{u\in Z_1}E_L(u)\right )$, and similarly we set $E^*_R=E_R\cup E'_R\cup \left(\bigcup_{u\in Z_1}E_R(u)\right )$. We define an ordering $\jset^*_L$ of the edges in $E^*_L$ as follows. We place the edges in the sets $E_L,E_L(u_1)$, $E_L(u_2),\ldots$, $E_L(u_{|Z_1|})$, $E'_L$ in this order, where the ordering of the edges within each set is identical to the ones we just defined: edges of $E_L$ and $E'_L$ are ordered according to the orderings $\jset_L$ and $\jset'_L$, respectively, and for each $1\leq i\leq |Z_1|$, edges of $E_L(u_i)$ are ordered according to $\jset^L(u_i)$. 
We define an ordering $\jset_R$ of the edges in $E^*_R$ similarly. The following observation follows from our discussion so far:

\begin{observation}
	Suppose we follow the curve $\zeta'$ from $v'$ to $v''$, immediately to the right of the curve, in drawing $\hat \phi$ of $G''$. Then we encounter the edges of $E^*_L$ in the order $\jset^*_L$. Similarly, if we follow the curve $\zeta'$ from $v'$ to $v''$ immediately to the left of the curve, then we encounter the edges of $E^*_R$ in the order $\jset^*_R$.
\end{observation}

\paragraph{Constructing the Two Subinstances.}
We define the set $E'_1$ of edges that we delete from $G^*$, as follows. Set $E'_1$ contains, for every vertex $u\in Z_2$, all edges incident to $u$. 
Recall that, if $\event$ does not happen, then $|E_1'|\leq \chi^{1-4\eps}$, from Fact \ref{fact: few edges incident to non-common vertices}.
Let $E'_2$ be a set of edges from $G^*$ that contains, for every vertex $u_i\in Z_1$, all edges that are incident to $u_i$, except for the edges of $E_L(u_i)\cup E_R(u_i)$. Notice that every edge in $E'_2$ is an interfering edge, that lies in $\eint$. From Fact \ref{fact: few interfering edges}, if $\event$ did not happen, then $|\eint|\leq \chi^{1-7\eps}+|\hat E(G'')|\leq 5\chi^{1-5\eps}$ (from Invariant \ref{inv: few difficult edges}).

Next, we define a graph $\hat G^*$, as follows. Initially, we let $\hat G^*=G^*\setminus (E'_1\cup E'_2)$. Next, we consider every vertex $u_i\in Z_1$ one-by-one. When vertex $u_i$ is considered, we split it into two vertices: vertex $u_i'$ that is incident to all edges in $E_L(u_i)$, and vertex $u_i''$, that is incident to all edges in $E_R(u_i)$. Once all vertices in $Z_1$ are processed, we obtain the final graph $\hat G^*$. We denote $S_L=\set{u'_i\mid u_i\in Z_1}\cup \set{v'_L,v''_L}$, and we denote $S_R=\set{u''_i\mid u_i\in Z_1}\cup \set{v'_R,v''_R}$.

Drawing $\hat \phi$ of $G''$ naturally defines a drawing of the resulting graph $\hat G^*$, that we denote by $\hat \phi^*$. It is easy to verify, using this drawing, that the removal of all interfering edges from $\hat G^*$, together with the edges of $E(Q_L)\cup E(Q_R)$, disconnects vertices of $S_L$ from vertices of $S_R$ in $\hat G^*$. 

Next, we let $E'_3$ be a minimum-cardinality set of edges, whose removal from $\hat G^*$ disconnects vertices of $S_L$ from vertices of $S_R$. We need the following observation:

\begin{observation}\label{obs: small cut 2}
If bad event $\event$ did not happen, then $|E'_3|\leq 5\chi^{1-5\eps}$.
\end{observation}

\begin{proof}
From the above discussion, the edge set $ \eint\cup E(Q_L)\cup E(Q_R)$ separates $v'_L$ from $v'_R$ in $G^*$. From Invariant \ref{inv: few difficult edges}, $|\hat E(G'')|\leq 4\chi^{1-5\eps}$. If $\event$ does not happen, then $|\eint\setminus \hat E(G'')|\leq \chi^{1-7\eps}$, and, since paths $Q_L$, $Q_R$ are short, $|E(Q_L)|,|E(Q_R)|\leq 10\chi^{2\eps}$. Overall, we get that $|E'_3|\leq 5\chi^{1-5\eps}$.
\end{proof}

Let $E'=E_1'\cup E_2'\cup E_3'$; we refer to the edges of $E'$ as \emph{deleted edges}. From the above discussion, if $\event$ did not happen, then $|E'|\leq 2\chi^{1-4\eps}$.
We let $G'_1\subseteq \hat G^*$ be the union of all connected components that contain vertices of $S^L$, and we let $G'_2=(\hat G^*\setminus E')\setminus G'_1$. We obtain instance $(G_1,\Sigma_1)$ as follows. Graph $G_1$ is obtained from graph $G_1'$, by contracting all vertices in $S_L$ into a single vertex that we denote it by $v^*_L$. Notice that the set of edges incident to $v^*_L$ in $G_1$ is $E^*_L\setminus E'$. The ordering $\oset_{v^*_L}$ of edges incident to $v^*_L$ is a circular ordering obtained from $\jset^*_L$. For every other vertex $u\in V(G_1)\setminus \set{v^*_L}$, vertex $u$ lies in $G''$, and the set of edges incident to $u$ in $G_1$ is a subset of $\delta_{G''}(u)$. We define the ordering $\oset_u$ of the edges in $\delta_{G_1}(u)$ to be consistent with the same ordering in $\Sigma''$. This defines the rotation system $\Sigma_1$ for graph $G_1$. We define instance $(G_2,\Sigma_2)$ similarly. We also define the sets of difficult edges: $\hat E(G_1)=(\hat E(G'')\cup \eint)\cap E(G_1)$, and $\hat E(G_2)=(\hat E(G'') \cup \eint)\cap E(G_2)$.

Next, we verify that all invariants hold for the resulting two instances.
First, recall that $|E_L|,|E_R|\geq  \chi^{1-2\eps}/3$, while, if bad event $\event$ did not happen, then $|E'|\leq 2\chi^{1-4\eps}$. Therefore, if bad event $\event$ did not happen, then $|E(G_1)|\geq |E_L\setminus E'|\geq \chi^{1-2\eps}/4$, and similarly $|E(G_2)|\geq \chi^{1-2\eps}/4$. Therefore, Invariant \ref{inv: many edges} holds for both instances.

If event $\event$ did not happen, then, from Fact \ref{fact: few interfering edges}, $|\eint\setminus \hat E(G'')|\leq \chi^{1-7\eps}$. Therefore, $|\hat E(G_1)|,|\hat E(G_2)|\leq \chi^{1-7\eps}+|\hat E (G')|$, and Invariant \ref{inv: few difficult edges} holds for both instances. Lastly, it remains to show that both resulting instances have special solutions. We show this for instance $(G_1,\Sigma_1)$, and the proof for the other instance is symmetric.  

We start from the drawing $\hat \phi^*$ of graph $\hat G^*$, that was obtained from the special solution $\hat \phi$ for instance $(G'',\Sigma'')$ by deleting the edges in $E'$ from it, and splitting the vertices of $Z_1\cup \set{v',v''}$. We delete from this drawing all edges except those in the set $E(G_1)\setminus \hat E(G_1)$. Notice that the only crossings in the current drawing are regular crossings of $\hat \phi$ -- that is, crossings that are present in the optimal drawing $\phi^*$ of $G$. In order to obtain a drawing of graph $G_1$, we need to contract all vertices in $S_L$ into a single vertex $v^*_L$. We show that we can do so without introducing any new crossings. Recall that we have defined a simple curve $\zeta'$ that is contained in the image of the path $Q$ in $\hat \phi$, and the images of the vertices $v',u_1,\ldots,u_{Z_1},v''$ appear on this curve in $\hat \phi$ in this order. Recall that the only edges whose images cross curve $\zeta'$ are edges of $\eint$. We slightly shift the curve $\zeta'$ to the left, so it contains the images of vertices $v'_L,u'_1,u'_2,\ldots,u'_{Z_1},v''_L$ in this order. Notice that no edge crosses this new curve $\zeta'$ in the current drawing, as we have deleted all edges of $\eint$ from it. Therefore, we can contract the vertices of $S_L$ into a single vertex without introducing any new crossings.

\paragraph{Combining the Solutions.}

We solve instances $(G_1,\Sigma_1),(G_2,\Sigma_2)$ recursively, obtaining solutions $\phi_{G_1}$ to the first instance and $\phi_{G_2}$ to the second instance. We now show how to combine these solutions in order to obtain a solution $\phi_{G'}$ to instance $(G',\Sigma')$.

Consider the drawing $\phi_{G_1}$ of graph $G_1$. Let $\eta(v^*_L)$ be a small disc around vertex $v^*_L$ in this drawing. For every edge $e\in E^*_L$, we let $p_e$ be the unique point lying in the intersection of the image of $e$ and the boundary of $\eta(v^*_L)$. Notice that we are guaranteed that the circular ordering of the points in $\set{p_e\mid e\in E^*_L}$ on the boundary of $\eta(v^*_L)$ is precisely $\oset_{v^*_L}$ -- a circular ordering obtained from $\jset^*_L$. We discard, for every edge $e\in E^*_L$, the part of the image of $e$ that lies inside disc $\eta(v^*_L)$. Next, we place a smaller disc $\eta'(v^*_L)$ inside $\eta(v^*_L)$, such that the boundaries of the two discs are disjoint, and the image of $v^*_L$ lies in the interior of  $\eta(v^*_L)$. We place images of vertices $v_L,u_1',u_2',\ldots,u_{|Z_1|}'$ on the boundary of  $\eta(v^*_L)$ in this order. We then extend the images of the edges $e\in E^*_L$ inside $\eta(v^*_L)\setminus \eta'(v^*_L)$, so that the image of each edge now terminates at the image of its endpoint. From the definition of the ordering $\jset^*_L$, we can do so without introducing any new crossings. 
We process the drawing $\phi_{G_2}$ of graph $G_2$ in exactly the same way. 

Consider now the sphere, and a simple closed curve $\sigma$ on the sphere, that splits it into two discs, $D$ and $D'$. We place the drawing that we have obtained from $\phi_{G_1}$ inside disc $D$, so that the boundary of $\eta'(v^*_L)$ coincides with the boundary of $D$. We also place the drawing that we have obtained from $\phi_{G_2}$ inside disc $D'$, so that the boundary of $\eta'(v^*_R)$ coincides with the boundary of $D'$, and, moreover, for all $1\leq i\leq |Z_1'|$, the images of vertices $u_i',u_i''$ coincide, and the images of vertices $v_L,v_R$ coincide as well (vertex $v_L$ represents the merger of vertices $v'_R,v''_R$ in $G_R$). So far we have not introduced any new crossings. It now remains to add the images of the edges in $E'$ to the resulting drawing. We do so using Lemma \ref{lem:add_back_discarded_edges new}, obtaining a drawing $\phi_{G'}$ of $G'$, with the number of crossings bounded by:

\[ \cro(\phi_{G_1})+\cro(\phi_{G_2})+|E'|\cdot |E(G')|. \]

Assuming that $\event$ did not happen, then $|E'|\leq 2\chi^{1-4\eps}$, and so the number of crossings is bounded by:

\[ \cro(\phi_{G_1})+\cro(\phi_{G_2})+2\chi^{1-4\eps}\cdot |E(G')|. \]

\subsection{Final Accounting}
\label{subsec:accounting}

Let $\phi_G$ be a solution to the instance $(G,\Sigma)$ of \CNwRS that our algorithm computes. We now bound the number of crossings in $\phi_G$. We start by summarizing the properties of our recursive algorithm. 
Let $(G',\Sigma')$ be an instance that was considered by the algorithm. Recall that we have denoted by $N(G')$ the total number of crossings in which the edges of $G'$ participate in the optimal solution $\phi^*$ to instance $(G,\Sigma)$.

\begin{itemize}
\item If $(G',\Sigma')$ is a base instance, then our algorithm is guaranteed to produce a solution to this instance with the number of crossings bounded by  $O\left(\left (N(G')+\chi^{1-\eps}\cdot |E(G')|\right )\poly(\log n)\right )$. We say that instance $(G',\Sigma')$ \emph{contributes} $\Upsilon(G')=O\left(\left (N(G')+\chi^{1-\eps}\cdot |E(G')|\right )\poly(\log n)\right )$ crossings.

%the current instance $G'$ is in the recursion base case, that is, there is no vertex in $G'$ with degree above $\eac^{1-\eps}$, and we can efficiently compute a drawing $\psi_{G'}\in \Phi(G',\Sigma)$ such that $\cro(\psi_{G'})\le \tilde{O}(\optcrors(G')+\eac^{1-\eps}\cdot|E(G')|)$.

\item If Case 1 happens when processing instance $(G',\Sigma')$, then 
our algorithm decomposes instance $(G',\Sigma')$ into two subinstances $(G_1,\Sigma_1)$ and $(G_2,\Sigma_2)$, such that $|E(G_1)|,|E(G_2)|\geq \chi^{1-\eps}/3$ (from Observation~\ref{RS_obs:case_1}). Our algorithm then computes a solution $\phi_{G'}$ to instance $(G',\Sigma')$, whose number of crossings is bounded by $\cro(\phi_{G_1})+\cro(\phi_{G_2})+\chi^{1-2\eps}\cdot |E(G')|$. We say that instance $(G',\Sigma')$ contributes
$\Upsilon(G')=\chi^{1-2\eps}\cdot |E(G')|$ new crossings.

%the current instance $G'$ is decomposed into sub-instances $G_1, G_2$ through Case 1, then we have $|E(G_1)|, |E(G_2)|\ge \eac^{1-\eps}/3$ (from Observation~\ref{RS_obs:case_1}). And given the drawings $\psi_{G_1}$ and $\psi_{G_2}$ of sub-instances $G_1$ and $G_2$, respectively, we can efficiently compute a drawing $\psi_{G'}\in \Phi(G',\Sigma)$ such that $\cro(\psi_{G'})\le \cro(\psi_{G_1})+\cro(\psi_{G_2})+\eac^{1-2\eps}\cdot |E(G')|$.

\item Lastly, if Case 2 happens when   processing instance $(G',\Sigma')$, then 
our algorithm decomposes instance $(G',\Sigma')$ into two sub-instances $(G_1,\Sigma_1)$ and $(G_2,\Sigma_2)$. Assuming that bad event $\event$ does not happen when processing $(G',\Sigma')$, we are guaranteed that $|E(G_1)|,|E(G_2)|\geq \chi^{1-2\eps}/4$. Moreover, our algorithm computes a solution $\phi_{G'}$ to instance $(G',\Sigma')$, whose number of crossings is bounded by $\cro(\phi_{G_1})+\cro(\phi_{G_2})+2\chi^{1-4\eps}\cdot |E(G')|$. We say that instance $(G',\Sigma')$ contributes
$\Upsilon(G')=2\chi^{1-4\eps}\cdot |E(G')|$ new crossings.

%\item If the current instance $G'$ is decomposed into sub-instances $G_1, G_2$ through Case 2, then with probability $1-O(\eac^{-3\eps})$, we have $|E(G_1)|, |E(G_2)|\ge \Omega(\eac^{1-2\eps})$ (from Observation~\ref{RS_obs:case_2.1} and Observation~\ref{RS_obs:case_2.2}). And given the drawings $\psi_{G_1}$ and $\psi_{G_2}$ of sub-instances $G_1$ and $G_2$, respectively, we can efficiently compute a drawing $\psi_{G'}\in \Phi(G',\Sigma)$ such that $\cro(\psi_{G'})\le \cro(\psi_{G_1})+\cro(\psi_{G_2})+O(\eac^{1-3\eps})\cdot |E(G')|$.
%\item If the current instance $G'$ is decomposed into smaller sub-instances $G_1, G_2$ through Case 2.2, the with probability $O(\cro^{-\eps})$, we have $|E(G_1)|, |E(G_2)|\ge \Omega(\cro^{1-2\eps})$ (from Observation~\ref{RS_obs:case_2.2}). and given the drawings $\psi'_{G_1}$ and $\psi'_{G_2}$ of sub-instances $G_1$ and $G_2$, respectively, we can efficient compute a drawing $\psi'_{G'}\in \Phi(G',\Sigma)$ such that $\cro(\psi_{G'})\le \cro(\psi_{G_1})+\cro(\psi_{G_2})+O(\cro^{1-3\eps})\cdot |E(G')|$.
\end{itemize}

\begin{figure}[h]
	\label{fig:recursion}
	\centering
	\includegraphics[scale=0.4]{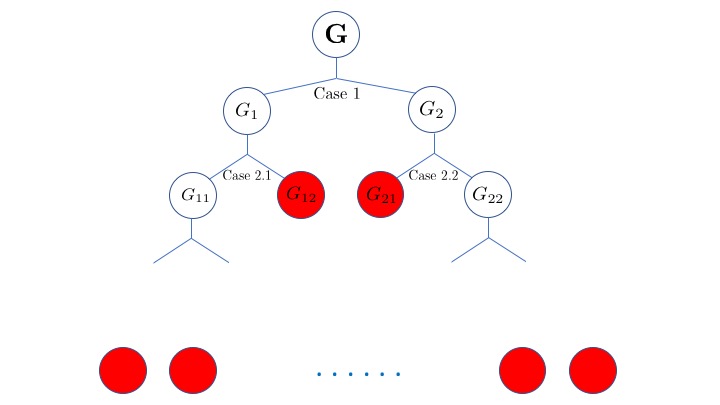}
	\caption{Decomposition tree associated with the algorithm. Vertices corresponding to base instances are shown in red.} 
\end{figure}

We can associate a \emph{decomposition tree} $T$ with the execution of the algorithm (see Figure~\ref{fig:recursion}). For every instance $(G',\Sigma')$ considered by the algorithm, the decomposition tree contains a vertex $v(G')$. If instance $(G',\Sigma')$ was decomposed into  subinstances $(G_1,\Sigma_1)$ and $(G_2,\Sigma_2)$, then we add edges $(v(G'),v(G_1))$ and $(v(G'),v(G_2))$ to the tree. The tree is rooted at vertex $v(G)$, and its leaves correspond to the base instances. We denote by $\hset$ the set of all graphs $G'$, such that instance $(G',\Sigma')$ was considered by the algorithm, and we let $\lset\subseteq \hset$ be the set of graphs $G'$ whose corresponding vertex $v(G')$ is a leaf of the tree $T$.

 Recall that, whenever instance $(G',\Sigma')$ is decomposed into  sub-instances $(G_1,\Sigma_1)$ and $(G_2,\Sigma_2)$, we have ensured that $E(G_1),E(G_2)\subseteq E(G')$ and $E(G_1)\cap E(G_2)=\emptyset$. 
%Let $\lset$ be the set of graphs $G'$, such that vertex $v(G')$ is a leaf of the tree $T$. 
Therefore, for all graphs $G',G''\in \lset$, $E(G')\cap E(G'')=\emptyset$.
Since, from Invariant \ref{inv: many edges}, for every instance $(G',\Sigma')$, $|E(G')|\geq \chi^{1-2\eps}/4$, and since $|E(G)|\leq \chi$, the number of the recursive levels is bounded by $4\chi^{2\eps}$, and $|\lset|\leq 4\chi^{2\eps}$. Moreover, since every inner vertex of tree $T$ has $2$ children, the total number of vertices in tree $T$ is at most $8\chi^{2\eps}$. Therefore, we have proved the following claim:

\begin{claim}\label{claim: rec levels}
	The number of the recursive levels is at most $4\chi^{2\eps}$, and the total number of instances that the algorithm considers is at most $8\chi^{2\eps}$. 
\end{claim}

Recall that we have defined, for every graph $G'\in \hset$, the number of crossings  $\Upsilon(G')$  that $G'$ contributes. Then the total number of crossings in $\phi_G$ is bounded by $\sum_{G'\in \hset}\Upsilon(G')$. Let $\hset_1\subseteq\hset$ be the set of all graphs $G'$, such that Case 1 happened when instance $(G',\Sigma')$ was considered, and define $\hset_2$ similarly for Case 2. We denote $\Upsilon_0=\sum_{G'\in \lset}\Upsilon(G')$, $\Upsilon_1=\sum_{G'\in \hset_1}\Upsilon(G')$, and $\Upsilon_2=\sum_{G'\in \hset_2}\Upsilon(G')$. We now bound each of the three terms separately.

\begin{claim}\label{claim: bound leaves}
$\Upsilon_0\leq O\left(\chi^{2-\eps}\poly(\log n)\right ).$
\end{claim}
\begin{proof}
Recall that for each graph $G'\in \lset$, $\Upsilon(G')=O\left(\left (N(G')+\chi^{1-\eps}\cdot |E(G')|\right )\poly\log n\right )$. Since the graphs in $\lset$ are disjoint in their edges, and since $\cro(\phi^*),|E(G)|\leq \chi$, we get that:
\[\begin{split} \Upsilon_0&=\sum_{G'\in \lset}\Upsilon(G')\\
&\leq \sum_{G'\in \lset}O\left(\left (N(G')+\chi^{1-\eps}\cdot |E(G')|\right )\poly\log n\right )\\
&\leq O\left(\left (\cro(\phi^*)+\chi^{1-\eps}\cdot |E(G)|\right )\poly\log n\right )\\
&\leq O\left(\chi^{2-\eps}\poly\log n\right ).\end{split}\]
\end{proof}

\begin{claim}\label{claim: bound case 2}
	With probability at least $1/2$, $\Upsilon_2\leq O(\chi^{2-\eps} )$.
\end{claim}
\begin{proof}
	For a graph $G'\in \hset_2$, let $\event(G')$ denote the bad event $\event$ for instance $(G',\Sigma')$. Recall that $\event(G')$ happens with probability at most $O(1/\chi^{3\eps})$. Since, from Claim \ref{claim: rec levels}, $|\hset_2|\leq 8\chi^{2\eps}$, from the union bound, with probability at least $1/2$, none of the events in $\set{\event(G')\mid G'\in \hset_2}$ happens.
	
For the remainder of the proof, we assume that none of the above events happens. Recall that for each graph $G'\in \hset_2$, $\Upsilon(G')=2\chi^{1-4\eps}\cdot |E(G')|$. We partition the graphs in $\hset_2$ into \emph{levels}, where two graphs $G_1,G_2$ lie in the same level iff the distances from $v(G_1)$ and from $v(G_2)$ to the root $v(G)$ of the tree $T$ are the same. Then the number of levels is bounded by $O(\chi^{2\eps})$ (from Claim \ref{claim: rec levels}), and the total contribution of graphs of $\hset_2$ that lie in the same level is bounded by $O(\chi^{1-4\eps}\cdot|E(G)|)\leq O(\chi^{2-4\eps})$. Therefore, overall, $\Upsilon_2\leq O(\chi^{2-\eps})$, with probability at least $1/2$.
\end{proof}

\begin{claim}\label{claim: bound case 1}
	$\Upsilon_1\leq O(\chi^{2-\eps} )$.
\end{claim}
\begin{proof}
Recall that for each graph $G'\in \hset_2$, $\Upsilon(G')=\chi^{1-2\eps}\cdot |E(G')|$. When instance $(G',\Sigma')$ is processed, we decomposed it into two sub-instances $(G_1,\Sigma_1)$ and $(G_2,\Sigma_2)$, such that $|E(G_1)|,|E(G_2)|\geq \chi^{1-\eps}/3$.
We mark, in the tree $T$, every vertex $v(G')$ with $G'\in \hset_2$. We say that instance $G'\in \hset_2$ \emph{lies at level $i$} iff there are exactly $i$ marked vertices on the unique path in tree $T$ from $v(G')$ to $v(G)$. 
\begin{observation}\label{obs: few levels}
The total number of non-empty levels is at most $3\chi^{\eps}$. 
\end{observation}
\begin{proof}
Consider any graph $G'\in \hset_2$, and assume that it lies at level $i$. Let $G=G_1,G_2,\ldots,G_i=G'$ be the marked vertices on the unique path $P$ connecting $v(G')$ to $v(G)$ in $T$. Notice that, for every vertex $v(G_j)$ on path $P$, there is a child $v(G'_j)$ of $v(G_j)$ that does not lie on $P$, such that $|E(G_j')|\geq \chi^{1-\eps}/3$. Since the edge sets of all graphs $G_1',G_2',\ldots,G_i'$ are mutually disjoint, and since $|E(G)|\leq \chi$, we get that $i\leq 3\chi^{\eps}$.
\end{proof}

For every $1\leq i\leq 3\chi^{\eps}$, the graphs $G'\in \hset_2$ that lie at level $i$ contribute in total at most $\chi^{1-2\eps}\cdot |E(G)|\leq \chi^{2-2\eps}$ to $\Upsilon_1$. Since the number of non-empty levels is bounded by $O(\chi^{\eps})$, $\Upsilon_1\leq O(\chi^{2-\eps})$.
\end{proof}

We conclude that, with probability at least $1/2$, the number of crossings in the drawing $\phi(G)$ is bounded by $O(\chi^{2-\eps}\cdot \poly\log n)$. In order to increase the probability of success, we repeat the algorithm multiple time, returning the best of the resulting solutions.

\appendix
\newpage
\section{Lower Bounds for the Relationship between \MCN and \MP}

\label{sec: appx-lower}

\begin{figure}[!h]
	\centering
	\subfigure[Graph $G$ and its optimal drawing; the edges of $E'$ are shown in red.]{\scalebox{0.40}{\includegraphics{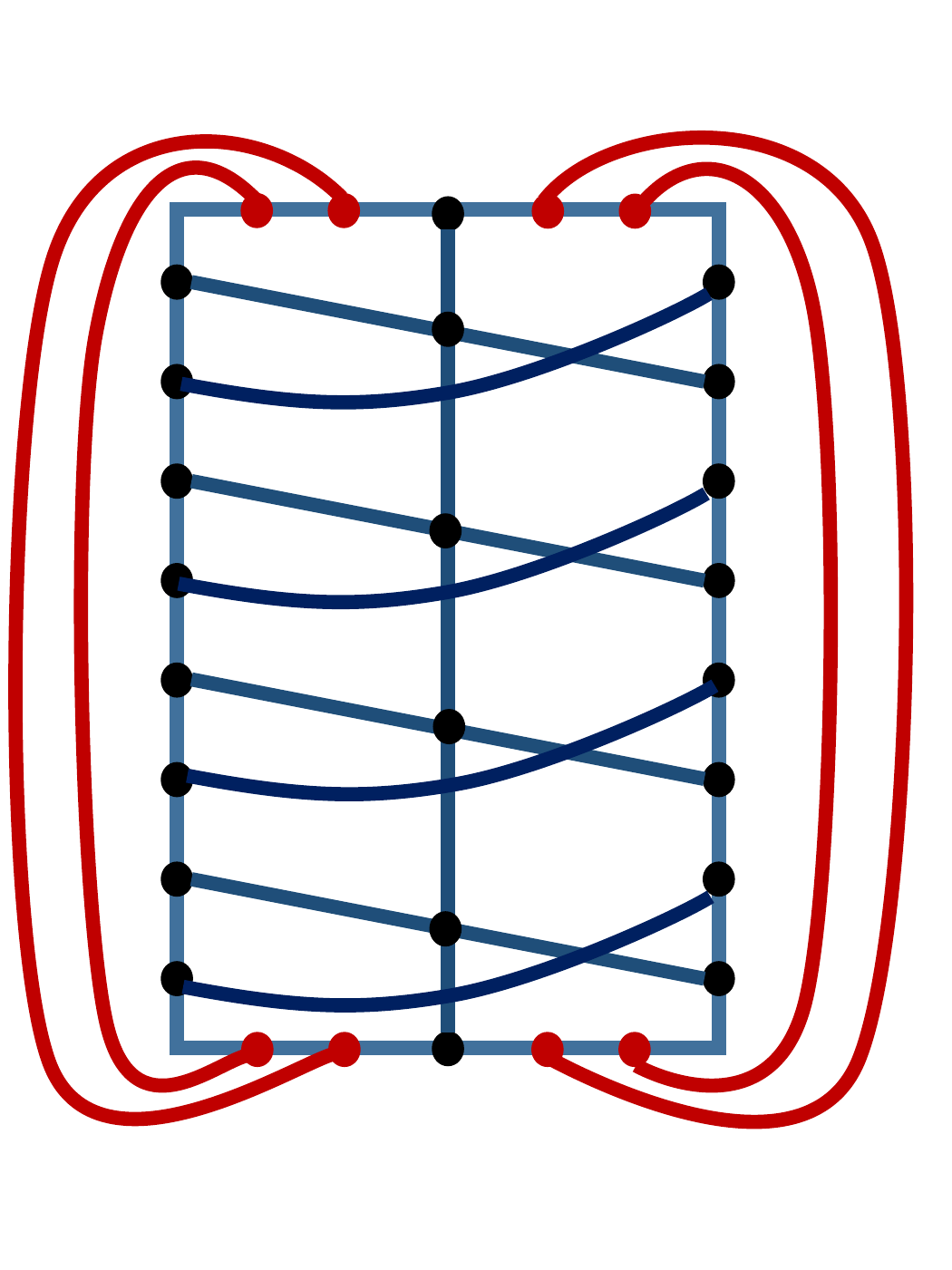}}}
	\hspace{.7cm}
	\subfigure[The unique planar drawing of graph $G\setminus E'$.]{
		\scalebox{0.40}{\includegraphics{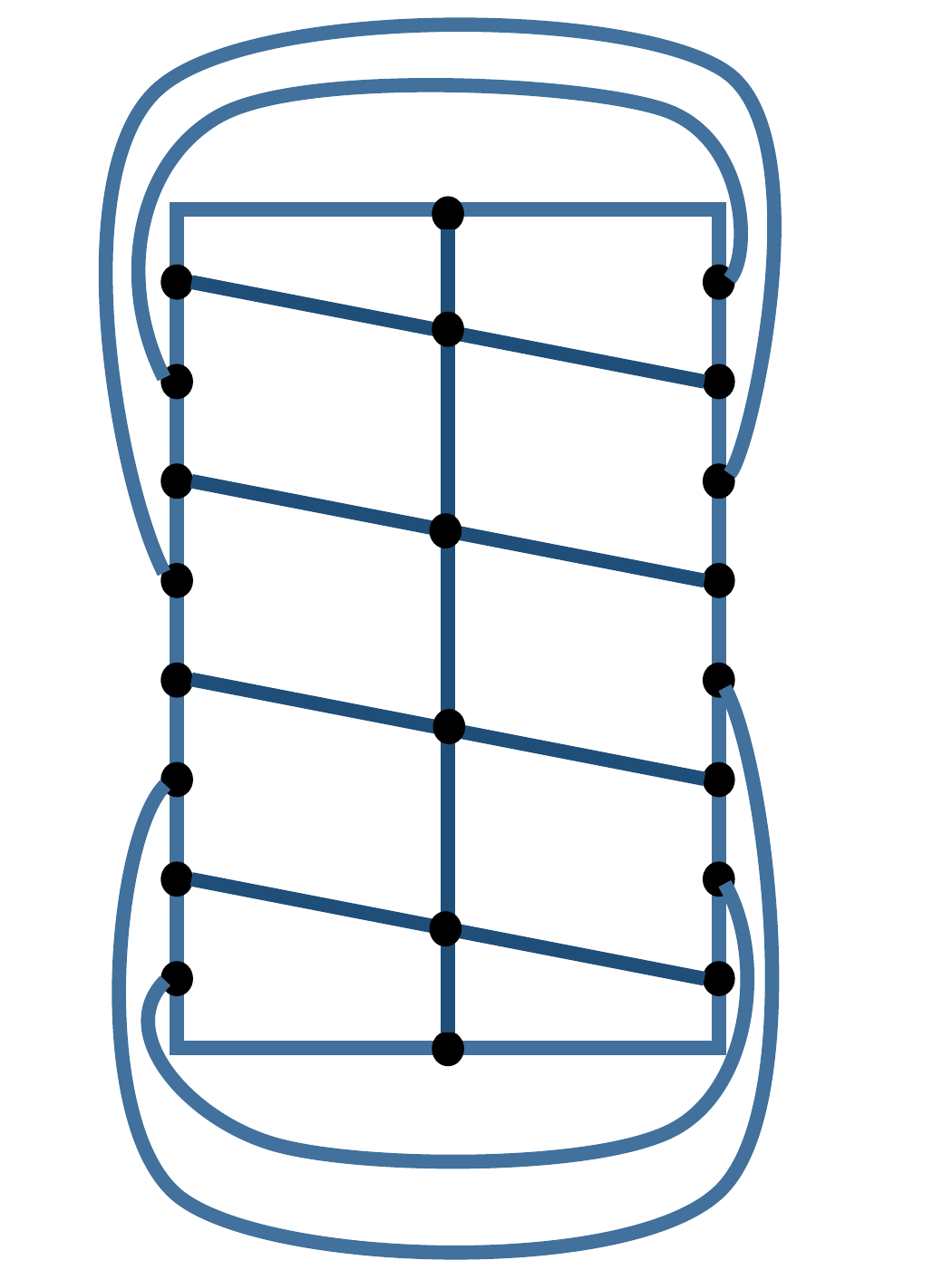}}}
	\hspace{.7cm}
	\subfigure[Adding images of edges of $E'$ to the planar drawing of $G\setminus E'$ must cause $\Omega(k\cdot\optcro(G))$ crossings.]{
		\scalebox{0.40}{\includegraphics{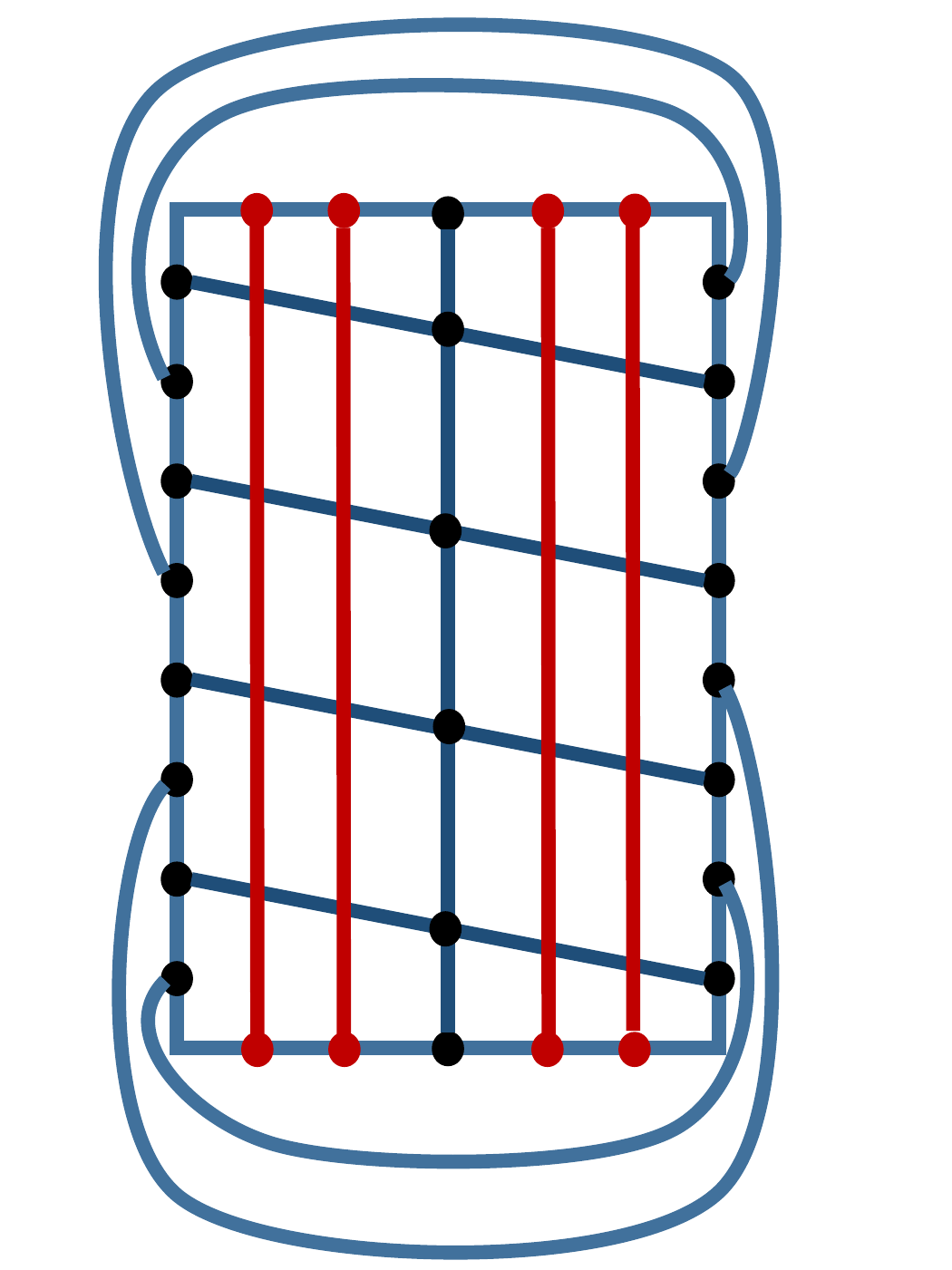}}}
	\caption{A negative example due to \cite{chuzhoy-lowerbound}, showing a lower bound of $\Omega(k\cdot \optcro(G))$ on the number of crossings that any algorithm using Paradigm $\Pi$ must incur. \label{fig:lower-bound-1}} %We note that this constructruction can be modified to work for arbitrary values of $k$ and $\optcro(G)$.\label{fig:lower-bound-1}}
\end{figure}

\begin{figure}[!h]
	\centering
	\subfigure[Graph $G$ and its optimal drawing; the edges of $E'$ are shown in red. ]{\scalebox{0.17}{\includegraphics{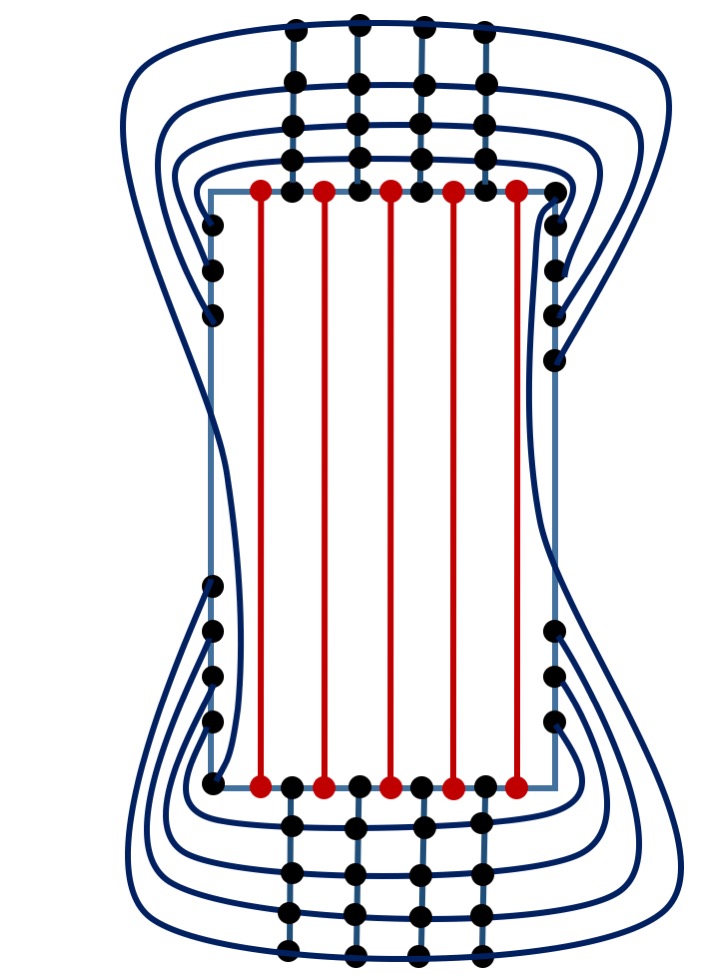}}}
	\hspace{.7cm}
	\subfigure[The unique planar drawing of graph $G\setminus E'$.]{
		\scalebox{0.17}{\includegraphics{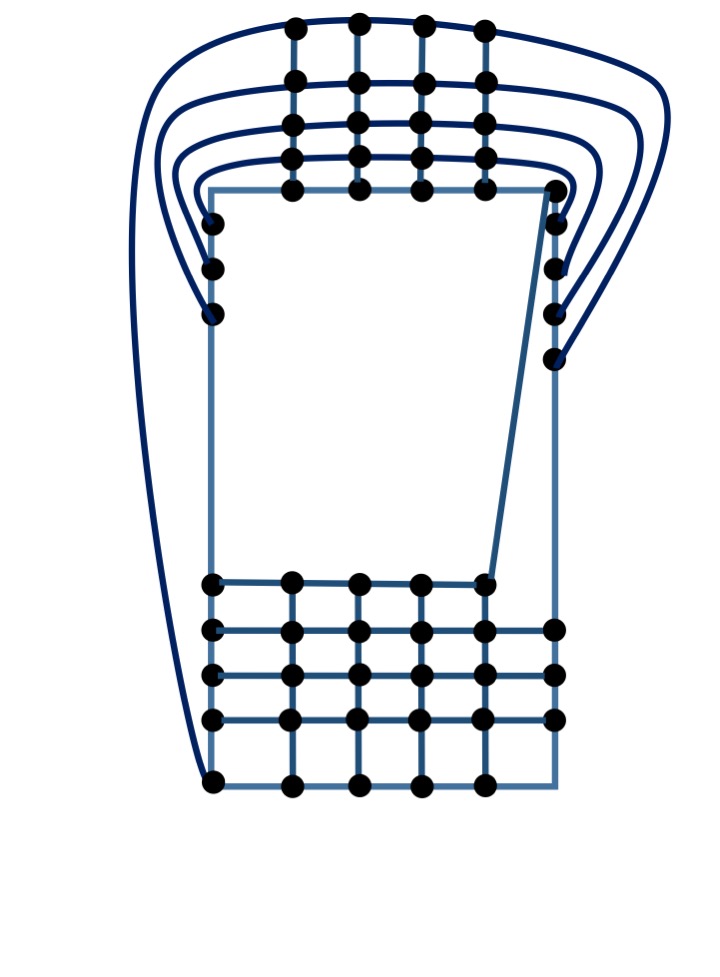}}}
	\hspace{.7cm}
	\subfigure[Adding images of edges in $E'$ to the planar drawing of $G\setminus E'$ must cause $\Omega(k^2)$ crossings.]{
		\scalebox{0.17}{\includegraphics{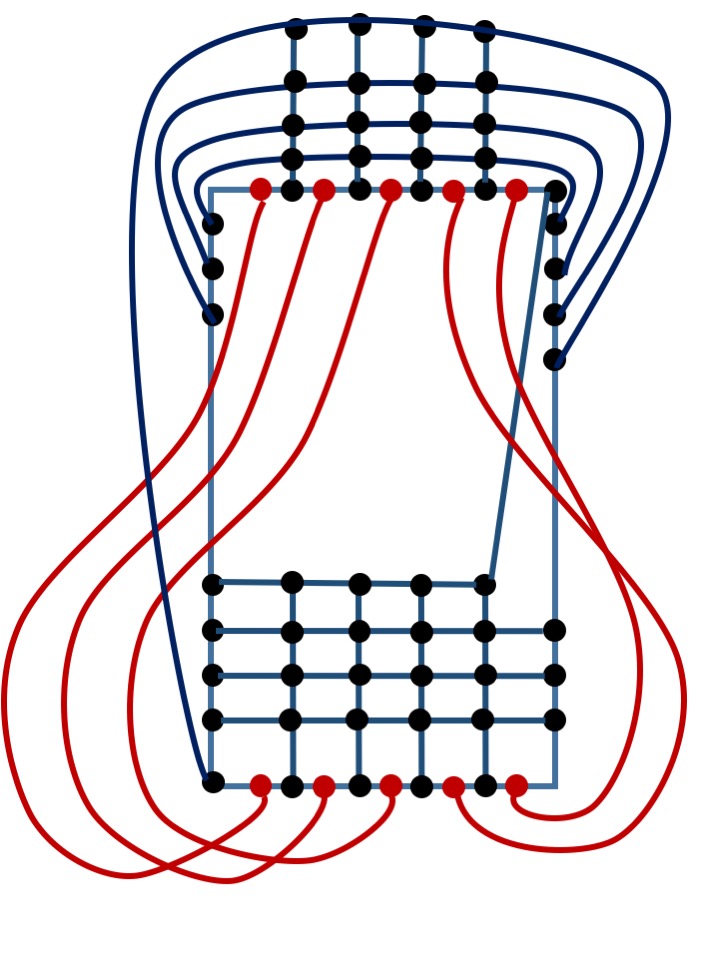}}}
	\caption{This example establishes the lower bound of $\Omega(k^2)$, in the case where $k\geq \optcro(G)$. We note that this construction can be modified to work for arbitrary values of $k$, while keeping $\optcro(G)=O(1)$.\label{fig:lower-bound-2}}
	%\znote{the thickness of lines in two figures are different, this seems weird to me, should we modify it?}
\end{figure}

Recall that \cite{chuzhoy2011graph} provided an efficient algorithm, that, given an input graph $G$, and a planarizing set $E'$ of $k$ edges for $G$, draws the graph with $O\left(\Delta^3\cdot k \cdot (\optcro(G) + k)\right)$ crossings, using Paradigm $\Pi$. %the paradigm outlined in Section \ref{sec: intro}: namely, the algorithm first computes a planar drawing of the graph $G\setminus E'$, and then adds the drawings of the edges of $E'$, to obtain the final drawing of $G$. 
In this section we provide two lower-bound constructions of \cite{chuzhoy-lowerbound},  that show that the algorithm of \cite{chuzhoy2011graph}  provides nearly best possible guarantees that can be achieved via this paradigm. 
Specifically, in Figure \ref{fig:lower-bound-1}, we show a graph $G$ and a planarizing set $E'$ of $k$ edges of $G$, such that graph $G\setminus E'$ is $3$-connected, and hence has a unique planar drawing. However, it is easy to see that, adding the images of the edges of $E'$ to this drawing must incur  $\Omega(k\cdot\optcro(G))$ crossings. %In this example, however, $k$ is close to $\optcro(G)$. 
In Figure \ref{fig:lower-bound-2}, we show another example, due to \cite{chuzhoy-lowerbound}, where $\optcro(G)$ may be arbitrarily small relatively to $k$, in which adding the images of edges of $E'$ to the unique planar drawing of $G\setminus E'$ must incur $\Omega(k^2)$ crossings.

\section{Proof of Corollary \ref{cor: approx alg}}\label{sec: cor main algo}

Recall that the algorithm of ~\cite{even2002improved} computes a drawing of the input graph $G$ with 
$O((n+\optcro(G)) \cdot \poly(\Delta\log n))$ crossings.
In order to obtain an $O(n^{1/2-\eps'}\cdot\poly(\Delta\log n))$-approximation algorithm
for \mcn, we apply the algorithm from Corollary \ref{cor: main_algo}, and the algorithm of \cite{even2002improved} to the input graph $G$, and output the better of the two solutions. 
If $\optcro(G)\geq n^{1/2+\eps/4}$, where $\eps$ is the constant from Corollary \ref{cor: main_algo}, then the algorithm of \cite{even2002improved} provides an $O(n^{1/2-\eps/4}\cdot\poly(\Delta\log n))$-approximation. Otherwise, $\optcro(G)<n^{1/2+\eps/4}$, and the algorithm from 
Corollary \ref{cor: main_algo} achieves an approximation factor of $O\left ((\optcro(G))^{1-\epsilon}\cdot\poly(\Delta\log n)\right ) \leq O\left (n^{1/2-\eps/4}\cdot\poly(\Delta\log n)\right )$. We note that we did not attempt to optimize the constant $\eps'=\eps/4$.

\section{Proofs Omitted from Section \ref{sec:overview} }\label{sec: appendix proofs from overview}
%--------------------------------------
%--------------------------------------
%--------------------------------------
\subsection{Proof of Theorem \ref{thm: solve cluster placement}}\label{sec: solving cluster placement problem} 
In order to simplify the notation, we denote the set $\hcset$ of clusters by $\cset$.
Consider some cluster $C\in \cset$.
For every face $F\in \fset_C$ of the drawing $\psi_C$ of $C$, we let $\hset_C(F)\subseteq \cset\setminus\set{C}$ contain all clusters $C'$ such that $F_C(C')=F$. We denote by $\fset'_C\subseteq \fset_C$ the set of all faces $F$ with $\hset_C(F)\neq\emptyset$.

The proof of the theorem a recursive algorithm. The base case is when, for every cluster $C\in \cset$, $|\fset'_C|=1$. For each cluster $C$, let $F_C$ be the unique face in $\fset'_C$. Consider a drawing of the clusters in $\bigcup_{C\in \cset}C$ in the plane, such that the drawing of each cluster $C$ is identical to $\psi_C$, with the face $F_C$ serving as the outer face of the drawing of $C$, and the images of all clusters are disjoint. Let $\phi'$ denote the resulting drawing of   $\bigcup_{C\in \cset}C$  in the plane, and let $\phi$ be the corresponding drawing on the sphere. Then it is easy to see that $\phi'$ defines a feasible solution for the input instance of the \CP problem.

Assume now that there is at least one cluster $C\in \cset$, with $|\fset'(C)|>1$. For every face $F\in \fset'(C)$, we define a new sub-instance of the current instance of the \CP problem, that only includes the clusters of $\hset_C(F)\cup \set{C}$; the faces $F_{C_1}(C_2)$ for pairs $C_1,C_2\in \hset_C(F)$ of clusters are defined exactly as before. Notice that, if the original instance of the \CP problem had a feasible solution, then each resulting sub-problem must also have a feasible solution. We solve each such sub-problem recursively, and obtain, for every face $F\in \fset'(C)$, a drawing $\phi_F$ of $\bigcup_{C'\in \hset_C(F)\cup C}C'$ on the sphere. We let $D_F$ denote a disc in this drawing that contains the images of all clusters if $\hset_C(F)$ but is disjoint form the image of $C$. In order to obtain a solution to the original instance of the problem, we start with the drawing $\psi_C$ of the cluster $C$ on the sphere. For every face $F\in \fset'(C)$, we copy the contents of the disc $D_F$ in the drawing $\phi_F$ to the interior of the face $F$. Once we process every face $F\in \fset'(C)$, we obtain a drawing of $\bigcup_{C'\in\cset}C'$. Moreover, it is easy to verify that, if the original instance of the \CP problem had a feasible solution, then we obtain a feasible solution for this instance.

\subsection{Proof of Lemma \ref{lem: cheap solution}} \label{sec: cheap soln for CNRS}
It is enough to show that, for every face $F\in \fset$, there is a solution $\phi^F$ to instance $(G^F,\Sigma^F)$ of \CNwRS, such that, if we denote by $\chi^F$ the set of all crossings of edges in $\phi^F$, then $\sum_{F\in \fset}|\chi^F|\leq O(\optcro(G)\cdot \poly(\Delta\log n))$.

We now fix a face $F\in \fset$ and construct a solution $\phi^F$ to instance $(G^F,\Sigma^F)$ of \CNwRS. Our starting point is the drawing $\phi$ of graph $G'$ given by Observation \ref{obs: drawing consistent with Cluster Placement}. We delete from $\phi$ the images of all vertices and edges, except for the vertices and edges of the clusters in $\hset(F)$, and the images of the edges in $E^F$. Recall that, in the resulting drawing, no edge of $\bigcup_{C\in \hset(F)}E(C)$ may participate in crossings. Since, for every ordered pair $(C,C')\in \hset(C)$ of clusters, the image of $C'$ in the resulting drawing must be contained in the face $F_C(C')$, there must be a face $F'$ in the resulting drawing that contains the images of every edge in $E^F$. Viewing face $F'$ as the outer face of a planar drawing of the resulting graph, we can contract the image of each cluster $C\in \hset(F)$ into a single point, that we view as the image of the corresponding vertex $v(C)$, without increasing the number of crossings. Therefore, we obtain a drawing $\tilde \phi^F$ of the graph $G^F$ on the sphere, and moreover:
\[\sum_{F\in \fset}\cro(\tilde \phi^F)\leq \cro(\phi)\leq O(\optcro(G)\cdot \poly(\Delta\log n)).\]
Consider again some face $F\in \fset$. Notice that the drawing $\tilde \phi^F$ of $G^F$ is not necessarily consistent with the rotation system $\Sigma^F$. Next, we modify the drawing $\tilde \phi^F$ in order to make it consistent with $\Sigma^F$, while only introducing a small number of crossings. Recall that we have defined a set $\Gamma\subseteq V(G')$ of vertices called terminals -- the set of all vertices that serve as endpoints of the edges in $E''$.

We process the vertices $v\in V(G^F)$ one-by-one. For every vertex $v\in V(G^F)$, we let $\eta(v)$ be a small disc around the drawing of $v$ in $\tilde \phi^F$.  When vertex $v$ is processed, we slightly modify the images of the edges of $\delta(v)$ in disc $\eta(v)$, so that the circular order in which the edges of $\delta(v)$ enter $v$ becomes identical to $\oset_v$.

Consider now some vertex $v=v(C)\in V(G^F)$. Assume first that $C\in \cset_1$. In this case, $|\delta(v)|\leq \poly(\Delta\log n)$. For every edge $e\in \delta(v)$, we replace the segment of the drawing of $e$ in disc $\eta(v)$, so that all resulting curves enter the image of the vertex $v$ in the order $\oset_v$, and every pair of curves intersects at most once.  This introduces at most $\poly(\Delta\log n)$ new crossings.

Assume now that $C\in \cset_2$. Let $\oset'_v$ be the order in which the images of edges of $\delta(v)$ enter $v$ in the current drawing $\tilde \phi^F$. Recall that we have defined a cycle $K^F(C)\subseteq C$, which is the intersection of the cluster $C$ and the boundary of the face $F$, and $\Gamma^F(C)\subseteq \Gamma(C)$ is the set of terminals that appear on $K^F(C)$. Recall that we have defined an ordering $\tilde \oset^F(C)$ of the terminals in $\Gamma^F(C)$ to be the circular order of the terminals of $\Gamma^F(C)$ along the cycle $K^F(C)$, and the ordering $\oset_v$ of $\delta(v)$ was defined based on $\tilde \oset^F(C)$. Since the edges of $C$ may not participate in crossings in $\phi$, and since the drawing of $C$ in $\phi$ is identical to $\psi_C$, for every terminal $t\in \Gamma^F(C)$, the edges of $\delta(v)$ that are incident to $t$ appear consecutively in the ordering $ \oset'_v$ (and from the definition, they also appear consecutively in the ordering $\oset_v$). The orderings of edges of $\delta(v)$ that are incident to different terminals in $\oset_v$ and $ \oset'_v$ must both be consistent with $\tilde \oset_v$. Therefore, the only difference between the orderings $\oset_v$ and $\oset'_v$ is that for every terminal $t\in \Gamma^F(C)$, the edges of $\delta(v)$ that are incident to $t$ may appear in different orders in $\oset_v$ and $\oset'_v$. Consider the small disc $\eta(v)$ around the vertex  $v$ in the current drawing $\tilde \phi^F$ of $G^F$. We can assume that this disc is small enough so it does not contain any crossings. For every edge $e\in \delta(v)$, let $p_e$ be the point that is the intersection of the current image of $e$ and the boundary of the disc $\eta(v)$ in $\tilde \phi^F$. Then for every terminal $t\in  \Gamma^F(C)$, the  points $p_e$ corresponding to the edges of $\delta(v)$ that are incident to $t$ appear consecutively on the boundary of $\eta(v)$. We rearrange the images of all edges of $\delta(v)$ that are incident to $t$ inside $\eta(v)$, so that they enter $v$ in the order consistent with $\oset_v$. This introduces, for every terminal $t\in \Gamma^F(C)$, at most $O(\Delta^2)$ new crossings. Once we process all vertices of $G^F$, we obtain the final drawing $\phi^F$ of $G^F$ that is consistent with the rotation system $\Sigma^F$. We now bound the total number of new crossings that this procedure has introduced.

Consider a cluster $C$. Note that for every terminal $t\in \Gamma(C)$, there can be at most $\Delta$ faces $F\in \fset$, such that $t$ lies on the boundary of $F$ in drawing $\tilde \phi$ of $\bigcup_{C\in \cset}C$. Therefore, there are at most $\Delta |\Gamma(C)|$ pairs $(t,F)$, where $t\in \Gamma(C)$ and $F\in \fset$, such that $t$ lies on the boundary of $F$. We denote the set of all such pairs for cluster $C$ by $\Pi(C)$.

If $C$ is a type-1 cluster, then the total increase in the number of crossings due to rearranging edges entering vertex $v(C)$ in all graphs $\set{G^F}_{F\in \fset}$ is at most $O(|\Pi(C)|^2\Delta^2)\leq O(\Delta^4 |\Gamma(C)|^2)\leq O(\poly(\Delta\log n))$.

If $C$ is a type-2 cluster, then every pair $(t,F)\in \Pi(C)$ contributes at most $\Delta^2$ crossings (by rearranging the images of edges of $\delta(v)$ in $\tilde \phi^F$ that are incident to $t$). 

Altogether, the number of new crossings is bounded by:
\[
\begin{split}
O(|\Gamma|\Delta^3)+O(|\cset_1|\poly(\Delta\log n))& \leq  O(|\Gamma|\poly(\Delta\log n))\\
&\leq O(|E''|\poly(\Delta\log n))\\
&\leq O(\optcro(G)\poly(\Delta\log n),\end{split}\]

and so the total number of crossings in all drawings in $\set{\phi^F}_{F\in \fset}$ is bounded by $ O(\optcro(G)\poly(\Delta\log n))$.

%-------------------------
%-------------------------
%-------------------------
%-------------------------
%-------------------------

\section{Proofs Omitted from Section \ref{sec:block_decompos}}\label{sec: appendix proofs from prelims}

\subsection{Proof of Lemma~\ref{lem:block_endpoints_routing}}
\label{subsec:block_endpoints_routing}

We assume w.l.o.g. that graph $G$ is $2$-connected, as otherwise we can prove the theorem for each of its super-blocks $Z\in \zset(G)$ separately. We denote by $\bset=\bset(G)$ the block decomposition of $G$, $\tilde\bset=\set{\tilde B\mid B\in \bset}$, and we let $\tbset^*\subseteq \tbset$ contain all graphs $\tilde B\in \tbset$ that are not isomorphic to $K_3$. We denote by $\tau=\tau(\bset)$ the decomposition tree corresponding to the block decomposition $\bset$ of $G$. If vertex $v(B_1)$ is a child of vertex $v(B_2)$ in tree $\tau$, then we say that $B_1$ is a \emph{child block} of $B_2$. For a block $B\in \bset$, we also denote by $\desc(B)$ the set of all \emph{descendant blocks} of $B$. The set $\desc(B)$ contains all blocks $B_1$ where vertex $v(B_1)$ is a descendant of vertex $v(B)$ in the tree $\tau$. We note that the set $\desc(B)$ also contains the block $B$.

Consider a block $B\in \bset$. For convenience, we call the fake edge $e^*_{\tilde B}$ connecting the endpoints of $B$ (if it exists) a \emph{bad} fake edge, and all other fake edges of $\tilde B$ are called good fake edges. We define a set $A'_{\tilde B}$ of fake edges as follows: if $B$ is isomorphic to $K_3$, then $A'_{\tilde B}=\emptyset$, and otherwise, $A'_{\tilde B}$ contains all good fake edges of $\tilde B$. %Note that, from the definition of a block decomposition, for every pair $(x,y)$ of vertices, edge $(x,y)$ may belong to at most one set $\set{A'_{\tilde B}}_{B\in \bset}$.   
Consider now some good fake edge $e=(u,v)\in A'_{\tilde B}$. Then there is some child block $B_1$ of $B$ with endpoints $u,v$. A \emph{valid embedding} of the fake edge $e=(u,v)\in A'_{\tilde B}$ is a path $P(e)$ that connects $u$ to $v$, is internally disjoint from $\tilde B$, and is contained in a block $B_1$ that is a child block of $B$, whose endpoints are $(u,v)$. A \emph{valid embedding} of the set $A'_{\tilde B}$ of fake edges is a collection $\qset(B)=\set{P(e)\mid e\in A'_{\tilde B}}$ of paths, were for each edge $e\in A'_{\tilde B}$, path $P(e)$ is a valid embedding of $e$. Note that from the definition, we are guaranteed that the paths in $\qset(B)$ are internally disjoint. The following lemma is central to the proof of Lemma~\ref{lem:block_endpoints_routing}.
\begin{lemma}\label{lem: find embeddings}
There is an efficient algorithm, that, given a block $B\in \bset\setminus\set{G}$, computes, for every descendant-block $B_1\in \desc(B)$, a valid embedding $\qset(B_1)$ of the set $A'_{\tilde B_1}$ %\snote{changed $A'_{\tilde B}$ to $A'_{\tilde B_1}$} 
of fake edges, and additionally a collection $\pset_1(B)$ of $6$ paths in $B$, connecting the endpoints of $B$, such that, if we denote by $\pset_2(B)=\bigcup_{B_1\in \desc(B)}\qset(B_1)$, then the paths in $\pset_1(B)\cup \pset_2(B)$ cause congestion at most $6$ in $B$.
\end{lemma}
We note that, from the definition of valid embeddings, all paths in $\pset_2(B)$ must be contained in $B$.

We prove Lemma \ref{lem: find embeddings} below, after we complete the proof of Lemma~\ref{lem:block_endpoints_routing} using it.
Recall that graph $G$ itself is a block in the decomposition $\bset$. Let $B_1,\ldots,B_r$ be the child blocks of $G$. We apply the algorithm from 
Lemma  \ref{lem: find embeddings} to each such block $B_i$ separately, obtaining the sets $\pset_1(B_i),\pset_2(B_i)$ of paths. 
Let $\qset^*=\bigcup_{i=1}^r\pset_2(B_i)$. Then set $\qset^*$ contains, for every block $B\in \bset\setminus \set{G}$, a set $\qset(B)$ of paths that defines a valid embedding of the set $A'_{\tilde B}$ of fake edges of $\tilde B$. 
For all $1\leq i\leq r$, let $(x_i,y_i)$ be the endpoints of the block $B_i$. %\snote{they might be repeated}. 
We embed the fake edge $(x_i,y_i)$ of $A'_{\tilde G}$ into any of the $6$ paths $P(x_i,y_i)\in \pset_1(B_i)$. We then set $\qset(G)=\set{P(x_i,y_i)\mid 1\leq i\leq r}$. Note that $\qset(G)$ is a valid embedding of the good fake edges of $G$. Adding the paths in $\qset(G)$ to the set $\qset^*$, we obtain a set of paths in $G$, that cause edge-congestion at most $6$, and contain, for every  pseudo-block $B\in \bset$, a valid embedding of the set $A'_{\tilde B}$ of fake edges. Lastly, consider any block $B\in \bset\setminus\set{G}$. Let $(x,y)$ be the endpoints of $B$, and let $B^c$ be its complement block. Then $B^c$ contains a path connecting $x$ to $y$. We let the embedding $P(e^*_{\tilde B})$ of the bad fake edge $e^*_{\tilde B}$ be any path in $B^c$ that connects $x$ to $y$.
We then set $\pset_{\tB}= \qset(B)\cup \set{P(e^*_{\tilde B})}$.
Observe that, from the definition of valid embeddings of bad fake edges, all paths in $\pset_{\tB}$ are mutually internally disjoint, and they are internally disjoint from $\tilde B$. As discussed above, the paths in set $\pset= \bigcup_{\tilde B\in \tilde\bset^*(G)}\left(\pset_{\tilde B}\setminus \set{P_{\tilde B}(e^*_{\tilde B})} \right)$ cause congestion at most $6$ in $G$. Therefore, in order to complete the proof of Lemma~\ref{lem:block_endpoints_routing}, it is now enough to prove  Lemma \ref{lem: find embeddings}.

\iffalse
\begin{lemma}
	\label{lem:block_endpoints_routing}
	Let $G$ be an $n$-vertex graph, and let $\bset(G)$ be its block decomposition.
	Denote $\tilde\bset(G)=\set{\tilde B\mid B\in \bset(G)}$, and let $\tilde \bset^*(G)\subseteq \tilde\bset(G)$ contain all graphs $\tilde B$ that are not isomorphic to $K_3$. Then we can efficiently compute, for each graph $\tilde{B}\in \tilde\bset^*(G)$, a collection $\pset_{\tilde B}=\set{P_{\tilde B}(e)\mid e\in \aset_{\tilde B}}$ of paths in $G$, such that:
	\begin{itemize}
		\item for each fake edge $e=(u,v)\in \aset_{\tilde B}$, the path $P_{\tilde B}(e)$ connects $u$ to $v$ in $G$ and it is internally disjoint from $\tilde B$; 
		\item all paths in $\pset_{\tilde B}$ are mutually internally disjoint; and
		\item if we denote $\pset=\bigcup_{\tilde B\in \tilde\bset^*(G)}\left(\pset_{\tilde B}\setminus \set{P_{\tilde B}(e^*_{\tilde B})} \right)$, then every edge of $G$ participates in at most $6$ paths in $\pset$.
	\end{itemize} 
\end{lemma}

\fi

\begin{proofof}{Lemma \ref{lem: find embeddings}}
	The proof is by induction on the length of the longest path from $v(B)$ to a leaf vertex of $\tau$ that is a descendant of $v(B)$ in $\tau$. 
The base case is when $v(B)$ is a leaf vertex of $\tau$. Let $(x,y)$ denote the endpoints of $B$. Observe that in this case, $\tilde B$ is obtained from block $B$ by adding the bad fake edge $e^*_{\tilde B}$ to it, and this is the only fake edge in $\tilde B$. In particular, $A'_{\tilde B}=\emptyset$, so we can set $\qset(B)=\emptyset$. Block $B$ must contain at least one path  $P$ connecting $x$ to $y$. We let $\pset_1(B)$ contain $6$ copies of this path. Setting $\pset_2(B)=\emptyset$, we get valid sets $\pset_1(B),\pset_2(B)$ of paths for $B$, with $\pset_1(B)\cup \pset_2(B)$ causing edge-congestion at most $6$ in $B$.

Consider now an arbitrary block $B\in \bset\setminus\set{G}$, such that $v(B)$ is not a leaf of $\tau$, and let $B_1,\ldots,B_r$ be its child blocks. Using the induction hypothesis, we compute, for all $1\leq i\leq r$, the sets $\pset_1(B_i),\pset_2(B_i)$ of paths that are contained in $B_i$, such that $\pset_1(B_i)\cup \pset_2(B_i)$ cause edge-congestion at most $6$ in $B_i$. We now consider three cases.

\paragraph{Case 1. $\tilde B=K_3$.} In this case, $A'_{\tilde B}=\emptyset$, and so $\qset(B)=\emptyset$ is a valid embedding of the edges in $A'_{\tilde B}$. %Note that $B$ has either one or two child blocks, so $r\leq 2$. 
We set $\pset_2(B)=\bigcup_{i=1}^r\pset_2(B_i)$. Clearly, set $\pset_2(B)$ contains, for each block $B^*\in \desc(B)$, a valid embedding $\qset(B^*)$ of the edges of $A'_{\tilde B^*}$. It now remains to define a set $\pset_1(B)$ of $6$ paths connecting the endpoints of $B$. Let $(x,y)$ denote the endpoints of $B$. Then there is a path $P$ in $\tilde B$, that is disjoint from the bad fake edge $e^*_{\tilde B}$, connecting $x$ to $y$. This path contains two edges, that we denote by $e_1$ and $e_2$. Initially, we let $\pset_1(B)$ contain six copies of the path $P$, that we denote by $P_1,\ldots,P_6$. Assume first that $e_1=(x_1,y_1)$ is a fake edge, and assume w.l.o.g. that the child block $B_1$ of $B$ has endpoints $x_1$ and $y_1$. Then we replace, for all $1\leq i\leq 6$, the edge $e_1$ on path $P_i$ by the $i$th path in set $\pset_1(B_1)$ (recall that this path connects $x_1$ to $y_1$ in $B_1$). If $e_2$ is a fake edge, then we proceed similarly. As a result, set $\pset_1(B)$ now contains $6$ paths that are contained in $B$, each of which connects $x$ to $y$. Moreover, it is easy to verify that the paths in $\pset_1(B)\cup \pset_2(B)$ cause edge-congestion at most $6$ in $B$.

\paragraph{Case 2. $\tilde B\ne K_3$, and block $B$ has a single child-block.} We denote by $B_1$ the child block of $B$. Let $(x,y)$ be the endpoints of $B$, and let $(x_1,y_1)$ be the endpoints of $B_1$. Let $e_1=(x_1,y_1)$ be the unique fake edge %\snote{can't it be a real edge?}
in $A'_{\tilde B}$. We let the embedding of this edge $P(e_1)$ be any of the $6$ paths in $\pset_1(B_1)$. We then let $\qset(B)=\set{P(e_1)}$ be a valid embedding of the edges of $A'_{\tilde B}$, and $\pset_2(B)=\qset(B)\cup \pset_2(B_1)$. Clearly, set $\pset_2(B)$ contains, for each block $B^*\in \desc(B)$, a valid embedding $\qset(B^*)$ of the edges of $A'_{\tilde B^*}$. It now remains to define a set $\pset_1(B)$ of $6$ paths connecting the endpoints of $B$. Since graph $\tilde B$ is $3$-connected, there are at least three internally disjoint paths connecting $x$ to $y$ in $\tilde B$. Since graph $\tilde B$ contains only two fake edges, at least one of these paths, that we denote by $P$, does not contain fake edges. We then let $\pset_1(B)$ contain $6$ copies of the path $P$. Clearly, the paths in $\pset_1(B)\cup \pset_2(B)$ cause edge-congestion at most $6$ in $B$.

\paragraph{Case 3. $\tilde B\ne K_3$, and block $B$ has at least two child-blocks.} Let $(x,y)$ be the endpoints of $B$, and, for all $1\leq i\leq r$, let $(x_i,y_i)$ be the endpoints of the child block $B_i$. We also denote by $e_i=(x_i,y_i)$ the fake edge in $A'_{\tilde B}$ connecting the endpoints of $e_i$ (if it exists). For each block $B_i$, we let $P^*_i\in \pset_1(B_i)$ be an arbitrary path in $\pset_1(B_i)$, and we let the embedding $P(e_i)$ of fake edge $e_i$ (if it exists) be $P^*_i$. We then define $\qset(B)=\set{P(e_i)\mid e_i\in A'_{\tilde B}}$. It is immediate to verify that $\qset(B)$ is a valid embedding of the fake edges in $A'_{\tilde B}$. We then set $\pset_2(B)=\qset(B)\cup \left (\bigcup_{i=1}^r\pset_2(B_i)\right )$, so set $\pset_2(B)$ contains, for each block $B^*\in \desc(B)$, a valid embedding $\qset(B^*)$ of the edges of $A'_{\tilde B^*}$. It now remains to define a set $\pset_1(B)$ of $6$ paths connecting the endpoints of $B$.

Since graph $\tilde B$ is $3$-connected, it contains at least three internally disjoint paths connecting $x$ to $y$. At least two of these paths, that we denote by $P$ and $P'$, are disjoint from the bad fake edge $e^*_{\tilde B}$. We let $\pset_1(B)=\set{P_1,P_2,P_3,P'_1,P'_2,P'_3}$, where initially, paths $P_1,P_2,P_3$ are copies of $P$, and paths $P'_1,P'_2,P'_3$ are copies of $P'$. Next, we consider every fake edge of $P$ and $P'$ one-by-one. Let $e_i\in P$ be a fake edge on path $P$, so $e_i=(x_i,y_i)$, and the child block $B_i$ of $B$ has endpoints $x_i,y_i$. Recall that $\pset_1(B_i)$ contains $6$ paths connecting $x_i$ to $y_i$; one of these paths has been used as $P^*_i$, and the remaining five paths have not been used yet. For each $1\leq j\leq 3$, we replace, on path $P_j$, the edge $e_i$, with the $j$th path in $\pset_1(B_i)\setminus\set{P^*_i}$. Once we process every fake edge on path $P$ and on path $P'$ in this fashion, we obtain a final set $\pset_1$ of $6$ paths connecting $x$ to $y$ in graph $B$. Moreover, it is easy to verify that the paths in $\pset_1(B)\cup \pset_2(B)$ cause edge-congestion at most $6$.
\end{proofof}

\subsection{Proof of Lemma~\ref{lem: block containment}}
\label{apd:block_containment}

In this section we provide the proof of Lemma~\ref{lem: block containment}.

Let $C_2\in \cset(H_2)$ be the connected component of $H_2$ containing $B_2$. Since $H_2\subseteq H_1$, there is a connected component $C_1\in \cset(H_1)$ containing $C_2$. Let $Z_2\in \zset(C_2)$ be the super-block containing $B_2$. Since $C_2\subseteq C_1$, it is easy to see that there is a super-block $Z_1\in \zset(C_1)$ containing $Z_2$. 
	
We now consider the block decomposition $\lset(Z_1)$ of $Z_1$. We let $B_1\in \lset(Z_1)$ be the pseudo-block containing all vertices of $\tilde B_2$, that minimizes $|V(B_1)|$. Since $B_2\subseteq Z_1$ and $Z_1\in \lset(Z_1)$, such a pseudo-block must exist. We now show that $B_1$ has all required properties.
	
We start by showing that $V(\tilde B_2)\subseteq V(\tilde B_1)$. Assume for contradiction that this is not the case. Since $V(\tilde B_2)\subseteq V(B_1)$, there must be a child block $B_3$ of $B_1$ that contains at least one vertex of $\tilde B_2$ as an inner vertex. However, $B_3$ does not contain all vertices of $\tilde B_2$, or  we would have chosen $B_3$ instead of $B_1$. Therefore, there are two vertices $u,v\in \tilde B_2$, with $u\in B_3$ and $v\not\in B_3$. The two endpoints of $B_3$ then disconnect $u$ from $v$ in graph $H_1$. We next show that this is impossible. 
Since $|V(\tilde B_2)|>3$ and $\tilde B_2$ is $3$-connected, there are three internally disjoint paths in $\tilde B_2$ connecting $u$ to $v$, that we denote by $P_1,P_2$ and $P_3$; note that these paths may contain fake edges. However, from Observation \ref{obs: paths for fake edges}, for every fake edge $e$ in $\tilde B_2$, there is a path $P(e)$ in $H_2$, connecting its endpoints, that is internally disjoint from $V(\tilde B_2)$. Moreover, if $e\neq e'$ are two distinct fake edges of $\tilde B_2$, then $P(e)$ and $P(e')$ are internally disjoint. By replacing every fake edge on paths $P_1,P_2$ and $P_3$ with the corresponding path in $H_2$, we obtain three internally disjoint paths $P_1',P_2',P_3'$ in graph $H_2$ connecting $u$ to $v$. Since $H_2\subseteq H_1$, these three paths belong to $H_1$ as well. However, from the definition of blocks, the endpoints of the block $B_3$ must disconnect $u$ from $v$ in $H_1$, a contradictiton.
We conclude that $V(\tilde B_2)\subseteq V(\tilde B_1)$.
	
Consider now some real edge $e=(u,v)\in E(\tilde B_2)$. Since $u,v\in \tilde B_1$, from the definition of graph $\tilde B_1$, edge $e$ must also belong to $\tilde B_1$, unless $u$ and $v$ are the endpoints of the block $B_1$.
	
Let $A$ be the collection of all fake edges in $\tilde B_2$. From Observation \ref{obs: paths for fake edges}, for every fake edge $e$ in $\tilde B_2$, there is a path $P(e)$ in $H_2$, connecting its endpoints, that is internally disjoint from $V(\tilde B_2)$. Moreover, the paths in $\pset=\set{P(e)\mid e\in A}$ are internally disjoint. Since $H_2\subseteq H_1$, the paths of $\pset$ also belong to graph $H_1$. Consider any such path $P=P(e)$. Note that not all edges of $P$ may lie in $\tilde B_1$. We partition the path $P$ into segments $\sigma_1,\eta_1,\sigma_2,\eta_2,\ldots,\eta_{r-1},\sigma_r$, where for all $1\leq i\leq r$, $\sigma_i$ is contained in $\tilde B_1$ (where all edges of $\sigma_i$ are real edges), and for all $1\leq i<r$, $\eta_i$ is internally disjoint from $\tilde B_1$. It is possible that a segment $\sigma_i$ only consists of a single vertex. We refer to the segments $\sigma_i$ as \emph{$\sigma$-segments}, and we refer to the segments $\eta_i$ as \emph{$\eta$}-segments.
We now consider two cases. The first case happens when there is a single $\eta$-segment, and both $\sigma_1,\sigma_2$ contain a single vertex. In this case we say that path $P$ is a \emph{type-1} path. Let $u,v$ be the endpoints of $P$. Then either $u,v$ are the endpoints of $B_1$, or there is a child block $B_3$ of $B_1$ whose endpoints are $u,v$. In either case, $(u,v)$ is a fake edge of $\tilde B_1$. %Notice that set $\set{P(e)\mid e\in A}$ of paths may contain several type-1 paths whose endpoints are $u,v$. %\snote{I am assuming it cannot be $u$ and $v$ at the same time because there are no parallel edges. however we can have $u$ or $v$ for different edges, right? if so we can replace $u,v$ by $u$ or $v$}
%However, since these paths are internally disjoint, the number of fake edges in $\tilde B_1$ connecting $u$ to $v$ must be at least as large as the number of such paths, so we can use a distinct fake edge for each such path. We then let $P'(e)$ contain a single fake edge of $\tilde B_1$ connecting $u$ to $v$.

Assume now that the first case does not happen; we then say that path $P$ is a type-2 path.
Consider now some segment $\eta_i$, for $1\leq i<r$, and let $(v_i,u_i)$ be its endpoints. Then either $(v_i,u_i)$ are the endpoints of the block $B_1$ (and then fake edge $(v_i,u_i)$ lies in $\tilde B_1$); or $(v_i,u_i)$ are the endpoints of some child block $B_3$ of $B_1$ (in which case the fake edge $(v_i,u_i)$ also lies in $\tilde B_1$). We replace each such segment $\eta_i$ with the corresponding fake edge of $\tilde B_1$, obtaining a new path $P'(e)$ that is contained in $\tilde B_1$. Consider now the resulting set $\pset'=\set{P'(e)\mid e\in A}$ of paths. Recall that the original paths in $\set{P(e)\mid e\in A}$ were internally disjoint. It is easy to see from our construction that the paths in $\pset'$ are internally disjoint. % (the following sentence is not good because internally disjoint means they can't share inner vertices, not edges) Indeed, if $P'(e),P'(e')\in \pset'$ share an edge $\hat e=(a,b)$, then both $a$ and $b$ must belong to both of the original paths $P(e),P(e')$. Since both paths are internally disjoint, $a$ and $b$ must be the endpoints of both paths, and both paths must be type-1 paths. But as we have observed above, a distinct fake edge can be used for each such path.

Finally, denote the endpoints of $B_1$ by $u$ and $v$, and assume that $\tilde B_2$ contains edge $(u,v)$ as a real edge. Recall that, by the definition, $\tilde B_1$ does not contain a real edge $(u,v)$, but it contains a fake edge $e=(u,v)$. We next show that this edge does not lie on any path in $\pset'$. Assume for contradiction that some path $P'(e')\in \pset'$ contains the edge $e$. 
Recall that path $P'(e')$ cannot contain any vertex of $\tilde B_2$ as an inner vertex.
Since both $u$ and $v$ belong to $\tilde B_2$, path $P'(e')$ must consist of a single edge -- the edge $e$.  But then the original path $P(e')$ consists of a single $\eta$-segment, that connects $u$ to $v$, and so the fake edge $e'$ of $\tilde B_2$ has the same endpoints as the edge $e$. This leads to a contradiction, since $e$ is a real edge of $\tilde B_2$, and $\tilde B_2$ contains no parallel edges between two vertices, where one of the edges is real and another edge is fake.

\section{Proofs Omitted from Section \ref{sec:canonical drawing}}
\subsection{Proof of Lemma~\ref{lem:uncrossing}}
\label{apd:non_interfering}
%In this section we provide the proof of Lemma~\ref{lem:uncrossing}.
In order to simplify the notation, we denote $H'$ by $H$.
%First we discard from $H$ all edges that do not belong to $E(\qset)$. 
%We replace each edge $e$ in $H$ with $c_{\qset}(e)$ copies, and let each path in $\qset$ that contains $e$ uses a distinct copy of $e$. 
%Therefore, the modified graph (that we still denote by $H$) may contain parallel edges, and the paths in $\qset$ now becomes edge-disjoint. 
%It is easy to see that the given drawing $\psi$ can be modified to a planar drawing $\psi'$ of the new graph $H$.
%We view the paths of $\qset$ as originating from vertices in $\Gamma$ and terminating at $u$, so each edge in $H$ is now a directed edge.
%The algorithm consists of two steps.
%In the first step, we compute a set $\qset_1$ of paths routing $\Gamma$ to $u$, such that (i) for each edge $e\in E(H)$, $c_{\qset_1}(e)\le c_{\qset_1}(e)$; and (ii) if we view the paths of $\qset_1$ as originating from vertices in $\Gamma$ and terminating at $u$, then there is no directed cycle in the graph induced by all paths in $\qset_1$.
%This defines a topological order $\oset$ on the vertices of $V(\qset_1)$.
%In the second step, we process the vertices of $H$ one by one and construct the desired set $\qset'$ of paths, together with its non-interfering representation. %We now describe each step in detail.

Let $\hat{H}$ be a graph that is obtained from $H$, by adding a new vertex $s$ to it, and connecting it to every vertex $v\in \Gamma$. We set edge capacities in graph $\hat H$ as follows. Each edge in $\set{(s,v)\mid v\in \Gamma}$ has capacity $1$, and each edge $e\in E(H)$ has capacity $\cong_{\qset}(e)$. Next, we compute maximum flow $F$ from $s$ to $u$ in graph $\hat H$. Since all edge capacities are integral, we can ensure that the flow $F$ is integral as well. It is immediate to verify that the value of the maximum $s$-$u$ flow in $\hat H$ is $|\Gamma|$. Moreover, we can ensure that for every edge $e=(u',v')$, the flow is only sent in one direction (either from $u'$ to $v'$ or from $v'$ to $u'$, but not both). We can also ensure that the flow is acyclic. Using the standard flow-path decomposition of the flow $F$, and deleting the first edge from each resulting flow-path, we obtain a set $\tilde \qset=\set{\tilde Q_v\mid v\in \Gamma}$ of directed paths in graph $\hat H$, where each path $\tilde Q_v$ connects $v$ to $u$. Moreover, for every edge $e\in E(\hat H)$, $\cong_{\tilde \qset}(e)\leq \cong_{\qset}(e)$.

Let $H^*$ be the directed graph that is obtained by taking the union of the paths in $\tilde \qset$, whose edges are directed towards the edge $u$; for every edge $e\in E(H)$, we add $\cong_{\tilde \qset}(e)$ copies of $e$ to this graph. Notice that $H^*$ is a Directed Acyclic Graph, and so there is an ordering $\oset$ of its vertices, such that, for every pair $x,y\in V(H^*)$ of vertices, if there is a path from $x$ to $y$ in $H^*$, then $x$ appears before $y$ in this ordering. In particular, ordering $\oset$ must be consistent with the paths in $\tilde \qset$. Notice that we can use the drawing $\psi$ of $H$ in order to obtain a drawing of graph $H^*$, as follows. First, we delete from $\psi$ all edges and vertices that do not participate in the paths in $\tilde \qset$. Next, every edge $e=(u,v)$ that remains in the current drawing, we create $\cong_{\tilde \qset}(e)$ copies of the edge $e$, each of which is drawn along the original drawing of $e$, with the copies being drawn close to each other. For each such edge $e$, we will view the different copies of $e$ as different edges, and we think of each of these copies as belonging to a distinct path in $\tilde \qset$. We denote this new drawing of graph $H^*$ by $\psi^*$. Note that $\psi^*$ is a planar drawing.

For every vertex $v\in V(H^*)$, we let $\eta(v)$ be a small disc around the image of $v$ in $\psi^*$. We denote by $\delta^+(v)$ the set of all edges that are leaving $v$ in $H^*$, by $\delta^-(v)$ the set of all edges entering $v$, and by $\delta(v)=\delta^+(v)\cup \delta^-(v)$. For every edge $e\in \delta(v)$, we denote by $p_e(v)$ the unique point on the boundary of $\eta(v)$ that the image of $e$ in $\psi^*$ contains. We use the following simple observation.

\begin{observation}
	There is an efficient algorithm to compute, for every vertex $v\in V(H^*)$, a perfect matching $M(v)\subseteq \delta^-(v)\times \delta^+(v)$ between the edges of $\delta^-(v)$ and $\delta^+(v)$, and, for every pair $(e,e')\in M(v)$ of edges, a curve $\zeta(e,e')$ that is contained in $\eta(v)$ and connects $p_e(v)$ to $p_{e'}(v)$, such that all curves in $\set{\zeta(e,e')\mid (e,e')\in M(v)}$ are disjoint from each other.
\end{observation}

\begin{proof}
	Observe first that the paths in $\tilde \qset$ define a perfect matching between edges of $\delta^-(v)$ and edges of $\delta^+(v)$, as each path $Q\in \tilde \qset$ that contains $v$ must contain exactly one edge from $\delta^-(v)$ and exactly one edge from $\delta^+(v)$. 
	
	We maintain a disc $\eta$ in the plane; originally, $\eta=\eta(v)$, and we gradually delete some areas of $\eta$, making it smaller. We also start with $M(v)=\emptyset$, and we perform $|\delta^-(v)|$ iterations. In every iteration, we select a pair $e\in \delta^-(v)$, $e'\in \delta^+(v)$ of edges that appear consecutively in the current drawing. We add $(e,e')$ to $M(v)$, and we delete both of these edges from the current drawing. Additionally, we select a curve $\zeta(e,e')$, that is contained in the current disc $\eta$, connecting $p_e(v)$ to $p_{e'}(v)$, such that curve $\zeta(e,e')$ is disjoint from the images of all edges that remain in the drawing, and is close to the boundary of the disc $\eta$. Curve $\zeta(e,e')$ splits the disc $\eta$ into two discs, $\eta'$ and $\eta''$, where exactly one of the discs (we assume that it is $\eta'$) contains the image of $v$, while the other disc is disjoint from the images of all edges that remain in the drawing. We set $\eta=\eta'$ and continue to the next iteration. It is immediate to see that, when the algorithm terminates, we obtain the desired matching $M(v)$, and a set $\set{\zeta(e,e')\mid (e,e')\in M(v)}$ of curves with the desired properties.
\end{proof}

We now gradually transform the paths in $\tilde \qset$ in order to turn them into a set of non-interfering paths, as follows. We process all vertices in $V(H^*)$ one-by-one, according to the ordering $\oset$. We now describe an iteration when vertex $v$ is processed. 
Let $\pset(v)\subseteq \tilde \qset$ be the set of paths containing the vertex $v$. For every path $\tilde Q_t\in \pset(v)$, we delete the unique edge of $\delta^+(v)$ that lies on this path, thereby decomposing $\tilde Q_t$ into two sub-paths: path $P^1_t$ connecting $t$ to $v$, and path $P^2_t$ connecting some vertex $v'$ that is incident to an edge of $\delta^+(v)$ to $u$. Let $\pset_1(v)=\set{P^1_t\mid \tilde Q_t\in \pset(v)}$, and let $\pset_2(v)=\set{P_t^2\mid \tilde Q_t\in \pset(v)}$. We will now ``glue'' these paths together using the matching $M(v)$. Specifically, we construct a new set $\tilde \qset'=\set{\tilde Q'_t\mid t\in \Gamma}$ of paths as follows. Consider any vertex $t\in \Gamma$. 
If the original path $\tilde Q_t\in \tilde \qset$ does not lie in $\pset(v)$, then we let $\tilde Q'_t=\tilde Q_t$. Otherwise, consider the unique path $P_t^1$ in $\pset_1(v)$ that originates at $t$, and let $e$ be the last edge on this path. Let $e'$ be the unique edge of $\delta^+(v)$ that is matched to the edge $e$ by the matching $M(v)$, and let $\tilde Q_{t'}\in \pset(v)$ be the unique path that contained $e'$.
We then let the new path $\tilde Q'_t$ be the concatenation of the path $P_1(t)$ and the path $P_2(t)$, thereby making $e$ and $e'$ consecutive on this path.

Once we process every vertex of $H^*$, we obtain the final collection $\qset'$ of paths, routing $\Gamma$ to $u$. Clearly, for every edge $e\in E(H)$, $\cong_{\qset'}(e)\leq \cong_{\qset}(e)$. It is also easy to verify that the ordering $\oset$ of the vertices of $V(H^*)$ is consistent with the paths in $\qset'$, since every path in $\qset'$ is a directed path in the directed acyclic graph $H^*$. We extend $\oset$ to an ordering that includes all vertices of $H$ arbitrarily. We next show that the paths in $\qset'$ are non-interfering, by providing a non-interfering representation of these paths. This representation exploits the drawing $\psi^*$ of the graph $H^*$, and the curves $\set{\zeta_{e,e'}(v)}_{v\in V(H^*), (e,e')\in M(v)}$ that we defined. Consider any path $Q\in \tilde \qset'$, and let $Q=(v_1,v_2,\ldots,v_r=u)$. For all $1\leq i<r$, we denote $e_i=(v_i,v_{i+1})$. The corresponding curve $\gamma(Q)$ is a concatenation of the following curves from the drawing $\psi^*$:

\begin{itemize}
	\item The image of edge $e_1$, from the image of $v_1$, to the point $p_{e_1}(v_2)$ on the boundary of the disc $\eta(v_2)$;
	\item For all $1\leq i<r$, curve $\zeta_{e_i,e_{i+1}}$, connecting the point $p_{e_i}(v_{i+1})$ to  the point $p_{e_{i+1}}(v_{i+1})$ in $\eta(v_{i+1})$;
	\item For all $1<i<r$, the image of the edge $e_i$, between points $p_{e_i}(v_i)$ and $p_{e_i}(v_{i+1})$; and
	\item Image of edge $e_{r-1}$, from $p_{e_{r-1}}(v_{r-1})$ to the image of $v_r$.
\end{itemize}

It is immediate to verify that we can draw all segments of $\gamma(Q)$, such that all resulting curves in $\set{\gamma(Q')}_{Q'\in \qset'}$ are disjoint from each other, and each of them is drawn in the thin strip $S_Q$ around the image of $Q'$ in $\psi^*$. From the definition of the drawing $\psi^*$, the resulting curves are a valid non-interfering representation of the paths in $\qset'$ with respect to the original drawing $\psi$ of $H$.

\subsection{Proof of Lemma~\ref{lem:find_confluent_paths}}
\label{apd:find_confluent_paths}

%In this section we provide the proof of Lemma~\ref{lem:find_confluent_paths}.
%Let $H$ be a graph. Given a set $\pset$ of paths in $H$ and a number $\beta>0$, we define the \emph{congestion} $c(\pset)$ of the set $\pset$ of paths to be the maximum number of paths that contain a same edge, namely $c(\pset)=\max_{e\in H}\set{c_{\pset}(e)}$.

%We use the following lemmas.

%\mynote{the choice of notation $C$ for contours is not good because throughout the paper we are using $C$ for clusters. Curves are usually denoted by $\lambda,\zeta,\gamma$ and such. But if it's too complicated to fix this right now we can leave it for later.}

%(1) In Corollary F.3 does $\hat \qset$ contain a path from each vertex on the interface to $u^*$? if so it should say so.
%\mynote{(2) in section F, when you use “mapping” h it should be “model” h. Lemma F.4 is saying “with h being the mapping”. But you mean that the algorithm finds the mapping, right? then you should say “together with the mapping” or “together with the model”. You don’t “find a minor R in G”. You compute a model of a minor of R in G.}

Throughout the section, we use $R$ to denote the $(r\times r)$ grid, for some parameter $r$ that is an integral power of $2$, %\mynote{it should be "the" $k\times k$ grid, because there is only one of those.} 
and we use $I$ to denote the set of vertices in its last row. 

We use the following lemma.
\begin{lemma}
\label{lem: paths with bounded congestion square}
There is an efficient algorithm, that, given 
an $n$-vertex planar graph $H$ and a subset $S$ of $|S|=r$ vertices of $V(H)$ that are $\alpha'$-well-linked in $H$, for some $0<\alpha'<1$, computes a distribution $\dset$ over pairs $(u^*,\qset)$, where $u^*$ is a vertex of $H$, and $\qset$ is a collection of %{\bf confluent}
paths in $H$ routing vertices of $S$ to $u^*$, such that the distribution $\dset$ has support size at most $O(r^2)$, and for each edge $e\in E(H)$, \[\mathbb{E}_{(u^*,\qset)\in \discset}[(\cong_{\qset}(e))^2]=O\left(\frac{\log r}{(\alpha')^4}\right).\]
\end{lemma}

We provide the proof of Lemma~\ref{lem: paths with bounded congestion square} later, after we complete the proof of Lemma~\ref{lem:find_confluent_paths} using it.
%As our second step, we compute the desired vertex $\hat u^*$ and the set $\hat \qset$ of paths. 
%We first apply the algorithm in Claim~\ref{lem:confluent_paths_in_grid} to compute a distribution $\discset$ on pairs $(u^*,\qset)$. 
Let $\dset$ be the distribution we get from the algorithm in Lemma~\ref{lem: paths with bounded congestion square} applied to graph $H$, set $S$ and parameter $\alpha'$.
%Recall that for each edge $\tilde e$ in $G$, we have $\mathbb{E}_{(\tilde u^*,\tilde\qset)\in\tilde\discset}[(\cong_{\tilde\qset}(\tilde e))^2]=O(\log r)$. Therefore, 
From linearity of expectation, $\mathbb{E}_{(u^*,\qset)\in \discset}[\sum_{e\in E(H)}w(e)\cdot(\cong_{\qset}(e))^2]=O\left(\frac{\log r}{(\alpha')^4}\cdot \sum_{e\in E(H)}w(e)\right)$.
Clearly, there exists a pair $(\hat u^*,\hat\qset)$ with non-zero probability in $\discset$, such that
$\sum_{e\in E(H)}w(e)\cdot(\cong_{\hat\qset}(e))^2=O\left(\frac{\log r}{(\alpha')^4}\cdot \sum_{e\in E(H)}w(e)\right)$.
Since the distribution $\discset$ has support size $O(r^2)$, such a pair can be found by checking all pairs $(u^*,\qset)$ with non-zero probability in $\dset$.

%\mynote{later on you denote by $I$ the last row, not the set of its vertices.}
The remainder of this section is dedicated to the proof of Lemma~\ref{lem: paths with bounded congestion square}.

We use the following claim from~\cite{chuzhoy2011algorithm} and its corollary. We note that
the claim appearing in \cite{chuzhoy2011algorithm} is somewhat weaker, but their proof immediately implies the stronger result that we state below.
%(Claim D.11 in~\cite{chuzhoy2011algorithm}). We note that the claim in~\cite{chuzhoy2011algorithm} is somewhat weaker. In particular, the confluency of the path set is not guaranteed. But the proof in~\cite{chuzhoy2011algorithm} immediately implies the stronger result that we state below.
\begin{claim}[Claim D.11 from \cite{chuzhoy2011algorithm}]
\label{lem:confluent_paths_in_grid}
There is a distribution $\discset$ over pairs $(u^*,\qset)$, where $u^*$ is a vertex of $R$, and $\qset$ is a collection of %{\bf confluent}
paths in $R$ connecting every vertex of $I$ to $u^*$, such that, for each edge $e\in E(R)$, $\mathbb{E}_{(u^*,\qset)\in \discset}[(\cong_{\qset}(e))^2]=O(\log r)$. Moreover, such a distribution with support size at most $O(r^2)$ can be computed efficiently.
\end{claim}

We say that a graph $J$ is a \emph{minor} of a graph $G$, iff there is a function $h$, mapping each vertex $v\in V(J)$ to a connected subgraph $h(v)\subseteq G$, and each edge $e=(u,v)\in E(J)$ to a path $h(e)$ in $G$ connecting a vertex of $h(u)$ to a vertex of $h(v)$, such that: (i) for all $u,v\in V(J)$, if $u\ne v$, then $h(u)\cap h(v)=\emptyset$; and (ii) the paths in the set $\set{h(e)\mid e\in E(J)}$ are mutually internally disjoint, and they are internally disjoint from $\bigcup_{v\in V(J)}h(v)$. 
A function $h$ satisfying these conditions is called a \emph{model} of $J$ in $G$. %We also say that a model $h$ \emph{embeds $H$ into $G$}.
We use the following corollary of Claim \ref{cor:confluent_paths_in_grid}.

\begin{corollary}
\label{cor:confluent_paths_in_grid}
There is an efficient deterministic algorithm, that, given a graph $G$ that contains $R$ as a minor, together with the model $h$ of $R$ in $G$, and for each vertex $x\in V(R)$, a vertex $v_x\in h(x)$, 
a distribution $\tilde\discset$ on pairs $(\tilde u^*,\tilde\qset)$, where $\tilde u^*$ is a vertex in $G$, and $\tilde\qset$ is a collection of %{\bf confluent}
paths in $G$ connecting every vertex of $\set{v_x\mid x\in I}$ to $\tilde u^*$, such that distribution with support size at most $O(r^2)$, for each edge $e\in E(G)$, $\mathbb{E}_{(\tilde u^*,\tilde\qset)\in \tilde\discset}[(\cong_{\tilde\qset}(e))^2]=O(\log r)$.
%computes a vertex $\hat u^* \in V(G)$ and a set $\hat\qset$ of $r$ paths in $G$, each connecting a distinct vertex of $\set{v_x\mid x\in I}$ to $\hat u^*$, such that
%$\sum_{e\in E(G)}w(e)\cdot (\cong_{\hat\qset}(e))^2\leq O\left(\log r\cdot\left(\sum_{e\in E(G)}w(e)\right)\right)$.
\end{corollary}
\begin{proof}
%For each $x\in V(R)$, we arbitrarily fix a vertex $v_x$ in $h(x)$. 
%As our first step, we show that there is a distribution $\tilde\discset$ on pairs $(\tilde u^*,\tilde\qset)$, where $\tilde u^*$ is a vertex in $G$, and $\tilde\qset$ is a collection of %{\bf confluent} paths in $G$ connecting every vertex of $\set{v_x\mid x\in I}$ to $\tilde u^*$, such that, for each edge $e\in E(G)$, $\mathbb{E}_{(\tilde u^*,\tilde\qset)\in \tilde\discset}[(\cong_{\tilde\qset}(e))^2]=O(\log r)$. We also show that such a distribution with support size at most $O(r^2)$ can be computed efficiently.
For each vertex $x\in V(R)$, we let $\delta(x)$ be the set of edges incident to $x$ in $R$, so $|\delta(x)|\le 4$. For each edge $e\in \delta(x)$, we denote by $b_x(e)$ the vertex in $h(x)$ that serves as the endpoint of the path $h(e)$. %Therefore, $|b(x)|\le 4$ as well. 
We denote $B(x)=\set{b_x(e)\mid e\in \delta(x)}$.
We now select: (i) for each pair $b_x(e),b_x(e')$ of distinct vertices of $B(x)$, a path $P^x_{e,e'}$ in $h(x)$ that connects $b_x(e)$ to $b_x(e')$; and
(ii) for each vertex $b_x(e)\in B(x)$, a path $W^x_e$ connecting $v_x$ to $b_x(e)$.
We call these paths \emph{auxiliary paths in $h(x)$}.

We now apply Claim~\ref{lem:confluent_paths_in_grid} to $R$. Let $\discset$ be the distribution over pairs $(u^*,\qset)$ that we get, where $u^*$ is a vertex of $R$, and $\qset$ is a collection of paths in $R$ connecting every vertex of $I$ to $u^*$. We now use the model $h$ of $R$ in $G$, and the auxiliary paths to transform the distribution $\discset$ into another distribution $\tilde\discset$ over pairs $(\tilde u^*, \tilde \qset)$, where where $\tilde u^*$ is a vertex in $\set{h(x)\mid x\in V(R)}$, and $\tilde\qset$ is a collection of %{\bf confluent}
paths in $G$ connecting every vertex of $\set{v_x\mid x\in I}$ to $\tilde u^*$, as follows.
For each pair $(u^*,\qset)$ with non-zero probability in $\discset$, we define a corresponding pair $(\tilde u^*,\tilde \qset)$ as follows. We set $\tilde u^*=v_{u^*}$. Let $Q=(x_1,\ldots,x_{r-1},x_r=u^*)$ be a path in $\qset$, where we denote $e_i=(x_i,x_{i+1})$ for each $1\le i\le r-1$.
We let $\tilde Q$ be the path obtained by concatenating the paths  $W^{x_1}_{e_1},h(e_1),P^{x_2}_{e_1,e_2},h(e_2),P^{x_3}_{e_2,e_3}, \ldots,h(e_{r-1}),W^{x_r}_{e_{r-1}}$. It is easy to verify that the path $\tilde Q$ is a path in $G$ that connects $v_{x_1}$ to $v_{u^*}$. We then let $\tilde \qset=\set{\tilde Q\mid Q\in \qset}$. Therefore, $\tilde\qset$ is a collection of paths in $G$ connecting every vertex of $\set{v_x\mid x\in I}$ to $\tilde u^*$. To define $\tilde \discset$, we simply assign, for every pair $(u^*,\qset)$ with non-zero probability in $\discset$, the same probability to the pair $(\tilde u^*, \tilde Q)$. %This completes the description of the distribution $\tilde \discset$.

It remains to show that, for each edge $\tilde e\in E(G)$, $\mathbb{E}_{(\tilde u^*,\tilde\qset)\in \tilde\discset}[(\cong_{\tilde\qset}(\tilde e))^2]=O(\log r)$. 
%For brevity we call this \emph{congestion property}.
We first consider an edge $\tilde e$ that does not belong to any subgraph of $\set{h(x)\mid x\in V(R)}$. Clearly either $\tilde e$ belongs to a unique path $h(e)$ in $\set{h(e)\mid e\in R}$, or it does not belong to any path in $\set{h(e)\mid e\in R}$. If the latter case happens, then $\mathbb{E}_{(\tilde u^*,\tilde\qset)\in \tilde\discset}[(\cong_{\tilde\qset}(\tilde e))^2]=0$. If the former case happens, then
$\mathbb{E}_{(\tilde u^*,\tilde\qset)\in \tilde\discset}[(\cong_{\tilde\qset}(\tilde e))^2]\le \mathbb{E}_{(u^*,\qset)\in\discset}[(\cong_{\qset}(e))^2]=O(\log r)$.
%Therefore, the edge $\tilde e$ satisfies the congestion property.
Consider now an edge $\tilde e$ in $h(x)$ for some vertex $x\in V(R)$. Note that, from the construction of $\tilde\discset$, whenever the edge $\tilde e$ is contained in some path $\tilde Q$, the corresponding path $Q$ in $R$ has to contain at least one edge of $\delta(x)$. Therefore, for each pair $(u^*,\qset)$ with non-zero probability in $\discset$, $\cong_{\tilde\qset}(\tilde e)\le \sum_{e\in \delta(x)}\cong_{\qset}(e)$. As a result,  
$\mathbb{E}_{(\tilde u^*,\tilde\qset)\in \tilde\discset}[(\cong_{\tilde\qset}(\tilde e))^2]\le \mathbb{E}_{(u^*,\qset)\in\discset}[(4\cdot\max_{e\in \delta(x)}\set{\cong_{\qset}(e)})^2]=O(\log r)$.
\iffalse
As our second step, we compute the desired vertex $\hat u^*$ and the set $\hat \qset$ of paths. 
%We first apply the algorithm in Claim~\ref{lem:confluent_paths_in_grid} to compute a distribution $\discset$ on pairs $(u^*,\qset)$. 
Recall that for each edge $\tilde e$ in $G$, we have $\mathbb{E}_{(\tilde u^*,\tilde\qset)\in\tilde\discset}[(\cong_{\tilde\qset}(\tilde e))^2]=O(\log r)$. Therefore, by linearity of expectation, we get that $\mathbb{E}_{(\tilde u^*,\tilde \qset)\in \tilde\discset}[\sum_{\tilde e\in G}w(\tilde e)\cdot(\cong_{\tilde \qset}(\tilde e))^2]=O(\log r\cdot\left(\sum_{\tilde e\in G}w(\tilde e)\right))$.
Clearly, there exists a pair $(\hat u^*,\hat\qset)$ with non-zero probability in $\tilde\discset$, such that
$\sum_{\tilde e\in G}w(\tilde e)\cdot(\cong_{\tilde \qset}(\tilde e))^2=O(\log r\cdot\left(\sum_{\tilde e\in G}w(\tilde e)\right))$.
Since the distribution $\tilde\discset$ has support size $O(r^2)$, such a pair can be found by checking all pairs $(\tilde u^*,\tilde\qset)$ with non-zero probability.
\fi
\end{proof}

We use the following lemma, whose proof is deferred to Section \ref{apd:proof_of_route_to_grid}.
\begin{lemma}\label{lem:route_to_grid}
There exists an efficient algorithm that, given a planar graph $G$ with maximum vertex degree $\Delta$ and a set $S$ of $r$ vertices that is $\alpha'$-well-linked in $G$ for some $0<\alpha'<1$, computes (i) an $R$-minor in $G$, where $R$ is the $(k\times k)$ grid and $k=\Omega(\alpha' r/\poly(\Delta))$ and $k$ is an integral power of $2$, together with a model $h$ of $R$ in $G$; (ii) for each vertex $x\in V(R)$, a vertex $v_x\in h(x)$ in $G$; and (iii) $k$ edge-disjoint paths in $G$, each connecting a distinct vertex of $S$ to a distinct vertex of $\set{v_x\mid x\in I}$.
\end{lemma}

We now prove Lemma~\ref{lem: paths with bounded congestion square} using Corollary~\ref{cor:confluent_paths_in_grid} and Lemma~\ref{lem:route_to_grid}.
Recall that we are given a planar graph $H$ and a set $S\subseteq V(H)$ of vertices that are $\alpha'$-well-linked for some $0<\alpha'<1$. 
Let $\phi$ be a planar drawing of $H$.
Since the maximum vertex-degree $H$ could be as large as $n$, the size of the grid minor obtained by directly applying Lemma~\ref{lem:route_to_grid} to $H$ may be too small for us.
We therefore construct a graph $H'$ from $H$, that has constant maximum vertex-degree.
%Let $V'\subseteq V(H)$ contain all vertices in $H$ with degree at least $5$.
We start from $H$, and process every vertex of $V(H)$ as follows. Let $v$ be a vertex of $V(H)$, let $d=\deg_H(v)$ and let $e_1,\ldots,e_{d}$ be the edges incident to $v$ in $H$, indexed according to the circular ordering in which they enter the image of $v$ in the drawing $\phi$. We let $R_v$ be the $(d\times d)$-grid, and we denote the vertices of its first row by $x_1(v),\ldots,x_d(v)$. We then replace the vertex $v$ by the graph $R_v$, and let, for each $1\le i\le d$, the edge $e_i$ be now incident to vertex $x_i(v)$. Let $H'$ be the graph obtained after all vertices in $V(H)$ are processed.
It is easy to see that the max vertex-degree of $H'$ is $4$, and $H$ can be simply obtained from $H'$ by contracting each cluster in $\set{R_v\mid v\in V(H)}$ back into the vertex $v$, so each edge of $H$ is also an edge of $H'$. We use the following simple observations whose proofs are straightforward and are omitted here.

\begin{observation}
\label{obs: convert path set}
Let $\qset'$ be a set of paths in $H'$.
For each path $Q'\in \qset'$, let $Q$ be the path obtained from $Q'$ by contracting, for each vertex $v\in V(H)$, every edge of $R_v$ that lies on path $Q'$. Define $\qset=\set{Q\mid Q'\in \qset'}$. 
Then for each edge $e\in E(H)$, $\cong_{\qset}(e)\le \cong_{\qset'}(e)$.
\end{observation}

\begin{observation}
The set $S'=\set{x_1(v)\mid v\in S}$ of vertices is $\alpha'$-well-linked in $H'$.
\end{observation}

%We denote $r=|S|$.
For a set $\qset'$ of paths in $H$, we denote
$\cong(\qset')=\max_{e\in E(H)}\set{\cong_{\qset'}(e)}$.
%We will construct three sets $\qset'_1, \qset'_2, \qset'_3$ of paths in $H'$, and then concatenate the paths in $\qset'_1, \qset'_2, \qset'_3$ to obtain the final set $\qset'$ of paths.

First, we apply the algorithm in Lemma~\ref{lem:route_to_grid} to graph $H'$ and the input vertex set $S$, to compute a model $h$ of an $R$-minor in $H'$, where $R$ is the $(k\times k)$-grid with $k=\Omega(\alpha' r)$. 
%Recall that $I$ is the set of vertices in the last row of $R$.
We also obtain, for each vertex $x\in V(R)$, a vertex $v_x\in h(x)$ in $H'$; and a set of $k$ edge-disjoint paths in $H'$, each connecting a distinct vertex of $S'$ to a distinct vertex of $\set{x_1(z)\mid z\in I}$.
We denote this set of paths by $\qset_1$, and for each vertex $v\in S$, we denote the path in $\qset_1$ that contains $v$ as one of its endpoints by $Q^1_{v}$.

Let $S'_1\subseteq S'$ be the set of endpoints of paths in $\qset_1$ that lie in $S$, so $|S_1|=k=\Omega(\alpha' r)$. We arbitrarily partition the set $S'\setminus S'_1$ of vertices into  groups $S'_2,\ldots,S'_t$, where each group of $S'_2,\ldots,S'_{t-1}$ contains exactly $|S'_1|$ vertices, and the last group $S'_t$ contains at most $|S'_1|$ vertices, so $t=O(1/\alpha')$. Since the set $S'$ of vertices is $\alpha'$-well-linked in $H'$, for each $2\le i\le t$, there exists a set $\pset_i$ of paths in $H'$, each connecting a distinct vertex of $S'_i$ to a distinct vertex of $S'_1$, such that $\cong(\pset_i)=O(1/\alpha')$. Additionally, let the set $\pset_1$ contain, for each vertex $v\in S'_1$, a path that only contains the single vertex $v$. Denote $\qset_2=\bigcup_{1\le i\le t}\pset_i$. Then set $\qset_2$ contains, for each vertex $v\in S'$, a path connecting $v$ to a vertex in $S'_1$, such that each vertex $u\in S'_1$ serves as the endpoint of at most $O(1/\alpha')$ paths in $\qset_2$, and $\cong(\qset_2)\leq O(1/(\alpha')^2)$. For each path $Q\in \qset_2$, let $s(Q)\in S'_1$ be the endpoint of $Q$ in $S'_1$.

Next, we use the algorithm from  Corollary~\ref{cor:confluent_paths_in_grid} to compute a distribution
$\tilde\discset$ on pairs $(\tilde u^*,\tilde\qset)$, where $\tilde u^*$ is a vertex in $H'$, and $\tilde\qset$ is a collection of %{\bf confluent}
paths in $H'$ routing vertices of $\set{x_1(z)\mid z\in I}$ to $\tilde u^*$, such that distribution with support size at most $O(r^2)$, for each edge $e\in E(H')$, $\mathbb{E}_{(\tilde u^*,\tilde\qset)\in \tilde\discset}[(\cong_{\tilde\qset}(e))^2]=O(\log k)=O(\log r)$.
%$\sum_{e\in E(H)}w(e)\cdot (\cong_{\qset_3}(e))^2=O\left(\log k\cdot \left(\sum_{e\in E(H)}w(e)\right)\right)$. For each $v\in I$ we denote the path in $\qset_3$ with one endpoint being $v$ by $Q^3_v$. 

%We now construct the desired distribution $\dset$ on pairs $(u^*,\qset)$, where $u^*$ is a vertex in $H$ and $\qset$ is a collection of paths in $H$ routing $S$ to $u^*$, as follows.

We now construct a distribution $\hat\dset$ on pairs $(\hat u^*,\hat \qset)$, where $\hat u^*$ is a vertex in $H'$ and $\hat \qset$ is a collection of paths in $H'$ routing $S'$ to $\hat u^*$, as follows.
%show how to concatenate the paths in $\qset_1, \qset_2, \qset_3$ to obtain the final set $\qset$ of paths. 
Consider a pair $(\tilde u^*,\tilde\qset)$ in distribution $\tilde\dset$ with non-zero probability.
We let the set $\hat \qset$ contain, for each path $Q\in \qset_2$, a path formed by the concatenation of (i) the path $Q\in \qset_2$; (ii) the path $Q^1_{s(Q)}\in \qset_1$ (the path in $\qset_1$ whose endpoint in $S$ is $s(Q)$); and (iii) the path in $\tilde \qset$ that connects $s(Q)$ to $\tilde u^*$. It is clear that the set $\hat \qset$ contains, for each $v\in S'$, a path that connects $v$ to $\tilde u^*$. We add the pair $(\tilde u^*,\hat\qset)$ to distribution $\hat\dset$ with the same probability as the pair $(\tilde u^*,\tilde\qset)$ in distribution $\tilde\dset$.

From the definition of the set $\hat \qset$ of paths, and the property that each vertex in $S'_1$ serves as the endpoint of at most $O(1/\alpha')$ paths in $\qset_2$, we get that each path in $\qset_1$ serves as a subpath of at most $O(1/\alpha')$ paths in $\hat\qset$, and the same holds for $\tilde\qset$.
Therefore, for each edge $e\in H'$, 
\[
\begin{split}
\cong_{\hat\qset}(e)= &\text{ } O(1/\alpha')\cdot \cong_{\qset_1}(e)+\cong_{\qset_2}(e)+O(1/\alpha')\cdot \cong_{\tilde\qset}(e)\\
= &\text{ } O(1/(\alpha')^2)+O(1/\alpha')\cdot \cong_{\tilde\qset}(e).
\end{split}
\]
Therefore, for each edge $e\in H'$, 
\[
\begin{split}
\mathbb{E}_{(\hat u^*,\hat\qset)\in \hat{\dset}}[(\cong_{\hat\qset}(e))^2]= &\text{ } O(1/(\alpha')^4)+O(1/(\alpha')^3)\cdot \mathbb{E}_{(\tilde u^*,\tilde\qset)\in \tilde{\dset}}[\cong_{\tilde\qset}(e)]
+O(1/(\alpha')^2)\cdot \mathbb{E}_{(\tilde u^*,\tilde\qset)\in \tilde{\dset}}[(\cong_{\tilde\qset}(e))^2]\\
= &\text{ } O(\log r/(\alpha')^4).
\end{split}
\]

Finally, we define the distribution $\dset$ on pairs $(u^*,\qset)$ where $u^*$ is a vertex in $H$ and $\qset$ is a collection of paths in $H$ routing $S$ to $u^*$, as follows.
%show how to concatenate the paths in $\qset_1, \qset_2, \qset_3$ to obtain the final set $\qset$ of paths. 
Consider a pair $(\hat u^*,\hat\qset)$ in $\hat\dset$ with non-zero probability.
We let $\qset$ contains, for every path $\hat Q\in \hat\qset$ connecting a vertex of $x_1(v_1)$ to a vertex of $x_1(v_2)$ for a pair $v_1,v_2$ of vertices of $S$, the corresponding path in $H$ connecting vertex $v_1$ to vertex $v_2$ (obtained from $Q'$ by contracting each cluster in $\set{R_v\mid v\in V(H)}$ back into the vertex $v$). 
From Lemma~\ref{obs: convert path set}, for each edge in $E(H)$, $\cong_{\qset}(e)\le \cong_{\hat\qset}(e)$.
Therefore, it follows immediately that
$\mathbb{E}_{(u^*,\qset)\in\dset}[(\cong_{\qset}(e))^2]= O(\log r/(\alpha')^4)$.
\iffalse
Therefore, %\mynote{please replace $\alpha'^x$ with $(\alpha')^x$ everywhere.}
\begin{equation}
\begin{split}
\sum_{e\in E(H)}w(e)\cdot (\cong_{\qset}(e))^2\leq  
&\text{ }O\left(\sum_{e\in E(H)}w(e)\cdot\poly(\Delta)\cdot \left(\frac{1+\alpha'\cong_{\qset_3}(e)}{(\alpha')^2}\right)^2\right)\\
\leq &\text{ } O\left(\sum_{e\in E(H)}w(e)\cdot\poly(\Delta)\cdot \left(\frac{1+2\alpha' \cong_{\qset_3}(e)+(\alpha'\cong_{\qset_3}(e))^2}{(\alpha')^4}\right)\right)\\
\le &\text{ } O\left(\frac{\poly(\Delta)}{(\alpha')^4}\cdot\sum_{e\in E(H)}w(e)+\frac{\poly(\Delta)}{(\alpha')^3}\cdot\sum_{e\in E(H)}w(e)\cdot (\cong_{\qset_3}(e))^2\right)\\
\le &\text{ } O\left(\frac{\poly(\Delta)\log n}{(\alpha')^4}\cdot\left (\sum_{e\in E(H)}w(e)+1\right )\right).
\end{split}
\end{equation}
\fi
%This completes the proof of Lemma~\ref{lem: paths with bounded congestion square}, and also completes the proof of Lemma~\ref{lem:find_confluent_paths}.
In order to complete the proof of Lemma~\ref{lem: paths with bounded congestion square}, it now remains to prove Lemma \ref{lem:route_to_grid}, which we do next.

\subsubsection{Proof of Lemma \ref{lem:route_to_grid}}
\label{apd:proof_of_route_to_grid}
In this section we provide the proof of Lemma \ref{lem:route_to_grid}.
Our proof uses techniques similar to those used in the proof of Theorem 3.1 of~\cite{chekuri2004edge}.
%For a graph $G$, we say that a set $V'\subseteq V(G)$ of vertices is \emph{node-well-linked} in graph $G$ iff for every pair $A,B$ of disjoint subsets of $V'$ such that $|A|=|B|$, there exist $|A|$ node-disjoint paths in $G$ connecting vertices of $A$ to vertices of $B$.

We assume that we are given some fixed drawing of $G$ on the sphere. We fix a point $\nu$ on the sphere that does not belong to the image of $G$. A \emph{contour} $\lambda$ with respect to this drawing is a simple closed curve that does not contain point $\nu$, and only intersects the drawing at the vertices of $G$. We denote by $V_{\lambda}$ the set of vertices of $G$ whose image lies on $\lambda$, and we refer to $|V_{\lambda}|$ as the \emph{length} of $\lambda$. We say that a subset $A\subseteq V_{\lambda}$ of vertices is \emph{contiguous} iff $A$ contains all vertices of $G$ that lie on a contiguous subcurve of $\lambda$.
Clearly, a contour $\lambda$ separates the sphere into two open regions. We define the interior $\text{ins}(\lambda)$ of $\lambda$ to be the region not containing the point $\nu$, and define the graph $G_{\lambda}$ to be the subgraph of $G$ consisting of all edges and vertices whose image lies in $\lambda\cup\text{ins}(\lambda)$. 
%\mynote{is $\text{ins}\lambda$ an open region? then you should write "lie on $\lambda\cup ins(\lambda)$, because otherwise an edge whose image lies partly in \lambda and partly in $ins(\lambda)$ is not included.}

The proof consists of two steps. Throughout the proof, we set $\beta=\lceil\alpha'r/(100\Delta)\rceil$.
In the first step, we will construct a contour $\lambda$ such that at least half of vertices of $S$ lie in $\text{ins}(\lambda)$, and the following additional properties hold: %\mynote{looks like this should be a statement of a lemma?}
\begin{properties}{P}
\item \label{Pro1} $|V_{\lambda}|=\beta$;
\item \label{Pro2} for each pair $A,B\subseteq V_{\lambda}$ of disjoint equal-cardinality contiguous subsets of vertices, there exists $|A|$ node-disjoint paths in $G_{\lambda}$ connecting vertices of $A$ to vertices of $B$; and
\item \label{Pro3} there exist a set of $\lfloor\beta/2\rfloor$ edge-disjoint paths in $G$, each connecting a distinct vertex of $S$ that lies inside the interior of $\lambda$ to a distinct vertex of $V_{\lambda}$.
\end{properties}
In the second step, we will use the contour constructed in the first step in order to compute a grid minor and the edge-disjoint paths connecting vertices of $S$ to it. Before we describe each step in details, we state and prove the following observation. %Denote $r=|S|$.
\begin{observation}
\label{obs:slim_or_fat}
If $\lambda$ is a contour such that $|V_{\lambda}|\le \beta$, then the number of vertices of $S$ that lie in the interior of $\lambda$ is either at most $r/10$ or at least $9r/10$.
\end{observation}
\begin{proof}
Let $r'$ be the number of vertices of $S$ that lie in the interior of the contour $\lambda$, and let $r''$ be the number of vertices of $S$ that lie in the exterior of the contour $\lambda$. %\mynote{please eliminate "resp"}
Assume first that $r'\le r''$. 
%Consider the set $V(G_{\lambda})\setminus V_{\lambda}$ of vertices. 
Since the vertices of $S$ are $\alpha'$-well-linked in $G$, $|\out_G(V(G_{\lambda})\setminus V_{\lambda})|\ge \alpha'r'$. Note that every edge in $\out_G(V(G_{\lambda})\setminus V_{\lambda})$ must be incident to a vertex of $V_{\lambda}$, so $\alpha'r'\le \Delta|V_{\lambda}|\le \alpha'r/25$. Therefore, $r'\le r/25$. Assume now that $r'>r''$. It is easy to see that we can derive that $r''\le r/25$ similarly. Therefore, $r'\ge r-r''-|V_{\lambda}|\ge 9r/10$.
\end{proof}

We say that a contour $\lambda$ is \emph{short} iff $|V_{\lambda}|\le \beta$, and we say that a contour $\lambda$ is \emph{fat} iff the number of vertices of $S$ that lie in $\text{ins}(\lambda)$ is at least $9r/10$.
%For two contours $\lambda,\lambda'$, we say that $\lambda'$ is \emph{smaller} than $\lambda$ iff $G_{\lambda'}\subsetneq G_{\lambda}$, namely the interior of $\lambda$ contains a strict superset of vertices and edges of $G$ than the interior of $\lambda'$. \mynote{I think this is not how $G_{\lambda}$ was defined. Also "smaller" is not a good word to use here. Why not just state the containment thing in the claim without giving this a name? Also, wouldn't it be easier just to say that $ins(\lambda')\subseteq ins(\lambda)$ instead of talking about graphs?}
We use the following claim that appears in the (first paragraph of) proof of Theorem 3.1 of \cite{chekuri2004edge}.
\begin{claim}
\label{clm:nudge_contour}
There is an efficient algorithm that, given a short and fat contour $\lambda$ of length less than $\beta$, computes another short and fat contour $\lambda'$ of length exactly $\beta$, such that $G_{\lambda'}\subsetneq G_{\lambda}$.
\end{claim}

\subsubsection*{Step 1. Computing a Contour}
We now describe the algorithm for the first step. The algorithm maintains a contour $\hat \lambda$, that is initialized to be the small circle around the point $\nu$ that does not intersect any vertices of $G$.
The algorithm will iteratively update $\hat \lambda$, and will continue to be executed as long as not all properties \ref{Pro1}, \ref{Pro2}, \ref{Pro3} %\mynote{you should use the properties environment so the numbering is associated with a letter and it is clear what you are talking about. It was used e.g. in defining invariants in Sec 9.} 
are satisfied by $\hat \lambda$. Note that each of these properties can be checked efficiently. 
Clearly, $\hat{\lambda}$ is short and fat initially. We will ensure that this is true for all curves $\hat{\lambda}$ that are considered over the course of the algorithm. Moreover, as we will see in the description, graph $G_{\lambda}$ will become smaller after each iteration. Therefore, the algorithm will eventually terminate and output a desired contour.
We now describe an iteration. We distinguish between the following three cases.

\paragraph{Case 1. Property \ref{Pro1} is not satisfied.} In this case we simply apply the algorithm in Claim~\ref{clm:nudge_contour} to $\hat{\lambda}$, update $\hat \lambda$ to be the contour $\hat\lambda'$ that we obtain, and then continue to the next iteration.
From Claim~\ref{clm:nudge_contour} and Observation~\ref{obs:slim_or_fat}, the new contour is short, fat, and satisfies that $G_{\hat\lambda'}\subsetneq G_{\hat\lambda}$.

\paragraph{Case 2. Property \ref{Pro2} is not satisfied.} In this case we let $A, B$ be a pair of disjoint contiguous subsets of $V_{\lambda}$ such that $|A|=|B|$, and there does not exist a set of $|A|$ node-disjoint paths connecting vertices of $A$ to vertices of $B$. From among all such pairs of subsets, we choose one where $|A|$ is minimized.
We use the following claim, which is an immediate corollary of Theorem 3.6 in \cite{robertson1986graph}.

%\mynote{you should use a bulletted and not a numbered list. numbered lists should only be used if it's a sequence of things, or if you are planning to refer to the elements of the list by the number.}
\begin{claim}
There is a simple non-closed curve $J$, such that:
\begin{itemize}
\item $J$ is entirely contained in $\hat \lambda\cup \text{ins}(\hat \lambda)$; %\mynote{you should either use $ins(\lambda)$ or the word interior, and you should declare from the beginning if it's open or closed.}
\item $J$ only intersects the drawing of $G_{\hat \lambda}$ at its vertices, and only intersects with $\hat \lambda$ at the endpoints $a,b$ of $J$ (and $a,b$ are not vertices of $G$);
\item if we denote by $U_1$ and $U_2$ the two subcurves of $\hat \lambda$ connecting $a$ and $b$, then either all vertices of $A$ lie on $U_1$ and all vertices of $B$ lie on $U_2$, or all vertices of $A$ lie on $U_2$ and all vertices of $B$ lie on $U_1$; and
\item the number of vertices lying on $J$ is at most $|A|-1$.
\end{itemize}
Moreover, such a curve $J$ can be found efficiently.
\end{claim}
We compute such a curve $J$, and assume with loss of generality that vertices of $A$ lie on $U_1$ and vertices of $B$ lie on $U_2$.
We define the new contour $\lambda_1$ to be the concatenation of $U_1$ and $J$, and the new contour $\lambda_2$ to be the concatenation of $U_2$ and $J$. %\mynote{Please eliminate "resp"} 
Clearly, at least one of $\lambda_1$ and $\lambda_2$ contains at least $r/3$ vertices of $S$ in its interior. Assume without loss of generality that it is $\lambda_1$. 
Then $|V_{\lambda_1}|\le |V_{\hat \lambda}|-|B|+|J|\le \beta-1$.
From Observation \ref{obs:slim_or_fat}, the interior of $\lambda_1$ contains at least $9r/10$ vertices of $S$. We update $\hat \lambda$ to be $\lambda_1$ and continue to the next iteration. From the above discussion, $\lambda_1$ is a short and fat contour, and satisfies that $G_{\lambda_1}\subsetneq G_{\hat\lambda}$.

\paragraph{Case 3. Property \ref{Pro3} is not satisfied.} 
Let $\tilde G_{\hat \lambda}$ be a graph that is obtained from $G_{\hat{\lambda}}$ by adding to it (i) two new vertices $s,t$; (ii) for each vertex $v$ of $S$ that lies in the interior of $\hat \lambda$, an edge $(s,v)$; and (iii) for each vertex $v'$ in $V_{\hat \lambda}$, an edge $(t,v')$. We assign capacity $1$ to each edge of $\tilde G_{\hat \lambda}$, and then compute the minimum cut separating $s$ from $t$ in $\tilde G_{\hat \lambda}$. Since the property \ref{Pro3} is not satisfied, the minimun cut has value at most $\lfloor \beta/2 \rfloor-1$.
Denote by $E'$ the set of edges in the cut. Let $E'_1\subseteq E'$ contain all edges in $E'$ that are incident to $s$, so $|E'_1|\le \lfloor \beta/2 \rfloor$. Let $S'\subseteq S$ be the set of vertices in $S$ that lie in the interior of $\hat{\lambda}$, and are not an endpoint of an edge in $E'_1$, so $|S'|\ge 4r/5$. 
Let $E'_2=E'\setminus E'_1$, so $|E'_2|\le \lfloor \beta/2 \rfloor$. Note that, in the dual graph of $G_{\hat{\lambda}}$ with respect to its drawing, the edges corresponding to the edges of $E'_2$ form a set of cycles that separate the faces corresponding to vertices of $S'$ from the faces corresponding to vertices of $V_{\hat \lambda}$, and these cycles naturally form a set of disjoint closed curves in the drawing of $G$, such that each vertex of $S'$ lie in the interior of one of the curves, and all vertices of $V_{\hat \lambda}$ lie in the exterior of each of these curves. It is not hard to see that, each of these closed curves can be further transformed into a contour, by shifting every intersection between the curve and an edge of $G$ to an endpoint of the edge, such that (i) the resulting contour contains the same set of vertices and edges in its interior as the closed curve; and (ii) the length of the contour is at most the number of intersections between the curve and the drawing of $G_{\hat \lambda}$.
Let $\lambda_1,\ldots, \lambda_{l}$ be the contours that we obtain. From the above discussion, the total length of $\lambda_1,\ldots,\lambda_l$ is at most $\lfloor \beta/2 \rfloor$. Using reasoning similar to that in the proof of Observation~\ref{obs:slim_or_fat}, one of these contours contains at least $9r/10$ vertices of $S'$ (otherwise, the removal of all vertices lying on the countours separates the graph into connected components, each of which contains fewer than $9r/10$ vertices of $S'$, contradicting the well-linkedness of $S$). Assume without loss of generality that it is $\lambda_1$. We then update $\hat {\lambda}$ to $\lambda_1$ and continue to the next iteration.
Clearly, $\lambda_1$ is short and fat, and satisfies that $G_{\lambda_1}\subsetneq G_{\hat\lambda}$.

\subsubsection*{Step 2. Constructing the Grid Minor}

Let $\lambda$ be the contour that is obtained from the first step. We denote $V(\lambda)=\set{v_1,\ldots,v_{\beta}}$, where the vertices in $V(\lambda)$ are indexed in the clockwise order of their appearance on $\lambda$.
Denote $\gamma=\lfloor\beta/4\rfloor$.
We partition the vertices on $\lambda$ into $4$ consecutive subsets of cardinality  %\mynote{please use segments instead of chunks. Chunks is informal word} 
$\gamma$ each: for each $0\le i\le 3$, $B_i=\set{v_j\mid i\gamma+1\le h\le (i+1)\gamma}$.
From property \ref{Pro2}, we can find a set $\pset_0$ of $\gamma$ node-disjoint paths connecting vertices of $B_0$ to vertices of $B_2$, and another set $\pset_1$ of $\gamma$ node-disjoint paths connecting vertices of $B_1$ to vertices of $B_3$.
We now compute a grid minor in $G_{\lambda}$ from the sets $\pset_0$ and $\pset_1$ of paths.

Let $H$ be the graph consisting of all paths in $\pset_0$ and $\pset_1$. We first iteratively modify $H$ as follows. If there is an edge $e$ in $H$, such that in the graph $H\setminus \set{e}$, there is a set of $\gamma$ node-disjoint paths connecting vertices of $B_0$ to vertices of $B_2$, and another set of $\gamma$ node-disjoint paths connecting vertices of $B_1$ to vertices of $B_3$, then we delete $e$ from $H$ and continue to the next iteration. We call such an edge $e$ an \emph{irrelevant} edge. We iteratively remove irrelevant edges from $H$ in this way until we are not able to do so. Let $\hat H$ be the remaining graph, so $\hat{H}$ does not contain any irrelevant edge.
Let $\hat{\pset}_0$ be a set of $\gamma$ node-disjoint paths connecting vertices of $B_0$ to vertices of $B_2$ in $\hat H$, and let $\hat{\pset}_1$ be a set of $\gamma$ node-disjoint paths connecting vertices of $B_1$ to vertices of $B_3$ in $\hat H$.

We claim that, for each path $P\in \hat{\pset}_0$ and each path $P'\in \hat{\pset}_1$, $P\cap P'$ is a path. Note that this implies that combining the sets $\hat{\pset}_0$ and $\hat{\pset}_1$ of paths yields a $(\gamma\times \gamma)$-grid minor in $G_{\lambda}$.
We now prove the claim.
We call the paths in $\hat{\pset}_0$ \emph{vertical paths} and view them as being directed from vertices of $B_0$ to vertices of $B_2$. We call the paths in $\hat{\pset}_1$  \emph{horizontal paths}, and view them as being directed from vertices of $B_1$ to vertices of $B_3$.
Denote $\hat{\pset}_0=\set{P_1,P_2,\ldots,P_{\gamma}}$, where for each $1\le i\le \gamma$, the endpoint in $B_0$ of path $P_i$ is $v_{i}$.
Denote $\hat{\pset}_1=\set{P'_1,P'_2,\ldots,P'_{\gamma}}$, where for each $1\le j\le \gamma$, the endpoint in $B_1$ of path $P'_j$ is $v_{\gamma+j}$.
Note that the planar drawing of $G$ induces a planar drawing of $\hat H$. Since all vertices of $B_0, B_1, B_2, B_3$ lie on the contour $\gamma$, the image of each path $P_i\in \hat{\pset}_0$ separates the interior of $\lambda$ into two regions, that we call the \emph{left region of $P_i$} and \emph{right region of $P_i$}, respectively. In particular, the left region of $P_i$ contains the image of $P_1,\ldots,P_{i-1}$, and right region of $P_i$ contains the image of $P_{i+1},\ldots,P_{\gamma}$. We define the left and right regions for each path $P'_j\in \hat{\pset_1}$ similarly.

Assume the claim is false, and assume without loss of generality that some vertical path $P$ visits some horizontal path $P'$ more than once. 
Therefore, either there is a subpath of $P$ whose image lies in the right region of $P'$ and does not contain any vertex of $P'$ as its inner vertex, that we call a \emph{bump}, or there is a subpath of $P$ whose image lies in the left region of $P'$ and does not contain any vertex of $P'$ as its inner vertex, that we call a \emph{pit}. We show that neither bumps nor pits may exist, thus completing the proof of the claim.
We now show that bumps do not exist. The arguments for pits are symmetric.
Assume for contradiction that there is a bump. 
Consider a bump that is created by a path $P\in \hat\pset_0$ and a path $P'\in\hat{\pset}_1$, and let $u,w$ be two vertices shared by $P$ and $P'$. 
We say that a bump is \emph{aligned} iff $u$ appears before $w$ on both paths $P,P'$, or $w$ appears before $u$ on both paths $P,P'$. Since we can reverse the direction of the paths in $\hat{\pset_0},\hat{\pset_1}$, we can assume without loss of generality that there exists an aligned bump.

We now take the aligned bump that, among all pairs $P_i\in \hat{\pset}_0, P'_j\in \hat{\pset}_1$ of paths that form an aligned bump that minimizes $j$, minimizes $i$, namely
\[j=\min\set{j'\mid\exists i', \text{ s.t. } P_{i'}, P'_{j'} \text{ form an aligned bump}},\]
and
\[i=\min\set{i\mid P_{i}, P'_{j} \text{ form an aligned bump}}.\]
%Let $Q\subsetneq P_i$ be the subpath of $P_i$ whose image lies in the right region of $P'_j$ and does not contain any vertex of $P'_j$ as its inner vertex. Clearly, either $j=1$, or the image of $Q$ lies entirely in the left region of $P'_{j+1}$, as otherwise the pair $P_i,P'_{j+1}$ of paths will create a bump, contradicting our choice of $P_i,P'_j$. 
Let $u,w$ be the vertices shared by $P_i$ and $P'_j$, with $u$ appearing before $w$ on both $P_i$ and $P'_j$.
Let $Q$ be the subpath of $P_i$ between $u$ and $w$, and let $Q'$ be the subpath of $P'_j$ between $u$ and $w$.
We now distinguish between the following two cases, depending on whether or not $Q'$ contains a vertex of some other path $P_{i'}\in \hat{\pset_0}$ with $i'\ne i$ as an inner vertex.
We first assume that $Q'$ does not contain such a vertex, then we claim that the first edge $e$ of $Q$ is irrelevant. 
To see this, observe first that no path of $\hat\pset_1$ may contain $e$, since otherwise the paths in $\hat\pset_1$ are not node-disjoint. 
Therefore, if we modify the path $P_i$ in $\hat\pset_0$ by replacing the segment $Q$ by $Q'$, then we obtain a new pair $\hat{\pset}_0', \hat{\pset}_1$ of sets of node-disjoint paths in $\hat H\setminus e$, where $\hat{\pset}_0'$ routes $B_0$ to $B_2$ and $\hat{\pset}_1$ routes $B_1$ to $B_3$. This contradicts the fact that $\hat H$ contains no irrelevant edges.
We now consider the case where $Q'$ does contain a vertex $u'$ from a path $P_{i'}\in \hat{\pset_0}$ with $i'\ne i$. 
From the definition of an aligned bump, $u'$ lies in the left region of $P_i$, and therefore $i'<i$.
Since we view the path $P_{i'}$ as being directed from a vertex of $B_0$ to a vertex of $B_2$, it is easy to see that the subpath of $P_{i'}$ between $u'$ and its endpoint in $B_2$ must contain another vertex of $P'_j$. Let $w'$ be the first such vertex on the subpath of $P_{i'}$ between $u'$ and its endpoint in $B_2$, and we denote by $Q''$ the subpath of $P_{i'}$ between $u'$ and $w'$. We claim that $Q''$ does not contain any vertex of another path $P'_{j'}$ with $j'\ne j$. To see this, observe first that the image of $Q''$ lies in the left region of $P'_{j}$, since otherwise the pair $P_{i'}, P'_j$ of paths also creates an aligned bump, contradicting to the choice of $i$. Observe next that $Q''$ cannot contain any vertex of another path $P'_{j'}$ with $j'< j$, since otherwise the pair $P_{i'}, P'_{j'}$ of paths also creates an aligned bump, contradicting to the choice of $j$.
Therefore, $Q''$ does not contain any vertex of another path $P'_{j'}$ with $j'\ne j$. We now show that the edge of $P'_j$ going out of $u'$ (that we denote by $e'$) is irrelevant.
First, since $u'\in P_{i'}\cap P'_j$, no other path of $\hat{\pset_0}\cup \hat{\pset_1}$ may contain $u'$, and therefore no other path of $\hat{\pset_0}\cup \hat{\pset_1}$ may contain the edge $e'$. We can then modify the path $P'_j$ in $\hat{\pset_2}$ by replacing its segment between $u'$ and $w'$ by $Q'$. Clearly, we obtain a new pair $\hat{\pset}_0, \hat{\pset}_1'$ of sets of node-disjoint paths in $\hat H\setminus e$, where $\hat{\pset}_0$ routes $B_0$ to $B_2$ and $\hat{\pset}_1'$ routes $B_1$ to $B_3$. This contradicts the fact that $\hat H$ contains no irrelevant edges.
Therefore, no bumps may exists.

Let $h'$ be the model that embeds the $(\gamma\times\gamma)$-grid into $G_{\lambda}$.
%and let $\hat\pset_0$ and  $\hat\pset_1$ be the sets of paths in $\hat H$ routing $B_0$ to $B_2$ and routing $B_1$ to $B_3$ respectively.

Since the property \ref{Pro3} is satisfied, we can efficiently find a set $\hat\pset$ of at least $\lfloor\beta/2\rfloor$ edge-disjoint paths, each connecting a distinct vertex of $S$ to a distinct vertex of $V_{\lambda}$. Let $V'\subseteq V_{\lambda}$ be the set of endpoints  of these paths lying in $V_{\lambda}$.
Recall that vertex set $V_{\lambda}$ is partitioned into four contiguous subsets $B_0,B_1,B_2,B_3$.
Therefore, at least one of these four vertex sets (say $B_0$) 
%\mynote{you should say what are $B_i$'s, type them to remind the reader what they are "the four vertex sets ...$\subseteq I$} 
contains at least $\lfloor\beta/2\rfloor/4$ vertices of $V'$. 
%Let $\pset'\subseteq\hat{\pset}$ contain all paths with an endpoint in $V'_0$.
We view the paths connecting vertices of $B_0$ to vertices of $B_2$ as forming the columns of the grid, and we view the paths connecting vertices of $B_1$ to vertices of $B_3$ as forming the rows of the grid. 
Therefore, each column of $R'$ corresponds to a vertex in $B_0$.
We let $V'_0=B_0\cap V'$, and let $\pset'\subseteq\hat{\pset}$ contain all paths with an endpoint in $V'_0$. 
Lastly, we delete from $R'$ all columns that do not correspond to vertices of $V'_0$ and delete arbitrary $\gamma-|V'_0|$ columns. Let $R$ be the resulting $|V'_0|\times |V'_0|$ grid and let $h$ be the model induced by $h'$. From the discussion, $|V'_0|=\Theta(\beta)=\Theta(\alpha'r/\Delta)$. For each vertex $x\in V(R)$, we select an arbitrary vertex of $h(x)$ as $v_x$.
Denoting by $I$ the set of vertices in last row of $R$, it is easy to see that each path of $\pset'$ that connects a vertex of $S$ to a vertex of $B_0$ can be extended to a path that connects a vertex of $S$ to a vertex of $\set{v_x\mid x\in I}$, by concatenating it with a subpath of $\hat\pset_0$. We denote by $\pset$ the resulting paths obtained from $\pset'$ and $\hat\pset_0$. It is clear that the paths in $\pset$ are edge-disjoint. %\mynote{that's two sentences with a comma between them.}
This completes the construction of the grid minor $R$ and the set $\pset$ of edge-disjoint paths connecting $S$ to the vertices of $\set{v_x\mid x\in I}$, thus completing the proof of Lemma~\ref{lem:route_to_grid}.

\iffalse
\znote{to be moved to intro:}
To the best of our knowledge, the problem \CNwRS was only previously studied by Pelsmajer et al in~\cite{pelsmajer2011crossing}. They first showed that the \CNwRS problem is NP-complete, and then gave approximation algorithms to some special cases. In particular, when the number of vertices is $k$, they presented an algorithm that approximates the $\optcrors(G,\Sigma)$ to within a factor of $O(k^4)$ in time $O(m^k\log m)$, where $m$ is the number of edges in the graph.
\fi

%\input{apd_drawing_type_1}
\section{Proofs Omitted from Section~\ref{sec:cr_rotation_system}}

\subsection{Proof of Lemma \ref{lem:base}}\label{subsec: proof of base of recursion}
We use the following theorem from~\cite{shahrokhi1994book}.

\begin{lemma}[Theorem 4.2 in \cite{shahrokhi1994book}]
	\label{lem:cr_potential}
	There is an efficient algorithm, that, given a simple graph $G$, finds a drawing of $G$ with $O\left((\optcro(G)+\sum_{v\in V(G')}\deg^2_G(v))\cdot\poly(\log n)\right)$ crossings. %Moreover, in the drawing, all vertices of $G$ are drawn on a line and all edges are drawn in the same halfspace with respect to the line.
\end{lemma}

Note that our input graph $G$ is not necessarily simple. We remedy this by subdividing every edge of $G$ with a new vertex, obtaining a simple graph $G'$. Applying the algorithm from Lemma \ref{lem:cr_potential} to graph $G'$, we obtain a drawing $\phi'$ of $G'$ with the number of crossings bounded by:
\[\begin{split}
&O\Bigg(\Big(\optcro(G')+\sum_{v\in V(G')}\deg^2_{G'}(v)\Big)\cdot  \poly(\log n)\Bigg)\\
&\quad\quad\quad\quad\quad\quad\quad\quad\leq O\Bigg(\Big(\optcro(G)+\Delta \cdot \sum_{v\in V(G')}\deg_{G'}(v)\Big)\cdot\poly(\log n) \Bigg)\\
&\quad\quad\quad\quad\quad\quad\quad\quad\leq O\left (\left (\optcro(G)+\Delta \cdot |E(G')|\right )\cdot\poly(\log n) \right )\\
&\quad\quad\quad\quad\quad\quad\quad\quad\leq O\left (\left (\optcrors(G)+\Delta \cdot |E(G)|\right )\cdot\poly(\log n) \right ).
\end{split}
\]
Here we use the fact that any solution to an instance $(G,\Sigma)$ of \CNwRS defines a solution to instance $G$ of \MCN of the same value, so $\optcro(G)\leq \optcrors(G)$.

Notice that drawing $\phi'$ of $G'$ naturally induces a drawing $\phi$ of $G$, with the number of crossings bounded by $O\left (\left (\optcrors(G)+\Delta \cdot |E(G)|\right )\cdot\poly(\log n) \right )$. However, this drawing may not respect the rotation system $\Sigma$. We slightly modify the drawing $\phi$ in order to rectify that.

We process every vertex $v\in V(G)$ one-by-one. When vertex $v$ is processed, we let $\eta(v)$ be a small disc around $v$ in the current drawing $\phi$. For every edge $e\in \delta_{G}(v)$ that is incident to $v$, we let $\sigma_e$ be the segment of the image of $v$ in $\phi$ that is contained in $\eta(v)$, and we let $p_e$ be the unique point on the intersection of $\sigma_e$ and the boundary of the disc $\eta(v)$. We define new curves $\set{\sigma'_e\mid e\in \delta_G(v)}$, that are contained in $\eta(v)$, such that each curve $\sigma'_e$ originates at $p_e$ and terminates at the image of $v$ in $\phi$. We ensure that every pair of the resulting curves crosses at most once, and moreover, the images of the curves in $\set{\sigma'_e\mid e\in \delta_G(v)}$ enter the vertex $v$ in  a circular order consistent with the order $\oset_v\in \Sigma$. 

Once every vertex of $V(G)$ is processed, we obtain a drawing of $G$ that respects $\Sigma$. The total number of new crossings that we introduced is bounded by $\sum_{v\in V(G)}(\deg_G(v))^2\leq \Delta\cdot |E(G)|$. Therefore, we obtain a final drawing of graph $G$ that respects $\Sigma$, and has at most $O\left (\left (\optcrors(G)+\Delta \cdot |E(G)|\right )\cdot\poly(\log n) \right )$ crossings.

\subsection{Proof of Lemma \ref{lem:add_back_discarded_edges new}}\label{subsec: proof of adding back edges}

%Note that it is sufficient to prove the lemma for the special case where $|E'|=1$, as we can then add the edges of $E'$ to the drawing one-by-one. We use the following claim.
The following claim is central to the proof of the lemma.

\begin{claim}\label{claim: add one edge}
	There is an efficient algorithm, that, given an instance $(H,\Sigma)$ of the \CNwRS problem, an edge $e\in E(H)$, and a drawing $\phi$ of graph $H\setminus \set{e}$ that respects $\Sigma$, computes one of the following:
	
	\begin{itemize}
		\item either a drawing $\hat \phi$ of $H\setminus \set{e}$ that respects $\Sigma$, with $\cro(\hat \phi)<\cro(\phi)$; or
		\item a drawing $\phi'$ of $H$ that respects $\Sigma$, with $\cro(\phi')\le \cro(\phi)+|E(H)|$.
	\end{itemize} 
\end{claim}

It is now easy to complete the proof of Lemma \ref{lem:add_back_discarded_edges new} using Claim \ref{claim: add one edge}. 
We denote $E'=\set{e_1,\ldots,e_r}$. The algorithm consists of $r$ phases, where in the $i$th phase we compute a drawing $\phi_i$ of graph $(H\setminus E')\cup \set{e_1,\ldots,e_i}$ that respects the rotation system $\Sigma$. We set $\phi_0=\phi$. We now describe the $i$th phase. 

Recall that, at the beginning of the $i$th phase, we are given a drawing $\phi_{i-1}$ of graph $(H\setminus E')\cup \set{e_1,\ldots,e_{i-1}}$.  We iteratively apply the algorithm from Claim \ref{claim: add one edge} to the current drawing $\phi_{i-1}$. If the outcome of the algorithm is another drawing  $\hat \phi_{i-1}$ of of graph $(H\setminus E')\cup \set{e_1,\ldots,e_{i-1}}$ that respects $\Sigma$ with $\cro(\hat \phi_{i-1})<\cro(\phi_{i-1})$, then we replace $\phi_{i-1}$ with $\hat \phi_{i-1}$ and continue to the next iteration. Otherwise, we obtain the desired drawing $\phi_i$ of graph $(H\setminus E')\cup \set{e_1,\ldots,e_{i}}$ that respects $\Sigma$, with $\cro(\phi_i)\le \cro(\phi_{i-1})+|E(H)|$ (recall that $E'\subseteq E(H)$). We then return this drawing as the outcome of the $i$th phase, and continue to the next phase.

Let $\phi'=\phi_r$ be the drawing of $H$ that is obtained after the $r$th phase. Then we are guaranteed that $\phi'$ respects the rotation system $\Sigma$, and moreover, from the above discussion, $\cro(\phi')\le \cro(\phi)+|E'|\cdot|E(H)|$.
Therefore, in order to complete  the proof of Lemma \ref{lem:add_back_discarded_edges new}, it is now enough to prove Claim \ref{claim: add one edge}.

\begin{proofof}{Claim \ref{claim: add one edge}}
Let $e=(u,v)$. We compute a curve $\gamma$ in the current drawing $\phi$ of $H\setminus\set{e}$ connecting $\phi(u)$ to $\phi(v)$, such that (i) $\gamma$ intersects the drawing of $H\setminus\set{e}$ at images of edges only; (ii) $\gamma$ does not contain any crossings of the drawing $\phi$; and (iii) if we add the edge $e$ to the drawing $\phi$ such that the image of $e$ is $\gamma$, then the resulting drawing of $H$ respects the rotation system $\Sigma$. Among all curves $\gamma$ with the above properties, we choose the one that has fewest crossings with the edges of $H\setminus \set{e}$, and does not cross itself. Such a curve $\gamma$ can be computed as follows.
Let $\fset$ be the set of all faces in the drawing $\phi$ of $H\setminus\set{e}$. Let $F$ be the face that is incident to $v$, such that, if edge $e$ is added to the drawing in a manner that respects the ordering $\oset_v$ of all edges incident to $v$, then there must be a segment of the image of $e$ that contains $v$, does not contain any crossings, and is contained in $F$. We define the face $F'$ incident to $u$ similarly. We then compute a graph that is dual to the drawing $\phi$ of $H\setminus\set{e}$, and in that graph we compute a shortest path between the two vertices that represent faces $F$ and $F'$. The resulting path then naturally defines the desired curve $\gamma$.
	
Assume first that curve $\gamma$ participates in at most $|E(H)|$ crossings. In this case we simply add the edge $e$ to the drawing $\phi$ so that the image of $e$ is $\gamma$, and return the resulting drawing of $H$. Clearly, this drawing has at most $\cro(\phi)+|E(H)|$ crossings, and it respects $\Sigma$. 

Assume now that $\gamma$ has more than $|E(H)|$ crossings. Then there is some edge $e'\in E(H)\setminus\set{e}$ whose image intersects $\gamma$ at least twice. Let $x,y$ be two intersections of $\gamma$ and the image of $e'$. We denote by $\sigma$ the segment of $\gamma$ between $x$ and $y$, and we denote by $\sigma'$ the segment of the image of $e'$ between $x$ and $y$. Let $N$ be the number of crossings in which $\sigma$ participates (excluding $x$ and $y$), and let $N'$ be defined similarly for $\sigma'$. Observe that $N< N'$ must hold. Indeed, if $N\geq N'$, then we can modify the curve $\gamma$ as follows. Assume that $\gamma$ is directed so that $x$ appears before $y$ on $\gamma$. Let $t$ be a point on $\gamma$ that appears just before point $x$ on it, and let $t'$ be a point on $\gamma$ that appears just after point $y$ on it. Let $\sigma^*$ be a curve that connects $t$ to $t'$, and follows the segment $\sigma'$. By replacing the segment of curve $\gamma$ between $t$ and $t'$ with $\sigma^*$, we obtain a new curve $\gamma'$ that has all the required properties, but has fewer crossings than $\gamma$, a contradiction. We conclude that $N<N'$ must hold.
We then modify the drawing of  the edge $e'$ in $\phi$, by replacing the segment $\sigma'$ with $\sigma$. This produces a new drawing $\hat \phi$ of $H\setminus\set{e}$ with $\cro(\hat \phi)<\cro(\phi)$, that respects the rotation system $\Sigma$.
\end{proofof}

\subsection{Proof of Lemma \ref{lem: initial path set}}
\label{apd:proof_of_initial path set}

%\mynote{todo: vertex $v$ should be replaced with $u$ throughout because we already use $v$ for a specific vertex}
Recall that the minimum-cardinality set of edges in $G''$ whose removal separates $v'$ from $v''$ has size at least $\eac^{1-2\eps}$. 
From Menger's theorem, $G''$ contains a collection of at least $\ceil{\eac^{1-2\eps}}$ edge-disjoint paths connecting $v'$ to $v''$. 
Our first step is to compute such a set $\tilde \pset$ of $\ceil{\eac^{1-2\eps}}$ edge-disjoint paths connecting $v'$ to $v''$, using standard max-flow techniques.
We view the paths in $\tilde\pset$ as being directed from $v'$ to $v''$. Let $H$ be the directed graph obtained by taking the union of the  directed paths in $\tilde \pset$. We can assume, from the properties of maximum flow, that $H$ is a directed acyclic graph, so there is an ordering $\oset$ of its vertices, such that, for every pair $x,y$ of vertices in $H$, if there is a path of $\tilde \pset$ in which $x$ appears before $y$, then $x$ also appears before $y$ in the ordering $\oset$. 
Note that the rotation system $\Sigma''$ for graph $G''$ naturally induces a rotation system for graph $H$, that we denote by $\tilde \Sigma$. For each vertex $u\in V(H)$, we denote by $\delta^+(u)$ the set of edges leaving $u$ in $H$, and by $\delta^-(u)$ the set of edges entering $u$ in $H$. Clearly, if $u\not\in\set{ v',v''}$, then $|\delta^+(u)|=|\delta^-(u)|$.
We use the following simple observation.

\begin{observation}
\label{obs:rerouting_matching}
There is an efficient algorithm to compute, for each vertex $u\in V(H)\setminus \set{v',v''}$, a perfect matching $M(u)\subseteq \delta^-(u)\times \delta^+(u)$ between the edges of $\delta^-(u)$ and the edges of $\delta^+(u)$, such that, for each pair of matched pairs $(e^-_1,e^+_1)$ and $(e^-_2,e^+_2)$ in $M(u)$, the intersection of the path that consists of the edges $e^-_1,e^+_1$ and the path that consists of edges $e^-_2,e^+_2$ at vertex $u$ is non-transversal with respect to $\tilde\Sigma$.
\end{observation}
\begin{proof}
We start with $M(u)=\emptyset$ and perform $|\delta^-(u)|$ iterations. In each iteration, we select a pair $e^-\in \delta^-(u), e^+\in \delta^+(u)$ of edges that appear consecutively in the rotation $\oset_v$ of $\tilde\Sigma$. We add $(e^-,e^+)$ to $M(u)$, delete them from $\delta^-(u)$ and $\delta^+(u)$ respectively, and then continue to the next iteration. It is immediate to verify that the resulting matching $M(u)$ satisfies the desired properties.
\end{proof}

We now gradually modify the set $\tilde \pset$ of paths in order to obtain a set $\pset$ of edge-disjoint paths connecting $v'$ to $v''$, that is non-transversal with respect to $\Sigma''$.
We process all vertices of $V(H)\setminus \set{v',v''}$ one-by-one, according to the ordering $\oset$. We now describe an iteration in which the vertex $u$ is processed. Let $\tilde\pset_u\subseteq\tilde \pset$ be the set of paths containing $u$. For each path $P\in \tilde\pset_u$, we delete the unique edge of $\delta^+(u)$ that lies on this path, thereby decomposing $P$ into two-subpaths: path $P^-$ connecting $v'$ to $u$; and path $P^+$ connecting some vertex that is the endpoint of an edge of $\delta^+(u)$ to $v''$. Define $\tilde\pset^-_u=\set{P^-\mid P\in \pset_u}$ and 
$\tilde\pset^+_u=\set{P^+\mid P\in \pset_u}$.
We will glue these paths together using the edges in $\delta^+(u)$ and the matching $M(u)$ produced in Observation \ref{obs:rerouting_matching}. Specifically, we construct a new path set $\tilde{\pset}'$ that contains, for each path $P\in \tilde \pset$, a new path $P'$, defined as follows. Consider a path $P\in \tilde\pset$. If $P\notin\tilde{\pset}_u$, then we let $P'=P$. Otherwise, consider the unique path $P^-\in \tilde{\pset}^-_u$ that is a subpath of $P$, and let $e^-_P$ be the last edge on this path. Let $e^+$ be the edge in $\delta^+(u)$ that is matched with $e^-_P$ in $M(u)$, and let $\hat P^+$ be the unique path in $\tilde\pset^+_u$ that contains $e^+$.
We then define the new path $P'$ to be the concatenation of the path $P^-$, the edge $e^+$, and the path $\hat P^+$. This finishes the description of an iteration.

Let $\pset$ be the final set of path that we obtain from the above process. It is easy to verify that the set $\pset$ of paths is non-transversal with respect to $\tilde \Sigma$, and is therefore non-transversal with respect to $\Sigma''$.

\newpage
\bibliographystyle{alpha}
\bibliography{REF}

\end{document}